\documentclass[twoadvisors,numrefs,final,noinfo]{nddiss2e}
\usepackage{revsymb}
\usepackage{graphicx}
\usepackage{dcolumn}
\usepackage{longtable}
\usepackage[T1]{fontenc}

\begin{document}

\frontmatter

%\title{Shape Changes at High Spin: Triaxiality and Reflection Asymmetry}
\title{EXOTIC COLLECTIVE EXCITATIONS AT HIGH SPIN:\\TRIAXIAL ROTATION AND OCTUPOLE CONDENSATION}
\author{Xiaofeng Wang}
\work{Dissertation}
\degaward{Doctor of Philosophy}
%\degaward{Doctor of Philosophy\\in\\Physics}
\degprior{B.S., M.S.}
\advisor{Umesh Garg}
\secondadvisor{Robert V. F. Janssens}
\department{Physics}
\degdate{December 2007}

\maketitle
%%%%%%%%%%%%%%%%%%%%%%%%%%%%%%%%%%%%%%%%%%%%%%%%%%%%%%%%%%%%%%%%%%%%%%%%
%
% Front stuff
%
%%%%%%%%%%%%%%%%%%%%%%%%%%%%%%%%%%%%%%%%%%%%%%%%%%%%%%%%%%%%%%%%%%%%%%%%

\copyrightholder{Xiaofeng Wang}
\copyrightyear{2007}
\makecopyright

\begin{abstract}
In this thesis work, two topics, triaxiality and reflection asymmetry, have been discussed. 
Band structures in $^{163}$Tm were studied in a ``thin'' target experiment as well as in a 
DSAM lifetime measurement. Two new excited bands were shown to be characterized by a deformation 
larger than that of the yrast sequence. These structures have been interpreted as 
Triaxial Strongly Deformed bands associated with particle-hole excitations, rather 
than with wobbling. Moreover, the Tilted-Axis Cranking calculations provide 
a natural explanation for the presence of wobbling bands in the Lu isotopes 
and their absence in the neighboring Tm, Hf and Ta nuclei. A series of so-called ``unsafe'' 
Coulomb excitation experiments as well as one-neutron transfer measurements was carried 
out to investigate the role of octupole correlations in the $^{238,240,242}$Pu isotopes. 
Some striking differences exist between the level scheme and deexcitation patterns seen in 
$^{240}$Pu, and to a lesser extent in $^{238}$Pu, and those observed in $^{242}$Pu and in 
many other actinide nuclei such as $^{232}$Th and $^{238}$U, for example. 
The differences can be linked to the strength of octupole correlations, which 
are strongest in $^{240}$Pu. Further, all the data find a natural explanation within the 
recently proposed theoretical framework of octupole condensation. 
\end{abstract}

\renewcommand{\dedicationname}{\mbox{}}
\begin{dedication}
  Dedicated to my wife $\it{Canli}$\\in heartful recognition of\\her love and encouragement.
\end{dedication}

\tableofcontents
\listoffigures
\listoftables

\begin{preface}
Shell structure is one of the cornerstones of our description of atomic nuclei. Systems 
with none or a small number of nucleons outside a closed shell are generally spherical, 
while those away from closed shells are usually deformed because of the long-range forces 
between valence nucleons. The purpose of this thesis is to explore two exotic modes of 
collectivity that have only been proposed recently. 

Direct evidence for triaxial nuclear shapes has, historically, been difficult to obtain. 
Nevertheless, early in the 21st century, evidence was found in nuclei with proton number 
$Z{\sim}70$ and mass $A{\sim}165$ for wobbling, a collective mode uniquely associated with 
triaxiality. In the present work, the properties of band structures discovered in a nucleus 
close to those where wobbling was reported are examined. It is shown that these bands are 
associated with triaxial rotation, but not wobbling. 

During the past year, the concept of octupole condensation has been proposed in order 
to account for band structures observed in some neutron-deficient actinide nuclei. In the 
present work, the strength of octupole correlations in plutonium isotopes is investigated. 
It is shown that the rotational sequences observed in $^{240}$Pu find a natural 
interpretation within the new concept. 

For clarity and ease of reading, this thesis is divided into five chapters. 
In the first, the theoretical concepts relevant to the problems under discussion 
are described. The second chapter is devoted to the various experimental techniques and data 
analysis methods. In the next chapter, the results obtained for Triaxial Strongly Deformed 
bands in $^{163}$Tm are discussed; a general introduction of triaxiality in nuclei is followed 
by the presentation of the data and a detailed interpretation. The fourth chapter discusses 
an investigation of octupole correlations in three even-even Pu isotopes ($A=238$, 240, 242). 
The first three sections of this chapter contain a general introduction on reflection 
asymmetry in nuclei, a motivation of the present work and relevant information about 
the experiments and the data analysis. The data for each nucleus are then presented 
one by one in the next three sections, and this is followed by the interpretation within 
the available models. Finally, this thesis ends with a brief summary of the present work 
and a perpective on possible future measurements. 
\end{preface}

\begin{acknowledge}
This thesis is a summation of my research efforts over the past four and one half years. 
In this long and hard period, I worked with as much wisdom and enthusiasm as I am capable of, 
but, honestly, I do not think that I could have finished this job without the help 
and support of many people. Here, I would like to deeply thank every 
contributor to this work from the bottom of my heart. Please forgive me if some 
of their names are not mentioned within the limited available space. 

Dr. Umesh Garg, my advisor at Notre Dame, opened the door for me and introduced me 
to the world of nuclear physics, which may become my lifetime career. He is 
a mentor for me not only in work, but also in life. 

Dr. Robert Janssens is my advisor at Argonne. I feel very lucky that I was able to work 
on my thesis in his group, at the side of a wonderful instrument -- Gammasphere. 
His guidance and enthusiasm throughout the course of this work were determinant 
for its accomplishment. The things that have made a great impact on me in the past years 
and will stay with me in the future are his profound knowledge of science as 
well as his positive attitude towards work and life. 

My gratitude also goes to Drs. Stefan Frauendorf and Takashi Nakatsukasa; their excellent 
theoretical work made a nice interpretation of the data possible. 

Drs. Frank Moore, Michael Carpenter, Torben Lauritsen, Shaofei Zhu, Constantin Vaman, 
Daryl Hartley and N. S. Pattabiraman will never be forgotten because all of them have 
initiated me to the many data analysis techniques used in this work. Dr. Ingo Wiedenh{\"o}ver, 
is thanked for the many fruitful discussions on the Pu data. These made me understand 
better the structure of the heavy elements. 

I thank Dr. Donald Peterson for his patience in teaching me how to use Latex, 
and Dr. Mario Cromaz, for his answers to my questions about the Blue database. 

I am indebted to Dr. Irshad Ahmad and John Greene for the high quality targets used 
in this work, and to Drs. Filip Kondev, Sean Freeman, Augusto Macchiavelli, 
Neil Hammond, R.S. Chakrawarthy, S.S. Ghugre and G. Mukherjee for their participation 
in the thesis experiments. 

I thank Drs. Birger Back, Susan Fischer, Kim Lister, Darek Seweryniak, Cheng-lie Jiang, 
Teng Lek Khoo, Sujit Tandel, Partha Chowdhury and Xiaodong Tang for 
involving me in their research projects. In this way, I obtained valuable research 
experience, different from that from my thesis work. 

Drs. Philip Collon, Gordon Berry, Ani Aparahamian, Michael Wiescher, Kathy Newman, 
and James Kaiser are thanked for their continuous support and help throughout the 
course of my graduate studies. 

I owe much to Dr. Walter Johnson, my former advisor. It was his kindness and 
generousness that made the transfer of my research interests from atomic theory 
to experimental nuclear physics smooth. The period of my first two years at Notre 
Dame when I worked with him left me with many good memories. 

My friends, Lou Jisona and Nate Hoteling, made my stay at Argonne easier and 
happier. 

My appreciation is also due to the Argonne Physics Division, the Notre Dame Physics Department 
and the administrative staffs therein (Allan Bernstein, Colleen Tobolic, Janet Bergman, 
Barbara Weller, Barbara Fletch, Jennifer Maddox, Shelly Goethals, Shari Herman, 
Sandy Trobaugh, Lesley Krueger $\it{et~al.}$) for all the help I received in the past years. 

Last, family is most important for everyone. I would not have accomplished anything 
without the love and support from my family. My Mom and Dad gave me birth, 
education and love, I cannot adequately express my gratitude to them in any word. 
This thesis is dedicated to my wife, Canli, for her love, support and encouragement. 

\end{acknowledge}

\mainmatter

\setcounter{chapter}{0}

%
% Chapter 1
%

%
% Chapter 1
%

\chapter{\label{chap:theoriBkgd}THEORETICAL BACKGROUND}

\section{Fundamental properties of nucleus}
\begin{figure}
\includegraphics[angle=270,width=\columnwidth]{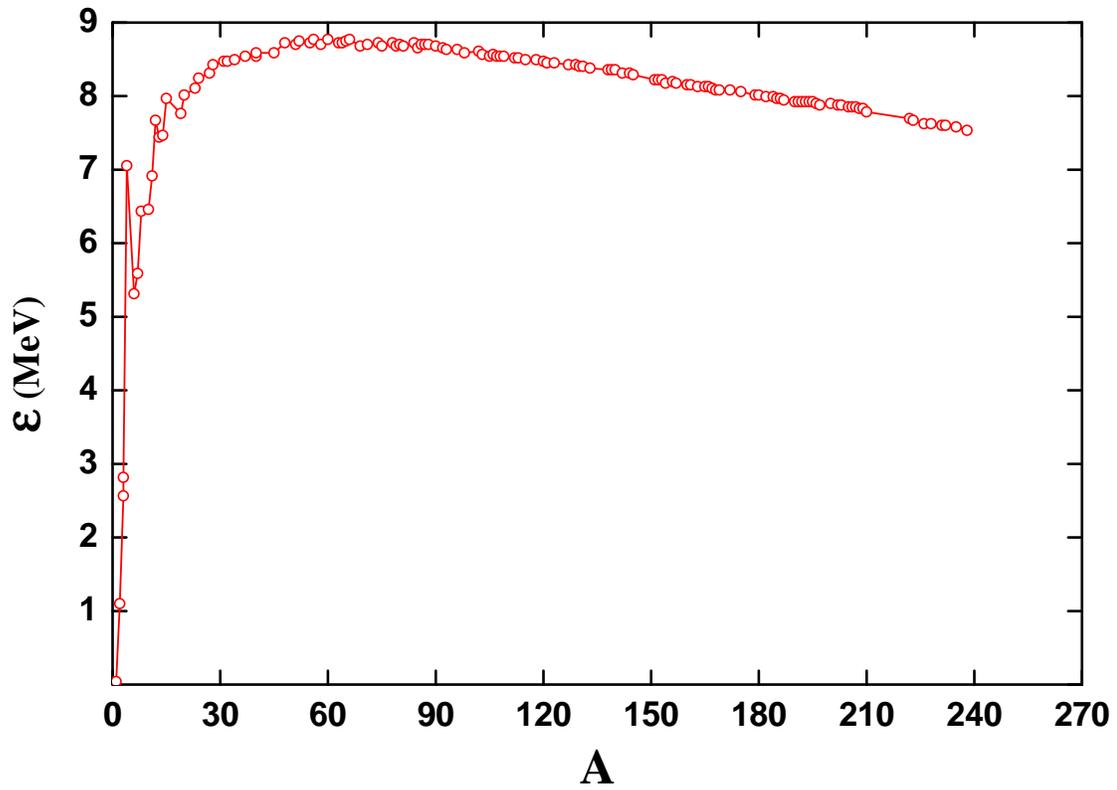}
\caption{Average binding energy per nucleon $\varepsilon$ as a function of atomic 
mass number $A$. Adapted from Ref.~\cite{Krane-88-book}.\label{fig:bind_energy}}
\end{figure}

Since the discovery of the atomic nucleus following the famous Rutherford 
scattering experiment~\cite{Rutherford-PM-6-21-11} in 1911, numerous facts 
regarding this small (10$^{-14}$ to 10$^{-15}$ m in diameter, only one 
ten-thousandth of an atom in size), but heavy (about $99.9\%$ of an atom in mass) 
object have been studied and characterized. The well-known properties of the nucleus 
include the fact that ($i$) it consists of protons, positively charged particles, and 
neutrons, electrically neutral particles; ($ii$) between nucleons a strong, 
but short-ranged nuclear force exists, which overcomes the Coulomb repulsion 
and results in a bound system; ($iii$) the binding energy per nucleon, which 
originates from the fact that the mass of a given nucleus is less than 
the sum of its constituent nucleons, keeps increasing as a function of mass 
until reaching a maximum of about 8 $MeV$ at mass number A $\sim$ 60, and 
above this value remains approximately constant~\cite{Krane-88-book} 
(see Figure~\ref{fig:bind_energy}); and, ($iv$) the nuclear force saturates as indicated 
by the trend of the binding energy per nucleon as a function of mass and by the 
fact that the nuclear density is almost constant. 

There is strong evidence that nuclei with certain numbers of protons (Z) or 
neutrons (N) are more stable than others. This is seen, for example, in the neutron 
and proton separation energy, the energy of first excited states, $\it{etc.}$ These specific 
N or Z numbers, called ``magic numbers'', provide an insight into the fact that the 
nucleons inside nucleus occupy shells, similar to those occupied by the electrons 
surrounding the nucleus of the atom. On the other hand, the existence of the 
collective motion of a large number of nucleons 
in a nucleus, $\it{e.g.}$, the collective rotation and vibration of the nucleus, $\it{etc.}$, to be 
discussed later in this chapter, is also firmly supported by a large number of experimental 
observations. It is the occupation of shells by two types of nucleons that gives 
the nucleus its special character. Such occupation is under some conditions responsible 
for the so-called single-particle aspects of nuclear structure and under some other 
accounts for its collective behavior. Understanding these two fundamental modes 
and the interplay between them is one of the most important goals of 
nuclear structure research. 

\section{Shell model and deformation}
\subsection{\label{subsec:NUSMD}The nuclear shell model}

\begin{figure}
\begin{center}
\includegraphics[angle=0,width=0.90\columnwidth]{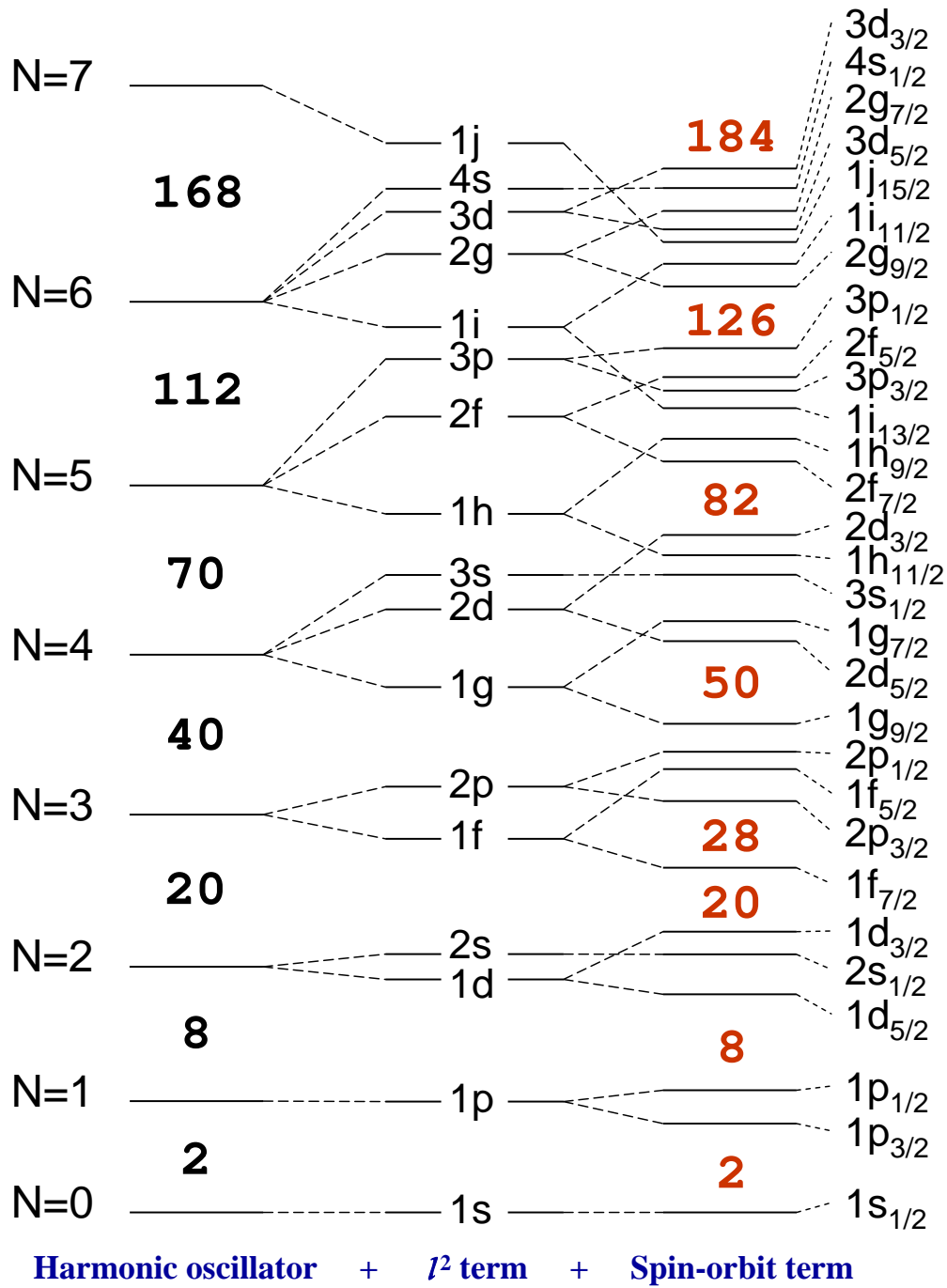}
\caption[Energy levels from calculations using a modified oscillator potential.]{Energy 
levels from calculations using a modified oscillator potential.
The three columns of states correspond to the three terms in the potential, and the numbers 
shown in the right column with red color are the ``magic numbers'' 
(see text for details). Adapted from Ref.~\cite{Krane-88-book}.\label{fig:ShellScheme}}
\end{center}
\end{figure}

In order to account for the shell structure found in the nucleus, the Nuclear Shell 
Model~\cite{Mayer-PR-75-1969-49,Haxel-PR-75-1766-49} was developed, 
and it has proved to be a most successful model. In the shell model, each 
nucleon is described as moving in an average potential generated by all the other 
nucleons, the so-called mean field potential. Hence, the ordering and energy of nuclear 
states can be calculated by choosing an appropriate form of the potential and solving 
the Schr{\"o}dinger equation: 
\begin{equation}
\left[-\frac{{\hbar}^2}{2m}{\nabla}^2+V(r)\right]{\psi}(r)=e\,{\psi}(r).\label{eq:SchrodingerEQ}
\end{equation}
One of the applicable potentials can be expressed as, 
\begin{equation}
V(r)=\frac{1}{2}m({\omega}r)^2+{\beta}\;{l^2}+{\alpha}\;{\vec{l}\cdot\vec{s}},\label{eq:MHOpotential}
\end{equation}
where $\vec{l}$ is the orbital angular momentum and $\vec{s}$ is the intrinsic spin. The 
first term, the harmonic oscillator potential, leads to the sequence of levels given in the left column 
of Figure~\ref{fig:ShellScheme}, where $N$ is the principal quantum number. 
This first term accounts only for the first three magic numbers. 
The addition of a $l^2$ term removes some of the degeneracy, as shown in 
the middle column of Figure~\ref{fig:ShellScheme}, but it still does not result in the correct 
magic numbers. Therefore, an additional spin-orbit coupling term 
$\left(\vec{l}\cdot\vec{s}\right)$~\cite{Haxel-PR-75-1766-49} is necessary to obtain 
the sequence of levels in the right column where the ``magic numbers'', $\it{e.g.}$, 2, 8, 20, 
28, 50, 82, 126, 184, can be understood. The states obtained in this way are occupied by the nucleons 
in an order of ascending energy starting from the lowest level while obeying the Pauli Exclusion Principle, 
$\it{i.e.}$, a maximum of two nucleons can fill into any single level. 

The intrinsic spin of a nucleon is $1/2$, so for a given $l$ there are two values of total 
angular momentum $j$, $j=l{\pm}1/2$, corresponding to different spin orientations 
with respect to the direction of the orbital angular momentum. In spectroscopic notation, 
the $j$ value is added as a subscript, for example, $1p_{1/2}$ and $1p_{3/2}$, and the 
multiplicity of the states is $2j + 1$. As can be seen in Figure~\ref{fig:ShellScheme}, 
for $l > 3$, the energy splitting between $j+1/2$ and $j-1/2$ states will be large enough
to lower the $j+1/2$ state from one oscillator shell ($N$) to the shell below ($N-1$). 
Such levels are known as intruder states and are of opposite parity, $\pi=(-1)^l$, to the shell
that they eventually occupy. 

It should be noted that the magic numbers mentioned here apply 
to nuclei close to stablity and recent evidence~\cite{Janssens-Nature-435-897-05} 
indicates that these numbers are modified for exotic, neutron-rich nuclei. 
It is also worth pointing out that the scheme of proton states is slightly different at 
high energy from that of neutrons because of Coulomb repulsion. 

Another often used and more realistic potential is the Woods-Saxon (WS) potential, 
\begin{equation}
V(r)=-{V_0}{[1+exp^{(\frac{r-{R_0}}{a})}]^{-1}}+{\alpha(r)}\,{\vec{l}\cdot\vec{s}},\label{eq:WSpotential}
\end{equation}
where $R_0={r_0}{A^{1/3}}$, $V_0~{\approx}~50~MeV$, $a~{\approx}~0.5~fm$ and $r_0~{\approx}~1.2~fm$. 
It can also reproduce the ``magic numbers'' and the shell structure. In the WS potential, the 
total angular momentum, $j=l+s$, and the parity, $\pi=(-1)^l$, are the only good 
quantum numbers~\cite{Mayer-55-book}. 

Until this point, the nuclear problem is considered as one where each nucleon 
is treated as an independent particle moving in an average potential representing the effective 
interaction of all other nucleons with the one being described. This description is often referred 
to as the mean-field approximation. The assumption above is not accurate and, in fact, 
the nuclear problem should be treated as a many-body problem, due to the mutual interactions between 
the nucleons. These types of interactions, called residual interactions, must be taken care of 
if an accurate description of the nucleus is to be achieved. 

\subsection{\label{subsec:Deformation}Deformation}

The Shell Model, using the nuclear potentials with spherical symmetry described above, has 
been successful in explaining many of nuclear phenomena and in predicting the properties 
of spherical or near-spherical nuclei, in which the number of nucleons outside a closed shell
is small. However, when considering nuclei away from closed shells, the residual 
interactions between the valence nucleons (nucleons beyond a closed shell) can not be 
described by the spherical Shell Model. In such nuclei, 
the long-range effective forces between valence nucleons will lead to collective
motion. In some cases, these collective effects can be strong enough to drive 
to a breaking of the spherical symmetry, and a permanent deformation of the nucleus 
is then established as the total energy of nuclear system with a deformed shape 
becomes lower than that associated with a spherical shape. The nuclear shape can be described using 
a radius vector in terms of a set of shape parameters $\alpha_{\lambda\mu}$ in the 
following way: 
\begin{equation}
R(\theta,\phi) = {R_0}\left(1+{\alpha_{00}}+\sum_{\lambda=1}^{\infty}
\sum_{\mu=-\lambda}^{\lambda}{\alpha_{\lambda\mu}}
{Y_{\lambda\mu}(\theta,\phi)}\right),\label{eq:NuclRadius}
\end{equation}
where $R(\theta,\phi)$ is the distance from the center of the nucleus to the surface at 
angles $(\theta,\phi)$, $R_0$ is the radius of a sphere having the same volume as the deformed 
nucleus, the factor $\alpha_{00}$ is due to nuclear volume conservation, and 
$Y_{\lambda\mu}(\theta,\phi)$ is a spherical harmonic function of $\theta$ and $\phi$. 

In expression~\ref{eq:NuclRadius}, the lowest multipole, $\lambda=1$, 
corresponds to a shift of the position of the center 
of mass. It can be easily eliminated by requiring the origin of the coordinate system 
to coincide with the center of mass. The terms associated with $\lambda=2$ represent the 
quadrupole deformation. In such a case, the nucleus is either of oblate deformation 
(with two equal semi-major axes) or of prolate deformation (having two equal semi-minor 
axes), or of triaxial deformation (having three unequal axes). The latter case is one of 
the two foci of this thesis work. The $\lambda=3$ terms introduce octupole deformation, which 
is reflection asymmetric with a pear shape as one of the typical shapes; 
this is the other emphasis of this work. For the issues addressed in the present thesis, 
the $\lambda=4$ (hexadecapole) and higher order terms are sufficiently small that they 
can be ignored. 

\begin{figure}
\begin{center}
\includegraphics[angle=0,width=0.60\columnwidth]{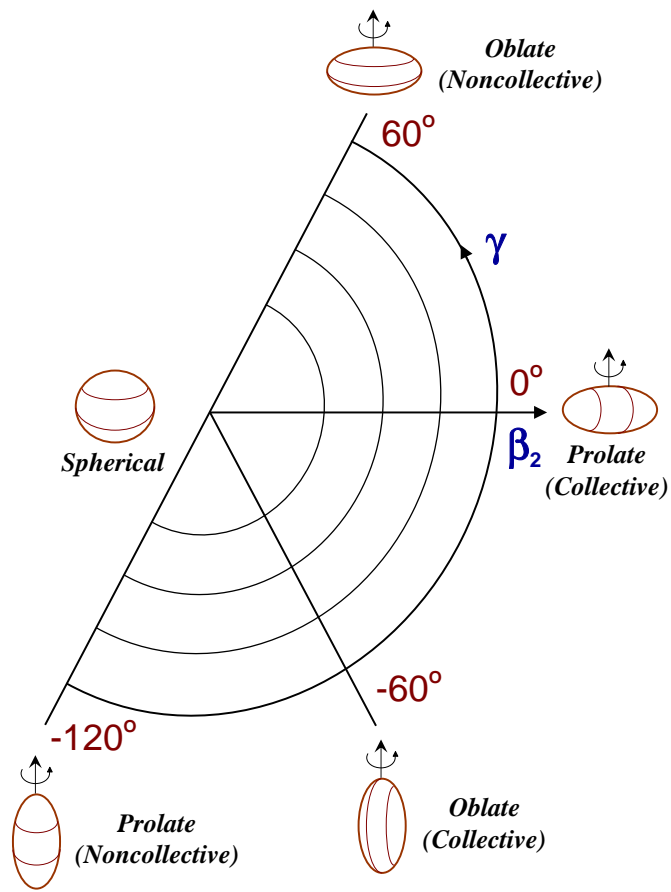}
\caption{The nuclear deformations described in the Lund convention. Adapted from Ref.~\cite{Zhu-03-thesis}.\label{fig:QuadDeform}}
\end{center}
\end{figure}

In the case of pure quadrupole deformation, Eq.~\ref{eq:NuclRadius} can be simplified to
\begin{equation}
R(\theta,\phi) = {R_0}\left(1+{\alpha}_{20}{Y}_{20}(\theta,\phi)
+{\alpha}_{22}{Y}_{22}(\theta,\phi)
+{\alpha}_{22}{Y}_{2-2}(\theta,\phi)\right).\label{eq:QuadNuclRadius}
\end{equation}
The choice of an appropriate coordinate system where the principal axis is lined up with 
the axis of symmetry of the nuclear shape leads to $\alpha_{21}=\alpha_{2-1}=0$, $\alpha_{22}=\alpha_{2-2}$. 
Using the so-called Lund convention (shown in Figure~\ref{fig:QuadDeform}), the coefficients 
$\alpha_{20}$ and $\alpha_{22}$ can be expressed as 
\begin{eqnarray}
\alpha_{20} & = & \beta_2\cos\gamma\label{eq:LundConvDef1} \\
\alpha_{22} & = & \beta_2\sin{\gamma},\label{eq:LundConvDef2}
\end{eqnarray}
where the parameters $\beta_2$ and $\gamma$ represent the excentricity and non-axiality of the nuclear 
shape, respectively (see Figure~\ref{fig:QuadDeform}), and are defined by the Lund convention as: 
\begin{eqnarray}
\frac{R_x-R_0}{R_0} & = & \sqrt{\frac{5}{4\pi}}{\beta_2}{\cos}(\gamma-\frac{2}{3}{\pi})\label{eq:LundConvDef3} \\
\frac{R_y-R_0}{R_0} & = & \sqrt{\frac{5}{4\pi}}{\beta_2}{\cos}(\gamma-\frac{4}{3}{\pi})\label{eq:LundConvDef4} \\
\frac{R_z-R_0}{R_0} & = & \sqrt{\frac{5}{4\pi}}{\beta_2}{\cos}{\gamma}.\label{eq:LundConvDef5}
\end{eqnarray}

For axially symmetric deformation, $\gamma=0$, $\beta_2$ can be derived from the equations above as: 
\begin{equation}
\beta_2 = \frac{4}{3}\sqrt{\frac{\pi}{5}}\frac{{\Delta}R}{R_0},\label{eq:BetaDefi}
\end{equation}
where ${\Delta}R$ $(=R_z-R_x)$ is the difference between the major ($R_z$) and minor ($R_x$) 
axis of the ellipsoid. It can be concluded from Eq.~\ref{eq:BetaDefi} that ${\beta_2}<0$ for 
oblate deformation, in which $R_z<R_x$, whereas ${\beta_2}>0$ for a prolate shape, in which $R_z>R_x$. 
Typical values for $\beta_2$ found in nuclei are: 0.2 -- 0.3 for normal deformation and 
0.4 -- 0.6 for superdeformation. 
Another popular deformation parameter $\varepsilon_2$ is often used in the literature. For small deformation, 
\begin{equation}
\varepsilon_2\; {\approx}\; \frac{{\Delta}R}{R_0} = \frac{3}{4}\sqrt{\frac{5}{\pi}}\beta_2 = 0.946\beta_2.\label{eq:EpslonDefi}
\end{equation}
The parameter $\gamma$ is the one to describe the degree of triaxiality. As can be seen in 
Figure~\ref{fig:QuadDeform}, the nuclei are axially deformed only when $\gamma$ is equal to 
multiples of $60^{\circ}$, while intermediate values of $\gamma$ describe various degrees of 
triaxiality with the maximum degree of triaxial deformation being reached when $\gamma$ is an odd multiple of 
$30^{\circ}$. 

\subsection{\label{subsec:DefmShelModl}The deformed shell model}

As was mentioned above, the spherical Shell Model has difficulies when dealing with 
issues regarding deformed nuclei. Therefore, the deformed shell model was introduced. 

\begin{figure}
\begin{center}
\includegraphics[angle=0,width=\columnwidth]{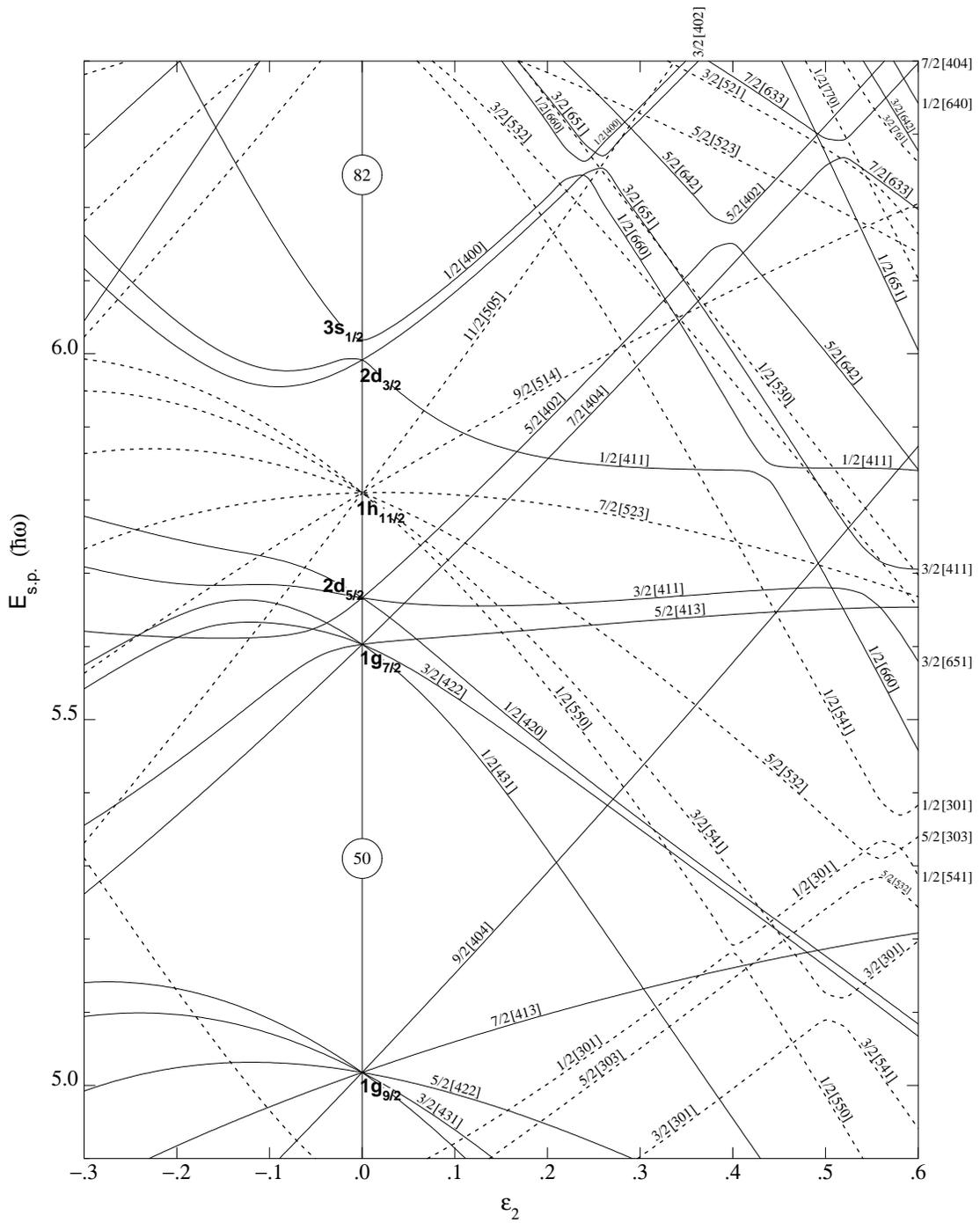}
\caption{Nilsson diagram for protons, $50\le{Z}\le82$. Adapted from Ref.~\cite{Bohr-98-book}.\label{fig:NilssonDiagm1}}
\end{center}
\end{figure}

\begin{figure}
\begin{center}
\includegraphics[angle=0,width=1.05\columnwidth]{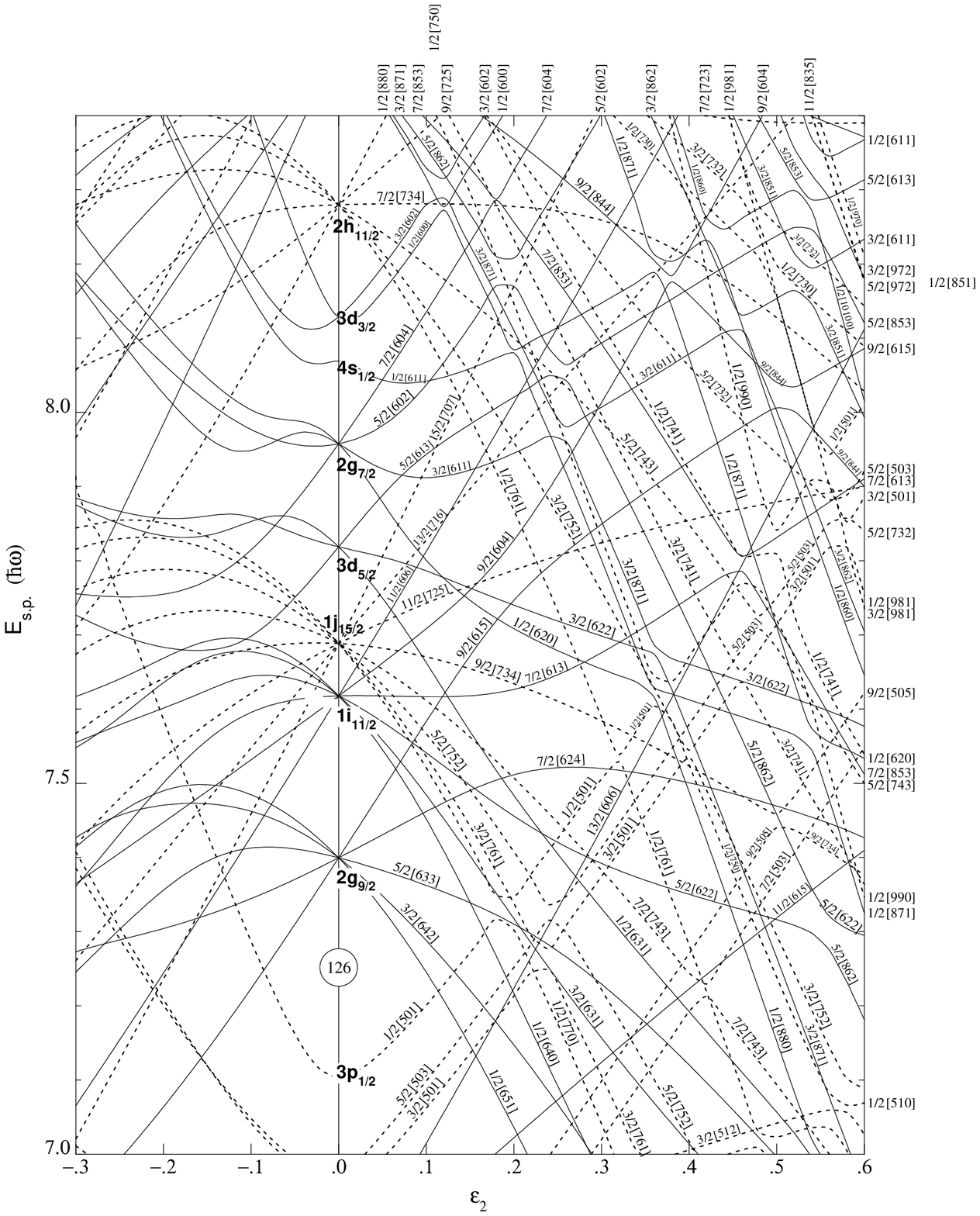}
\caption{Nilsson diagram for neutrons, $N\ge126$. Adapted from Ref.~\cite{Bohr-98-book}.\label{fig:NilssonDiagm2}}
\end{center}
\end{figure}

The modified harmonic oscillator potential, $\it{i.e.}$, Nilsson potential~\cite{Nilsson-DMFM-29-16-55}, 
allows to take deformation into account. The Hamiltonian in this case can be written as: 
\begin{equation}
H_{Nilsson}=\frac{-{\hbar}^2}{2m}{{\nabla}^2}+\frac{m}{2}\left(\omega_x^2x^2+
\omega_y^2y^2+\omega_z^2z^2\right)-2\kappa\hbar{\omega_0}\left[\vec{l}\cdot\vec{s}-
\mu\left(l^2-{\langle{l^2}\rangle}_N\right)\right],\label{eq:NilsnHamil}
\end{equation}
where the $\left(\vec{l}\cdot\vec{s}\right)$ term represents the spin-orbit force, and the 
$\left(l^2-{\langle{l^2}\rangle}_N\right)$ term was introduced by Nilsson to simulate the 
flattening of the nuclear potential at the bottom of the well (as obtained with a WS potential). 
The factors $\kappa$ and $\mu$ determine the strength of the spin-orbit and $l^2$ term, 
respectively. The $\omega_{x,y,z}$ terms are the one-dimensional oscillator frequencies 
which can be expressed as a function of the deformation. In the axially-symmetric case, 
\begin{equation}
\omega_x^2 = \omega_y^2 = \omega_0^2\left(1+\frac{2}{3}{\varepsilon_2}\right),~\omega_z^2 = \omega_0^2
\left(1-\frac{4}{3}{\varepsilon_2}\right),\label{eq:NilsnParamts}
\end{equation}
where $\omega_0$ is the oscillator frequency ($\hbar\omega_0=41A^{-1/3}MeV$) in the spherical 
potential, for which ${\varepsilon_2}=0$. Using the deformation-dependent Hamiltonian, the 
single-particle energies can be calculated as a function of the deformation ${\varepsilon_2}$. 
A plot of single-particle energies versus deformation is known as a Nilsson diagram; two examples 
of which are given in Figures~\ref{fig:NilssonDiagm1} and \ref{fig:NilssonDiagm2} for 
$50~{\le}~Z~{\le}~82$ and $N~{\ge}~126$, respectively. 

\begin{figure}
\begin{center}
\includegraphics[angle=270,width=0.60\columnwidth]{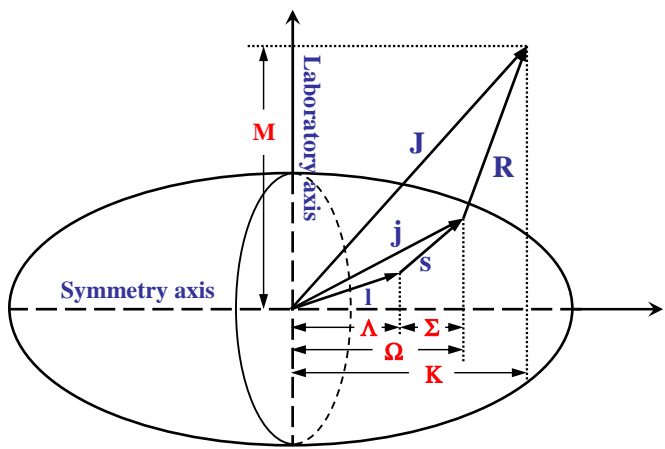}
\caption[Asymptotic quantum numbers for the Nilsson model.]{Asymptotic quantum numbers 
for the Nilsson model: $R$ is the angular momentum of 
collective rotation; $M$ and $K$ are the projections of the total angular momentum $J$ onto 
the laboratory axis and the symmetry axis respectively, and $K$ is equal to $\Omega$ if no collective 
rotation occurs or the axis of collective rotation is perpendicular to the symmetry axis. 
Adapted from Ref.~\cite{Zhu-03-thesis}.\label{fig:AsympQuantNums}}
\end{center}
\end{figure}

The Nilsson orbitals can be characterised by the so-called asymptotic quantum numbers
\begin{equation}
\Omega[Nn_{z}\Lambda]\label{eq:AsymQuanNumbs}
\end{equation}
where $N$ is the principal quantum number from the harmonic oscillator, $\Omega$ is the 
projection of the single-particle angular momentum onto the symmetry axis ($z$), 
$\Lambda$ is the projection of the orbital 
angular momentum onto the symmetry axis and $n_{z}$ is the number of oscillator quanta 
along the symmetry axis. While $N$ and $\Omega$ are strictly valid quantum numbers for the Hamiltonian 
(Eq.~\ref{eq:NilsnHamil}), $n_{z}$ and $\Lambda$ become good quantum numbers only for large 
deformations and are approximate quantum numbers otherwise. 
The parity of the state, $\pi$, is determined by $(-1)^N$. The 
projection of the intrinsic spin of the nucleon onto the symmetry axis is $\Sigma(=\pm\frac{1}{2})$, 
thus we can define $\Omega=\Lambda\pm\frac{1}{2}$. The asymptotic quantum numbers for 
the Nilsson model are shown schematically in Figure~\ref{fig:AsympQuantNums}. 

It should also be remembered that if $N$ is even, then ($n_z+\Lambda$) must also be even. 
Similarly if $N$ is odd, then ($n_z+\Lambda$) must be odd. It can be seen in Figure~\ref{fig:NilssonDiagm1} and 
Figure~\ref{fig:NilssonDiagm2} that at zero deformation, the ($2j+1$)-fold degeneracy 
of a given $j$ state is not lifted. When the deformation is introduced, the $j$ states 
split into two-fold degenerate levels, the number of which for a state $j$ is 
$(j+\frac{1}{2})$. 

Many properties of nuclear excitations based on orbitals in the Nilsson model can be 
understood with these quantum numbers. For example, in Figure~\ref{fig:NilssonDiagm2}, it can be seen 
that the $j_{15/2}$ shell, with negative parity, $\pi$, lies in a region of predominantly 
positive-parity orbits. As a result, the various trajectories of the orbits of $j_{15/2}$ 
parentage are rather straight and the associated wavefunctions are rather pure, while 
the orbits in the neighboring shells, $\it{e.g.}$, $d_{5/2}$, $g_{7/2}$, are bent and changing 
slopes much more often. Two levels with the same $\Omega$ and $\pi$ quantum numbers can not cross 
because the deformed potential couples them and causes a repulsion. In contrast, only levels 
with different $\Omega$ or $\pi$ cross, because the axial symmetric, reflection symmetric potential 
has no matrix elements due to its symmetry. Hence, it is just the 
high-$K$ values and negative parities of $j_{15/2}$ orbits, being different from those 
of the neighboring orbits, that lead to the above observation in the Nilsson diagram. 

In order to calculate the total energy of the nucleus, a summation of all populated 
single-particle energies can be made. The big shell gaps at finite values of $\varepsilon_2$, 
seen in the Nilsson diagram, suggest the existence of stable deformations. Thus, within the 
framework of the Nilsson model, it is possible to predict the magnitude of the deformation for 
nuclei away from closed shells. This is only a very rough estimate. An accurate method is described 
in the following section. 

\subsection{\label{subsec:StrutinskyMethod}The Strutinsky-shell correction}

The method to predict the existence of stable, deformed nuclei by calculating the total 
energy of the nuclear system with the shell model, described earlier, has proved to be 
successful in interpreting many microscopic aspects of the nucleus, mostly properties of 
excited states relative to the ground state. However, it fails to 
accurately reproduce some of the bulk properties of the nucleus, such as the total 
binding energy. In contrast, another approach, the liquid drop model~\cite{Nilsson-95-book}, 
where the nucleus is described in analogy to a liquid drop, has difficulty in predicting 
properties related to shell structure, but 
is often able to provide an adequate interpretation of the macroscopic properties of the nucleus. 
A new approach that can incorporate the advantages of both of these models was proposed by 
Strutinsky~\cite{Strutinsky-NPA-95-420-67,Strutinsky-NPA-112-1-68} to accurately reproduce, 
for example, the observed nuclear ground-state energies. In the Strutinsky approach, the total 
energy $E_{tot}$ is split into two terms: the first is a macroscopic term, $E_{ldm}$, derived 
from the liquid drop model, and the second is the microscopic term $E_{shell}$, which accounts 
for the fluctuations in the shell energy,
\begin{equation}
E_{tot}=E_{ldm}+E_{shell}(protons)+E_{shell}(neutrons).\label{eq:StrutinskyEnergy}
\end{equation}
In Eq.~\ref{eq:StrutinskyEnergy}, the $E_{shell}$ quantity is calculated independently 
for protons and neutrons. It is defined by the difference between the actual discrete 
level density and a ``smeared'' level density. The actual discrete level density $g$ 
consists of a sequence of $\delta$-functions, and the smeared density $\tilde{g}$ 
uses a Gaussian distribution instead. The respective definitions are: 
\begin{equation}
g(e)=\sum_{i}\delta(e-e_i),\label{eq:ActualDensity}
\end{equation}
and
\begin{equation}
\tilde{g}(e)=\frac{1}{\gamma\sqrt{\pi}}\sum_{i}f_{corr}\left(\frac{e-e_i}{\gamma}\right)
exp\left(-\frac{(e-e_i)^2}{{\gamma}^2}\right).\label{eq:SmearedDensity}
\end{equation}
Here, $\gamma$ is an energy of the order of the shell spacing $\hbar\omega_{0}$, 
$\sim~8MeV$, and $f_{corr}$ is a correction function for keeping unchanged the long-range 
variation over energies much larger than $\hbar\omega_{0}$. The shell energy $E_{shell}$ 
can thus be calculated using
\begin{equation}
E_{shell}=2\sum_{i}{e_{i}}-2\int{e}\tilde{g}(e)\,de,\label{eq:ShellEnergy}
\end{equation}
where the factor 2 arises because of the double degeneracy of the deformed levels. 
Calculations using this method have predicted well, for example, the existence of stable 
reflection-asymmetric deformation in nuclear ground 
states~\cite{Moller-NPA-361-117-81,Nazarewicz-NPA-429-269-84}. 

\section{\label{sec:rotateCSM}Rotation and cranked shell model}
\subsection{\label{subsec:NuclRotation}Nuclear rotation and rotational band}

\begin{figure}
\begin{center}
\includegraphics[angle=270,width=0.60\columnwidth]{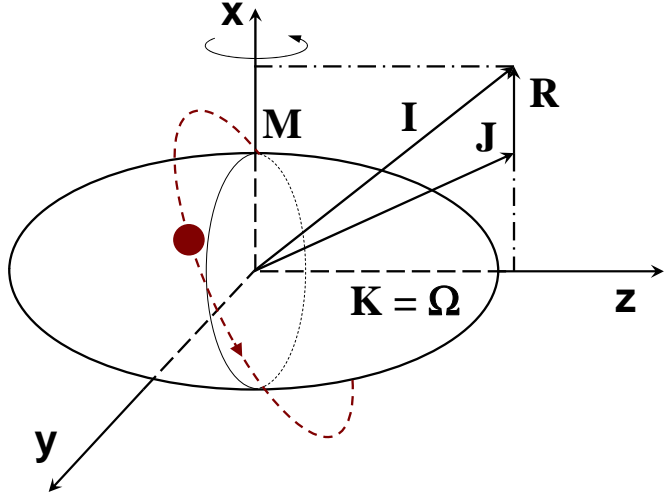}
\caption[Schematic coupling of angular momenta for collective rotation of an axially deformed 
nucleus.]{Schematic coupling of angular momenta for collective rotation of an axially deformed 
nucleus: $\vec{R}$ is the collective angular momentum, $\vec{J}$ is the intrinsic angular 
momentum, and $K$ is the projection of the total angular momentum, $\vec{I}$ $(=\vec{R}+\vec{J})$, 
onto the symmetry axis $z$. Adapted from Ref.~\cite{AbuSaleem-02-thesis}.\label{fig:DefNuclRot}}
\end{center}
\end{figure}

Because of deformation, discussed earlier, the collective rotation of the nucleus 
becomes possible. This started attracting people's attention early in 
the 1950s~\cite{Rainwater-PR-79-432-50,Bohr-PR-81-134-51,Bohr-DMFM-27-16-53}. 
In a quantum mechanical description, a system with a symmetry axis (conveniently 
named as the z-axis) is given by a wave function which is an eigenfunction of the angular 
momentum operator $\vec{J_z}$, and any rotation about this axis produces only a phase 
change. The rotating system has, therefore, the same wave function and the same 
energy as the ground state. This simply means that this system can not rotate about 
the symmetry axis collectively~\cite{Bengtsson-84-book}. The spherical nuclei are 
symmetric with respect to any axis, therefore, it is not possible to observe collective 
rotation in them. In the case of an axially deformed nucleus, there is a set of axes of 
rotation, perpendicular to the symmetry axis. A rotation around such an axis is 
presented schematically in Figure~\ref{fig:DefNuclRot}, and it gives rise to a 
distinct rotational pattern. 

Here, the rotational angular momentum $\vec{R}$ is generated by the collective motion of 
many nucleons about the axis $x$, which is perpendicular to the symmetry axis $z$. The intrinsic 
angular momentum, $\vec{J}$, is the sum of the angular momenta of the nucleons, 
$\it{i.e.}$, $\vec{J}=\sum_{i=1}^{A}\vec{j_i}$. The total angular momentum $\vec{I}$ 
is then $\vec{I}=\vec{R}+\vec{J}$, and its projection onto the symmetry axis $z$, $K$, is 
equal to the sum of the projection of the angular momentum $\vec{j}$ of the individual nucleons 
onto the symmetry axis, $\it{i.e.}$, ${\Omega}=\sum_{i=1}^{A}{\Omega}_i$, in this case. 

The classical kinetic energy of the rotating rigid body is, $E=\frac{L^2}{2\Im}$, where $L$ is the 
angular momentum and $\Im$ is the moment of inertia. In analogy, for a quantum system, the 
rotational energy is the expectation value of the Hamiltonian of rotation. For the rotating nuclear 
system schematically described in Figure~\ref{fig:DefNuclRot}, the Hamiltonian of rotation 
is given by:
\begin{equation}
H_{rot}=\frac{{\hbar}^2}{2\Im}{{I}_x^2}\; {\approx}\; \frac{{\hbar}^2}{2\Im}\left[{I}^2-{I}_z^2\right],\label{eq:RotHamiltonian}
\end{equation}
where $I_x$, $I_z$ are the projection of the total angular momentum $\vec{I}$ onto the axis of 
rotation $x$ and onto the symmetry axis $z$, respectively. The approximation assumes that the $x$ and $y$ components 
of $\vec{J}$ can be neglected (strong coupling), which is often the case. 

The state of the rotating system can be described in terms of three quantum numbers, the total 
angular momentum ($I$), its projection onto the axis of rotation ($M$), and its projection onto 
the symmetry axis ($K$). Hence, the energy of the rotating system can be obtained:
\begin{equation}
E_{rot}=\frac{{\hbar}^2}{2{\Im}}\left[I(I+1)-{K^2}\right].\label{eq:RotEnergy}
\end{equation}
It can be seen in Figure~\ref{fig:DefNuclRot} that the quantum number $K$ is associated with the intrinsic 
degrees of freedom of the valence nucleons, thus, in the energy of Eq.~\ref{eq:RotEnergy}, one term 
depends on intrinsic degrees of freedom, and the other depends on the total angular momentum of the 
system. The latter, generally called the rotational energy, can be written as: 
\begin{equation}
E=\frac{{\hbar}^2}{2{\Im}}I(I+1).\label{eq:RotEnergyP}
\end{equation}

The total wavefunction of the rotating system ${\Psi}^{I}_{MK}$ is the combination of the rotational 
wavefunction ($D^{I}_{MK}$) and the single-particle wavefunction (${\phi}_{K}$), and can be expessed as:
\begin{equation}
{\Psi}^{I}_{MK}={\left(\frac{2I+1}{16{\pi}^2}\right)}^{1/2}\left[{\phi}_{K}D^{I}_{MK}+(-)^{I+K}{\phi}_{-K}D^{I}_{M-K}\right].\label{eq:RotWaveFunc}
\end{equation}
In this expression, the second term reflects the property that a rotation by $\pi$ around the 
axis of rotation leaves the system unchanged. This rotational invariance results in two degenerate 
states, ${\phi}_{K}$ and ${\phi}_{-K}$, which form a single series of rotational states with spins 
give by:
\begin{equation}
I=K,~K+1,~K+2,...\label{eq:SpinRotSequce}
\end{equation}
The phase factor $(-)^{I+K}$ is called the signature. If the $J_x$ and $J_y$ components are taken into 
account,
\begin{eqnarray}
H_{rot} & = & \frac{{\hbar}^2}{2\Im}\left[(I_x-J_x)^2+(I_y-J_y)^2\right] \nonumber \\
 & = & \frac{{\hbar}^2}{2\Im}(I^2-I_z^2-2{I_x}{J_x}-2{I_y}{J_y}+J_x^2+J_y^2).\label{eq:RotHamiltonianGenXY}
\end{eqnarray}
The new terms compared with Eq.~\ref{eq:RotHamiltonian}, ${I_x}{J_x}$ and ${I_y}{J_y}$, are called Coriolis 
interactions. They represent the influence of rotation on the motion of the individual nucleons. Among other 
things, they disturb the regular sequence (Eq.~\ref{eq:SpinRotSequce}). According to the signature quantum number, 
$\alpha=(-1)^{I+K}$, the states of Eq.~\ref{eq:SpinRotSequce} can be divided into two distinct sets 
with an opposite value of the signature:
\begin{equation}
I=K,~K+2,~K+4,...\label{eq:SpinRotBandFav}
\end{equation}
and
\begin{equation}
I=K+1,~K+3,~K+5,...,\label{eq:SpinRotBandUnfav}
\end{equation}
and, each set of states is just a so-called rotational band. This means that the rotational bands 
are restricted to favored bands and unfavored partners with opposite signature. In an odd-A nucleus, 
for example, the levels in the favored bands possess spins, $I=\frac{1}{2},~\frac{5}{2},...$, while 
the unfavored partner bands are characterized by spins, $I=\frac{3}{2},~\frac{7}{2},...$, and opposite 
signature. 

In the excitation mode of rotation, a nucleus deexcites mostly in the form of emitting 
$\gamma$ rays, therefore, it is necessary to briefly introduce the fundamental 
properties of the $\gamma$ rays here. 

\begin{figure}
\begin{center}
\includegraphics[angle=270,width=0.40\columnwidth]{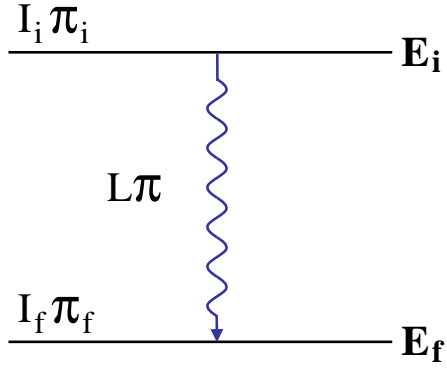}
\caption{Scheme of a $\gamma$ ray. Adapted from Ref.~\cite{Krane-88-book}.\label{fig:IllustGamRay}}
\end{center}
\end{figure}

As shown in Figure~\ref{fig:IllustGamRay}, the energy of a $\gamma$ ray that decays 
from an initial level with energy $E_i$ to a final level with energy $E_f$ is: 
\begin{equation}
E_{\gamma}=E_i-E_f.\label{eq:GamRayEner}
\end{equation}
Since each nuclear state has a definite angular momentum $I$, and parity $\pi$, a 
photon must take out angular momentum $\vec{L}$ (its eigenvalue is $L$) and 
parity $\pi$ in accordance with the conservation laws: 
\begin{eqnarray}
\vec{I_i}-\vec{L}=\vec{I_f},\label{eq:GamAMRule} \\
{\pi_i} \times {\pi}={\pi_f}.\label{eq:GamPatyRule}
\end{eqnarray}
The angular momentum of the photon, $L$, is called its multipolarity. 
For each multipolarity, two types of $\gamma$ transitions are possible: the electric 
transition (EL) or the magnetic transition (ML). Electric 
transitions have angular momentum $L$ and parity ${\pi}^{E}=(-)^{L}$, while 
the magnetic ones are characterized by angular momentum $L$ and parity ${\pi}^{M}=(-)^{L+1}$. 
Therefore, the selection rules for any $\gamma$ ray are: 
\begin{eqnarray}
|I_i-I_f|~{\le} & {L} & {\le}~(I_i+I_f), \nonumber \\
1~{\le} & {L} & {\le}~(I_i+I_f)~~~for~I_i=I_f~{>}~0; \nonumber \\
{\pi_i}{\pi_f} & = & {(-)^{L}}~~~for~EL, \nonumber \\
{\pi_i}{\pi_f} & = & {(-)^{L+1}}~~~for~ML.\label{eq:GamRaySelectRule}
\end{eqnarray}
Since the photon has an intrinsic spin of 1, a $\gamma$ transition from $I_i=0$ state to 
$I_f=0$ state can not occur. Often, the so-called stretched $E2$ transitions, $\it{i.e.}$, 
$\gamma$ decays from levels with an angular 
momentum $I$ to levels with an angular momentum ($I-2$) and the same parity, dominate 
in a rotational band, while the $E1$ transitions, $\it{i.e.}$, $\gamma$ decays 
from levels with the angular momentum $I$ to levels with the angular momentum ($I{\pm}1$) and the opposite 
parity, dominate inter-band deexcitations, especially in the case of octupole bands 
discussed in detail in this thesis work. 

It is also worth to note that the real nucleus is intermediate between two extremes, a rigid body and 
a superfluid, as the measured moments of inertia are less than the rigid body values at low spin 
and larger than those calculated for the rotation of a superfluid. Furthermore, experimentally the 
moment of inertia of nucleus is found to change as a function of spin. For the rotating nucleus, 
the important angular rotational frequency, $\omega$, can be written as
\begin{equation}
\hbar\omega = \frac{dE(I)}{d{I_x}} = \frac{dE(I)}{d(\sqrt{I(I+1)-{K^2}})},\label{eq:RotFrequen}
\end{equation}
where $I_x$ is called the aligned angular momentum and is the projection of the total angular momentum 
$I$ onto the rotation axis. In the simplest case, $K=0$. For a rotational band, where states are linked 
by $E2$ transitions, the angular rotational frequency can be approximated as
\begin{equation}
\hbar\omega=\frac{E(I)-E(I-2)}{\sqrt{I(I+1)}-\sqrt{(I-2)(I-1)}}~{\simeq}~\frac{E_{\gamma}}{2},\label{eq:RotFrequenKZo}
\end{equation}
where $E_{\gamma}$ is the energy of $\gamma$ ray between two consecutive levels in the rotational band. 
For a rotational band, two spin-dependent moments of inertia, which are related to two different 
aspects of nuclear dynamics, have been introduced in terms of the derivatives of the excitation energy 
with respect to the aligned angular momentum. The kinematic moment of inertia is the first order 
derivative
\begin{equation}
\Im^{(1)}={I_x}{\left({\frac{dE}{d{I_x}}}\right)}^{-1}{{\hbar}^2}=\hbar\frac{I_x}{\omega},\label{eq:KineMomtInert}
\end{equation}
and can be used to express the transition energy, $E_{\gamma}$, in a rotational band with Eq.~\ref{eq:RotEnergyP} as:
\begin{equation}
E_{\gamma}=E(I)-E(I-2)=\frac{{\hbar}^2}{\Im^{(1)}}(2I-1)\label{eq:E2GamRayEner}
\end{equation}
through Eq.~\ref{eq:RotEnergy}; while the dynamical moment of inertia is the second order derivative:
\begin{equation}
\Im^{(2)}={\left(\frac{{d^2}E}{d{{I_x}^2}}\right)}^{-1}{\hbar}^2=\hbar\frac{d{I_x}}{d\omega},\label{eq:DynaMomtInert}
\end{equation}
and can be related to the energy spacing of consecutive $\gamma$ rays in a rotational band
\begin{equation}
\Delta{E_{\gamma}}=\frac{4{\hbar}^2}{\Im^{(2)}}.\label{eq:DeltGamRayEner}
\end{equation}
Moreover, the two moments of inertia have the following relation, 
\begin{equation}
\Im^{(2)}=\frac{d}{d{\omega}}\left(\omega\Im^{(1)}\right)=\Im^{(1)}+\omega\frac{d{\Im^{(1)}}}{d\omega},\label{TwoMomtInertRelt}
\end{equation}
and $\Im^{(1)}\simeq\Im^{(2)}$ if $\Im^{(1)}$ is constant in a band. 

\subsection{\label{subsec:PairInteract}Pairing interaction}

The pairing interaction is a force responsible for binding together two identical nucleons with opposite intrinsic spins in the same 
orbit, and this interaction is such that the energy of the configuration of opposite spins for the two nucleons is much 
lower than the one of any other configuration. The existence of pairing forces in the nucleus 
is firmly supported by many experimental results, for example: (1) the ground state of even-even 
nuclei always has $0^{+}$ spin and parity, (2) the ground-state spin of odd-mass nuclei is always determined by the spin
of the last nucleon, which is the only unpaired one, and (3) the binding energy of an odd-mass nucleus is found to be always 
smaller than the average values for two neighboring even-even nuclei. The strength of the pairing interaction, $G$, which favors 
the maximum spatial overlap between the wave functions of nucleons, is lower for protons ($G_p=\frac{17}{A}MeV$) than for neutrons 
($G_n=\frac{23}{A}MeV$). The Hamiltonian describing pairing is usually written in the form:
\begin{equation}
H_{pair} = -G{P^{+}P}-{\mu}{\hat{N}},\label{eq:PairHamilt}
\end{equation}
where $P^{+}$ and $P$ are pair creation and annihilation operators, respectively, $\mu$ is the 
chemical potential, and $\hat{N}$ is the number operator. 

\begin{figure}
\begin{center}
\includegraphics[angle=270,width=0.50\columnwidth]{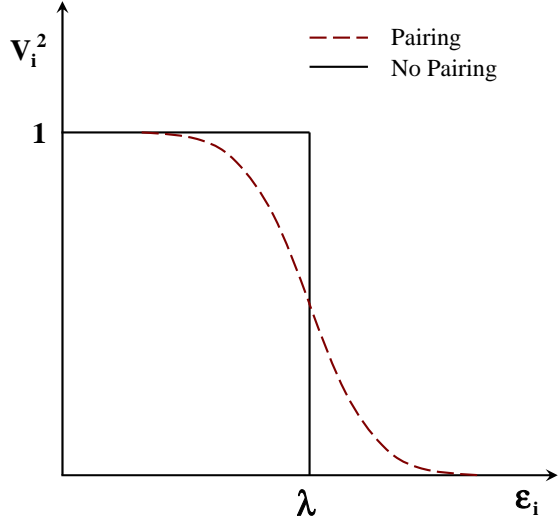}
\caption{Schematic representation of the smearing of the Fermi surface due to the pairing interaction. 
Adapted from Ref.~\cite{Greenlees-99-thesis}.\label{fig:PairFermiSurface}}
\end{center}
\end{figure}

Near to the Fermi surface, $\it{i.e.}$, near the last filled level, some unoccupied orbits are present. 
The pairing interaction scatters pairs of nucleons with $J^{\pi}=0^{+}$ from occupied states $j$ into empty 
states $j'$ and this will result in a ``smearing'' of the Fermi surface. In the absence of pairing, the 
Fermi surface would be a sharp rectangle (see Figure~\ref{fig:PairFermiSurface}). 

The smearing of the Fermi surface leads to the concept of 
quasi-particles~\cite{Bardeen-PR-108-1175-57,Bohr-PR-110-936-58}, where particle and hole wave functions are combined. 
The probability that a state $i$ is occupied by a hole is given by the expression:
\begin{equation}
U_i^2=\frac{1}{2}\left[1+\frac{({\varepsilon}_{i}-{\lambda})}{\sqrt{({\varepsilon}_{i}-{\lambda})^2+{\Delta}^2}}\right],\label{eq:ProbHole}
\end{equation}
while the corresponding expression for the occupation by a particle is given as:
\begin{equation}
V_i^2=\frac{1}{2}\left[1-\frac{({\varepsilon}_{i}-{\lambda})}{\sqrt{({\varepsilon}_{i}-{\lambda})^2+{\Delta}^2}}\right],\label{eq:ProbParticle}
\end{equation}
where ${\varepsilon}_{i}$ is the single particle energy, and $\lambda$ is the average Fermi energy associated with a certain particle number 
(see Ref.~\cite{Casten-90-book} for a detailed discussion of these quantities). The probabilities are normalised such that 
$U_i^2+V_i^2=1$. It can be seen that, far below the Fermi surface (${\varepsilon}_{i}~{\ll}~{\lambda}$) $V_i^2=1$, and far above 
the Fermi surface (${\varepsilon}_{i}~{\gg}~{\lambda}$) $U_i^2=1$. Close to the Fermi surface, the occupation probabilities are 
mixed. Following the treatment described in Ref.~\cite{Bardeen-PR-108-1175-57}, the quasi-particle energy can be expressed by:
\begin{equation}
E_{qp}(n,p)=\sqrt{({\varepsilon}_{i}-{\lambda})^2+{\Delta}^2}.\label{eq:QuasiPartEner}
\end{equation}

As the nucleus rotates, the induced Coriolis force, in analogy to the one of the classical rotations, competes with the 
pairing interaction and attempts to break the pair and align the individual angular momenta of the two nucleons with 
the rotation axis. More generally, the rotational motion weakens the pairing interaction in the nucleus, $\it{i.e.}$, while 
some pairs of nucleons are broken and align at specific rotational frequencies, pairing is affected for all pairs. 
This is known as the Coriolis anti-pairing effect (CAP)~\cite{Mottelson-PRL-5-511-60}. 

\subsection{\label{subsec:CSMintro}The cranked shell model}

In order to understand the interplay between the collective and intrinsic degrees of freedom of the nucleons, 
the cranked shell model (CSM) was developed by Bengtsson and Frauendorf~\cite{Bengtsson-NPA-327-139-79}, built on 
the original cranking concepts introduced by Inglis in 1954~\cite{Inglis-PR-96-1059-54}. 
In this model, the nucleons can be viewed as particles independently moving in an average potential, which rotates 
around the principal axis ($x$), which is perpendicular to the symmetry axis of the nucleus 
(an example is shown in Figure~\ref{fig:DefNuclRot}). 

The cranking model is formulated in the body-fixed frame. The transformation from the laboratory frame to the 
body-fixed frame can be made easily using the the rotation operator, 
$\Re=e^{-i{\omega}t{j_x}/{\hbar}}$, where $j_x$ is the projection of the total angular momentum onto the rotational axis $x$. 
The time-dependent Schr{\"o}dinger equation of a single particle in the laboratory system can be written as:
\begin{equation}
i{\hbar}\frac{{\partial}{\phi}^{l}}{{\partial}t}=h^{l}{{\phi}^{l}}.\label{eq:SinlLabSchodingerEQ}
\end{equation}
Using the rotation operator $\Re$, the wavefuction in the laboratory frame ${\phi}^{l}$ can be expressed in terms of the 
intrinsic wavefunction ${\phi}^{0}$, 
\begin{equation}
{\phi}^{l}=\Re{\phi}^{0},\label{eq:RtWvFunc}
\end{equation}
and the Hamiltonian in the laboratory frame $h^{l}$ can be expressed in terms of the intrinsic Hamiltonian $h^{0}$, $\it{i.e.}$, 
the non-rotating Hamiltonian expressed in the body-fixed frame, 
\begin{equation}
h^{l}=\Re{h^{0}}{\Re}^{-1}.\label{eq:RtHamilt}
\end{equation}
Hence, the time-dependent Schr{\"o}dinger equation of a single particle in the intrinsic (body-fixed) frame can be obtained, 
\begin{equation}
i{\hbar}\frac{{\partial}{\phi}^{0}}{{\partial}t}=(h^{0}-{\omega}{j_x}){{\phi}^{0}},\label{eq:SinlCSMschodingerEQ}
\end{equation}
by replacing ${\phi}^{l}$ and $h^{l}$ with the expressions~\ref{eq:RtWvFunc} and \ref{eq:RtHamilt}, respectively, and 
computing the time derivation in Eq.~\ref{eq:SinlLabSchodingerEQ}. 
The single-particle cranking Hamiltonian $h^{\omega}$ becomes:
\begin{equation}
h^{\omega}=h^{0}-{\omega}{j_x},\label{eq:SinlCrankHamiltn}
\end{equation}
where the term ${\omega}{j_x}$ represents the Coriolis and centrifugal forces resulting from the rotating frame. 
The eigenvalue of the single-particle cranking Hamiltonian, $e_{\nu}^{\omega}$, derived from the Schr{\"o}dinger equation, 
\begin{equation}
h^{\omega}{|{\nu}^{\omega}>}=e_{\nu}^{\omega}{|{\nu}^{\omega}>},\label{eq:SinSchodingerEQ}
\end{equation}
is the single-particle Routhian, where $|{\nu}^{\omega}>$ is the single-particle eigenfunction in the rotating frame. 
Taking into account the pairing interaction, the single-particle (quasi-particle) cranking Hamiltonian becomes:
\begin{equation}
h^{\omega}=h^{0}-{\omega}{j_x}-{\Delta}(P^{+}+P)-{\mu}{\hat{N}},\label{eq:RotQuasiPartEner}
\end{equation}
where $\Delta$ is the pair gap. For a given configuration, the total Routhian $e^{\prime}$ can be deducted by diagonalizing as:
\begin{equation}
e^{\prime}=\sum_{i}e_{\nu}^{\omega}(i).\label{eq:TotalRouthian}
\end{equation}
The single-particle (quasi-particle) aligned angular momentum (alignment), which is the projection of angular momentum onto the axis of 
rotation, can be obtained from the slope of the single-particle (quasi-particle) Routhian versus the rotational frequency, $\it{i.e.}$, 
$i_{x}^{\omega}=-\frac{de_{\nu}^{\omega}}{d{\omega}}$. 
%\begin{equation}
%i_{x}^{\omega}=-\frac{de_{\nu}^{\omega}}{d{\omega}}=-{\hbar}<{\nu}^{\omega}|j_{x}|{\nu}^{\omega}>.\label{eq:SingAlignAngMomt}
%\end{equation}
Similar to the Routhian, the total alignment, $i_{x}$, is given as:
\begin{equation}
i_x=\sum_{i}i_{x}^{\omega}(i).\label{eq:TotalAlign}
\end{equation}

Therefore, after being appropriately transformed into the rotating frame (to be discussed in Sec.~\ref{subsec:LabTrnsfBody}), the measured 
Routhian $e^{\prime}$ and alignment values $i_x$ as a function of the rotational frequency $\omega$ can be compared with the 
results of calculations from Eqs.~\ref{eq:TotalRouthian} and \ref{eq:TotalAlign}. 

\begin{figure}
\includegraphics[angle=270,width=\columnwidth]{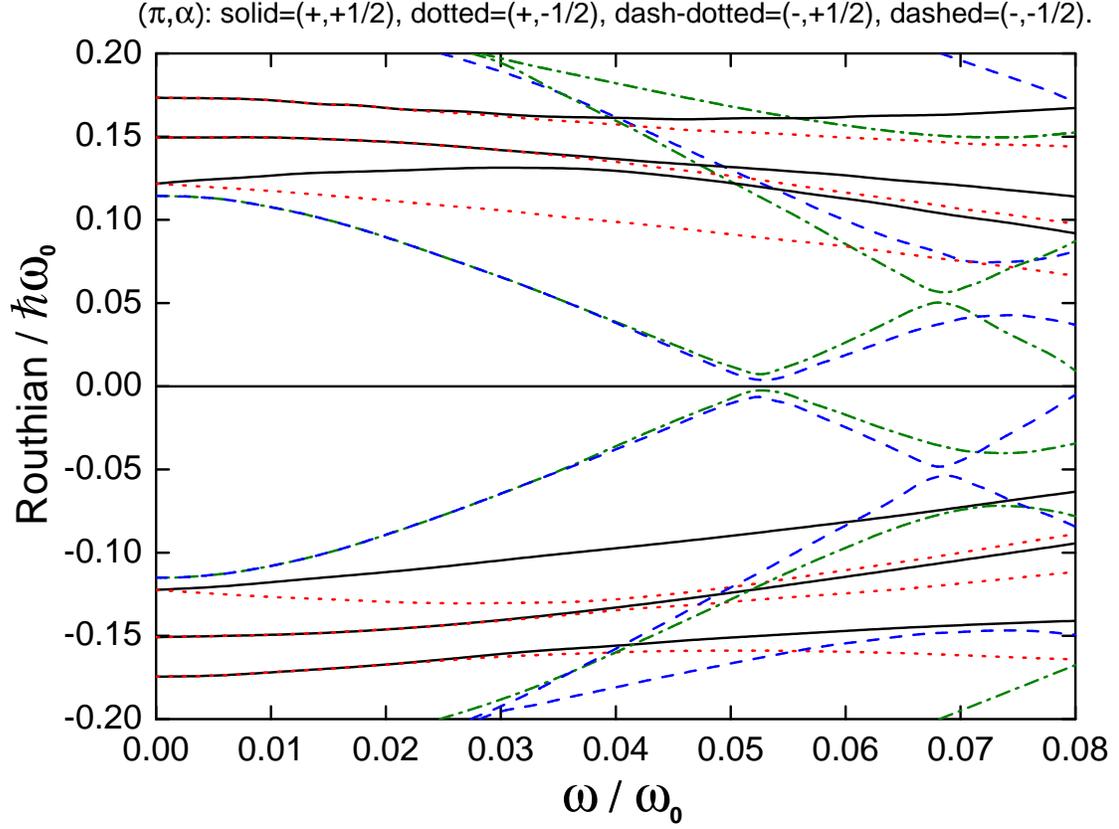}
\caption[Quasi-proton Routhians of $^{164}$Er.]{Quasi-proton Routhians of $^{164}$Er calculated with the following parameters: the quadrupole deformation, ${\varepsilon_2}=0.258$; 
the hexadecapole deformation (a high-order term in Eq.~\ref{eq:NuclRadius} in Sec.~\ref{subsec:Deformation}), 
${\varepsilon_4}=0.001$; the triaxiality, ${\gamma}=0^{\circ}$; the pairing gap, ${\Delta}=0.111~MeV$; 
and the Fermi energy, ${\lambda}=5.812~MeV$. 
Further details about these parameters may be found in Refs.~\cite{Nilsson-NPA-131-1-69,Bengtsson-NPA-327-139-79}, and the 
energy scale $\hbar{\omega_0}$ is 
defined in Ref.~\cite{Nilsson-NPA-131-1-69} (for a spherical shape, commonly, $\hbar{\omega_0}=41A^{-1/3}~MeV$). The figure is taken 
from Ref.~\cite{Bengtsson-PLB-135-358-84}.\label{fig:164Er-quasi-Routh}}
\end{figure}

Since the non-rotating single-particle wavefunctions are not eigenfunctions of $j_x$, the rotation leads to 
a mixing of the single-particle states and breaks the time-reversal symmetry. Thus, for the single-particle states the only 
remaining good quantum numbers are the parity, $\pi$, which is a conserved quantum number as long as the shape of the 
potential can be expanded in even multipoles, and the signature, $\alpha$, which is related to the properties of a 
nucleonic state under a rotation by $180^{\circ}$ around an axis ($x$) perpendicular to the symmetry axis. The signature is defined by:
\begin{equation}
{\Re}_{x}(\pi){\phi}_{\alpha}=e^{-i{\pi}j_x}{\phi}_{\alpha}=e^{-i{\pi}{\alpha}}{\phi}_{\alpha},\label{eq:SignatureDef}
\end{equation}
where $\phi_{\alpha}$ denotes a wavefunction with signature $\alpha$. While the parity is $+$ or $-$, the signature of a 
single particle state can be written as $+\frac{1}{2}$ or $-\frac{1}{2}$ conventionally. In a non-rotating potential 
(if $\omega=0$, $h^{\omega}=h^{0}$), the time-reversed states with the quantum number $+\Omega$ and $-\Omega$, $\it{i.e.}$, the projection 
of spin onto the symmetry axis ($z$), are energetically degenerate. Although they do not have a good signature with respect 
to a rotation perpendicular to the symmetry axis, it is always possible to form linear combinations of $\pi$ and $\alpha$. 
These linear combinations can then be used as basis states when solving the cranking equation~\ref{eq:SinSchodingerEQ}. 
which is then split into four independent sets of equations, each one corresponding to a particular combination of the 
parity, $\pi$, and the signature, $\alpha$. The solutions, $\it{i.e.}$, Routhians of quasi-particles, can therefore be classified 
by the quantum numbers ($\pi,\alpha$), which have four available values: $(+,+\frac{1}{2})$, $(+,-\frac{1}{2})$, $(-,+\frac{1}{2})$, 
and $(-,-\frac{1}{2})$. The Routhians are calculated as a function of the rotational frequency, $\hbar\omega$, for a given 
deformation and pairing gap using the cranking equation. They are usually summarized through quasi-particle diagrams. An example 
is given in Figure~\ref{fig:164Er-quasi-Routh} which presents the quasi-proton diagram calculated for $^{164}$Er~\cite{Bengtsson-PLB-135-358-84}. 
In the figure, the trajectories (orbitals) are labeled by ($\pi,\alpha$), and it is especially noticeable that orbitals with the same 
($\pi,\alpha$) do not cross; they rather come within some energy and then repel each other. The interaction regions can be 
interpreted as virtual crossings between different quasi-particle configurations, resulting in changes in alignment and energy. The experimental 
observation associated with a virtual crossing between the occupied and unoccupied quasi-particle orbitals is characterized by a sudden, 
large increase of the angular momentum along with a decrease in rotational frequency; $\it{i.e.}$, the curve bends back and up. 
The same happens in a plot of the moment of inertia vs. the rotational frequency. This phenomenon has been called ``backbending''. 
It was first observed in the ground state rotational bands of $^{162}$Er and $^{158,160}$Dy~\cite{Johnson-PLB-34-605-71}. 
The underlying physical explanation is the decoupling of a pair of high-$j$ quasi-particles from time reversed orbitals, where 
they have opposite intrinsic spins, and the alignment of their spins with the rotational axis (x) due to the increase of the 
Coriolis force with rotation~\cite{Stephens-NPA-183-257-72}. Hence, the rearrangement of the quasi-particle configuration of the nucleus 
represents the rotational alignment of a pair of quasi-particles. 

\subsection{\label{subsec:LabTrnsfBody}Transfering the experimental data to the intrinsic frame of nucleus}
As described earlier in this section, the Cranking shell model provides an opportunity 
to make predictions about the properties of a nuclear system, particularly the alignment $i_x$ and the quasi-particle energy 
(Routhian) $e^{\prime}$, in the rotating frame of reference. On the other hand, the measured values of the alignment and 
Routhian can be extracted from experimental data. Hence, after transferring the data 
from the laboratory frame to the rotating frame, a comparison between experiment and theory can be made to give the data 
an appropriate theoretical interpretation as well as to test how well the model predicts the experimental observations. 

For a rotational band, where states are linked by $E2$ transitions, the angular rotational frequency $\omega$ can be derived 
by using the transformed expression of Eq.~\ref{eq:RotFrequenKZo} in Sec.~\ref{subsec:NuclRotation}, 
\begin{equation}
\hbar\omega(I+1)=\frac{E_{\gamma}(I)}{\sqrt{(I+2)(I+3)-K^2}-\sqrt{I(I+1)-K^2}},\label{eq:measedAngFrqu}
\end{equation}
from the measured $\gamma$-ray energy $E_{\gamma}(I)$ (for the transition $(I+2){\rightarrow}I$), the spin $I$, and the 
known $K$ value which represents the projection of $I$ onto the axis of symmetry. The two spin-dependent 
moments of inertia can then be deduced from the measured $E_{\gamma}$ and spin $I$ values for three consecutive levels in the band, $\it{e.g.}$, 
$I+2$, $I$, and, $I-2$, through applying the following formula which are deduced from Eqs.~\ref{eq:KineMomtInert} and \ref{eq:DynaMomtInert}: 
\begin{eqnarray}
\Im^{(1)}(I+1) & = & {{\hbar}^2}\frac{(I+1)_x}{\hbar\omega(I+1)};\label{eq:CalKintMomt} \\
\Im^{(2)}(I) & = & {{\hbar}^2}\frac{(I+1)_x-(I-1)_x}{\hbar\omega(I+1)-\hbar\omega(I-1)},\label{eq:CalDynaMomt}
\end{eqnarray}
where $I_x=\sqrt{I(I+1)-K^2}$, $(I+1)_x$ and $(I-1)_x$ are in the same form with only $I$ substituted by $I+1$ and $I-1$, respectively, 
and, $\hbar\omega(I-1)$ is deduced by replacing $I$ with $I-2$ in Eq.~\ref{eq:measedAngFrqu}. 
For the simplest case, in which $K=0$, $\omega=\frac{E_{\gamma}}{2\hbar}$, and, $\Im^{(2)}(I)=\frac{4{\hbar}^2}{\Delta{E_{\gamma}}}$, 
where $\Delta{E_{\gamma}}$ ($=E_{\gamma}(I)-E_{\gamma}(I-2)$) is the energy spacing of consecutive $\gamma$ rays in the band. 

The experimental Routhian is given in terms of the excitation energy of level $E(I)$, the angular frequency $\omega$, and the 
angular momentum on the symmetry axis $I_x$ as:
\begin{equation}
E_{exp}^{\omega}(\omega)=\frac{1}{2}[E(I)+E(I+2)]-\hbar\omega(I+1)(I+1)_x.\label{RoutnBfSubRef}
\end{equation}
To evaluate the quasi-particle contribution alone, a reference energy must be subtracted in order to eliminate the contribution of the rotating 
core. This is commonly done by parameterizing the moments of inertia $\Im^{(1)}$, $\Im^{(2)}$ as functions of the rotational frequency $\omega$, 
as described in Ref.~\cite{Hackman-PRL-79-4100-97}:
\begin{eqnarray}
\Im^{(1)}(\omega) & = & J_0+J_1{\omega}^2;\label{eq:FitKintMomt} \\
\Im^{(2)}(\omega) & = & J_0+3J_1{\omega}^2,\label{eq:FitDynaMomt}
\end{eqnarray}
$J_0$ and $J_1$ are called Harris parameters. Hence, the fits of measured $\Im^{(1)}$ and $\Im^{(2)}$ values, 
obtained from Eqs.~\ref{eq:CalKintMomt} and \ref{eq:CalDynaMomt}, in the form of Eqs.~\ref{eq:FitKintMomt} and \ref{eq:FitDynaMomt}, where $\omega$ 
values are extracted from Eq.~\ref{eq:measedAngFrqu}, will give the value of the Harris parameters. The reference aligned angular momentum 
$I_x^{ref}(\omega)$ can be deduced by using the expression as:
\begin{equation}
I_x^{ref}(\omega)={J_0}\,{\omega}+{J_1}\,{\omega}^3,\label{eq:RefAlign}
\end{equation}
and the reference energy is given by:
\begin{equation}
E_{ref}^{\omega}(\omega) = -\int{I_x^{ref}(\omega)}\,d{\omega}
~{\simeq}~-\frac{1}{2}{J_0}{\omega}^2-\frac{1}{4}{J_1}{\omega}^4+\frac{{\hbar}^2}{8J_0},\label{eq:RefEnerg}
\end{equation}
omitting the last term, if $J_0=0$. 
Finally, the experimental Routhian $e^{\prime}(\omega)$ in the rotating frame, refering to the quasi-particle contribution alone, is written as:
\begin{equation}
e^{\prime}(\omega)=E_{exp}^{\omega}(\omega)-E_{ref}^{\omega}(\omega),\label{eq:FinalRouthn}
\end{equation}
and, in the same manner, the experimental alignment $i_x(\omega)$ in the rotating frame, which reflects the quasi-particle contribution only, can 
be given as:
\begin{equation}
i_x(\omega)=I_x(\omega)-I_x^{ref}(\omega),\label{eq:FinalAlign}
\end{equation}
where $I_x(\omega)$ is just $(I+1)_x$ defined above. 

\subsection{Shape vibrations}
\begin{figure}[h]
\includegraphics[angle=270,width=\columnwidth]{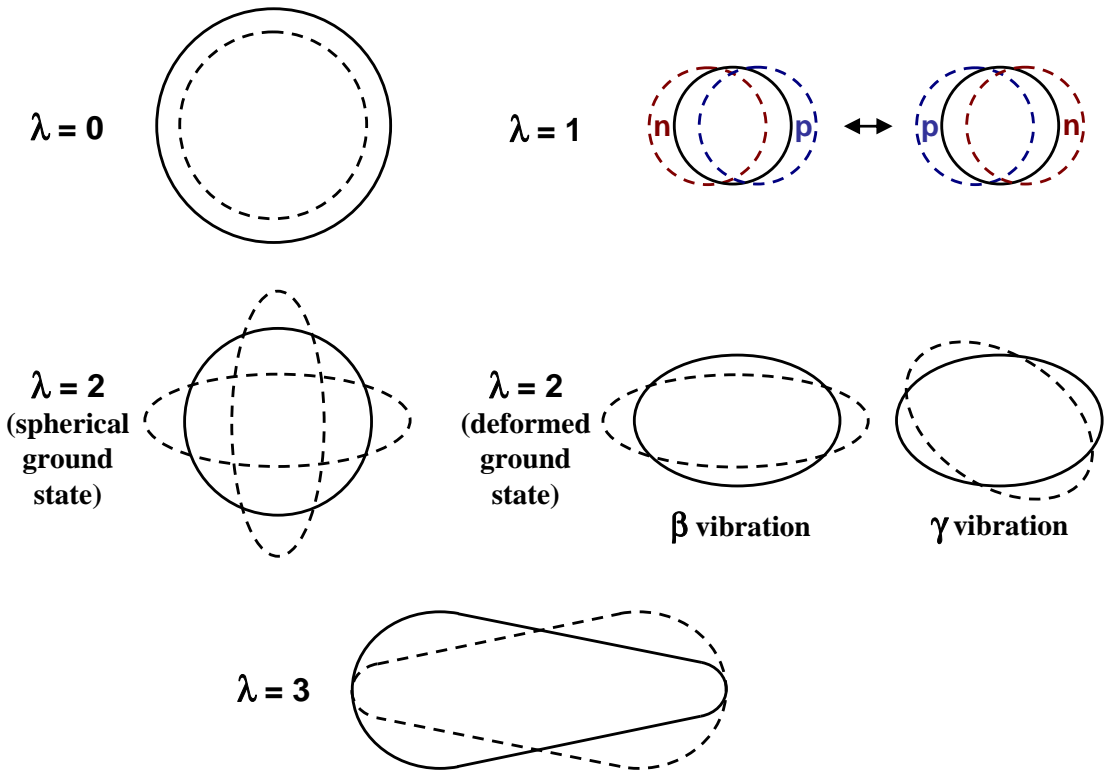}
\caption{A schematic illustration of the different modes of the nuclear vibration.\label{fig:VibrationModes}}
\end{figure}

In addition to rotation, vibration is also one of the collective excitation modes of the nucleus. 
One of the foci of this work is the nature of octupole vibrations in the Pu isotopes. It corresponds to a type of oscillation of the shape 
of the nucleus. When a spherical nucleus absorbs small amounts of energy, its density distribution can start to vibrate around the 
spherical shape. The magnitude of this vibration can be described by the coefficients $\alpha_{{\lambda}{\mu}}$, defined in 
Eq.~\ref{eq:NuclRadius} in Sec.~\ref{subsec:Deformation}. For small amplitude vibrations, the Hamiltonian for a vibration of 
multipole order $\lambda$, which is actually the difference between the energy of the deformed shape corresponding to 
the vibration and the energy of the nucleus at rest, can be written as:
\begin{equation}
H_{\lambda} = \frac{1}{2}C_{\lambda}\sum_{\mu}|\alpha_{\lambda\mu}|^2+\frac{1}{2}D_{\lambda}\sum_{\mu}
\left|\frac{d{\alpha}_{{\lambda}{\mu}}}{dt}\right|^2.\label{eq:VibHamilt}
\end{equation}
%\begin{equation}
%C_{\lambda} = \frac{1}{4\pi}(\lambda-1)(\lambda+2){\alpha}_2{A}^{2/3}-\frac{5}{2\pi}
%\frac{(\lambda-1)}{2\lambda+1}{\alpha}_3\frac{Z(Z-1)}{A^{1/3}},\label{eq:CLambdaDef}
%\end{equation}
%\begin{equation}
%D_{\lambda}=\frac{\rho{R_0}^5}{\lambda},\label{eq:DLambdaDef}
%\end{equation}
With the assumption that the different modes of vibrational excitation are independent from one another, the 
classical equation of motion can be obtained from the above Hamiltonian,
\begin{equation}
D_{\lambda}\frac{d^2{\alpha}_{\lambda\mu}}{dt^2}+C_{\lambda}{\alpha}_{\lambda\mu} = 0.\label{eq:EqOFVibMotn}
\end{equation}
Therefore, a small vibration can be considered as an harmonic oscillation with the amplitude, $\alpha_{\lambda\mu}$, 
and the angular frequency, $\omega_{\lambda}=\left(\frac{C_{\lambda}}{D_{\lambda}}\right)^{1/2}$. 
The vibrations are quantized. The quanta are called phonons, and $\hbar\omega_{\lambda}$ is quantity of vibrational 
energy for the multipole $\lambda$. Each 
phonon is a boson carrying angular momentum $\lambda\hbar$ and a parity $\pi=(-1)^{\lambda}$. The different 
modes of low order vibrational excitation ($\lambda=0,1,2,3$) are illustrated in Figure~\ref{fig:VibrationModes}. 

\section{Electromagnetic properties of deformed nuclei}
\subsection{Electric quadrupole moment}

The nuclear quadrupole moment is one of the most important properties of a deformed nucleus, and the observation of 
large quadrupole moments in nuclei away from closed shells is one of the direct evidences for the existence of stable 
nuclear deformation. The intrinsic quadrupole momemt, $Q_{0}$, in the body fixed frame of a deformed nucleus rotating 
about its z-axis can be defined in terms of the charge distribution in the nucleus, ${\rho}_{e}(r)$, and, hence, of the 
nuclear shape, as:
\begin{equation}
Q_{0}=\int(3z^2-r^2){\rho}_{e}(r)\,d^{3}r~{\approx}~\frac{8Z}{5}\frac{a-b}{a+b}\,{r_{0}^{2}},\label{eq:QuadMomt}
\end{equation}
where $a$ and $b$ are the lengths of the major and minor axes of nucleus, respectively, and $r_{0}=\frac{a+b}{2}$. 
Therefore, the nuclear quadrupole moment is a direct measure of the nuclear deformation, $\it{i.e.}$, for a spherical shape, 
$Q_0=0$; for a prolate shape, $Q_0>0$; and for an oblate nucleus, $Q_0<0$. The $Q_0$ moment can also be related to 
the deformation parameter, ${\beta}_2$. In axially symmetric nuclei with quadrupole deformation only, the first order 
expression can be given as: 
\begin{equation}
Q_{0}=\frac{3Z}{\sqrt{5}\pi}{r_0^2}{{\beta}_2}.\label{eq:Q0Beta2Relt}
\end{equation}

Generally, the experimental quadrupole moments measured in the laboratory frame are the spectroscopic quadrupole moments, 
$Q_{spec}$, which will be discussed in Chapter~\ref{chap:163Tm-lifetime}. As shown in Ref.~\cite{Olani-NPA-403-572-83}, 
the intrinsic quadrupole moment, $Q_{0}$ 
can be obtained by projecting the spectroscopic quadrupole moment onto the frame of reference fixed on the nucleus 
through the following relation:
\begin{equation}
Q_0=\frac{(I+1)(2I+3)}{3K^2-I(I+1)}{Q_{spec}},\label{eq:Q0QsRelt}
\end{equation}
where $K$ is the projection of $I$ onto the symmetry axis, as described in Sec.~\ref{subsec:NuclRotation}. 
For a $K=0$ band, such as the ground state band in even-even nuclei, this relation has the simpler form:
\begin{equation}
Q_0=\frac{(I+1)(2I+3)}{I(2I-1)}{Q_{spec}}.\label{eq:Q0QsGSRelt}
\end{equation}

Moreover, as shown in Ref.~\cite{Jolos-PRC-72-024310-05}, the experimental transition quadrupole moment, $Q_t$, 
which can be derived from the measurement of the lifetime of a state (see below), is related to the $Q_0$ moment 
by the relation:
\begin{equation}
Q_t(I+1)=\sqrt{Q_0(I)Q_0(I+2)}.\label{eq:Q0QtRelt}
\end{equation}

\subsection{\label{subsec:MagMomt}Magnetic moment}
In contrast to the nuclear electric moment, the nuclear magnetic moment reflects the contribution of the individual nucleons 
inside the nucleus. It is convenient to separate the orbital and spin contributions of the neutrons and protons. 
The magnetic moment operator can be expressed as:
\begin{equation}
\hat{\mu}={\mu_{N}}\sum_{i=1}^{A}[g_{li}l_{i}+g_{si}s_{i}],\label{eq:MagMontOpt}
\end{equation}
where $\mu_{N}$ is the nuclear magneton, $g_{li}$ and $g_{si}$ are the orbital and the spin gyromagnetic ratios 
(the gyromagnetic ratio is the ratio of the magnetic dipole moment to the angular momentum of a nucleus), respectively. 
Besides this contribution, the rotation of the core as a whole, $\it{i.e.}$, the collective rotation, 
contributes to the nuclear magnetic moments. In units of the nuclear magneton, the latter contribution is proportional 
to the angular momentum of rotation, $R$. Combining all of the contributions together, the magnetic moment operator can 
be written after some mathematical treatment as:
\begin{equation}
\hat{\mu}={g_R}I+[{g_K}-{g_R}]\frac{K^2}{I+1}.\label{eq:MagMontOptTl}
\end{equation}
The observed nuclear magnetic moment is the expectation value of the magnetic moment operator on a nuclear state $|I,K>$:
\begin{equation}
\mu={\langle}I,K|{\hat{\mu}_z}|I,K{\rangle},\label{eq:ObvMagMont}
\end{equation}
where $I$ is the total angular momentum, $K$ is the projection of $I$ onto the symmetry axis, and the 
z-axis is the axis of rotation. 

\subsection{\label{subsec:Gamma-rayProperty}Gamma-ray transition probability and branching ratio}

As described in Sec.~\ref{subsec:NuclRotation}, in the process of nuclear deexcitation between two levels of energy 
$E_i$ and $E_f$ (${E_i}>{E_f}$), a $\gamma$ ray or a conversion electron is emitted, which carries the energy and angular 
momentum difference between initial and final states. The angular momentum of the $\gamma$ ray (photon) always has 
an integer value of at least 1, and the photon can be of electric or magnetic character. 

As described in Ref.~\cite{Blatt-52-book,Alder-RMP-28-432-56}, the transition probability for a $\gamma$ ray of 
multipolarity $L$ from an excited state $a$ to a final state $b$ is given by, 
\begin{equation}
T(L)=\frac{8\pi(L+1)}{L[(2L+1)!!]^2}\left(\frac{1}{\hbar}\right)
\left(\frac{E_{\gamma}}{{\hbar}c}\right)^{2L+1}B(L),\label{eq:GamTransProb}
\end{equation}
where the reduced transition probability $B(L)$ for $a{\rightarrow}b$ is
\begin{equation}
B(L)=(2J_a+1)^{-1}|{\langle}{\psi}_b\|\varkappa(L)\|{\psi}_a{\rangle}|^2,\label{eq:GamReduTransProb}
\end{equation}
and $\varkappa$ is the electromagnetic operator. The reduced transition probability, $B(L)$, 
represents a sum of squared $\varkappa(L,\mu)$ matrix elements over the $m$ substates of 
the final state and an average over the $m$ substates of the initial state~\cite{Cerny-74-book}. 
An approximation to single-particle matrix elements is often used to calculate an approximate unit of strength, 
which is called Weisskopf unit~\cite{Blatt-52-book}, 
\begin{eqnarray}
B(EL)_{W} & = & (1/{4\pi})[3/(3+L)]^2(1.2A^{1/3})^{2L}~~~~[{e^2}{fm^{2L}}],\label{WeiskfUnitA} \\
B(ML)_{W} & = & (10/{\pi})[3/(3+L)]^2(1.2A^{1/3})^{2L-2}~~~~[{{\mu_0}^2}{fm^{2L-2}}],\label{WeiskfUnitB}
\end{eqnarray}
where $A$ is the mass number, $fm$ is femtometer ($10^{-15}$ m), and the units $e^2$ and ${\mu_0}^2$ are 
$e^2=1.44~MeV\;fm$ and ${\mu_0}^2=(e{\hbar}/2{M_p}c)^2=0.01589~MeV\;fm^3$. 

In this work, the observed $\gamma$ rays are of E1, E2 or M1 character, and the reduced transition 
probabilities for these three basic cases can be written according to Eq.~\ref{eq:GamTransProb}, 
\begin{eqnarray}
B(E1) & = & 6.288{\times}10^{-16}(E_{\gamma})^{-3}{\lambda}(E1)~~~~[{e^2}{fm^{2}}],\label{eq:E1ReduTranProb} \\
B(E2) & = & 8.161{\times}10^{-10}(E_{\gamma})^{-5}{\lambda}(E2)~~~~[{e^2}{fm^{4}}],\label{eq:E2ReduTranProb} \\
B(M1) & = & 5.687{\times}10^{-14}(E_{\gamma})^{-3}{\lambda}(M1)~~~~[{\mu_0}^2],\label{eq:M1ReduTranProb}
\end{eqnarray}
where $E_{\gamma}$ is the $\gamma$-ray energy in $MeV$, and $\lambda(XL)$ (X is E or M) can be measured experimentally using the relation 
with the $\gamma$-ray intensity, $I(XL)$ (the summation goes over all of the emitted $\gamma$ rays from the initial state $a$),
\begin{equation}
\lambda(XL)\propto\frac{I(XL)}{\sum{I(XL)}}.\label{eq:XLTransProb}
\end{equation}
Furthermore, within a rotational band, it is possible to express the reduced transition probabilities of E1, E2 and M1 
transitions in terms of the dipole moment (unit: $e\;fm$), $D_0$, the intrinsic quadrupole moment (unit: $e\;(fm)^2$), 
$Q_0$, and the gyromagnetic ratios, $g_K$ and $g_R$, respectively, as:
\begin{eqnarray}
B(E1) & = & \frac{3}{4\pi}{e^2}{D_0}^2|{\langle}{J}_{a}{K}10|{J}_{b}{K}{\rangle}|^2,\label{eq:E1ReltQ0B} \\
B(E2) & = & \frac{5}{16\pi}{e^2}{Q_0}^2|{\langle}{J}_{a}{K}20|{J}_{b}{K}{\rangle}|^2,\label{eq:E2ReltQ0B} \\
B(M1) & = & \frac{3}{4\pi}\left(\frac{e\hbar}{2Mc}\right)^2|{\langle}{J}_{a}{K}20|{J}_{b}{K}{\rangle}|^2(g_K-g_R)^2{K}^2,\label{eq:M1ReltQ0B}
\end{eqnarray}
where ${J}_{a}$, ${J}_{b}$ and $K$ are the total angular momenta described earlier and the projection onto symmetry axis. 

Hence, it is possible to derive from the $\gamma$-ray intensities the experimental branching ratios of the reduced probabilities by 
using equations~\ref{eq:E1ReduTranProb} to \ref{eq:M1ReltQ0B} in this section:
\begin{equation}
\frac{B(E1)}{B(E2)}=7.705{\times}10^{-7}\;\frac{E_{\gamma}^5(E2)}{E_{\gamma}^3(E1)}\;\frac{I(E1)}{I(E2)}~~~~[{fm}^{-2}],\label{eq:BrRtoE1E2}
\end{equation}
and
\begin{equation}
\frac{B(M1)}{B(E2)}=6.9685{\times}10^{-5}\;\frac{E_{\gamma}^5(E2)}{E_{\gamma}^3(M1)}\;\frac{I(M1)}{I(E2)}~~~~[{{\mu_0}^2}{e^{-2}}{fm^{-4}}].\label{eq:BrRtoM1E2}
\end{equation}
These measured branching ratios can be compared with the results of theoretical calculations, and the information brought by the 
comparison may help us understand some important properties of the nucleus studied. The branching ratios will be discussed 
in further details in the following chapters.

%
% Chapter 2
%

%
% Chapter 2
%

\chapter{\label{chap:exp_techs}EXPERIMENTAL TECHNIQUES}

It is well known to nuclear physics experimentalists that there are two basic 
phases to every research project in nuclear structure: 
producing the nuclei in the desired conditions and 
collecting all useful physical signals from the experiment. In order to achieve the first 
aspect an appropriate reaction must be selected, while 
the second requires proper instrumentation and a powerful data acquisition system. 

\section{Reaction and target}
\subsection{Fission barrier and angular momentum}
\begin{figure}
\begin{center}
\includegraphics[angle=0,width=0.65\columnwidth]{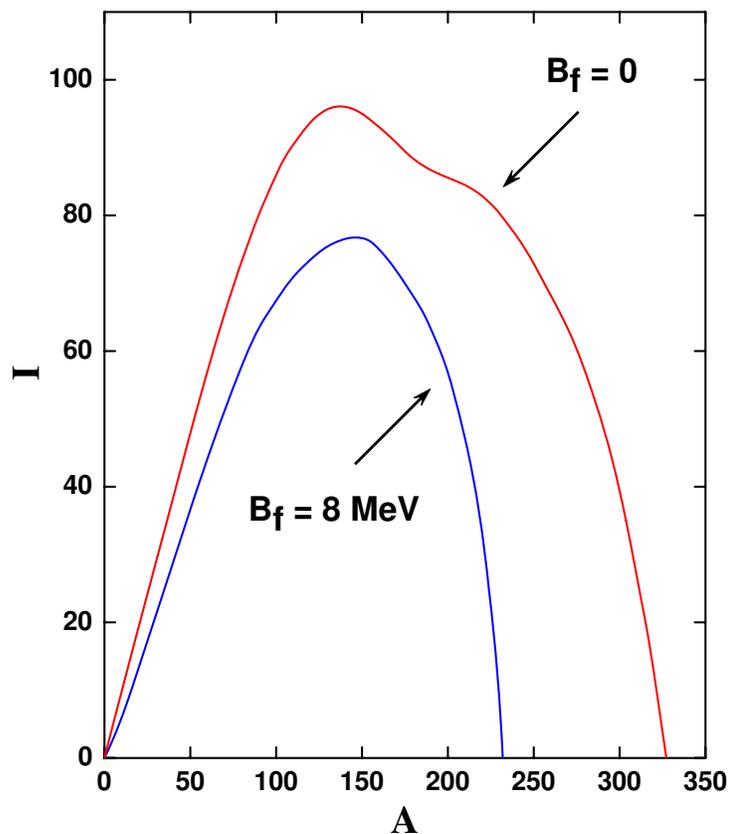}
\caption[The maximum angular momentum, as a function of mass number, which a nucleus 
can acquire before fission occurs in the liquid drop model.]{The maximum 
angular momentum, as a function of mass number, which a nucleus 
can acquire before fission occurs in the liquid drop model. The upper curve 
represents the angular momentum at which the fission barrier vanishes, while the 
lower curve represents the angular momentum for which the fission barrier is 8 $MeV$, $\it{i.e.}$, 
a condition where fission does not compete effectively with neutron evaporation. 
Adapted from Ref.~\cite{AbuSaleem-02-thesis}.\label{fig:MaxAnguMomtMass}}
\end{center}
\end{figure}

The work in this thesis focuses on phenomena in nuclei at high spin. Therefore, 
nuclei with large amounts of angular momentum must be produced with the chosen 
reaction. Often, a fusion-evaporation reaction with two heavy ions is the best option 
since a light projectile would only bring a small amount of angular momentum in the 
nucleus. However, this method is limited by the maximum amount of angular momentum 
a nucleus can accomodate while remaining stable against fission. This number is given 
qualitatively in Figure~\ref{fig:MaxAnguMomtMass}, from a calculation~\cite{Cohen-AP-82-557-74} 
done within the liquid drop model. 

In terms of the size and energy of the projectile, the maximum amount of angular momentum 
can be calculated using the following approximate relation:
\begin{equation}
l_{max}=\sqrt{2A_{B}E_{B}b},\label{eq:GenlMaxAngMomt}
\end{equation}
where $A_{B}$ and $E_{B}$ are the mass and energy of the projectile, and $b$ is the classical 
impact parameter. This parameter plays an important role in the fusion process and can be defined 
in terms of the distance between the centers of the beam and target nuclei. 

\subsection{Fusion-evaporation reaction}
\begin{figure}
\begin{center}
\includegraphics[angle=270,width=0.90\columnwidth]{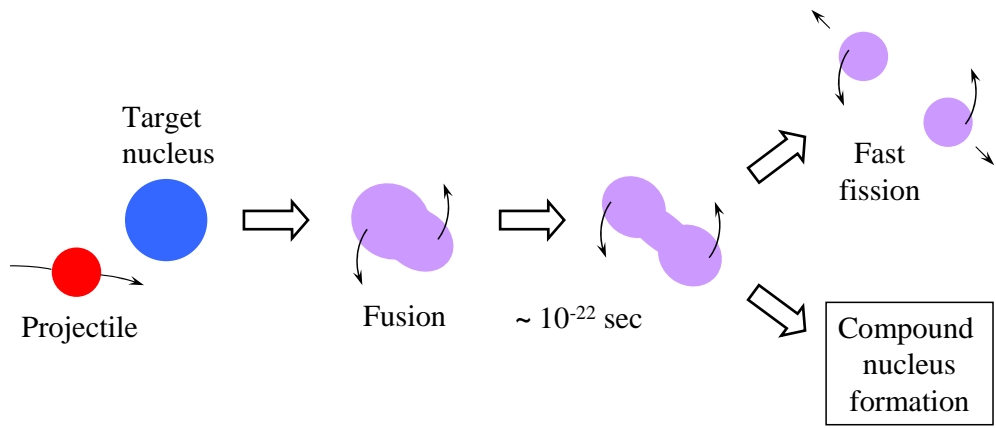}
\caption{A schematic illustration of fusion-evaporation reaction (process of compound 
nucleus formation).\label{fig:FusnEvpReactIlulst}}
\end{center}
\end{figure}

In the first part of this thesis work, the nucleus $^{163}$Tm was populated with high angular 
momentum via an heavy-ion induced fusion evaporation reaction, $\it{i.e.}$, $(HI,xn)$ 
reaction. In such a reaction, 
the target and projectile nuclei collide and fuse together. In a very short time 
($\sim$ $10^{-22}$ $s$), they either separate via fast fission or form a compound nucleus; this 
is schematically illustrated in Figure~\ref{fig:FusnEvpReactIlulst}. The idea of compound 
nucleus formation was first suggested by Niels Bohr in 1936~\cite{Bohr-Nature-344-137-36}. 

The compound nucleus is in a state of extreme excitation, and typically it may have an 
excitation energy of about $40~MeV$ and very large angular momentum, up to $70~{\hbar}$. 
However, it can only exist for $\sim$ $10^{-20}$ $s$~\cite{Blatt-52-book} before starting 
to get rid of its excess angular momentum and excitation energy. At first, the compound 
nucleus will deexcite in the most efficient way, particle emission, in which charged-particle 
emission (proton and alpha particles) is hindered by the Coulomb barrier and neutron evaporation 
usually dominates, until the process is no longer energetically possible. Each of the 
emitted neutrons carries away an excitation energy of on average 8 -- 10 $MeV$. The 
centrifugal barrier inhibits neutrons with considerable amount of orbital angular momentum 
from escaping~\cite{Amro-99-thesis}, thus, most of the evaporated neutrons are in $l=0$ or 1 
states ($l$ is the quantum number of orbital angular momentum). Hence, through the process 
of neutron evaporation, the nucleus loses most of the excitation energy, but only little 
angular momentum. Following particle emission, the nucleus, which is still in a state 
with a rather high excitation energy and a correspondingly large level density, will 
continue to deexcite through the emission of statistical $\gamma$ rays. These $\gamma$ rays are 
usually high energy dipole transitions, carrying away large amounts of excitation energy, 
but again very little angular momentum. As the nucleus approaches the yrast line, $\it{i.e.}$, 
the line which connects the states with the lowest energy for a given spin, the decay proceeds 
mostly through stretched quadrupole transitions (though other multipolarities also contribute) 
which remove the bulk of the angular momentum. When the level density is still high, these 
transitions form an unresolved continuum of $\gamma$ rays. Finally, as soon as the level density 
becomes low enough, the nucleus continues to decay by discrete transitions carrying lower energy 
(compared to statistical $\gamma$ rays) until the ground state is reached. These $\gamma$ 
rays will form cascades along, or parallel to, the yrast line. 

%\begin{figure}
%\begin{center}
%\includegraphics[angle=0,width=0.85\columnwidth]{FusnEvpEvIfig}
%\caption[A schematic of the decay of the compound nucleus.]{A schematic 
%of the decay of the compound nucleus. In this example, four neutrons 
%are first evaporated, carrying away a large amount of excitation energy, but very little 
%angular momentum. When below the particle evaporation threshold, deexcitation continues 
%to the ground state through $\gamma$-ray emission.\label{fig:FusnEvpEvIfig}}
%\end{center}
%\end{figure}

In order to have the compound nucleus formed and obtain the particular final nucleus of 
interest in an $(HI,xn)$ reaction, some factors have to be considered before the experiment:
(1) the target and beam should be available; 
(2) the energy of the projectile must be larger than the Coulomb barrier (in $MeV$), 
$E_{CB}$, 
\begin{equation}
E_{CB}=\frac{1.44{Z_1}{Z_2}}{1.16(A^{1/3}_1+A^{1/3}_2+2)},\label{eq:CoubBarir}
\end{equation}
and can often be estimated (particularly for $A~>~100$) appropriately using the empirical 
expression: 
\begin{equation}
E_{pk}(x)=\frac{1+{A_1}/{A_2}}{-Q_x+{\alpha}x},\label{eq:BeamEnerEst}
\end{equation}
where $Q_x$ is the $Q$ value for the given $(HI,xn)$ reaction, $\alpha$ is about 
$6~MeV$~\cite{Alex-PR-133-93-64}, and $A_1$ and $A_2$ refer to the mass of projectile 
and target, respectively; 
(3) as few as possible open reaction channels; 
(4) the maximum angular momentum, $l_{max}$, which can be estimated classically by the formula: 
\begin{equation}
l_{max}=\frac{\sqrt{\mu({E_0}-{E_{CB}})}}{4}(A^{1/3}_1+A^{1/3}_2){\hbar}\label{eq:MaxAngMomtEst}
\end{equation}
with $\mu={A_1}{A_2}/(A_1+A_2)$ and $E_0$ being the incident beam energy, should be comparable 
to the spins of levels of interest. 

\subsection{\label{subsec:UnsafeCulxTech}``Unsafe'' Coulomb excitation}
For the second part of this thesis work, the study of the properties of octupole correlations at high spin 
in the $^{238,240,242}$Pu isotopes, we need to carefully examine the feasibility of populating 
the desired states via fusion-evaporation reactions. In the framework of the liquid drop model, 
it is found empirically that no nucleus can survive fission if the condition 
$\frac{Z^2}{A}$ $\ge$ $50$ is satisfied. Obviously, the Pu isotopes of interest nearly satisfy 
this criterion. Moreover, the height of the fission barrier is inversely proportional to the 
value of the angular momentum. As can be seen in Figure~\ref{fig:MaxAnguMomtMass}, the high-spin 
states in the Pu isotopes of interest, $A$ $\sim$ $240$, are located beyond the curve of stability 
against fission, where ${B_f}=8~MeV$. Therefore, fusion-evaporation reactions can not be used 
effectively in work on these Pu isotopes, and another approach is required. 

The so-called ``unsafe'' Coulomb excitation (Coulex) technique, which was pioneered by 
D. Ward, $\it{et~al.}$~\cite{Ward-NPA-600-88-96} and proved to be successful in earlier work on actinide 
nuclei~\cite{Hackman-PRC-57-R1056-98,Wiedenhover-PRL-83-2143-99,Saleem-PRC-70-024310-04}, 
was exploited in this thesis. In this technique, a thick target, consisting of a 
target layer and a thick stopper foil is bombarded by a heavy beam (high $Z$) of energies $\sim$ 10 -- 15$\%$ 
above the Coulomb barrier (Eq.~\ref{eq:CoubBarir}). Thus, the dominant process is Coulomb excitation 
of the target and the projectile, but by raising the energy above the barrier, higher spin states 
are populated more strongly. In addition, transfer reactions will populate neighboring nuclei and 
provide an opportunity to investigate their structural properties as well. Unfortunately, the high beam 
energy will also generate fission and this process introduces background in the spectra. As 
shown in Ref.~\cite{Ward-NPA-600-88-96}, the deexcitation from states fed in unsafe Coulex can 
be selectively studied with detection systems comprising a number of Compton-suppressed Ge 
spectrometers plus a high-efficiency sum energy/multiplicity inner array, $\it{e.g.}$, Gammasphere. 
By gating on the $\gamma$-ray multiplicity, the longest rotational cascades, $\it{i.e.}$, the sequences 
involving the highest spin levels, can be selected. Under such conditions, most of the $\gamma$ 
rays are emitted after the excited nucleus has come to rest in 
the thick target, and the majority of transitions in a collective cascade are measured with the 
intrinsic resolution of the Ge detectors. This feature is especially useful in the actinide nuclei 
where many collective bands are characterized by nearly degenerate transition energies. In this 
technique, weak cascades that are not seen in the traditional particle-$\gamma$ Coulex experiments 
can be resolved in $\gamma$-$\gamma$ coincidence measurements. This technique has no limit on the 
thickness of the target material, and in some special cases, $\it{e.g.}$, $^{238}$U~\cite{Ward-NPA-600-88-96}, 
a thick foil of the material can be used both as a target and as a stopper. 

One of the deficiencies of this unsafe Coulex technique is that it is not possible to reliably 
measure absolute transition probabilities from the $\gamma$ yields. However, for the present work, 
this is not critical since only level schemes and relative $\gamma$-ray intensities are discussed. Another 
potential drawback of this technique originates in the Doppler broadening of transitions emitted from the 
highest spin states with lifetimes shorter than the stopping time. This effect makes such transitions harder 
to resolve. Fortunately, this drawback can also be overcome to some extent by obtaining and analyzing the 
angular spectra, as will be discussed in detail in Chapter~\ref{chap:PuIsotopes}. 

\subsection{Target preparation}
A crucial precondition for a successful experiment is the making of good targets. The quality of the 
targets may affect much the quality of final data. In the first part of this thesis work, the lifetime 
measurement of $^{163}$Tm, a thick target of isotopically enriched ($\ge$ $99\%$) $^{130}$Te backed 
by Au and Pb was used. The preparation of this target was fairly routine for the ANL target maker. 
In contrast, in the other part of this thesis work, the targets, $^{239,240,242}$Pu (backed by thick 
$^{197}$Au), are radioactive ($T_{1/2}$ $\sim$ $10^3$ -- $10^5$ years). As a result, the making and 
handling of these Pu targets were carried out very carefully in accordance with radiation safety concerns. 

\section{ATLAS and accelerating ions}
All experiments in this thesis work were performed at ATLAS (Argonne Tandem Linear 
Accelerator System) in the Physics Division at Argonne National Laboratory. ATLAS is the world's first 
superconducting linear accelerator for heavy ions at energies in the vicinity of the Coulomb barrier, 
and it consists of a sequence of sections of superconducting rf cavities where 
each accelerates charged atoms and then feeds them into the next section for additional energy gain. 
The beams are provided by one of two ``injector'' accelerators, either a 9 million volts ($MV$) electrostatic 
tandem Van de Graaff, or a new 12 $MV$ low-velocity linac coupled to an electron cyclotron resonance 
(ECR) ion source. The beam from one of these injectors is sent onto the 20 $MV$ ``booster'' 
linac, and then finally into the 20 $MV$ ``ATLAS'' linac section. High precision heavy-ion beams, with the 
size of $\le$ 1 $mm$ in diameter and pulses of $\le$ 500 $ps$ separated by 82 $ns$ intervals, 
ranging over all possible elements, from Hydrogen to Uranium, can be accelerated to energies of 7 -- 17 $MeV$ 
per nucleon and delivered to one of three target areas. 

\section{Gamma-ray detection}

\subsection{\label{subsec:GamRayIntactMatt}Interactions of gamma rays with matter}

With the appropriate beam and target, the nuclei to be studied are produced in a high-spin experiment, hence, 
the next step must be detecting the emitted $\gamma$ rays, which carry most of the useful physical information, 
and, possibly, charged particles as well. It is the interaction of electromagnetic radiation with matter 
(detector material) that makes the detection of $\gamma$ rays possible. For the energy range of $\gamma$ rays 
in high-spin research, $10~keV~<~E_{\gamma}~<~10~MeV$, only three major processes need to be considered, 
neglecting other small effects. These processes are the photoelectric absorption, Compton scattering, and 
pair production. 

In the case of photoelectric absorption, an incident photon is completely absorbed by an atom in the material, 
and one of the atomic electrons is ejected because of the energy deposited by photon. In Compton scattering, 
an incident $\gamma$ ray is inelastically scattered by an electron over an angle $\theta$, and a portion of 
$\gamma$-ray energy is transferred to the electron. The third process, pair production, is effective when the 
incident $\gamma$-ray energy is larger than 1.022 $MeV$, $\it{i.e.}$, twice of the rest mass of an electron. An 
electron/positron pair may be generated in the material, and, after slowing down, the positron will 
annihilate with one of the atomic electrons producing two $\gamma$ rays of energy 511 $keV$. 

\subsection{\label{subsec:GeDetIntro}Germanium detector}
\begin{figure}[h]
\begin{center}
\includegraphics[angle=270,width=0.60\columnwidth]{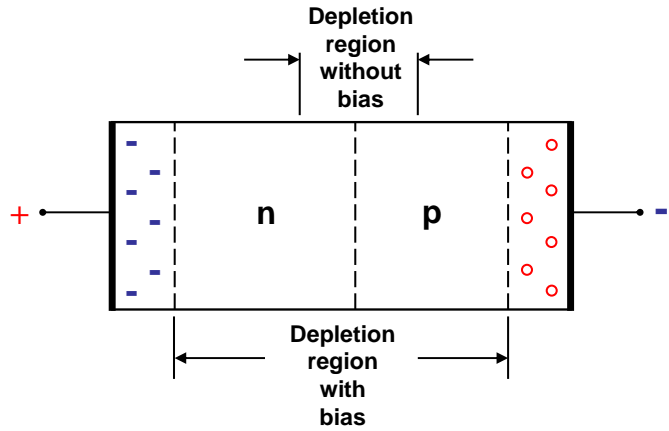}
\caption{Reverse-biased $p$-$n$ junction.\label{fig:BiasRev-p-n-junction}}
\end{center}
\end{figure}

In order to detect the $\gamma$ rays efficiently and accurately using the three processes just mentioned, 
the material of the detector must have a good enough absorption efficiency. This can be provided by a material 
of high atomic number ($Z$). The best energy resolution is generally provided by semiconductors, $\it{i.e.}$, materials 
which can to first order be viewed as a reservoir of loosely bound electrons. Of all types of 
semiconductor detectors, the High-Purity Germanium (HPGe) detectors are the best ones to satisfy the above 
requirements. They are widely applied in modern $\gamma$-ray spectroscopy experiments. An HPGe detector 
is a large reverse-biased $p$-$n$ diode junction, as shown in Figure~\ref{fig:BiasRev-p-n-junction}. The depletion 
region is a region of net zero charge in a $p$-$n$ junction. The reverse high voltage has the effect of enlarging 
the depletion region and, thus, the active volume for the radiation detection. 

Any $\gamma$ ray interacting with the germanium crystal, through the three processes described in 
Sec.~\ref{subsec:GamRayIntactMatt}, produces electron-hole pairs in the depletion region. These electron-hole pairs 
will then be swept to the edges of the detector, because of the electric gradient, and constitute an electric current. 
In the detector material, multiple processes, $\it{e.g.}$, typically a Compton scattering followed by another Compton event 
or by a photoelectric absorption, occur for an incident $\gamma$ ray. Moreover, the photopeak efficiency increases 
considerably with the rising of the active volume (width of depletion region), therefore, the use of larger volume 
detectors is desirable. 

The energy required to create an electron-hole pair in germanium is only about 3 $eV$. Thus, an incident $\gamma$ ray with a 
typical energy of hundreds of $keV$ can produce a large number of such pairs, leading to good resolution and small statistical 
fluctuations. The energy resolution $\Delta{E_{in}}$ (FWHM, $\it{i.e.}$, full width at half maximum) of any 
HPGe detector due to statistical fluctuations is given in the unit of $keV$ as:
\begin{equation}
\Delta{E_{in}}=2.36\sqrt{F{\epsilon}{E_{\gamma}}}=4.06\sqrt{F{E_{\gamma}}},\label{eq:GeDetEnerReslStatis}
\end{equation}
where $\epsilon=2.96~eV$ is the average ionization energy, $E_{\gamma}$ is the $\gamma$-ray energy in units of $MeV$, 
and, $F$ is a constant, called the Fano factor (for the interested reader, please see Chapter 5 of Ref.~\cite{Leo-93-book}). 
The Fano factor for a semiconductor has been studied experimentally in the work of Ref.~\cite{Pehl-NIM-81-329-70} and was 
shown to be between $0$ and $1$. 

\begin{figure}
\begin{center}
\includegraphics[angle=270,width=0.70\columnwidth]{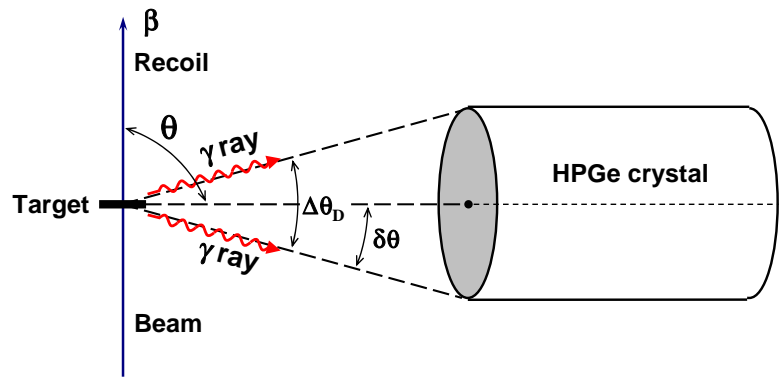}
\caption{Illustration for the discussion of Doppler broadening.\label{fig:DetResolDopplerBrd}}
\end{center}
\end{figure}

For an in-beam $\gamma$-ray experiment, the energy resolution of the Ge detector is often dominated by the Doppler broadening due 
to the motion of the recoiling nucleus and the opening angle ${\Delta}\theta_{D}$ of the Ge detector. The energy of a Doppler shifted 
$\gamma$ ray emitted from a recoiling nucleus with the velocity $\beta={v}/c$ ($c$ is the speed of light) and observed at an angle $\theta$ 
relative to the beam direction, $\it{i.e.}$, detector angle, can be written as:
\begin{equation}
E_{\gamma}=E_{{\gamma}0}\frac{\sqrt{1-\beta^2}}{1-\beta\cos{\theta}},\label{eq:DopplerShiftEnergy}
\end{equation}
where $E_{\gamma}$ is the enery of the $\gamma$ ray collected by a detector at the angle $\theta$ and $E_{{\gamma}0}$ is 
the nominal energy of the $\gamma$ ray (as emitted by recoils at rest). 
As shown in Figure~\ref{fig:DetResolDopplerBrd}, according to Eq.~\ref{eq:DopplerShiftEnergy}, the Doppler 
broadening $\Delta{E_{D}}$ can be given in terms of $\beta$, 
$\theta$ and ${\Delta}\theta_{D}$ ($=2\delta{\theta}$; $\delta{\theta}$ is the 
variation of the detector angle $\theta$) as: 
\begin{equation}
\Delta{E_{D}}=\left|E_{\gamma}(\theta+\delta\theta)-E_{\gamma}(\theta-\delta\theta)\right|
=E_{{\gamma}0}{\beta}{\sqrt{1-{\beta}^2}}\frac{\sin{\theta}}{(1-\beta\cos{\theta})^2}{{\Delta}\theta_{D}}.\label{eq:DopplerBrdTheta}
\end{equation}
It can be easily concluded that, for a given detector opening angle ${\Delta}\theta_{D}$, the energy of detected $\gamma$ rays will be 
maximally broadened at the angle $\theta=90^{\circ}$. Besides $\Delta{E_{in}}$ and $\Delta{E_{D}}$, the energy resolution of a Ge detector is 
also affected by $\Delta{E_{R}}$, $\it{i.e.}$, the Doppler broadening due to the angle spread of the recoils, and $\Delta{E_{V}}$, $\it{i.e.}$, the 
Doppler broadening due to the velocity distribution of the recoils. 

Based on the above discussion, the FWHM of the photon peaks, $\it{i.e.}$, the total energy resolution of a Ge detector $\Delta{E_{\gamma}}$, 
is expressed as (unit used generally: $keV$):
\begin{equation}
\Delta{E_{\gamma}^2}=\Delta{E_{in}^2}+\Delta{E_{D}^2}+\Delta{E_{R}^2}+\Delta{E_{V}^2}.\label{eq:GeDetEnerReslTl}
\end{equation}

The detection efficiency of HPGe detectors depends on the energy of the collected $\gamma$ ray, and reaches a maximum in the energy range of 
200 -- 400 $keV$. The relation of the efficiency with the $\gamma$-ray energy will be discussed in detail in Sec.~\ref{subsec:calibration}. 

HPGe detectors are operated at temperatures of around 77 $K$, in order to reduce noise from electrons which may be thermally 
excited across the small band gap in Ge (0.67 $eV$) at room temperature. This is achieved through thermal contact of the Ge 
crystal with a dewar of liquid nitrogen ($LN_2$), using a copper rod. 

\subsection{\label{subsec:CSupresBGO}Compton suppression and BGO detector}
\begin{figure}
\begin{center}
\includegraphics[angle=270,width=0.55\columnwidth]{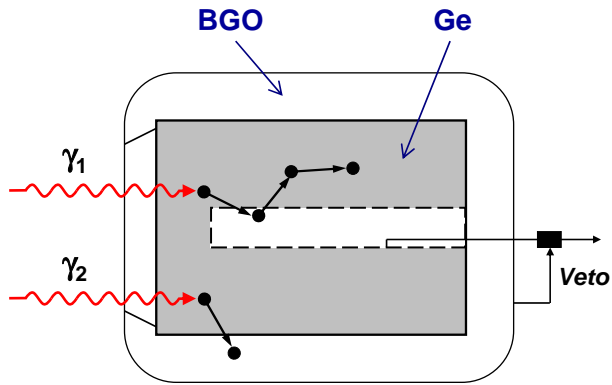}
\caption{Working principle of Compton suppression.\label{fig:ComptonSupresIllust}}
\end{center}
\end{figure}

Compton scattering, as discussed in Sec.~\ref{subsec:GamRayIntactMatt}, is a major process in $\gamma$-ray detection. Many 
$\gamma$ rays which enter the Ge detector will not deposit their full energy, leading to a large Compton continuum. One way to 
reduce the contribution of scattered $\gamma$ rays in the spectrum is to surround the Ge detector with a BGO (Bismuth Germanate) 
detector which detects $\gamma$ rays Compton-scattered out of the Ge crystal and then provides a signal to electronically 
suppress the partial energy pulse left in the Ge detector. 
Figure~\ref{fig:ComptonSupresIllust} schematically indicates the working principle of Compton 
suppression. With suppression shielding (the Ge and BGO are operated in anti-coincidence, which means that if an event occurs at 
the same time in both detectors, the event is rejected), the $\gamma$ ray which deposits all of its energy in the Ge detector, 
for example $\gamma_1$, is only detected by the Ge detector and, hence, will be accepted, while the $\gamma$ ray which deposits 
only part of its energy and is scattered out of the Ge detector, for example $\gamma_2$, is detected by both the Ge and BGO detector 
(a signal is generated to veto the HPGe readout), and, hence, will be discarded. 

\begin{figure}
\begin{center}
\includegraphics[angle=270,width=0.80\columnwidth]{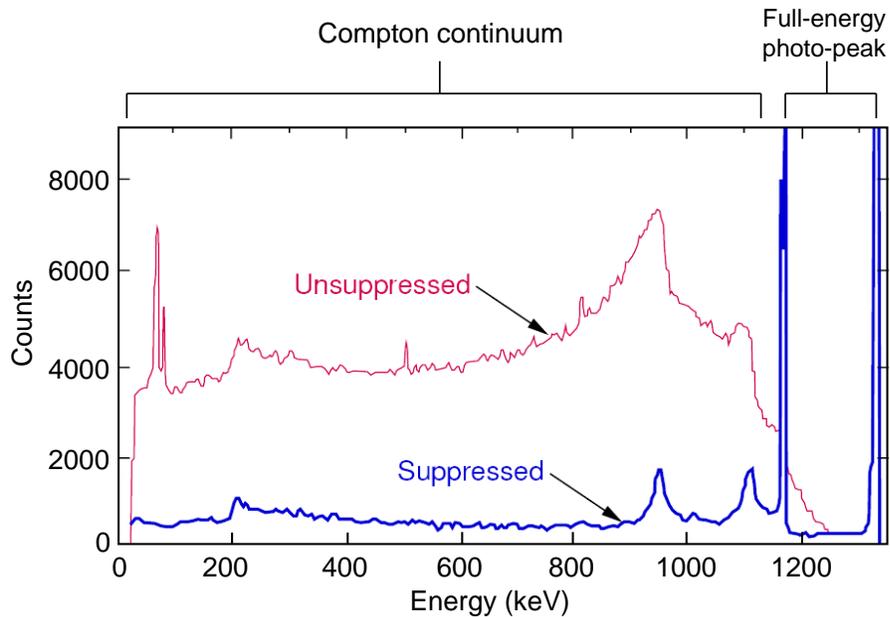}
\caption[Effect of Compton suppression.]{Effect of Compton suppression on the spectrum from the decay of a $^{60}$Co source. 
Taken from Ref.~\cite{Riley-98-book}.\label{fig:ComptnSupresImpPTratio}}
\end{center}
\end{figure}

BGO is a pure inorganic scintillator crystal. The reason to choose BGO detectors as Compton suppression shields, in spite 
of the notoriously low energy resolution of this material, is that, they have good timing properties (like other scintillator detectors), 
which is desirable for coincidence work, and high density (7.3 $g/{cm^3}$) and, hence, high efficiency (almost $100\%$ efficiency due 
to the large atomic number of $_{83}$Bi), which is suitable for shielding closely and tightly in large detector arrays. For 
$\gamma$-ray spectroscopy, the better the ratio of full-energy to partial-energy events (called the peak-to-total, 
or P/T ratio), the cleaner the $\gamma$-ray spectra will be. It has been proved that the P/T ratio can be improved considerably 
by Compton suppression. An example of the effect of Compton suppression on the spectrum from the decay of a $^{60}$Co 
source is shown in Figure~\ref{fig:ComptnSupresImpPTratio}. In this figure, it can be seen that the P/T ratio is 
increased from about $0.25$ for the bare crystal to about $0.60$ when Compton suppression is activated, while the photopeak is not 
affected appreciably~\cite{Riley-98-book}. 

\subsection{\label{subsec:GammasphereArray}Gammasphere detector array}
In an experimental work studying nuclear phenomena associated with high spin states, an average of 
20 -- 30 $\gamma$ rays are emitted in any single coincidence event (a $\gamma$-ray cascade). 
Therefore, a detection system capable of dealing simultaneously with a large number of $\gamma$ rays with 
good resolution and good efficiency is required. This has led to the development of large scale $\gamma$-ray 
detector arrays, such as Gammasphere~\cite{Janssens-NPN-6-9-96}, which was constructed early in 1990s, 
and is presently the most powerful $\gamma$-ray spectrometer in the world.

Gammasphere is an array of spherical shape (the diameter is about 6 feet and the weight is about 12 tons), 
consisting of up to 110 (101 in operation when running the experiments for this thesis work) 
Compton-suppressed HPGe detectors (HPGe + BGO) which cover almost $4\pi$ in solid angle around the target. 
The angle with respect to the beam direction of each ring of detectors and the maximum number of detectors 
that can be put in each ring can be found in Table~\ref{tab:GSdetRings}. 

\begin{table}
\begin{center}
\caption{ANGLE AND MAXIMUM NUMBER OF DETECTORS (DETS) FOR EACH RING OF GAMMASPHERE\label{tab:GSdetRings}}
\begin{tabular}{ccc}
\hline \hline
Ring No. & Angle & Maximum number of dets\\ \hline
1 & $17.27^{\circ}$ & 5\\
2 & $31.72^{\circ}$ & 5\\
3 & $37.38^{\circ}$ & 5\\
4 & $50.07^{\circ}$ & 10\\
5 & $58.28^{\circ}$ & 5\\
6 & $69.82^{\circ}$ & 10\\
7 & $79.19^{\circ}$ & 5\\
8 & $80.71^{\circ}$ & 5\\
9 & $90.00^{\circ}$ & 10\\
10 & $99.29^{\circ}$ & 5\\
11 & $100.81^{\circ}$ & 5\\
12 & $110.18^{\circ}$ & 10\\
13 & $121.72^{\circ}$ & 5\\
14 & $129.93^{\circ}$ & 10\\
15 & $142.62^{\circ}$ & 5\\
16 & $148.28^{\circ}$ & 5\\
17 & $162.73^{\circ}$ & 5\\
\hline 
\end{tabular}
\end{center}
\end{table}

\begin{figure}[h]
\begin{center}
\includegraphics[angle=270,width=0.60\columnwidth]{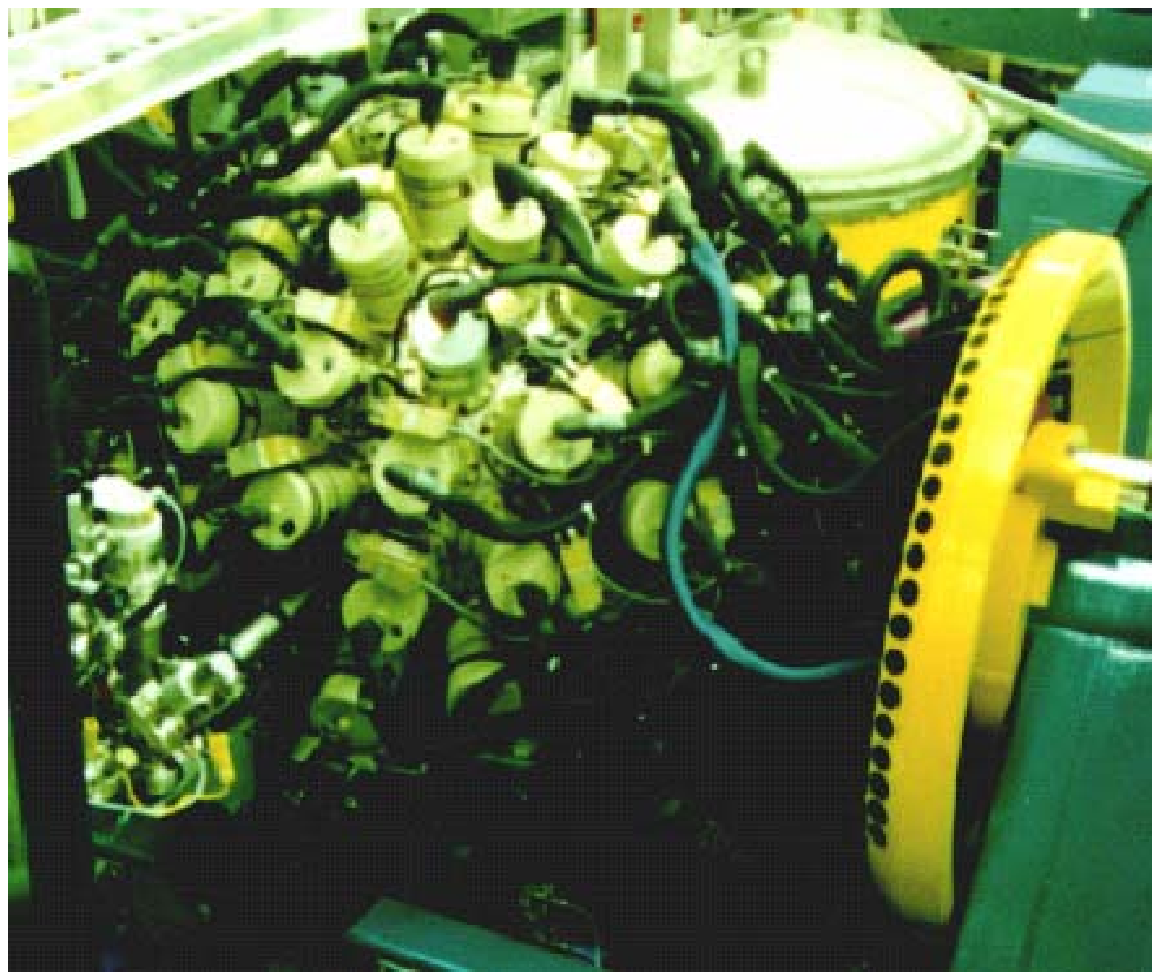}
\caption{Gammasphere detector array at ANL.\label{fig:Gammasphere}}
\end{center}
\end{figure}

The performance of the Gammasphere array depends on four factors~\cite{Baxter-NIMA-317-101-92}: the peak-to-total ratio, the 
energy resolution $\Delta{E_{\gamma}}$, the effective solid angle, and the resolving power $R$. The energy resolution 
is defined in terms of the FWHM of a $\gamma$ ray of average energy, as discussed in Sec.~\ref{subsec:GeDetIntro}, 
and $\Delta{E_{in}}$ is about 2 $keV$ for 1.3 $MeV$ $\gamma$ rays, which is good for high-precision spectroscopy. 
With such large solid angle coverage (almost $4\pi$), Gammasphere is perfectly suited to measure 5 -- 10 coincident $\gamma$ 
rays in a high multiplicity cascade with 20 -- 30 transitions emitted in a nuclear reaction. As described in Sec.~\ref{subsec:CSupresBGO}, 
the P/T ratio in a $\gamma$-ray spectrum is considerably improved by the Compton suppression. For a $\gamma$-ray event with 
fold $n$ ($\it{i.e.}$, with $n$ prompt coincident $\gamma$ rays), the resolving power of Gammasphere $R$ is proportional to the 
quantity ${\left(\frac{P/T}{\Delta{E_{\gamma}}}\right)}^{n}$. The total efficiency of Gammasphere is $10\%$ for 1.3 $MeV$ 
$\gamma$ rays. The characteristics of Gammasphere mentioned above make it an ideal device for studying high-spin phenomena 
in atomic nuclei. A picture of Gammasphere located at ATLAS is shown in Figure~\ref{fig:Gammasphere}. 

\subsection{Gammasphere electronics}
In order to detect two or more coincident $\gamma$ rays using the Gammasphere array, the electronic circuitry, which contains a 
number of amplifiers and discriminators for each Compton suppressed unit, must be employed to determine whether the detected 
$\gamma$-ray events satisfy the preset minimum coincidence requirement, $\it{i.e.}$, the trigger condition. 

Gammasphere (GS) is usually set to have a three-fold trigger for a common high-spin experiment without the use of any external 
detector. The first level of triggering is called the pre-trigger, of which the value is often set to be ``4''. This means 
that, once four or more Ge detectors of GS have fired simultaneously, $\it{i.e.}$, each of at least four Ge detectors detect a $\gamma$ ray 
within a time window of 200 -- 800 nanoseconds ($ns$) (the $\gamma$ rays are required to be in coindence not only with one other, but 
also with the beam burst), the logic will block any further acquisition. Then, the next 1 microsecond ($\mu{s}$) will be used to check 
the master trigger, set to be ``3'' typically, which guarantees that at least three ``clean'' (no veto signal generated) Ge detectors have 
fired in this event; $\it{i.e.}$, at least three Ge energies remain after checking the Compton suppression through the coincidence with the 
respective BGO detectors. Finally, a late trigger will generally spend 6 $\mu{s}$ to inspect the pile-up status with the purpose 
of excluding the possibility that more than one $\gamma$ ray was absorbed by a single Ge detector in the event. If all these 
conditions are fulfilled, the event is a valid one, and it will take another 8 $\mu{s}$ for the GS acquisition system to read out 
all the relevant information. If in any one of the above steps, the minimum coincidence requirement is not met, 
the acquisition is reset within $\sim$ 1 $\mu{s}$ to be available for incoming signals. More details about the GS electronics 
can be found in Ref.~\cite{Nisius-96-thesis}. 

\section{\label{sec:lifemeasretech}Lifetime measurements}
\subsection{Introduction}
The lifetimes of nuclear states, or more fundamentally, the electromagnetic transition matrix elements extracted from them, 
are vitally important to studies of the structure of nuclei. These electromagnetic matrix elements, of which the relation 
with the corresponding $\gamma$-ray probability was described in Eq.~\ref{eq:GamReduTransProb} in 
Sec.~\ref{subsec:Gamma-rayProperty}, provide one of the most important connections between theoretical model wave functions 
and data. It has become routine to ask of a nuclear model its prediction for the lifetimes of the nuclear states of interest 
and to judge the model's success by how well these reproduce the experimental data~\cite{Cerny-74-book}. The feasibility of 
nuclear lifetime measurements has greatly improved, due to technological advances in electronics, computers, 
accelerators, and above all, $\gamma$-ray detectors. The appearance of large scale $\gamma$-ray spectrometers, such as the 
Gammasphere array, exemplifies this spectacular progress. Without such a powerful device, lifetime measurements of short-lived 
states in the triaxial stongly deformed (TSD) bands of $^{163}$Tm (to be discussed 
in Chapter~\ref{chap:163Tm-lifetime}) using the Doppler shift attenuation method (DSAM) technique (Sec.~\ref{subsec:DSAMtech}) 
would hardly be conceivable. 

\begin{figure}
\begin{center}
\includegraphics[angle=270,width=0.90\columnwidth]{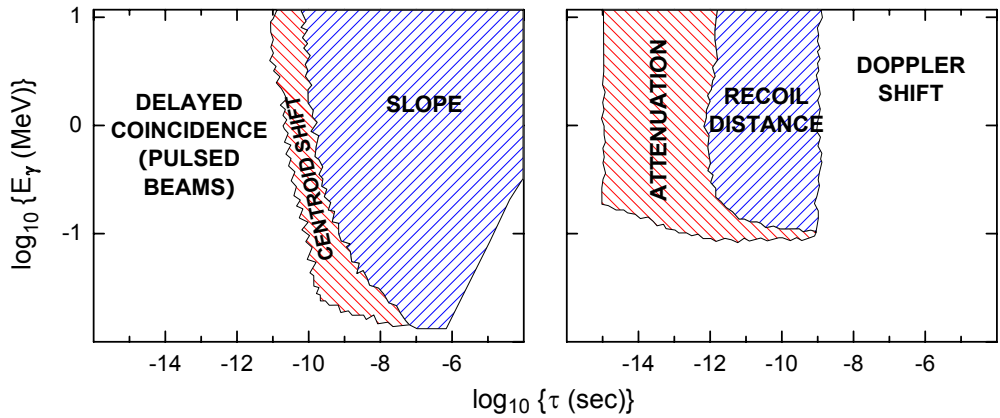}
\caption[Schematics of the applicable regions of lifetime and $\gamma$-ray energy for the three basic direct 
techniques of lifetime measurements.]{Schematics of the applicable regions of 
lifetime ($\tau$) and $\gamma$-ray energy ($E_{\gamma}$) for the three basic direct 
techniques of lifetime measurements. The left figure is for the electronic techniques, and the right one is for the RDM and DSAM 
(see Ref.~\cite{Cerny-74-book} for the detailed explanation of labels used in the figure). The regions are only crudely delineated 
and the boundaries indicated should not be taken as absolutely excluding nearby regions, but rather as an historical indication 
of the past use of these methods. Taken from Ref.~\cite{Cerny-74-book}.\label{fig:LifeTechsApplyRanges}}
\end{center}
\end{figure}

It is worth to mention the three most basic and widely applicable direct experimental techniques: the electronic technique, the recoil 
distance method (RDM) and the Doppler shift attenuation method. The time and $\gamma$-ray energy domains in which these techniques 
can be applied are illustrated in Figure~\ref{fig:LifeTechsApplyRanges}. The electronic technique, RDM and DSAM are used 
typically in the indicated time regions with time accuracies of 1, 10, and $15\%$, respectively, but with consirable variation 
depending on the detailed experimental conditions. Although there are lifetimes ($\tau$) for which more than one 
of these techniques can be used, each technique has its favorable timing region, and these regions are approximately~\cite{Cerny-74-book}: 
\begin{eqnarray}
  {\tau}~ & > & ~{10^{-10}}~sec~~~~~~electronic~technique;\nonumber \\
5\times{10^{-12}}~sec~<~& {\tau} &~<~{10^{-10}}~sec~~~~~~RDM;\nonumber \\
  {\tau}~ & < & ~5\times{10^{-12}}~sec~~~~~~DSAM.\label{eq:LifeTechsTimRanges}
\end{eqnarray}

In addition to these three basic direct techniques, there are several indirect methods and other special direct techniques which 
are often competitive with the three basic direct ones for the determination of lifetimes. The indirect techniques include: 
resonance fluorescence, capture cross-section measurements, Coulomb excitation, inelastic electron scattering, inelastic 
particle scattering, $\it{etc.}$ Other direct ones include: microwave and channeling techniques, $\it{etc.}$ The interested reader 
can refer to Ref.~\cite{Cerny-74-book} for further details. 

According to the selection rule defined in Eq.~\ref{eq:LifeTechsTimRanges}, and the theoretical prediction of lifetimes, we selected 
the DSAM technique to measure the lifetimes of TSD bands in $^{163}$Tm, 
which is one of the foci of this thesis work and will be discussed in detail in Chapter~\ref{chap:163Tm-lifetime}. 

\subsection{\label{subsec:DSAMtech}DSAM technique}
\begin{figure}
\begin{center}
\includegraphics[angle=270,width=0.80\columnwidth]{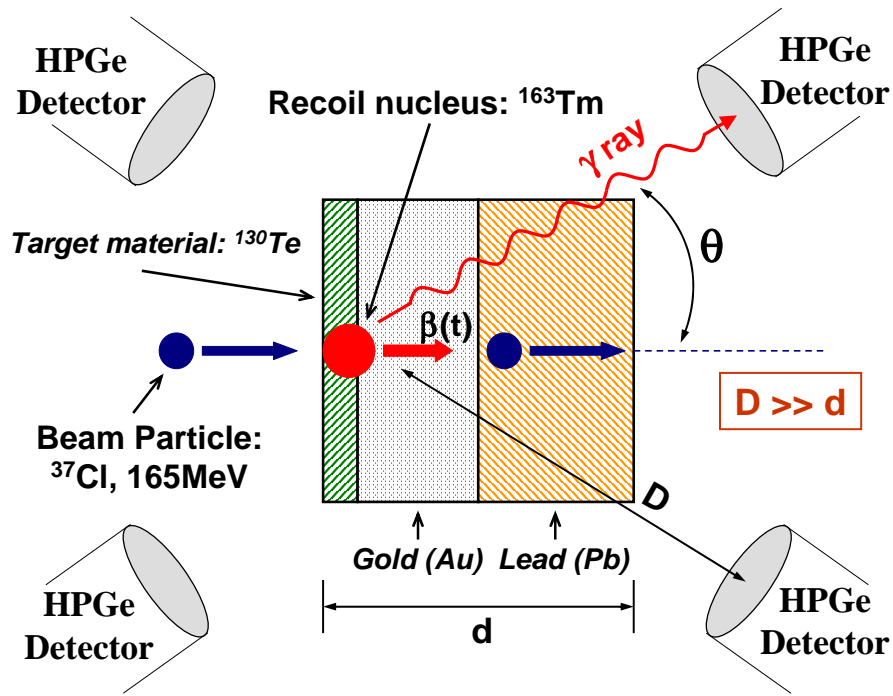}
\caption[Schematic illustration of the DSAM principle.]{Schematic 
illustration of the DSAM principle (see text for the interpretation). $D$ $\sim$ 20 $cm$, is the 
distance between target and detector, while $d$ $\sim$ $10^{-6}$ $m$ -- $10^{-3}$ $m$, is the thickness of target.\label{fig:DSAMtarget}}
\end{center}
\end{figure}

In accordance with the boundaries described in Figure~\ref{fig:LifeTechsApplyRanges}, to determine the lifetimes of nuclear 
states between $10^{-12}$ $s$ and $10^{-15}$ $s$, a Doppler Shift Attenuation Method (DSAM) measurement is performed. 
Typically, a target used in a DSAM experiment has the configuration of multiple layers, for example, as shown in Figure~\ref{fig:DSAMtarget}, 
the target for the $^{163}$Tm lifetime measurements consisted of a 0.813 mg/cm$^2$-thick layer of the actual target material, $^{130}$Te, 
evaporated on a 15 mg/cm$^2$ thick Au foil backed by a 15 mg/cm$^2$ layer of Pb. The thickness of the target was chosen to 
fully stop the recoiling evaporation residues in the Au layer, while the projectiles came to rest in the additional Pb foil. 

In Figure~\ref{fig:DSAMtarget}, the principle of the DSAM technique is schematically illustrated for the 
$^{163}$Tm lifetime measurements. A $^{37}$Cl beam particle, with an energy of 165 $MeV$, reacts with a $^{130}$Te 
target nucleus via a fusion-evaporation reaction, and, hence, a recoiling $^{163}$Tm residue with an initial 
velocity $\beta_0={v_0}/c$ is formed after four neutrons are evaporated from the compound nucleus. The nucleus 
$^{163}$Tm will then travel at a velocity $\beta(t)=v(t)/c$, which decreases with time as the nucleus slows down 
in the $^{130}$Te target layer, and the Au backing, until it is completely stopped. During the slowing down process 
in the thick target, a $\gamma$ ray of nominal energy $E_{\gamma{0}}$ emitted from the recoiling nucleus will be 
measured by a HPGe detector at the angle $\theta$, because of the Doppler shift, with the actual energy:
\begin{equation}
E_{\gamma}=E_{\gamma{0}}\frac{(1-{\beta}^2)^{1/2}}{1-\beta\cos{\theta}}.\label{eq:DopplerEffectGamRay}
\end{equation}
For small values of $\beta$, $\it{e.g.}$, $\beta$ $\sim$ $2\%$ in the $^{163}$Tm case, Eq.~\ref{eq:DopplerEffectGamRay} 
is commonly substituted by its first order approximation:
\begin{equation}
E_{\gamma}=E_{\gamma{0}}(1+\beta\cos{\theta}).\label{eq:DopplerEffectGam1stOdApprx}
\end{equation}
The ability of a given material to decelerate the recoiling nuclei is parameterized in terms of the stopping 
power. The appropriate calculation of the stopping power of the target and backing material leads to the 
determination of the velocity of the nucleus as a function of time by using the relation:
\begin{equation}
\frac{d\,E}{d\,x}=-M\frac{d\,v}{d\,t},\label{eq:StopPowerVelocity}
\end{equation}
where $\frac{d\,E}{d\,x}$ is the stopping power of the material that the recoil is traveling in, $M$ is the mass 
of the recoiling nucleus and $\frac{d\,v}{d\,t}$ is the change rate of the velocity $v$. The stopping power of 
material depends mainly on two processes: Coulomb collisions with the electrons of the atom of the target 
or backing material and nucleus-nucleus collisions with the target or backing material. A 
detailed discussion about stopping powers can be found in Ref.~\cite{Nisius-96-thesis}, and the value of the 
stopping power for a certain material can be obtained using computer codes and tabulations such as SRIM 2003~\cite{Ziegler-85-book}. 

Experimentally, there are three different scenarios to consider. First, for states with very short lifetimes and 
feeding times (the lifetimes of feeder states), as is the case for the TSD bands in $^{163}$Tm, 
the velocity distribution of the nucleus is narrow and all $\gamma$ rays associated with these levels 
are emitted before the complete stopping of the nucleus. Hence, several Doppler-shifted, well-defined 
peaks will be visible in a spectrum obtained for a specific angle. Second, for lifetimes and feeding times 
approaching the stopping time of the nucleus in the backing, the associated peaks in the spectrum present 
distinct lineshapes with the $\gamma$-ray intensity distributed between energies corresponding to the stopped 
and full velocity components. Finally, for states with long 
lifetimes or long feeding times, all associated $\gamma$ rays have no Doppler shift as the emission of these 
transitions occurs after the nucleus has come to rest in the backing material. For the first case, $\it{i.e.}$, 
the fully Doppler-shifted transitions, it is 
straightforward that the average energy of such a transition, $\overline{E_\gamma}$, $\it{i.e.}$, the centroid of the 
corresponding $\gamma$ peak, can be determined experimentally from the spectrum obtained at a specific angle $\theta$, 
and the measured value of the average velocity $\beta_{ave}$ will then be extracted by using 
Eq.~\ref{eq:DopplerEffectGam1stOdApprx}. The important fraction of full Doppler shift, $F(\tau)$, is defined 
in terms of $\beta_{ave}$ as:
\begin{equation}
F(\tau)=\beta_{ave}/{\beta}_{0}.\label{eq:FtauDef}
\end{equation}

On the other hand, Eq.~\ref{eq:StopPowerVelocity} has established a ``clock'' to measure time from the instant 
of fusion until the final stopping of the recoiling nucleus in the backing material. The velocity $v(t)$ as a function 
of time can be calculated using SRIM 2003~\cite{Ziegler-85-book} containing the necessary information on stopping powers. 
Thus, the experimental $F(\tau)$ values for the states of interest and the calculated velocity $v(t)$ for the recoiling 
nucleus are then used to calculate the states lifetimes from the expression:
\begin{equation}
F(\tau)=\int^\infty_0 \frac{N_{k}(t)}{N_{0}}\;\frac{v(t)}{v_{0}}\;d\,t,\label{eq:FtauVeloctyRel}
\end{equation}
where $N_k(t)$ is the population of state $k$ at time $t$, which is a function of all the lifetimes of all 
its feeding states and $N_0$ is the initial number of nuclei at state $k$ or a state that will decay to state $k$ 
at time $t=0$. The analytical solution to the population of state $k$ is given by the Bateman equation:
\begin{equation}
N_k(t)=N_{0}\prod^{k-1}_{i=1}{\lambda_i}\times\sum^{k}_{j=1}\frac{e^{-{\lambda_j}t}}{\prod\nolimits^{k}_{l{\neg}j}
({\lambda_l}-{\lambda_j})},\label{eq:StateKpopult}
\end{equation}
where ${\lambda_i}=1/{\tau_i}$, and $\tau_i$ is the mean lifetime of state $i$. For more precise results, the 
branching ratios, internal convertion coefficients, and side feeding cascades should be considered also, but these 
effects are small compared to the uncertainty $\sim$ $15\%$ associated with the limited knowledge of stopping powers. The 
simultaneous solution of all lifetimes in the band is accomplished with a $\chi^{2}$ minimization procedure 
in a program that attempts to fit the experimental $F(\tau)$ values with the expression of Eq.~\ref{eq:FtauVeloctyRel}. 
Error bars for the lifetimes can be determined by allowing the minimum of the total Chi-square $\chi^{2}_{min}$ of the fit to 
change by $\pm1$~\cite{Bevington-92-book}. Similarly, if the $F(\tau)$ values of states in a rotational band are measured, 
the transition quadrupole moment $Q_t$ and its error can be derived via running a set of fitting codes, which will be 
discussed in detail in Chapter~\ref{chap:163Tm-lifetime}. 

\section{\label{sec:DataAnytech}Data analysis techniques}
Once all the experimental data have been recorded on a storage medium, $\it{e.g.}$, a magnetic tape 
or a hard drive, the off-line analysis process begins. For a typical Gammasphere experiment, for example 
the six-day run of the $^{163}$Tm lifetime measurement, Gammasphere would collect more than 1.5 $\times$ 10$^9$ 
coincidence events with fold $\ge$ 3 ($\it{i.e.}$, with at least three prompt coincident $\gamma$ rays), and, hence, 
the total size of recorded data is huge, typically larger than 100 Giga Bytes ($GB$). On the other hand, the raw 
data commonly contains a large number of contaminant $\gamma$ rays from several kinds of background, which have 
to be subtracted in order to obtain the useful information from the data (the background subtraction will be 
discussed in detail in Sec.~\ref{subsec:GenSpecBKsub}). Moreover, even in the records of the ``clean'' $\gamma$ 
rays emitted from the nuclei of interest, some of the information that is part of standard GS data stream, for example, 
the energy and time information from the individual BGOs, $\it{etc.}$, is not necessary for our purpose and can be eliminated 
in a presort of the data. 

\subsection{\label{subsec:calibration}Calibration}
Before dealing with the actual data of experiments, the calibration data should be analyzed. These were obtained by placing 
$\gamma$-ray sources (nuclei with $\gamma$ rays of well-known energy, intensity, $\it{etc.}$), at the target position of 
Gammasphere. The purpose of these calibrations is to determine the relation between the channel number and the actual 
energy value, and to establish the detection efficiency of the GS detectors. The $^{56}$Co, $^{152}$Eu, $^{182}$Ta, and 
$^{243}$Am sources were chosen in the present work to cover the energy range from about 50 $keV$ to 
around 1.4 $MeV$. 

For any single detector, the recorded energy channel number, $x$, can be related to the actual energy of a $\gamma$ ray, 
$E_{\gamma}$, by using the polynomial form:
\begin{equation}
E_{\gamma}={a_0}+{a_1}x+{a_2}{x^2}+....\label{eq:EnerCalibration}
\end{equation}
For most cases, the first- or the second-order approximation is sufficient for GS data. In this thesis work, the 
second-order relation was employed and the coefficints $a_0$, $a_1$ and $a_2$ for each detector were determined by 
using the ``ENCAL'' codes from the Radware software package~\cite{Radford-NIMA-361-297-95}. 

\begin{figure}
\begin{center}
\includegraphics[angle=270,width=0.80\columnwidth]{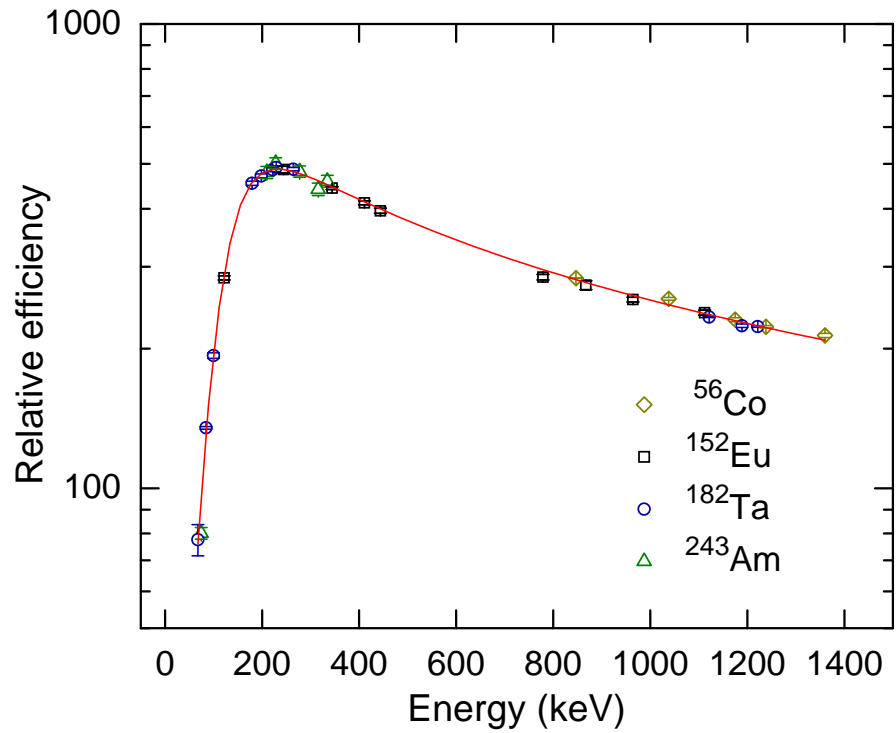}
\caption[Relative efficiency curve for all GS detectors as a function of energy.]{Relative 
efficiency curve for all GS detectors as a function of energy for the experiment on the Pu 
isotopes (Chapter~\ref{chap:PuIsotopes}). The fitting line, obtained by running the the ``EFFIT'' codes, represents the 
best-fit values of the seven parameters described in the text: $A=5.7531214$, $B=2.5643179$, $C=0.0$, $D=5.9341712$, $E=-0.62089050$, 
$F=-0.048151560$, and $G=9.3074818$.\label{fig:GSFMA35_eff_fit_curve}}
\end{center}
\end{figure}

In order to determine the relative intensity of the transitions within a cascade, the spectrum has to be corrected for 
the detection efficiency. This efficiency can be obtained by fitting a function to a set of measured data points (photopeak 
intensities) from a calibration source or to a combination of several sets of normalized data points from different calibration 
sources, as the relative intensity of each calibration transition is well known. The efficiency of GS detectors can be 
described by the following relation:
\begin{equation}
\epsilon=\exp\left[(A+Bx+C{x^2})^{(-G)}+(D+Ey+F{y^2})^{(-G)}\right]^{(-1/G)},\label{eq:EffCalibration}
\end{equation}
where $\epsilon$ is the relative efficiency, $x$ and $y$ are given by:
\begin{equation}
x=\ln({E_{\gamma}}/{E_1})~~~~~~~~y=\ln({E_{\gamma}}/{E_2}),\label{eq:EffCaliCoeff}
\end{equation}
and $E_{\gamma}$ is in $keV$, $E_1=100~keV$, $E_2=1000~keV$. The seven parameters A, B, C, D, E, F and G can be determined 
by fitting the experimental data points with the ``EFFIT'' codes in Radware. 
A, B and C describe the efficiency at low energies (typically $E_{\gamma}$ $\le$ 200 $keV$), while D, E and F have the same 
role for higher energies. In the fitting procedure, the parameter C is in general not required, and is by default fixed 
to be zero by the ``EFFIT'' codes. The last parameter G is an interaction parameter between the two regions. As shown in 
Figure~\ref{fig:GSFMA35_eff_fit_curve}, which is the total relative efficiency of all GS detectors for the experiment on the Pu 
isotopes (to be discussed in Chapter~\ref{chap:PuIsotopes}), the efficiency curve turns over around 
200 $keV$. The parameter G determines the efficiency in the region where the curve turns over: the larger G is, the sharper 
the turnover at the top will be. 

In Figure~\ref{fig:GSFMA35_eff_fit_curve}, it is also worth noting the sharp decrease of the curve in the low-energy region. 
This phenomenon arises from two main factors: (i) the absorption of low-energy $\gamma$ rays in the target and the 
materials between the target and the Ge crystal in the detector; (ii) cut-offs due to the electronics. Low-energy 
$\gamma$ rays are characterized by relatively poor timing in large Ge crystals~\cite{Leo-93-book}. As a result, the signal 
may be too late to be accepted either by the GS acquisition system or in the subsequent analysis. 

Similarly, the relative detection efficiency of a GS detector (or all detetors in a detector ring) for a $\gamma$ ray of specific 
energy can be extracted from the calibration data. These relative efficiencies will be useful later in measuring the 
lifetimes of states in Chapter~\ref{chap:163Tm-lifetime} and deriving the multipolarities of transitions in Chapter~\ref{chap:PuIsotopes}. 

\subsection{Blue database and Radware software package}
For a large array of HPGe detectors such as GS, a dramatic increase in its effective sensitivity is realized 
through the dispersion of less-correlated background $\gamma$ rays over a space of increased coincidence fold. Moreover, 
it has been shown in Ref.~\cite{Radford-92-proceed} that there exists an optimal analysis fold which is sufficiently 
high to dilute background $\gamma$-ray correlations and yet sufficiently low to maintain adequate statistics in the 
strong correlations observed in nuclear de-excitations. For GS, this fold has been found to be 4 in most cases, and, hence, 
the best sensitivity of the array is achieved with coincidence events of fold 4 or higher. For the analysis 
of coincidence data at these folds, the traditional histogram-based techniques, such as the Cube or Hypercube in 
Radware, becomes impractical because of the storage constraints. With the purpose to overcome the difficulty of efficient 
storage of and access to high-fold coincidence data, a new tool, the so-called Blue database, was developed by M. Cromaz, 
$\it{et~al.}$~\cite{Cromaz-NIMA-462-519-01} in 2000. 

\begin{figure}
\begin{center}
\includegraphics[angle=270,width=0.70\columnwidth]{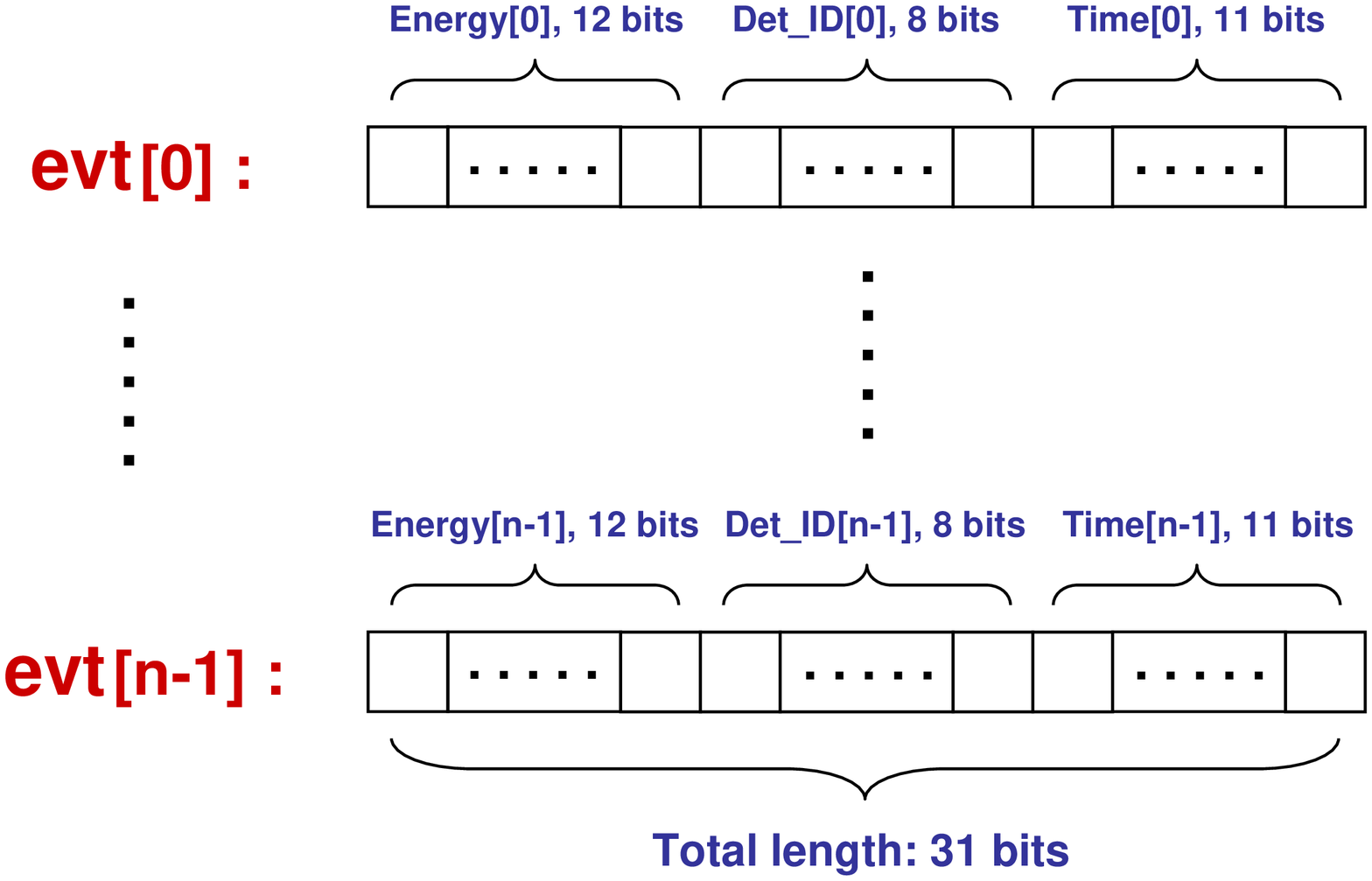}
\caption[A schematic illustration of a $n$-fold $\gamma$-ray event stored in Blue database file.]{A schematic 
illustration of a $n$-fold $\gamma$-ray event $evt$ stored in Blue database file. The total length of 
each element in the array can be as large as 32 bits (it is 31 bits in this example), while the length of individual 
parameter, $\it{i.e.}$, energy, or time or Det-ID, can be changed according to user's requirements.\label{fig:BlueDBIllust}}
\end{center}
\end{figure}

Blue is a library of routines which enables one to create and query a database specifically tailored to the storage of 
$\gamma$-ray coincidence events. A specialized indexing scheme which adapts to the density distribution of each particular 
dataset allows data to be retrieved quickly. Unlike traditional data-storage techniques for high-fold $\gamma$-ray 
coincidence events, the BLUE database stores the data in its original fold without unfolding. With Blue, the raw data 
can be sorted into several database files. Each BLUE database file corresponds to the ensemble of all coincidence events 
of a given fold $n$ ($n=1,2,3,...$), each event in the file for $n$-fold data is stored as an array of $n$ dimension, 
$\it{e.g.}$, the array $evt[n]$, and each element in the array ($evt[0]$, ..., $evt[n-1]$) remains encoded, not only with the 
$\gamma$-ray energy and time information, but also with the auxiliary information, $\it{e.g.}$, the detector identification 
(Det-ID). The data structure of $\gamma$-ray events stored in the Blue database is schematically illustrated in 
Figure~\ref{fig:BlueDBIllust}. Nevertheless, for certain experiments, $\it{e.g.}$, the Pu measurements to be discussed in 
Chapter~\ref{chap:PuIsotopes}, some auxiliary information of events, for example: the sum of energy absorbed by all 
detectors (H: sum energy), the total number of all detectors which fired promptly (K: multiplicity), $\it{etc.}$, need to 
be saved for the subsequent analysis. In such cases, the ($n+1$)-dimension arrays are used for storing the $n$-fold 
events instead of the regular $n$-dimension ones in the Blue database, since, with the new data structure, the 
last element in each array, being free of storing individual $\gamma$ rays, is perfect to contain a number generated 
by encoding the necessary auxiliary information. 

The specific data structure of BLUE is such that producing background-subtracted spectra at a given detector angle under 
specific coincidence requirements can be achieved efficiently with the method described in the next section. 
Such spectra are essential for studying the important Doppler-shifted or Doppler-broadened $\gamma$ transitions 
in Chapters~\ref{chap:163Tm-lifetime} and \ref{chap:PuIsotopes}, as they are analyzed in detail by using the 
``GF3'' codes in Radware. 

Moreover, the Cube or Hypercube files of Radware format (regular ones or ones under specific pregating conditions) can 
also be generated quickly through querying the corresponding Blue database instead of sorting the raw data that has 
much larger size and less organized structure. The Cube or Hypercube obtained in this way can then be viewed 
with the help of the ``LEVIT8R'' or ``4DG8R'' codes in Radware to, for example, search for new transitions or 
band structures. 

\subsection{\label{subsec:GenSpecBKsub}Background subtraction}
The proper identification of background is one of the key issues in the analysis of $\gamma$-ray 
spectroscopic experiments. The background contamination in such experiments is mainly of two types. 
The first one is the so-called ``random'' coincidence events, in which two or more detectors fire promptly with 
respect to one another but the $\gamma$ rays they receive are not correlated. An example of this situation 
would be the $\gamma$ rays emitted from a $^{240}$Pu nucleus of the target and from a $^{197}$Au nucleus from the 
backing material in the ``unsafe'' Coulex experiment described in Chapter~\ref{chap:PuIsotopes}. 
The other contamination arises from Compton scattering, in which the $\gamma$ ray only deposits a portion of 
its full energy in the detector as discussed in Sec.~\ref{subsec:GamRayIntactMatt}. 

\begin{figure}
\begin{center}
\includegraphics[angle=270,width=0.80\columnwidth]{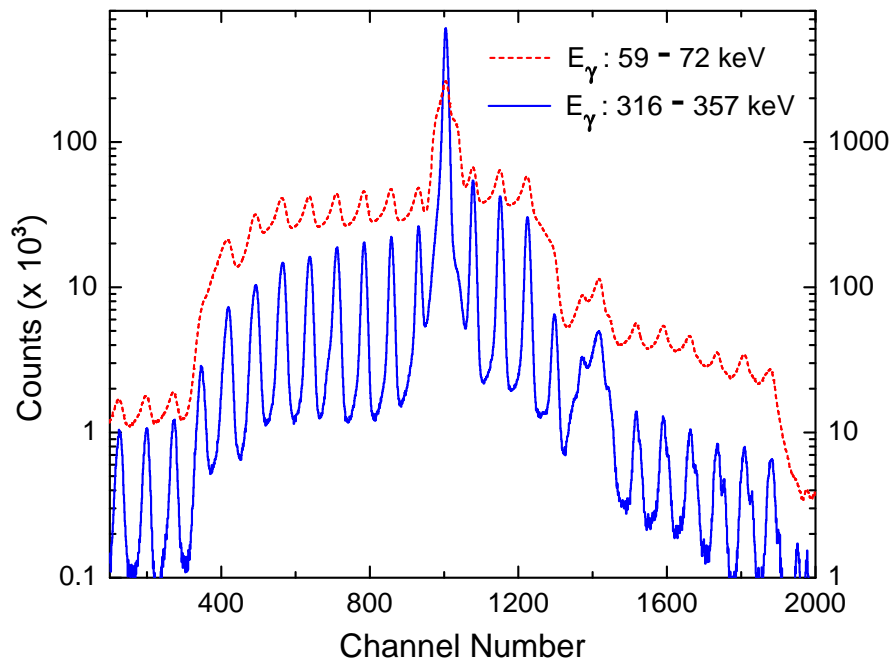}
\caption[Samples of time distribution for $\gamma$ rays.]{The 
distributions of time for $\gamma$ rays in the energy range 316 -- 357 $keV$ 
and $\gamma$ rays in the energy range 59 -- 72 $keV$, respectively, taken from the data on the $^{238,240,242}$Pu 
Coulex experiments. The channel number on the horizontal axis can be translated into real time, the time 
difference between consecutive peaks being 82 $ns$.\label{fig:Pu_T_gate}}
\end{center}
\end{figure}

These ``random'' $\gamma$ peaks in spectra can be removed by using complex gating conditions for sorting 
the data into the spectra (the 1-dimension histograms), or matrices (the 2-dimension histograms), 
or Cubes (the 3-dimension histograms), or Hypercubes (the 4-dimension histograms). Generally, such gating 
conditions include a window on the $\gamma$-ray time relative to the beam clock $RF$, a window on the multiplicity 
in the events, K, and a window on the sum energy in the events, H, in addition to the regular coincidence requirement 
with known transitions. The time window can be determined by investigating the distribution of $\gamma$-ray 
time as a function of $\gamma$-ray energy. As can be seen in Figure~\ref{fig:Pu_T_gate}, the curves represent 
sample data from $^{238,240,242}$Pu unsafe Coulex measurements. For $\gamma$ rays in the energy range 316 -- 357 $keV$, 
the time window can easily be set within a narrow range around the middle, strongest peak (the centroid is at about 
channel 1024), $\it{e.g.}$, channels 885 -- 1054. In contrast, for $\gamma$ rays in the energy range 59 -- 72 $keV$, a 
broad time window, $\it{e.g.}$, channels 500 -- 1054, should be chosen as the contrast of intensity between the middle 
peak and its neighboring peaks is low. This observation is due to the relatively poor timing response of Ge 
detectors for low-energy ($E_{\gamma}~<~200~keV$) $\gamma$ rays. Additionally, for the energy range 
$E_{\gamma}~>~200~keV$, the $\gamma$-ray spectra can be even cleaner. The possibility that those random contaminant 
$\gamma$ rays are distributed in any single beam burst is even. Hence, if the qualifying $\gamma$ rays with time 
positioned within the time window of the next beam burst (relative to the middle one), $\it{e.g.}$, channels 805 -- 884 in 
Figure~\ref{fig:Pu_T_gate}, under the exactly same gating conditions are subtracted with proper normalization 
from the gated spectrum containing the $\gamma$ rays with time positioned within the time window of the middle peak, $\it{e.g.}$, 
channels 885 -- 1054 in Figure~\ref{fig:Pu_T_gate}, in principle, the resulting spectra will be free of the random 
contaminations after the above processing. With the gating condition of time window only (or plus gates on some 
known transitions), the preliminary histograms can be produced and some strong transitions will be observed in the analysis. 

\begin{figure}
\begin{center}
\includegraphics[angle=270,width=0.80\columnwidth]{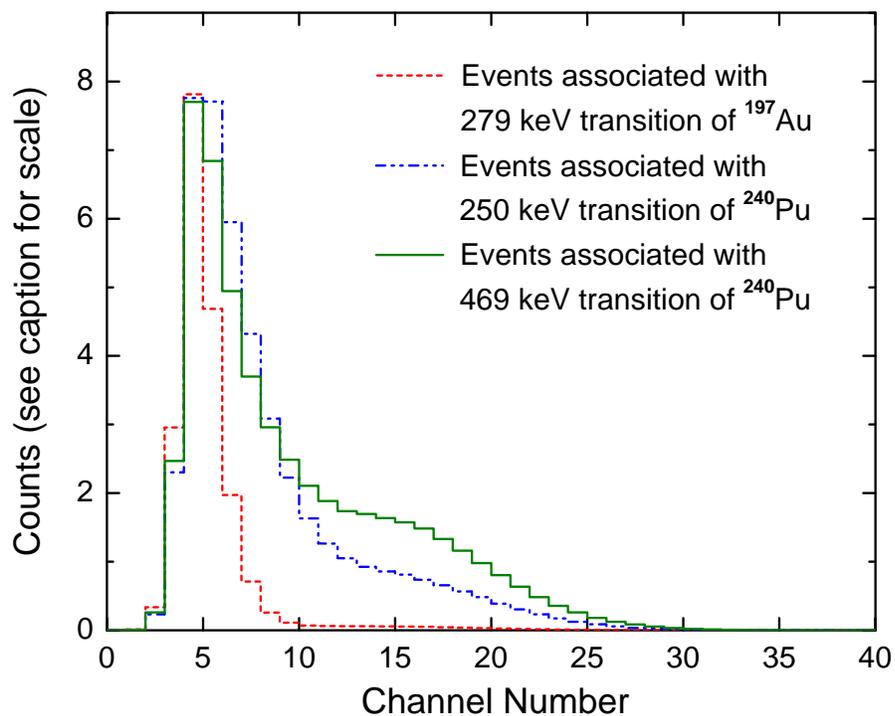}
\caption[Samples of multiplicity distribution for $\gamma$-ray events.]{Multiplicity distributions for 
events associated with the contaminant 279 $keV$ transition in 
$^{197}$Au (scale of counts: $1\times{10^6}$), with the 250 $keV$ $\gamma$ ray ($10^{+}~\rightarrow~8^{+}$ in the yrast band) 
of $^{240}$Pu (scale of counts: $4\times{10^4}$), and with the 469 $keV$ line ($22^{+}~\rightarrow~20^{+}$ in the 
yrast band) of $^{240}$Pu (scale of counts: $1.5\times{10^4}$). The channel number on the horizontal axis 
represents the total number of detectors firing in prompt coincidence.\label{fig:Pu_K_gate}}
\end{center}
\end{figure}

\begin{figure}
\begin{center}
\includegraphics[angle=270,width=0.80\columnwidth]{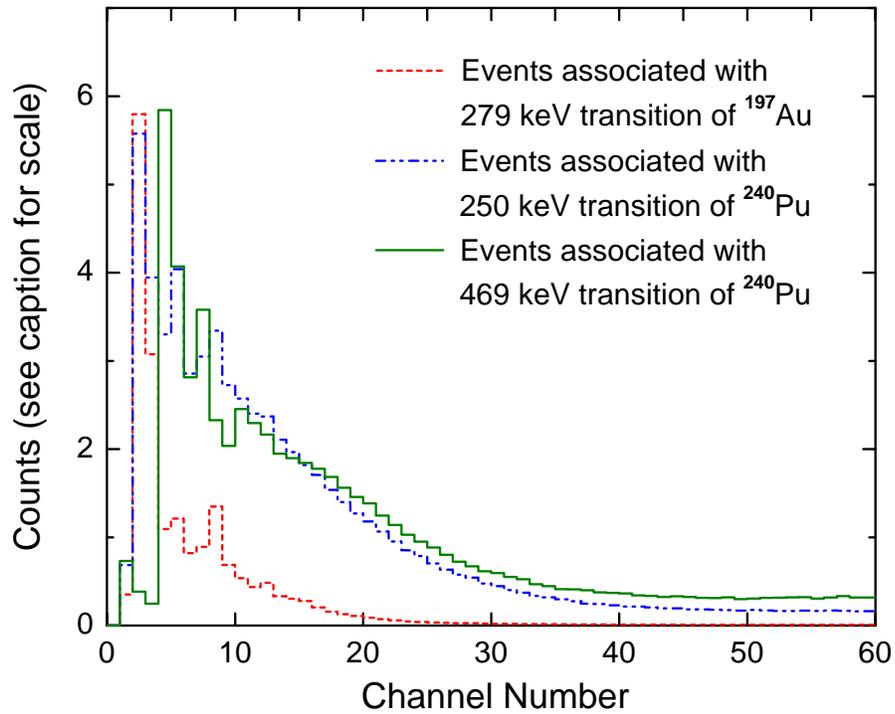}
\caption[Samples of sum-energy distribution for $\gamma$-ray events.]{Sum-energy 
distributions for events associated with the contaminant 279 $keV$ line in 
$^{197}$Au (scale of counts: $1\times{10^6}$), with the 250 $keV$ transition ($10^{+}~\rightarrow~8^{+}$ in the yrast band) 
in $^{240}$Pu (scale of counts: $2\times{10^4}$), and with the 469 $keV$ $\gamma$ ray ($22^{+}~\rightarrow~20^{+}$ in the 
yrast band) in $^{240}$Pu (scale of counts: $7\times{10^3}$). The channel number on the horizontal axis can be 
translated into energy units.\label{fig:Pu_H_gate}}
\end{center}
\end{figure}

The process to determine the appropriate K and H coincidence windows is explained with the example of the 
$^{240}$Pu Coulex data. As illustrated in Figs.~\ref{fig:Pu_K_gate} and \ref{fig:Pu_H_gate}, the distributions in 
multiplicty K and sum energy H for the events associated with the contaminant 279 $keV$ transition of $^{197}$Au, 
with the 250 $keV$ $\gamma$ ray ($10^{+}~\rightarrow~8^{+}$ in the yrast band) of $^{240}$Pu, and with the 469 $keV$ 
line ($22^{+}~\rightarrow~20^{+}$ in the yrast band) of $^{240}$Pu, were obtained from the same data. 
Based on the comparisons from Figure~\ref{fig:Pu_K_gate}, the following conclusions can be drawn: 
(i) the absolute intensity of the transition from the Coulex of $^{197}$Au is much larger than the one of any transition 
from the Coulex of the $^{240}$Pu target (considering that the scales of counts associated with the 279 $keV$, 250 $keV$ 
and 469 $keV$ lines are $1\times{10^6}$, $4\times{10^4}$ and $1.5\times{10^4}$, respectively); (ii) the 
contaminant $^{197}$Au Coulex transition favors low multiplicity, in a narrow range, $1~\le~K~\le~9$, 
while the events containing the transitions from the nucleus of interest, $^{240}$Pu, have much higher multiplicities, 
in a broad range, $1~\le~K~\le~30$; (iii) a trend can be clearly seen, $\it{i.e.}$, the higher the spin of the transition 
involved, the larger the multiplicity. A similar phenomenon can be observed in the comparison of the sum energies 
in Figure~\ref{fig:Pu_H_gate}, even though the discrimination is less pronounced than in Figure~\ref{fig:Pu_K_gate}. Therefore, 
taking into account also that the probability of fission goes up with multiplicity, 
the optimal K and H windows were set to be 6 -- 20 and 4 -- 35, respectively, in order to eliminate as much as 
possible background while keeping adequate statistics for true coincidence events. 

The second source of background, Compton scattering, results in a continuous spectrum for each $\gamma$ ray 
transition. As described in Sec.~\ref{subsec:CSupresBGO}, 
the efforts to minimize background using active Compton suppression shielding have led to significant improvements 
of the P/T ratios in $\gamma$-ray spectra. However, it is important to note that even with the best performance 
of current shields, only $\sim~60\%$ of $\gamma$ rays from a monochromatic 1 $MeV$ source are detected as photopeaks leaving 
a $\sim~40\%$ probability for a $\gamma$ ray to be detected as background. Therefore, such background is still an 
important factor affecting data analysis, and it has to be subtracted properly in order to acquire more useful information 
from the spectrum. 

If the data is stored in the traditional symmetric Cube or Hypercube file with decomposing each event with higher 
multiplicity into several 3- or 4-fold arrays, $\it{i.e.}$, unfolding, the background will be subtracted 
automatically in the analysis with the codes ``LEVIT8R'' or ``4DG8R'' of Radware which adopts the method described in 
Ref.~\cite{Radford-NIMA-361-306-95}. However, in some cases, $\it{e.g.}$, in a DSAM or angular distribution (to be discussed in 
the next section) analysis, the spectra generated at certain angles with complex gating conditions are needed, and, thus, it is 
convenient to store the data including the auxiliary information in its native fold via the Blue database technique. 
For the data of such unsymmetric format, the algorithm of subtracting background in Ref.~\cite{Radford-NIMA-361-306-95} 
is not valid. Hence, a new method was developed by K. Starosta, $\it{et~al.}$~\cite{Starosta-NIMA-515-771-03}, which has 
proven to be successful in the analysis of a number of DSAM measurements, 
including the $^{163}$Tm lifetime measurement in this thesis work. 
The principle of the new method is introduced briefly below by using an example of double gating on 
events of fold three and higher (for the interested reader, please refer to Ref.~\cite{Starosta-NIMA-515-771-03} 
for the details of this method). Any $\gamma$ ray detected in the spectrum may be either a peak (P) or 
a background (B). Denoting the first and the second gating transitions by ${\gamma}_1$ and ${\gamma}_2$, respectively, 
the events can be grouped into four classes, $\it{i.e.}$, PP, PB, BP and BB, which are defined in Table~\ref{tab:GamEvtsClass}. 

\begin{table}
\begin{center}
\caption{CLASSES OF EVENTS FOR DOUBLE GATING (SEE TEXT FOR DETAILS)\label{tab:GamEvtsClass}}
\begin{tabular}{ccc}
\hline \hline
Class & ${\gamma}_1$ & ${\gamma}_2$\\ \hline
(1) & Peak & Peak\\
(2) & Peak & Background\\
(3) & Background & Peak\\
(4) & Background & Background\\
\hline 
\end{tabular}
\end{center}
\end{table}

The objective is to obtain the spectrum $S_{pp}(j)$ which corresponds to class (1) from the double-gated spectrum 
$S_{tt}(j)$ which corresponds to the sum of all four classes. The task is accomplished following the subtraction of 
the appropriate background spectrum in the expression:
\begin{equation}
S_{pp}(j)=S_{tt}(j)-[N_{pb}S_{pb}(j)+N_{bp}S_{bp}(j)+N_{bb}S_{bb}(j)],\label{eq:DoubGtSpeWtBkSub}
\end{equation}
where $N_{pb}$, $N_{bp}$ and $N_{bb}$ represent the number of events in classes (2), (3) and (4), respectively, 
and $S_{pb}(j)$, $S_{bp}(j)$ and $S_{bb}(j)$ are the normalized spectra extracted for these classes:
\begin{equation}
\sum_{j}\;S_{pb}(j)=\sum_{j}\;S_{bp}(j)=\sum_{j}\;S_{bb}(j)=1.\label{eq:NormSpecClass}
\end{equation}
With some appropriate assumptions, the Eq.~\ref{eq:DoubGtSpeWtBkSub} can then be simplified into: 
\begin{equation}
S_{pp}(j)=S_{tt}(j)-{N}\left(\frac{b_2^{g1}}{N^{g1}}{S^{g1}(j)}+\frac{b_1^{g2}}{N^{g2}}{S^{g2}(j)}-
\frac{{b_1}b_2^{g1}+{b_2}b_1^{g2}}{2}\frac{1}{T}P(j)\right),\label{eq:DoubGtSpeWtBSFl}
\end{equation}
where $b_1^{g2}$ denotes the background to total (B/T) ratio for $\gamma_1$ in the spectrum single gated on $\gamma_2$, 
$\it{i.e.}$, the conditional probability that $\gamma_1$ is detected in background while $\gamma_2$ is detected in either manner, 
similarly, $b_2^{g1}$ denotes the B/T ratio for $\gamma_2$ in the spectrum single gated on $\gamma_1$; $b_1$ and $b_2$ 
denote the B/T ratios for $\gamma_1$ or $\gamma_2$ in the total projection spectrum $P(j)$ ($T=\sum_{j}P(j)$), 
respectively; $S^{g1}(j)$ and $S^{g2}(j)$ represent the spectrum single gated on $\gamma_1$ and the one single 
gated on $\gamma_2$, respectively, and, $N^{g1}=\sum_{j}S^{g1}(j)$, $N^{g2}=\sum_{j}S^{g2}(j)$. However, for $\gamma$ 
rays in the background subtracted spectrum $S_{pp}(j)$ there is a certain probability for peak detection and a 
certain probability for background detection as a result of Compton scattering; therefore, the smooth background 
in true coincidence with gating transitions is not removed by the algorithm defined in Eqs.~\ref{eq:DoubGtSpeWtBkSub} 
and \ref{eq:DoubGtSpeWtBSFl}. The shape of the smooth background $B(j)$ is then defined based on the total 
projection of the data according to the prescription of Ref.~\cite{Radford-NIMA-361-297-95}, and then the average 
probability for a $\gamma$ ray being detected in the background can be expressed as: $b^{T}=\frac{\sum_{j}B(j)}{T}$. 
Hence, the final clean spectrum with background subtraction can be written as:
\begin{eqnarray}
S_{pp}^{F}(j) & = & S_{pp}(j)-{b^T}\frac{\sum_{j}S_{pp}(j)}{N^B}B(j)\nonumber \\
 & = & S_{tt}(j)-{N}(\frac{b_2^{g1}}{N^{g1}}{S^{g1}(j)}+\frac{b_1^{g2}}{N^{g2}}{S^{g2}(j)}-
\frac{{b_1}b_2^{g1}+{b_2}b_1^{g2}}{2}\frac{1}{T}[P(j)-B(j)]\nonumber \\
 & & +[1-b_1^{g2}-b_2^{g1}]\frac{1}{T}B(j)).\label{eq:FinalClnDoubGtSpe}
\end{eqnarray}

\begin{figure}
\begin{center}
\includegraphics[angle=0,width=0.80\columnwidth]{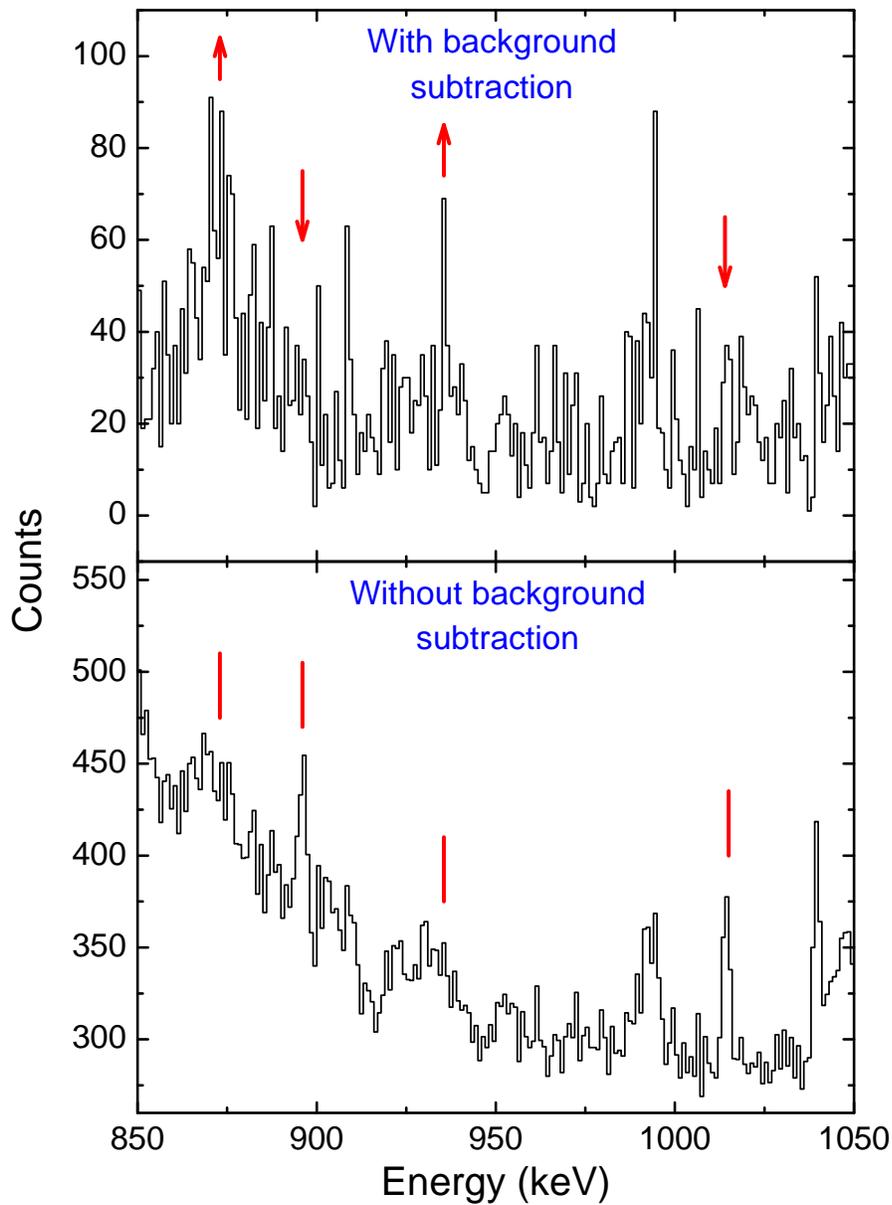}
\caption[Comparison between a background-subtracted spectrum and the spectrum 
without background subtraction.]{Comparison between 
a background-subtracted spectrum and the spectrum without background subtraction double-gated on 
the same transitions of the band TSD1 in $^{163}$Tm (see Sec.~\ref{subsec:163Tm_struct_expe_result} in 
Chapter~\ref{chap:163Tm-lifetime} for the band configuration). The method  presented in 
Ref.~\cite{Starosta-NIMA-515-771-03} was used for background subtraction. The spectra are 
summed over Gammasphere rings 2 to 5. Note the appearance of the true peaks (marked with 
$\uparrow$ symbols) and the disappearance of the background peaks (marked with $\downarrow$ symbols) 
in the background-subtracted spectrum; see text for details.\label{fig:sub_BK_effect}}
\end{center}
\end{figure}

The effect of proper background subtraction with this method is demonstrated by the comparison 
of a background-subtracted spectrum with data without subtraction in 
Figure~\ref{fig:sub_BK_effect}. It can be seen clearly that some 
peaks which can be observed in the spectrum without background 
subtraction disappear in the spectrum with background subtraction, 
while some peaks which are absent or too weak to be determined in the spectrum 
without background subtraction appear in the properly subtracted spectrum. 

\subsection{\label{subsec:LS-AngDist}Level scheme and gamma-ray angular distribution}
By analyzing a large number of spectra gated on transitions of interest, the various coincidence relations of $\gamma$ rays 
emitted in the deexcitation of a nucleus are studied. Hence, with the help of Radware, the level scheme of the nucleus can 
generally be constructed using the techniques described in the literature, such as Refs.~\cite{Zhu-03-thesis} 
and \cite{Radware-webpage}. Nevertheless, a specific method for identifying those high-spin transitions which are very weak 
and perhaps Doppler-shifted or -broadened will be introduced in Chapter~\ref{chap:PuIsotopes}. In a level scheme, transitions 
within a rotational band are assumed to be of $E2$ character, while transitions between bands which carry one or two 
units of angular momentum normally correspond to either $E2$, $E1$, or $M1$ multipolarity. In order to determine 
the multipolarities of $\gamma$ rays, the angular distribution of each $\gamma$ ray involved will be studied. The angular 
distribution is defined in the expression of $\gamma$-ray intensity ($W(\theta)$) as a function of the angle $\theta$ 
with respect to beam direction as:
\begin{equation}
W(\theta)=A_0[1+{A_2}P_2(\cos{\theta})+{A_4}P_4(\cos{\theta})],\label{eq:angu_dist}
\end{equation}
where $P_2(\cos{\theta})$ and $P_4(\cos{\theta})$ are the Legendre polynomials, $A_0$ represents the $\gamma$-ray intensity, 
and the coefficients $A_2$ and $A_4$ depend on the multipolarity ($L$) of the $\gamma$ ray. Typical values of the $A_2$ and 
$A_4$ coefficients are summarized in Table~\ref{tab:angu_dist_coeff_values}. 

With the help of a powerful detector array, such as GS, and of the proper data storage technique, the Blue database, these angular 
distribution coefficients $A_0$, $A_2$, $A_4$ can be extracted from data conveniently for transitions with sufficient statistics. 
Hence, the multipolarities of these $\gamma$ rays can be determined through the comparison of measured $A_2$ and $A_4$ coefficients 
with typical values. In most of cases, the band heads, $\it{i.e.}$, the lowest-spin level in a band, have been observed 
in particle-decay (for example: $\beta$-decay) experiments. Sometimes, the yrast cascade 
was also established previously. Normally, the spins and parities of these known states have been determined. 
Therefore, if the multipolarities of the transitions connecting the states of interest to those known levels (the yrast or low-spin 
states with known spins and parities) are obtained in an angular distribution measurement, 
the spins and parities of associated states will be assigned in accordance with Eqs.~\ref{eq:GamAMRule} 
and \ref{eq:GamPatyRule} in Chapter~\ref{chap:theoriBkgd}. 

\begin{table}
\begin{center}
\caption[TYPICAL VALUES OF ANGULAR DISTRIBUTION COEFFICIENTS $A_2$, $A_4$]{TYPICAL VALUES 
OF ANGULAR DISTRIBUTION COEFFICIENTS $A_2$, $A_4$ FOR $\gamma$ RAYS WITH DIPOLE ($D$), QUADRUPOLE ($Q$), OR 
MIXED DIPOLE-QUADRUPOLE ($D+Q$) MULTIPOLARITY. NOTE THAT $\Delta{I}$ DENOTES THE DIFFERENCE OF SPIN BETWEEN INITIAL AND FINAL STATES, 
WHILE $L$ REPRESENTS THE MULTIPOLARITY. TAKEN FROM REF.~\cite{Singh-ang-dist-URL}\label{tab:angu_dist_coeff_values}}
\begin{tabular}{cccccc}
\hline \hline
$\Delta{I}$ & $L$ & Sign of $A_2$ & Sign of $A_4$ & Typical value of $A_2$ & Typical value of $A_4$\\ \hline
2 & $Q$ & $+$ & $-$ & $+0.3$ & $-0.1$\\
1 & $D$ & $-$ &  & $-0.2$ & $0.0$\\
1 & $Q$ & $-$ & $+$ & $-0.1$ & $+0.2$\\
1 & $D+Q$ & $+/-$ & $+$ & $-0.8$ to $+0.5$ & $0.0$ to $+0.2$\\
0 & $D$ & $+$ &  & $+0.35$ & $0.0$\\
0 & $Q$ & $-$ & $-$ & $-0.25$ & $-0.25$\\
0 & $D+Q$ & $+/-$ & $-$ & $-0.25$ to $+0.35$ & $-0.25$ to $0.0$\\
\hline 
\end{tabular}
\end{center}
\end{table}

%
% Chapter 3
%

%
% Chapter 3
%

\chapter{\label{chap:163Tm-lifetime}TRIAXIAL STRONGLY DEFORMED BANDS IN $^{163}$Tm}

\section{\label{sec:163Tm_introduce}Triaxiality in nuclei}

\subsection{\label{subsec:TriaxialIntro}Introduction}
It is well known that the shape of a nucleus can be either deformed or spherical (see the discusssion 
in Sec.~\ref{subsec:Deformation} of Chapter~\ref{chap:theoriBkgd}). In most deformed nuclei the 
quadrupole deformation is dominant, and, so far, nuclear spectra have been associated mostly with 
axially symmetric deformed shapes, $\it{i.e.}$, with either prolate or oblate deformation. 
However, triaxial deformation (for example, a nuclear shape with parameters: 
${\varepsilon}_2>0$ and $|\gamma|{\sim}20^{\circ}$), 
$\it{i.e.}$, deformation implying the breaking of axial symmetry, has attracted much 
attention over the past decades since it opens a new dimension to the study of collective nuclear rotation 
in the sense that the rotation of axially symmetric nuclei becomes a limit to a more general description. 
Triaxiality relates to a nucleus with a shape characterized by three unequal principal body-fixed axes. 
Its occurence in nuclei has been a longstanding prediction of nuclear structure theory. In such triaxial 
nuclei, the mass distribution and, therefore, the moment of inertia is different along each of the 
three principal axes. 

\begin{figure}
\begin{center}
\includegraphics[angle=270,width=0.75\columnwidth]{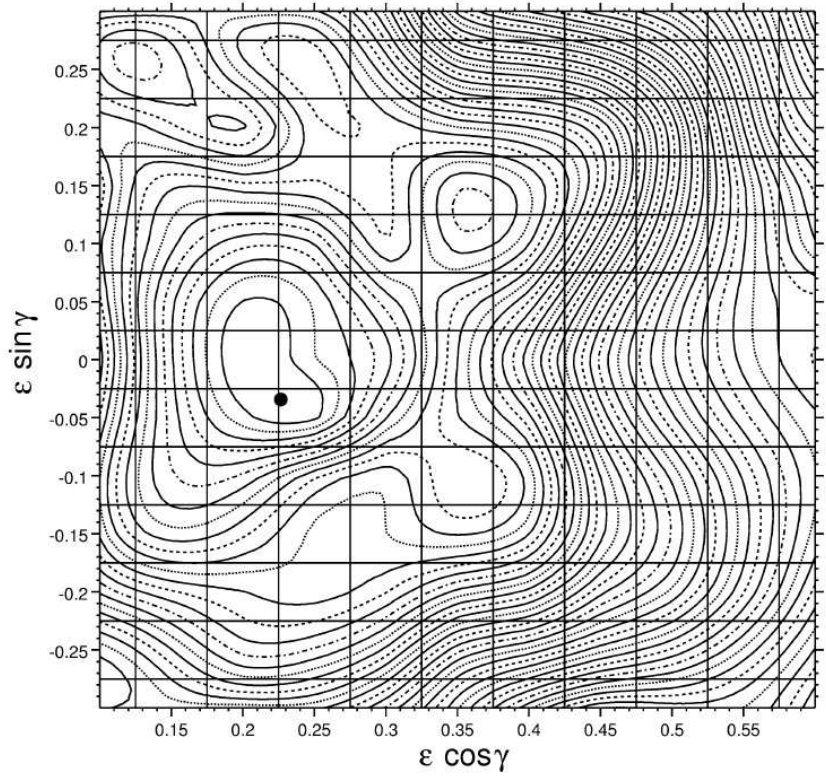}
\caption[Total Routhian Surface (TRS) for the $N$ $\sim$ 94 and $Z$ $\sim$ 71 region.]{Total 
Routhian Surface (TRS) for the $N$ $\sim$ 94 and $Z$ $\sim$ 71 region calculated by means of 
the ``Ultimate Cranker'' (UC) model~\cite{Bengtsson-UC-code-URL} at $I^{\pi}=61/2^{+}$. A 
normal deformed minimum at ${\varepsilon}_2=0.23$ and two triaxial strongly deformed minima 
at $({\varepsilon}_2,\gamma)=(0.38,{\pm}20^{\circ})$ are clearly seen. Adapted from 
Ref.~\cite{Schonwa-PLB-552-9-03}.\label{fig:165Lu-TRS-UC}}
\end{center}
\end{figure}

As discussed in Sec.~\ref{subsec:StrutinskyMethod} of Chapter~\ref{chap:theoriBkgd}, the calculation 
of the total energy of the nuclear system with the Strutinsky-shell correction, $\it{i.e.}$, the Strutinsky 
method, has proved to be successful in predicting where in the nuclear landscape deformation occurs. 
However, in order to account for collective rotation, the Strutinsky method must be transferred to 
the rotating frame of reference. Hence, when calculating the total energy of a rotating nucleus, 
Routhians are often used which represent the energy in the rotating frame. The results 
of such calculations are called Total Routhian Surfaces (TRS)~\cite{Dudek-87-book,Nazarewicz-NPA-467-437-87}. 
On the other hand, in the classical book of nuclear theory by Bohr and Mottelson~\cite{Bohr-98-book}, 
it was suggested that: ``...there is no clear-cut evidence for the occurence of stable 
nuclear deformation without axial symmetry... It is possible, however, that such 
shapes may be encountered for states with high angular momentum...''. 
Therefore, the more recent TRS calculations predict the occurrence of stable triaxial 
deformation at high spin, $\it{i.e.}$, the rotation has a direct impact on global nuclear deformation. 
For example, as can be seen in Figure~\ref{fig:165Lu-TRS-UC}, the TRS calculations for the 
$N$ $\sim$ 94 and $Z$ $\sim$ 71 region by the means of the so-called ``Ultimate Cranker'' (UC) 
model~\cite{Bengtsson-UC-code-URL}, based on a modified harmonic oscilator potential, predict 
two stable triaxial minima $({\varepsilon}_2,\gamma)=(0.38,{\pm}20^{\circ})$ 
at high spin. The calculations have by now been validated by experimental results - 
see Ref.~\cite{Schonwa-PLB-552-9-03}. Triaxiality in this region will be 
discussed further in Sec.~\ref{subsec:TSDwobbling}. 

Experimental signatures for a triaxial shape are difficult to establish, and, as a 
result, conclusive evidence has only appeared in the last few years, although the phenomenon was predicted 
more than 25 years ago. Triaxiality has now been invoked to describe various phenomena, including 
so-called chiral bands and wobbling bands. Both types of collective structures are now widely accepted 
as unique fingerprints for triaxiality. 

\subsection{\label{subsec:ChiralBands}Triaxiality and chiral bands}
\begin{figure}
\begin{center}
\includegraphics[angle=0,width=0.50\columnwidth]{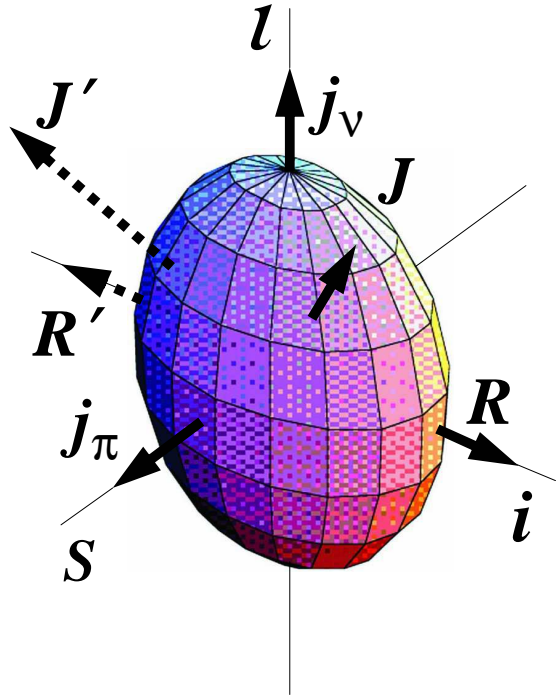}
\caption[Chirality in nuclei.]{Chirality in nuclei. $l$, $s$ and $i$ represent long, short and intermediate axes, respectively. 
$j_{\pi}$ and $j_{\mu}$ are angular momenta originating from the protons and neutron holes near to the Fermi 
surface. $R$ is angular momentum of the collective rotation. The total angular momenta $J$ and $J'$, 
corresponding to two opposite angular momenta, $R$ and $R'$, respectively, introduce the left-handed and 
right-handed configurations. The figure is adapted from Ref.~\cite{Zhu-03-thesis}.\label{fig:ChiralSymmty}}
\end{center}
\end{figure}

One of the best examples for chirality in nature is our hands -- the left hand can not be superimposed upon the 
right hand by means of rotation. More generally, any object with chirality can not be superimposed 
on its mirror image by means of a rotation only. Chirality often appears in molecules and is typical of 
biomolecules. Nuclei had originally been thought to be chiral symmetric, $\it{i.e.}$, not exhibiting chirality, 
because they consist of only two species of nucleons and have relatively simple associated shapes as compared to 
molecules. However, if the angular momentum $\vec{J}$ does not lie on any of the three principal axes in 
a triaxial nucleus, the combination of the three axes with $\vec{J}$ becomes chiral~\cite{Frauendorf-RMP-73-463-01}. 
Figure~\ref{fig:ChiralSymmty} illustrates how chirality emerges from the combination of dynamics 
(the angular momentum) and geometry (the triaxial shape). When nuclei are triaxially deformed 
with $\gamma~\sim~30^{\circ}$, the moments of inertia along the three principal axes will satisfy 
the relation $\Im_{s}\sim\Im_{l}<\Im_{i}$, where $l$, $s$ and $i$ represent the long, short and 
intermediate axes, respectively. Let's consider a nucleus with one (or more) proton in an orbital 
just above the Fermi surface (particle) and one (or more) neutron just below the Fermi surface (hole). 
The protons align their angular momentum $j_{\pi}$ along the axis $s$. This orientation maximizes the 
overlap of their orbitals with the triaxial density, which corresponds to minimal energy, since the 
core-particle interaction is attractive. The orientation of the angular momentum of neutron holes is 
aligned along the axis $l$ and minimizes the overlap of their orbitals with the triaxial density, 
reflecting the fact that the core-hole interaction is repulsive. The collective angular momentum of the core $R$ is 
preferably oriented along the intermediate axis $i$, which has the largest moment of inertia, because 
the density distribution achieves the largest deviation from rotational symmetry with respect to this 
axis. The total angular momentum, $J$, therefore, is out of any of the planes determined by the three principal 
axes. With respect to the frame defined by the three axes, a system with such total angular momentum has 
either a left-handed configuration ($J'$) or a right-handed one ($J$). Reverting the direction of 
the component of the angular momentum on the axis $i$ changes the chirality. 

The left-handed and right-handed configurations are related to each other by 
$|l{\rangle}={\mathcal{T}}{{\mathcal{R}}_y}(\pi)|r{\rangle}$~\cite{Frauendorf-RMP-73-463-01}. 
The operator ${\mathcal{T}}{{\mathcal{R}}_y}(\pi)$ is the time reversal operator combined 
with rotation about the y-axis by $180^{\circ}$. The simplest model case was considered as a high-$j$ particle 
and a high-$j$ hole coupled to a triaxial rotor with ratios ${\Im}_i=4{\Im}_s=4{\Im}_l$ between the 
moments of inertia. It was first studied by Frauendorf and Meng~\cite{Frauendorf-NPA-617-131-97,Frauendorf-ZPA-356-263-97}. 
They showed that the experimental signature of chiral rotation consists of two ${\Delta}I = 1$ 
bands of the same parity with nearly degenerate energy, $\it{i.e.}$, the so-called chiral doublet bands, which are 
connected by $E2$ and $M1$ mixed transitions. The interband $E2$ transitions are strongly 
suppressed, as compared to the inband $E2$ transitions, whereas the interband and inband $M1$ transitions 
are comparable in strength. Of course, there is always the possibility of two planar bands that are accidentally 
degenerate and would mimic chirality. Hence, evidence for the phenomenon should rest on all the 
signatures described above. 

So far, a number of pairs of nearly degenerate ${\Delta}I = 1$ sequences of the same parity have been 
seen in odd-odd nuclei with mass A $\sim$ 100~\cite{Vaman-PRL-92-032501-04,Timar-PRC-73-011301-06,Alcant-PRC-69-024317-04} 
and A $\sim$ 130~\cite{Koike-PRC-67-044319-03,Starosta-PRL-86-971-01} and have been 
interpreted as chiral doublets. However, it is also worth to point out that the argument of chirality in nuclei 
is based only on the symmetry of the rotating triaxial nucleus, and is independent of how the three components 
of the angular momentum are composed~\cite{Frauendorf-RMP-73-463-01}. Therefore, chirality is expected for all 
configurations that have substantial angular momentum components along the three principal axes, no matter 
how the individual components are composed. It is of fundamental importance to find a composite chiral pair 
of rotational bands in a system with more than two quasi-particles in order to clearly establish the general 
geometric character of this phenomenon. Such is the case in $^{135}$Nd~\cite{Zhu-PRL-91-132501-03}, 
a nucleus studied by the Notre Dame group, for which strong experimental evidence is available at the present 
time both in terms of energies of the chiral partner bands and in terms of the inband and interband transitions 
rates. 

\subsection{\label{subsec:TSDwobbling}Trixial strongly deformed band and wobbling}
Unlike the chiral bands, which are associated with a special arrangement of the angular momenta with 
respect to the triaxial shape and the small deformation of the nucleus (${\varepsilon}_2{\sim}0.2$), rotational 
sequences measured in A $\sim$ 160 -- 175 nuclei~\cite{Schmitz-NPA-539-112-92,Schmitz-PLB-303-230-93,%
Schnack-NPA-594-175-95,Bringel-EPJA-16-155-03,Odegard-PRL-86-5866-01,Jensen-NPA-703-3-02,Jensen-PRL-89-142503-02,%
Tormanen-PLB-454-8-99,Schonwa-PLB-552-9-03,Amro-PLB-553-197-03,%
Amro-PLB-506-39-01,Neuber-EPJA-15-439-02,Djong-PLB-560-24-03,Hartley-PLB-608-31-05} 
have been associated with stable triaxial deformation in a strongly deformed well, as predicted by 
theory~\cite{Schnack-NPA-594-175-95,Bengtsson-UC-code-URL} for this mass region. Because these observed sequences 
have relatively large deformation (${\varepsilon}_2{\sim}0.4$), which agrees with various theoretical 
predictions~\cite{Bengtsson-NPA-496-56-89,Bengtsson-NPA-512-124-90,Aberg-NPA-520-35-90,Schnack-NPA-594-175-95}, 
they are often referred to as triaxial strongly deformed (TSD) bands. In the literature, these TSD bands are 
interpreted as based on the deformation driven either by the $i_{13/2}$ proton intruder orbital, $\it{e.g.}$, for 
cases in several even-$N$ Lu isotopes~\cite{Schmitz-NPA-539-112-92,Schmitz-PLB-303-230-93,%
Schnack-NPA-594-175-95,Bringel-EPJA-16-155-03,Odegard-PRL-86-5866-01,Jensen-NPA-703-3-02,Jensen-PRL-89-142503-02,%
Schonwa-PLB-552-9-03,Amro-PLB-553-197-03}, or by the $i_{13/2}$ proton orbital coupled to different neutron orbitals, 
$\it{e.g.}$, for cases in odd-$N$ Lu nuclei~\cite{Bringel-EPJA-16-155-03,Tormanen-PLB-454-8-99}. 
Further, very recently, some of the observed TSD bands in nuclei of the region, $\it{i.e.}$, $^{161,163,165,167}$Lu, have 
been identified as so-called ``wobbling bands''~\cite{Bringel-EPJA-24-167-05,Odegard-PRL-86-5866-01,Jensen-PRL-89-142503-02,%
Schonwa-PLB-552-9-03,Amro-PLB-553-197-03}, the spectroscopic representation of the wobbling mode (excitation) which 
is generally considered as the other unambiguous fingerprint for nuclear triaxiality. 

\begin{figure}
\begin{center}
\includegraphics[angle=270,width=0.65\columnwidth]{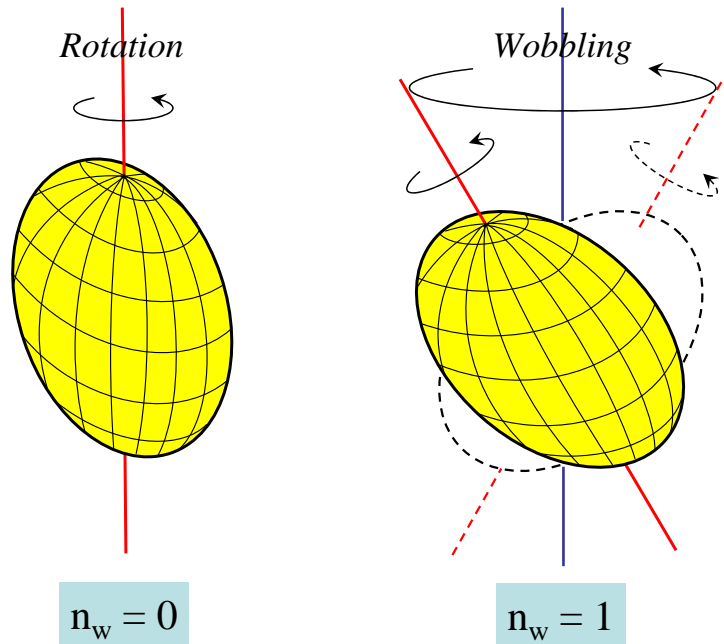}
\caption[A schematical illustration of the wobbling motion of a rotating nucleus with triaxial shape in the classical 
view.]{A schematical illustration of the wobbling motion of a rotating nucleus with triaxial shape in the classical 
view. $n_w$ is the wobbling quantum number. The left ($n_w=0$) and right ($n_w=1$) panels correspond to the yrast 
and the first excited wobbling bands, respectively.\label{fig:wobbling-illust}}
\end{center}
\end{figure}

It was first suggested by Bohr and Mottelson that a natural and unique consequence of a rotating triaxial 
nucleus is the occurrence of the wobbling excitation. In triaxial nuclei, where different moments of inertia 
(${\Im}_{x}>{\Im}_{y}~{\neq}~{\Im}_z$) are associated with the three principal axes, rotation about the three 
axes is quantum mechanically possible. The wobbling mode, analogous to the classical motion of an 
asymmetric spinning top in which perturbations are superimposed on the main rotation 
around the principal axes (as shown in Figure~\ref{fig:wobbling-illust}), is indicative of the 
three-dimensional nature of collective nuclear rotation~\cite{Bohr-98-book}. In the quantum 
picture, the low-spin spectrum of such a system corresponds to that of the well-known Davydov asymmetric rotor. 
However, the low spin data do not allow a clear distinction between a rigid rotor and a system that is soft 
with respect to triaxial deformation. In the high-spin limit ($I{\gg}1$), although rotation about the axis 
with the largest moment of inertia ($\Im_{x}$) is favored, the contributions from rotations about the other two 
axes can force the rotation angular momentum vector (R) off the principal axis to create a precession or wobbling 
mode. As a result, a triaxialy-deformed nucleus will exhibit a family of rotational cascades, called wobbling bands, 
each of which is associated with a wobbling phonon number ($n_w = 0,1,2,...$)~\cite{Hagemann-NPN-13-20-03}. So far, 
wobbling bands with $n_w$ up to 2 have been observed in Lu nuclei~\cite{Jensen-PRL-89-142503-02}. 
The energy of levels in a wobbling band can be given by:
\begin{equation}
E_R(I,{n_{\omega}})=\frac{I(I+1)}{2{\Im}_x}+{\hbar}{\omega}_{w}(n_w+\frac{1}{2}),\label{eq:WoblngLevlEner}
\end{equation}
where $\hbar{\omega}_w=\hbar{\omega}_{rot}\sqrt{({\Im}_x-{\Im}_y)({\Im}_x-{\Im}_z)/({\Im}_y{\Im}_z)}$ with 
$\hbar{\omega}_{rot}=I/{{\Im}_x}$~\cite{Bohr-98-book}. In the intrinsic system $(x,y,z)$, the quadrupole moments 
can be written as~\cite{Bohr-98-book}:
\begin{eqnarray}
Q_0{\equiv}{\langle}\sum_k(2z^2-x^2-y^2)_k{\rangle};\label{eq:WobbleQ0} \\
Q_2{\equiv}{\langle}\sqrt{3/2}\sum_k(x^2-y^2)_k{\rangle},\label{eq:WobbleQ2}
\end{eqnarray}
and $\tan{\gamma}=\sqrt{2}(Q_2/Q_0)$. The strength of interband transitions, 
\begin{equation}
B(E2;n_w,I{\rightarrow}n_w-1,I-1)=\frac{5}{16\pi}{e^2}\frac{n_w}{I}(\sqrt{3}{Q_0}x-\sqrt{2}{Q_2}y)^2,\label{eq:WobbleIntrBdBE2}
\end{equation}
is smaller than the one of inband transitions, 
\begin{equation}
B(E2;n_w,I{\rightarrow}n_w,I-2)\;{\approx}\;\frac{5}{16\pi}{e^2}{Q_2^2},\label{eq:WobbleInBdBE2}
\end{equation}
by a factor of $n_w/I$, which represents the square of the amplitude of the precessional (wobbling) motion~\cite{Bohr-98-book}.

The characteristics of a wobbling band family have been summarized~\cite{Odegard-PRL-86-5866-01,Jensen-NPA-703-3-02} as: 
(1) the bands are associated with very similar intrinsic structure, $\it{i.e.}$, alignment, moment of inertia, quadrupole moment, $\it{etc.}$ 
Each excited band with the quantum number $n_w=1,2,...$ can be seen as a wobbling phonon excitation built on the yrast 
$n_w=0$ band; (2) the $\Delta{I}={\pm}1$ interband transitions, $n_w{\rightarrow}n_w-1$ with $n_w=1,2,...$, possess large values 
of $B(E2)_{out}$ (in contrast, small values of $B(M1)$), in competition with $B(E2)_{in}$ of the inband $\Delta{I}=2$ transitions. 
The quantal phonon rule for transition probability implies that $B(E2;n_w=2{\rightarrow}n_w=1)=2B(E2;n_w=1{\rightarrow}n_w=0)$ 
and the transitions with ${\Delta}n_w=2$ are forbidden; (3) the bands are all associated with large $\varepsilon_2$ values and, 
hence, large quadrupole moments, $Q_t$, because of being built on a TSD minimum. 

The following sections in this chapter will focus on the investigation of the TSD bands in the $^{163}$Tm nucleus and the 
possibility of finding wobbling in this nucleus. 

\section{\label{sec:163Tm-struct}TSD band structures in $^{163}$Tm}

\subsection{\label{subsec:163Tm_struct_motivate}Motivation}
Until very recently, the fact that wobbling bands had only been observed in several ($A=161,163,165,167$) Lu 
isotopes, and not in any other element remained somewhat of a puzzle. Indeed, a number of TSD bands have been reported in 
many of the neighboring Ta and Hf nuclei of the region~\cite{Amro-PLB-506-39-01,Neuber-EPJA-15-439-02,Djong-PLB-560-24-03,%
Hartley-PLB-608-31-05,Fetea-NPA-690-239-01,Roux-PRC-63-024303-01}, (up to 8 TSD bands in case of $^{174}$Hf!), many 
of which may be grouped into possible families based on similarities of their dynamic moments of inertia, but none 
of them was found to exhibit deexcitation properties characteristic of the wobbling mode. In particular, the 
important $\Delta{I}=1$, $E2$ interband transitions that provide a clear signature for wobbling in the Lu isotopes are absent. 

A possible resolution of this issue has been proposed in our recent work of 
exploring the TSD band structures in the $^{163}$Tm nucleus, which was recently published~\cite{Pattabi-PLB-163Tm-07}. 
In this work, two strongly interacting TSD bands were identified. However, the linking transitions between TSD 
bands did not exhibit properties similar to the ones characteristic of wobbling. Rather, they seemed akin 
to what would be expected for collective structures associated with particle-hole (p-h) excitations 
in a TSD well. Still, this $^{163}$Tm case represents the first time that two TSD bands with 
interconnecting transitions have been observed in any element other than Lu. The results of the work of 
Ref.~\cite{Pattabi-PLB-163Tm-07} are discussed in some detail in the next two subsections. 

\subsection{\label{subsec:163Tm_struct_expe_result}Experiment and data}
\begin{figure}
\begin{center}
\includegraphics[angle=270,width=\columnwidth]{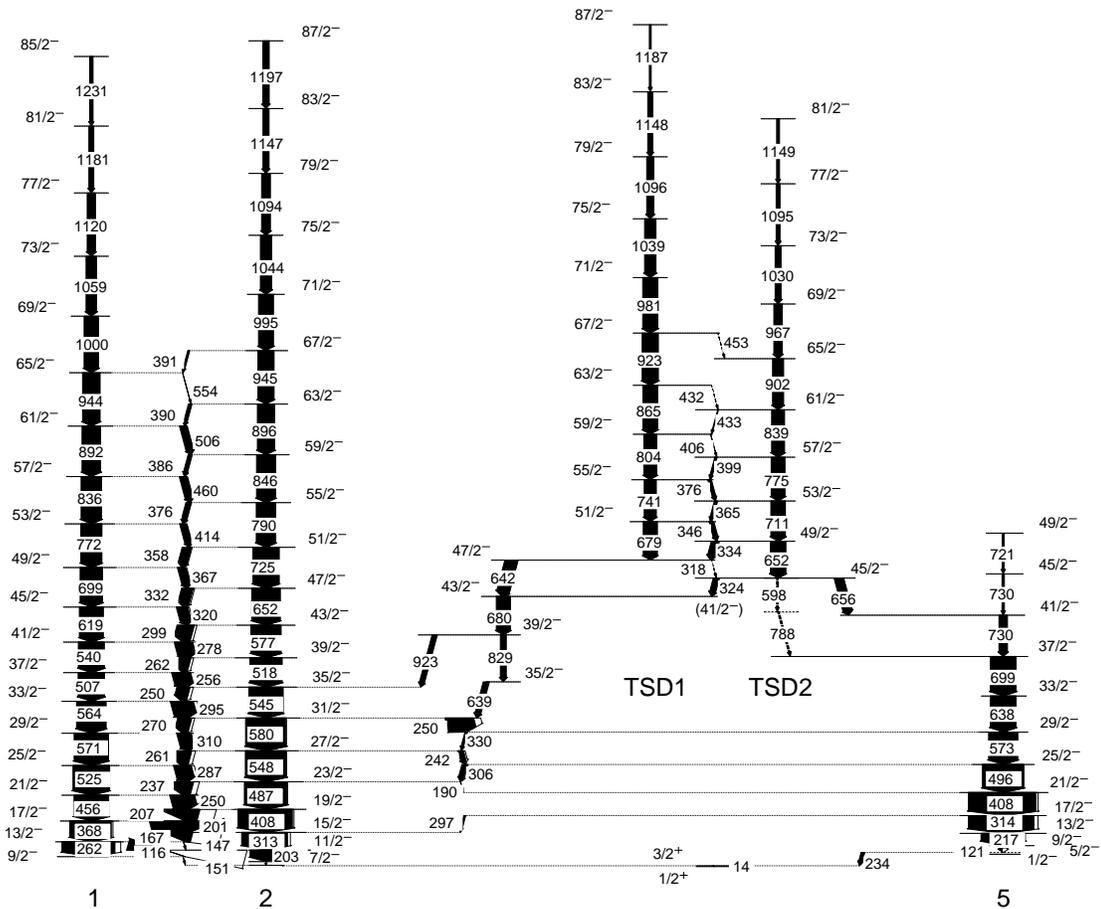}
\caption[Partial level scheme of $^{163}$Tm.]{Partial 
level scheme of $^{163}$Tm, showing the TSD bands as well as the yrast bands. Adapted from 
Ref.~\cite{Pattabi-PLB-163Tm-07}.\label{fig:163Tm_level_scheme_PRC}}
\end{center}
\end{figure}

A typical Gammasphere ``stand-alone'' experiment was carried out with the 170-$MeV$ $^{37}$Cl beam, provided by the 
88-inch cyclotron facility at the LBNL, and an isotopically enriched $^{130}$Te target foil of about 0.5-mg/cm$^2$ 
thickness, $\it{i.e.}$, a so-called ``thin'' target. A total of about one billion events was accumulated. The data 
analysis was performed using the standard techniques, introduced in Sec.~\ref{sec:DataAnytech} of Chapter~\ref{chap:exp_techs}, 
and resulted in a partial level scheme of $^{163}$Tm expanded with respect to earlier work~\cite{Jensen-ZPA-340-351-91}. As can be 
seen in this level scheme (Figure~\ref{fig:163Tm_level_scheme_PRC}), the two known yrast cascades~\cite{Jensen-ZPA-340-351-91}, 
$\it{i.e.}$, bands 1 and 2, are now extended up to spins $85/2^{-}$ and $87/2^{-}$, respectively, and two new excited bands 
(labeled as TSD1 and TSD2) are observed as well. The relevant supporting spectra can be found in Fig. 2 in 
Ref.~\cite{Pattabi-PLB-163Tm-07}.

\begin{figure}
\begin{center}
\includegraphics[angle=0,width=0.65\columnwidth]{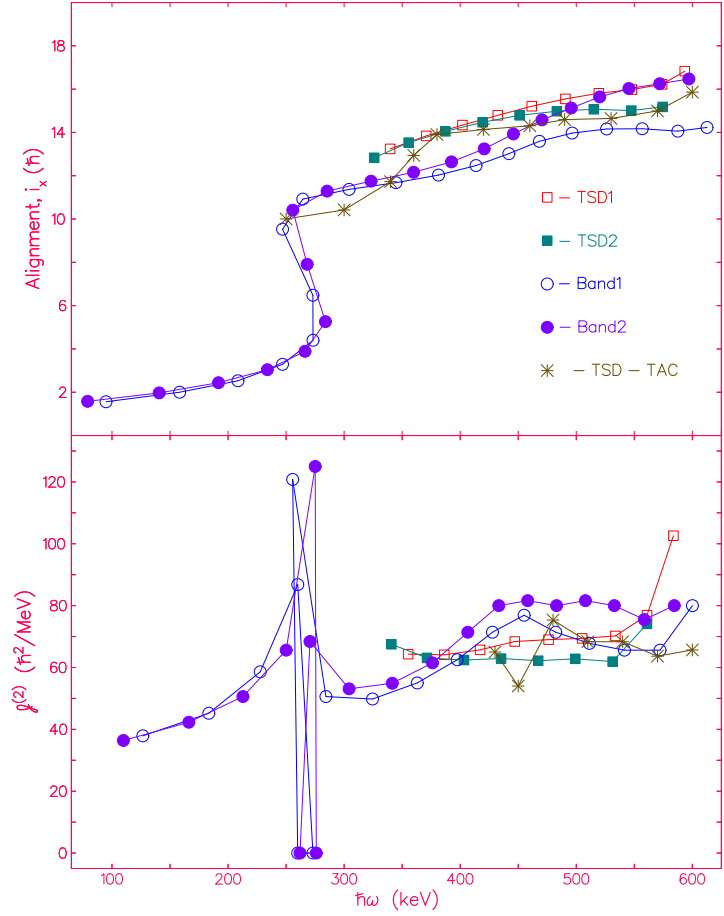}
\caption[Aligned spins and the experimental dynamic moments of inertia for the two TSD bands in $^{163}$Tm.]{Aligned 
spins $i_x$ (upper panel) and the experimental dynamic moments of inertia ${\Im}^{(2)}$ (lower panel) 
for the two TSD bands in $^{163}$Tm as a function of rotational frequency. The subtracted reference for the alignment 
is $I_{ref}={J_0}{\omega}+{J_1}{\omega^3}$ with the Harris parameters $J_0=30{\hbar^2}{MeV^{-1}}$ and 
$J_1=40{\hbar^4}{MeV^{-3}}$. The theoretical values for alignments and moments of inertia were obtained from 
the calculations in the TAC model. Adapted from Ref.~\cite{Pattabi-PLB-163Tm-07}.\label{fig:163Tm_align_moi}}
\end{center}
\end{figure}

From the established level scheme, the alignment, $i_x$, and the dynamic moment of inertia, ${\Im}^{(2)}$, 
for each band have been extracted and are plotted as a function of rotational frequency, $\hbar\omega$, in 
Figure~\ref{fig:163Tm_align_moi}. Both the alignments and the dynamic moments of inertia indicate that these 
four bands can be grouped into two families, $\it{i.e.}$, bands 1 and 2 in one, and, bands TSD1 and TSD2 in the other. 
As discussed below, the TSD1 and TSD2 bands are proposed to be associated with Triaxial Strongly Deformed (TSD) 
structures. 

\begin{figure}
\begin{center}
\includegraphics[angle=270,width=0.65\columnwidth]{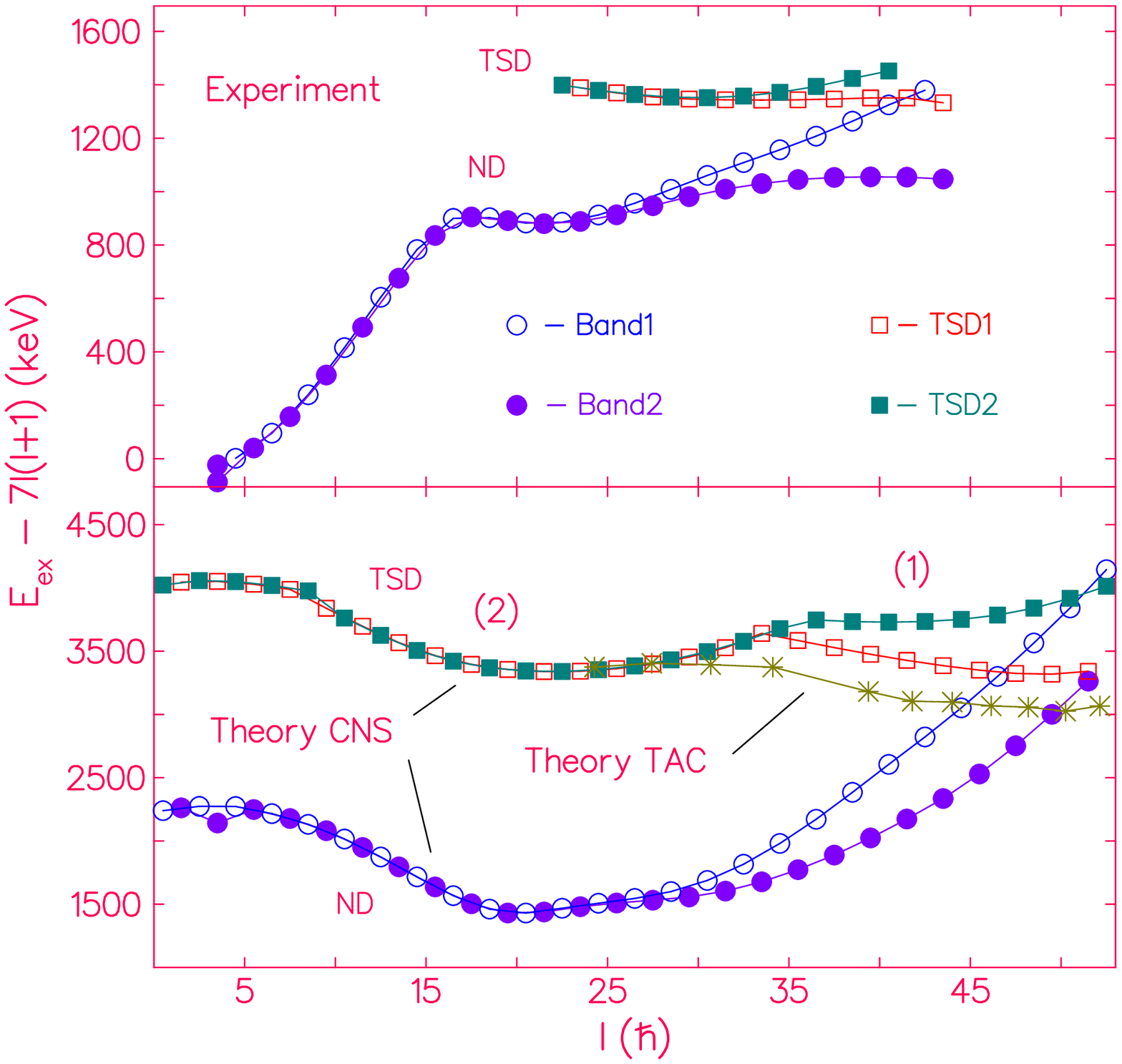}
\caption[Excitation energies for the two TSD bands and the two ND bands in $^{163}$Tm.]{Excitation 
energies relative to a rotational reference for the two TSD bands and the two 
ND bands (1 and 2) in $^{163}$Tm: the experimental data (top) and the calculated result in the CNS 
and the TAC models (bottom). The numbers (1) and (2) with the CNS calculation refer to the two associated 
TSD minima, respectively (presented in Figure~\ref{fig:163Tm_TRS}). Adapted from 
Ref.~\cite{Pattabi-PLB-163Tm-07}.\label{fig:163Tm_ex_ener}}
\end{center}
\end{figure}

\begin{figure}
\begin{center}
\includegraphics[angle=0,width=0.50\columnwidth]{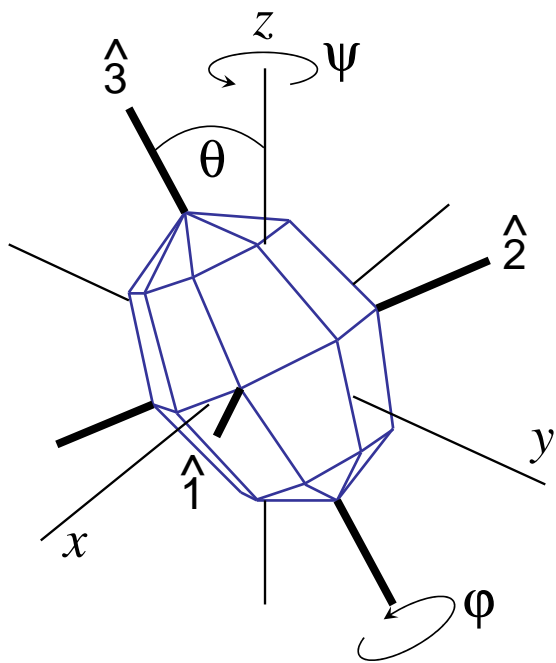}
\caption[The Euler angles specifying the orientation of the triaxial density distribution in the laboratory 
frame.]{The Euler angles specifying the orientation of the triaxial density distribution in the laboratory 
frame. A polyeder shape is shown, which makes the geometry more visible. The principal axes $\hat{1}$, 
$\hat{2}$ and $\hat{3}$ are the thick lines and the laboratory axes $x$, $y$, $z$ are thin ones. Adapted from 
Ref.~\cite{Frauendorf-NPA-677-115-00}.\label{fig:TACeulerAng}}
\end{center}
\end{figure}

In the top panel of Figure~\ref{fig:163Tm_ex_ener}, the experimental excitation energies of 
these proposed TSD bands as well as the yrast bands are shown, while the corresponding values calculated 
in the Cranked Nilsson-Strutinsky (CNS) model~\cite{Bengtsson-PSC-T5-165-83,Bengtsson-NPA-436-14-85,Afanasjev-PREP-322-1-99} 
and the Tilted-Axis Cranking (TAC) model~\cite{Frauendorf-NPA-557-259-93,Frauendorf-NPA-677-115-00,Frauendorf-RMP-73-463-01} 
are presented in the bottom panel. It is worth pointing out here that the TAC model is an approach seeking the solutions 
to the time-dependent Schr{\"o}dinger equation of a single particle in the intrinsic frame of nucleus ($\it{i.e.}$, 
Eq.~\ref{eq:SinlCSMschodingerEQ} in Sec.~\ref{subsec:CSMintro} of Chapter~\ref{chap:theoriBkgd}) in the case of 
uniform rotation about an axis that is tilted with respect to the principal axes of the deformed density 
distribution~\cite{Frauendorf-RMP-73-463-01}. Usually, three Euler angles $\theta$, $\phi$ and $\psi$ 
are used to specify the orientation of the principal axes in the body-fixed frame, denoted by $\hat{1}$, 
$\hat{2}$ and $\hat{3}$, with respect to the laboratory system, denoted by $x$, $y$ and $z$. As can be seen in 
Figure~\ref{fig:TACeulerAng}, $\psi(=\omega{t})$ is the angle that increases as the nucleus rotates around the 
$z$-axis, while the angles $\theta$ and $\phi$ determine the orientation of the rotation axis, $\it{i.e.}$, the $z$-axis. The 
CNS model is a special case of the TAC approach with the rotation axis fixed along one of the principal axes, $\it{i.e.}$, 
the interpolation technique into the cranked shell correction approach for the so-called Principal-Axis Cranking (PAC) 
solutions~\cite{Frauendorf-RMP-73-463-01} (the correspnding orientation angles satisfy $\theta=0,\pi/2$ and $\phi=0,\pi/2$). 

The measured excitation energies of 
TSD bands in $^{163}$Tm are quite different from the ones of the wobbling bands seen in Lu (see for example, 
Fig. 14 in Ref.~\cite{Jensen-NPA-703-3-02}). This observation seems to suggest a different nature for the 
two TSD bands in $^{163}$Tm, which is further supported by the structure of the linking transitions. These 
transitions connecting the two TSD bands in $^{163}$Tm are not similar to the ones characteristic of wobbling, 
$\it{i.e.}$, the linking transitions go both ways between the two TSD bands in $^{163}$Tm, whereas the 
connecting transitions always proceed only from a wobbling band to another with a lower 
$n_w$ value in Lu. 

\subsection{\label{subsec:163Tm_struct_discuss}Interpretation and discussion}
\begin{figure}
\begin{center}
\includegraphics[angle=0,width=0.65\columnwidth]{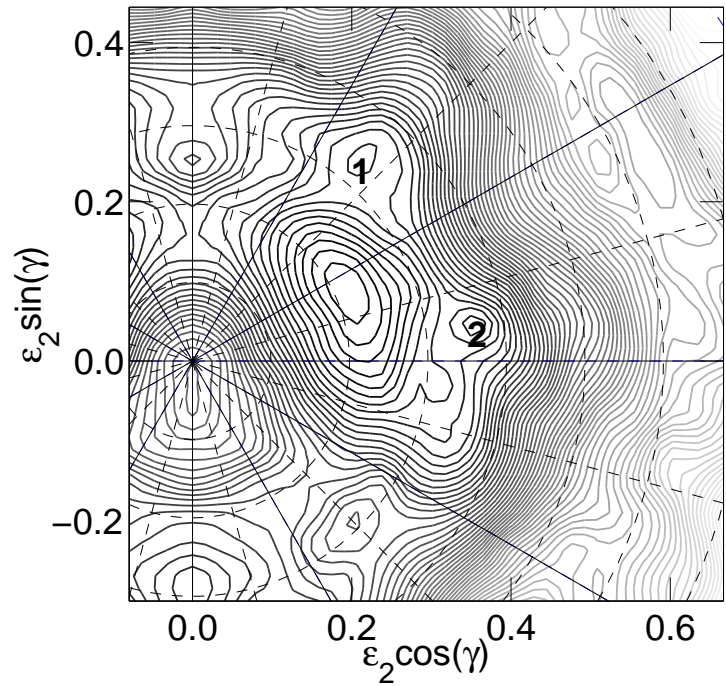}
\caption[Potential energy surface for $^{163}$Tm.]{Potential 
energy surface for $^{163}$Tm calculated in the CNS model at 
spin $63/2^{-}$. The two TSD minima are marked by 1 and 2, consistent with the 
labels used in Figure~\ref{fig:163Tm_ex_ener}. Adapted from 
Ref.~\cite{Pattabi-PLB-163Tm-07}.\label{fig:163Tm_TRS}}
\end{center}
\end{figure}

In order to understand the observed properties of these bands in $^{163}$Tm, the calculations in 
both the CNS and the TAC models were carried out. The potential energy surface (see Figure~\ref{fig:163Tm_TRS}), 
resulting from the CNS calculations for a configuration with $(\pi,\alpha)=(-,-1/2)$ (see details below) 
at spin $I^{\pi}=63/2^{-}$, indicates a prolate minimum at normal deformation (ND) 
$({\varepsilon}_2~{\approx}~0.21)$ and two triaxial strongly deformed (TSD) minima. The two TSD 
minima have nearly the same energy. It can be seen clearly in the bottom panel of Figure~\ref{fig:163Tm_ex_ener} 
that minimum 2 is energetically favored at low spin while minimum 1 becomes lower at high spin. The two minima have almost the 
same value of ${\varepsilon}_2$ and $|\gamma|$, indicating that both are associated with the same shape. 
The axis of rotation is the short one in minimum 1 (with $\gamma>0$), while it is the intermediate one for 
minimum 2 (with $\gamma<0$). Thus, the CNS calculations suggest that at $I=24$, where minimum 1 goes below 
minimum 2, the orientation of the rotational axis flips from the intermediate axis to the short one. 
This sudden flip is caused by the inherent assumption in the CNS model that the rotational axis must 
be a principal one, and in fact indicates that this assumption of rotation about a principal axis is 
inappropriate in this case. Therefore, the TAC calculations, which do not restrict the orientation of the rotational 
axis, should be adopted to interpret the observations in this experiment. As expected, a tilted solution 
with lower energy was found, which smoothly connects minimum 2 with minimum 1. For $I>23$, the angular 
momentum vector moves away from the intermediate axis toward the short axis. It does not quite reach 
it within the considered spin range. (For $I{\sim}50$, the angle with the intermediate axis is still 
about $20^{\circ}$.) This solution is assigned to bands TSD1 and TSD2. In accordance with the experiment, 
it corresponds to a ${\Delta}I=1$ band without signature splitting. As can be seen in 
Figure~\ref{fig:163Tm_ex_ener}, the observed onset of signature splitting at the highest spins is 
consistent with the calculated approach of the TAC solution in minimum 1 of the CNS result. At 
large frequency, the calculated TSD bands have a lower energy than the ND ones, while the data show 
that only band 1 crosses band TSD1 above $I=40$ (band 2 remains yrast up to the highest spins), $\it{i.e.}$, 
the experimental energy difference between the ND and TSD minima is somehwat larger than the calculated 
values at the highest spins. Moreover, Figure~\ref{fig:163Tm_align_moi} demonstrates that the calculated 
alignments and the dynamic moments also agree with the measured ones. Hence, all experimental observables 
for the TSD bands in $^{163}$Tm are well reproduced by the TAC calculations. 

\begin{figure}
\begin{center}
\includegraphics[angle=0,width=0.78\columnwidth]{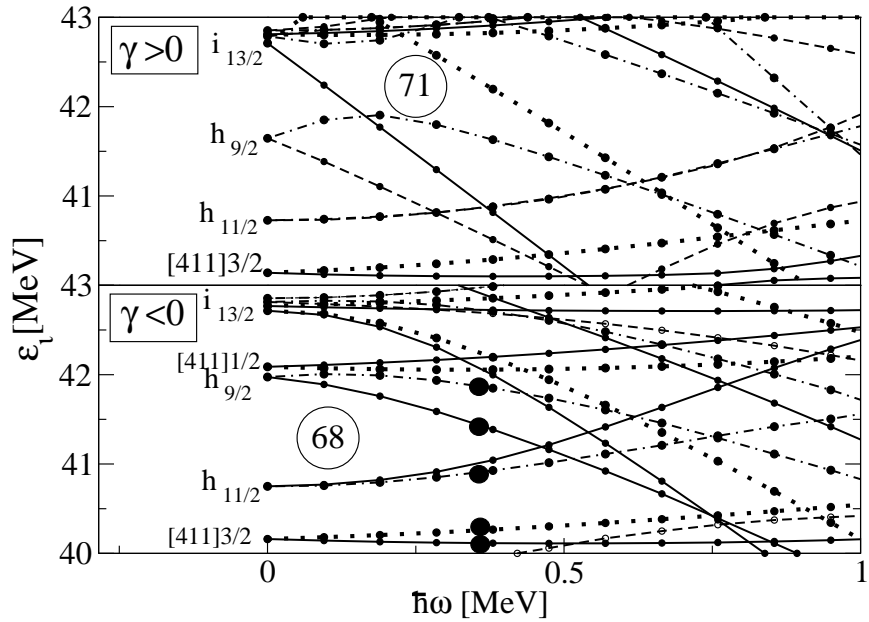}
\caption[Single-proton routhians as function of 
rotational frequency in TSD minima 1 and 2.]{Single-proton routhians as function of 
rotational frequency in TSD minima 1 (top) and 2 (bottom). The line convention is: 
$(\pi,\alpha)=$ (+,1/2) full, (+,-1/2) dot, (-,1/2) dash, (-,-1/2) 
dash dot. The deformation parameters used in the calculations are: 
$\epsilon_{2}= 0.39$, $\epsilon_{4} = 0.05$, $|\gamma|= 17^{\circ}$. Adapted from 
Ref.~\cite{Pattabi-PLB-163Tm-07}.\label{fig:163Tm_proton_routhian}}
\end{center}
\end{figure}

Further, the single proton routhians in both of the TSD minima were calculated and 
are shown in Figure~\ref{fig:163Tm_proton_routhian}. The configurations that we assign to 
the obseved TSD bands (indicated by the large filled circles in Figure~\ref{fig:163Tm_proton_routhian}) 
are the lowest with negative parity and small signature splitting, in agreement with the experiment. 
As discussed in Ref.~\cite{Pattabi-PLB-163Tm-07}, the CNS calculations also predict 
four competing TSD configurations (named TSD3, TSD4, TSD5, and 
TSD6 in the discussion below) at somewhat lower energy than TSD1 and TSD2. The 
positive-parity configurations TSD3 and TSD4 have the odd proton on one of the 
$[411]1/2$ routhians and have both $h_{11/2}$ signatures occupied; they are 
predicted by the CNS calculations to lie about 500 $keV$ below TSD1 at spin 20 
and have a larger energy above spin 50. However, as can be seen in Fig.~10 of 
Ref.~\cite{Jensen-ZPA-340-351-91}, some residual proton pair correlations 
in the lower-spin part will disfavor the configurations TSD3 and TSD4 with respect 
to TSD1 and TSD2. The configurations TSD5 and TSD6, with both signatures of the 
$h_{11/2}$ orbital occupied and the odd proton on one of the two $h_{9/2}$ routhians, 
would correspond to two well-separated $\Delta{I}=2$ sequences with little resemblance 
to the experiment. The favored signature branch, TSD5, is predicted by the CNS 
calculations to lie about 500 $keV$ below TSD1 at spin 20 and to have a larger 
energy than TSD1 above spin 40. We note here that the location of the $h_{9/2}$ 
orbital has been a longstanding open problem in calculations using the modified 
oscillator potential (cf. the discussion in Ref.~\cite{Jensen-ZPA-340-351-91}).

\begin{figure}
\begin{center}
\includegraphics[angle=0,width=0.75\columnwidth]{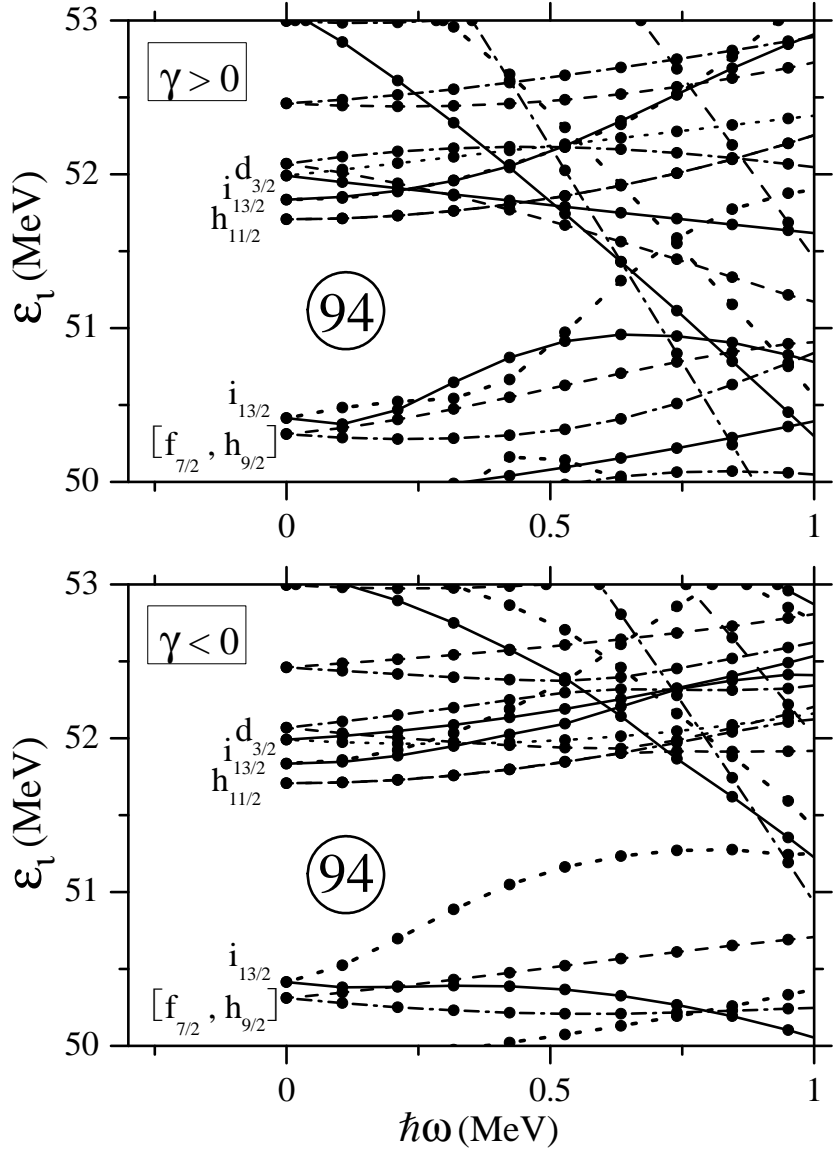}
\caption[Single-neutron routhians as function of 
rotational frequency in TSD minima 1 and 2.]{Single-neutron routhians as function of 
rotational frequency in TSD minima 1 (top) and 2 (bottom). The line convention is: 
$(\pi,\alpha)=$ (+,1/2) full, (+,-1/2) dot, (-,1/2) dash, (-,-1/2) 
dash dot. The deformation parameters used in the calculations are: 
$\epsilon_{2}= 0.39$, $\epsilon_{4} = 0.05$, $|\gamma|= 17^{\circ}$. 
Adapted from Ref.~\cite{Wang-PRC-163Tm-07}.\label{fig:163Tm_neutron_routhian}}
\end{center}
\end{figure}

The relevant single-proton routhians in the ND minimum can be found in 
Ref.~\cite{Bengtsson-EPJA-22-355-04}. Comparing them with the routhians in 
the TSD minima (see Figure~\ref{fig:163Tm_proton_routhian}), one finds that 
the $h_{11/2}$ orbital has a larger splitting between the two signatures, 
which reflects the proposed smaller deformation. Further, the $h_{11/2}$ 
levels are shifted up by about 2 $MeV$. Hence, bands 1 and 2 are interpreted 
as the signature partners of a configuration with an odd proton on the $h_{11/2}$ 
level at the Fermi surface and a pair of protons on the $[411]1/2$ orbitals. 
This assignment is further supported by the fact that the observed signature 
splitting between bands 1 and 2 is consistent with the CNS calculation based 
on the above configuration (shown in Figure~\ref{fig:163Tm_ex_ener}). 
It is also worth pointing out that, in contrast with previous calculations with $Z > 69$ and 
$N{\sim}94$~\cite{Odegard-PRL-86-5866-01,Jensen-NPA-703-3-02,Jensen-PRL-89-142503-02,%
Schonwa-PLB-552-9-03,Amro-PLB-553-197-03,Bringel-EPJA-24-167-05}, the 
$i_{13/2}$ proton level is empty in $^{163}$Tm, which means that this level is not 
essential in forming the TSD minima. Rather, it is the $N=94$ gap in the neutron routhians 
at ${\varepsilon}_2=0.39$, $|{\gamma}|=17^{\circ}$ 
(see Figure~\ref{fig:163Tm_neutron_routhian}) that stabilizes the TSD shape. 
We will return to this point further below. 

A natural explanation is provided by the calculated single-proton routhians in Figure~\ref{fig:163Tm_proton_routhian} 
for the presence of collective wobbling excitations in the Lu isotopes with $Z = 71$ and their 
absence in $^{163}$Tm with $Z = 69$. The TSD configurations of nuclei with $Z > 69$ belong 
typically to minimum 1 with ${\gamma}>0$~\cite{Bengtsson-EPJA-22-355-04}. For $Z = 71$, the Fermi level is the 
signature ${\alpha}=1/2$ routhian of $i_{13/2}$ parentage in the frequency range 
250 $keV$ $<$ $\hbar\omega$ $<$ 450 $keV$. The lowest TSD band is observed in this frequency 
range and has parity and signature $(+,1/2)$. The lowest particle-hole (p-h) excitation of the same parity lifts 
the odd proton onto the other signature, $\alpha=-1/2$ of this $i_{13/2}$ level, which lies 
at a relatively high energy (${\sim}~1~MeV$ at $\hbar\omega=0.4~MeV)$). This brings the collective 
wobbling excitation, which has an energy of about 0.3 $MeV$, well below the lowest p-h 
excitations. For $Z = 69$, however, the two signatures of the $h_{11/2}$ state are quite close 
together (see Figure~\ref{fig:163Tm_ex_ener}). Therefore, the wobbling excitation lies above the 
p-h excitations, likely too high in excitation energy to be populated with observable strength in 
the reaction employed in the present study. It is also worth mentioning that the relative energy 
of the collective wobbling mode and of the p-h excitations in $^{163}$Lu has been studied by means 
of the particle-rotor model~\cite{Hamamoto-PLB-193-399-87}, where the p-h excitations have been called the 
``cranking mode''~\cite{Odegard-PRL-86-5866-01}. They are found to be located well above the one-phonon 
wobbling excitation. With the level order suggested in Figure~\ref{fig:163Tm_proton_routhian}, one can expect, 
for $Z = 69$, that a band structure similar to the wobbling bands seen in $Z = 71$ is positioned 
at somewhat higher energy. It may be obtained by lifting the last proton from 
the $h_{11/2}$ into the $i_{13/2}$ orbital. For $Z = 73$, several TSD bands of 
both parities with similar energy are also expected. The possibility to experimentally 
identify a wobbling band is restricted by the competition of this collective excitation with the p-h 
excitations, $\it{i.e.}$, the wobbling band may become the first excitation mode above the yrast line, if its 
energy is lower than the energy of the p-h excitations. Such a case appears to occur in several odd-A 
Lu isotopes. Moreover, a large gap at $N = 94$, which is found in the neutron diagrams 
(Figure~\ref{fig:163Tm_neutron_routhian}) prevents the neutron p-h excitations from 
competing with the wobbling mode in the Lu isotopes. The opposite scenario occurs in 
$^{163}$Tm, where the energy for the p-h excitations between the signature 
partners of the $h_{11/2}$ orbital is smaller than the wobbling energy. In our experiment, only the 
first excited band, which corresponds to the p-h excitations, appears to have received sufficient intensity 
to be observed. However, the apparent absence of the wobbling mode does not necessarily 
imply a near-axial shape. If this was the case, there would be a conflict with the above calculations 
as well as with earlier ones~\cite{Bengtsson-EPJA-22-355-04}. 

\section{\label{sec:163Tm-lifework}Lifetime measurements of TSD bands in $^{163}$Tm}

\subsection{\label{subsec:163Tm_life_motivate}Motivation}
In the earlier structure work (see Sec.~\ref{sec:163Tm-struct} as well as 
Ref.~\cite{Pattabi-PLB-163Tm-07}), the interpretation of the two $^{163}$Tm 
sequences as TSD bands rested solely on indirect experimental indications (such as 
the magnitude and evolution with frequency of the moments of inertia) and on 
the agreement with the calculations. Therefore, with the purpose to verify directly that 
the proposed TSD bands are associated with a larger deformation than the yrast sequences 
as well as to further test how well the TAC model predicts or reproduces the experimental 
observations in the region, another experiment with the 165 $MeV$ $^{37}$Cl beam and a 
``thick'' $^{130}$Te target was performed using the ATLAS facility at ANL. The detailed 
configuration of the target used has been described in Sec.~\ref{subsec:DSAMtech} 
of Chapter~\ref{chap:exp_techs}. In this experiment, the lifetimes and, hence, the quadrupole 
moments for the four $^{163}$Tm bands were measured by means of the DSAM technique 
(see Sec.~\ref{subsec:DSAMtech} of Chapter~\ref{chap:exp_techs}). 

\subsection{\label{subsec:163Tm_life_expe_result}Experiment and data}

The details of the performed DSAM experiment have been given in 
Sec.~\ref{subsec:DSAMtech} of Chapter~\ref{chap:exp_techs}. 
In the six-day run, over 1.5 $\times$ 10$^9$ coincidence events with fold $\ge$ 3 
were collected by Gammasphere. Since the DSAM technique 
involves the detection of $\gamma$ rays during the slowing down process 
in the thick target, the relation between the average energy shifts and 
detector angles needs to be determined. For this purpose, the raw data 
were sorted into several BLUE database files. Subsequently, the background-subtracted 
spectra at a given detector angle under specific coincidence requirements, 
required by the DSAM mesurement, were achieved conveniently with the 
method introduced in Sec.~\ref{subsec:GenSpecBKsub} in Chapter~\ref{chap:exp_techs}. 
It is worth to point out that the proper method of subtracting background 
is a key factor in the success of such DSAM measurements. Recalling the effect 
of background subtraction in the process of generating coincidence spectra, as illustrated 
by Figure~\ref{fig:sub_BK_effect} in Chapter~\ref{chap:exp_techs}, it becomes rather 
obvious that accurate background subtraction is essential in a process where the crucial information, 
$\it{i.e.}$, the Doppler shift at a given angle, is derived from the determination of peak centroids. 
In addition, a thorough understanding of the spectra is very important as well since 
a measured centroid could potentially be in error if the peak is an unresolved doublet 
of two transitions. 

\begin{figure}
\begin{center}
\includegraphics[angle=270,width=\columnwidth]{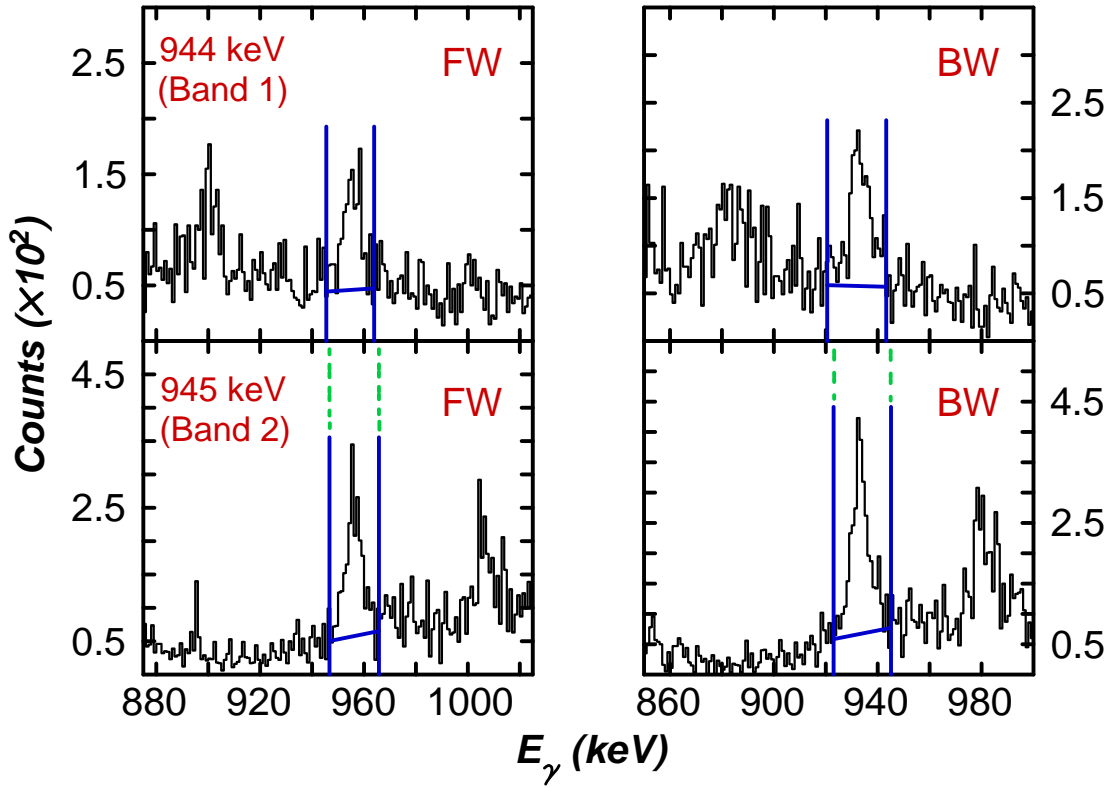}
\caption[Representative summed spectra for accurately determining the peak positions of distinguishable 
contaminated transitions.]{Representative summed spectra for accurately determining the peak positions of distinguishable 
contaminated transitions, the 944-$keV$ line in band 1 and 945-$keV$ line in band 2, at 2 detector 
angles: 35$^{\circ}$ (FW) and 145$^{\circ}$ (BW). The double gates used for producing the two 
spectra at the top are the combinations of the 1000-$keV$ line in band 1 and any one of the other inband 
transitions of band 1, while the gates for the two spectra at the bottom are the 896-$keV$ transition in 
band 2 plus any one of the other inband lines of band 2. The drawn low- and high-limit 
mark lines for each peak indicate the tiny shift of position from the 944-$keV$ transition to the 945-$keV$ 
transition.\label{fig:sp_b1_b2_dblt_944_945}}
\end{center}
\end{figure}

The four rotational bands of interest in the present measurements are the 
bands 1 ($85/2^{-}$ -- $9/2^{-}$ sequence) 
and 2 ($87/2^{-}$ -- $7/2^{-}$ cascade) associated with the $[523]7/2^{-}$ 
configuration and the bands TSD1 ($87/2^{-}$ -- $47/2^{-}$ sequence) and TSD2 
($81/2^{-}$ -- $45/2^{-}$ cascade), shown in Figure~\ref{fig:163Tm_level_scheme_PRC}. 
From a first inspection of the coincidence data, it was established 
that the transitions with energy $E_{\gamma}$ $\le$ 600 $keV$ in 
bands 1 and 2 did not exhibit any measurable shift or broadening 
as a function of detector angle. In other words, these 
deexcitations must have occurred after the recoiling nuclei 
have come to rest in the Au layer of the target. These ``stopped'' 
transitions could thus be used as a starting point to obtain 
coincidence spectra for each band at ten detector angles, from 
which energy shifts would subsequently be determined. The use of "stopped" 
transitions alone proved to be insufficient. Hence, angle-dependent 
coincidence gates had to be placed on band members in an iterative procedure 
starting with the lowest $\gamma$ ray exhibiting a shift and 
moving up in the band one transition at each step. This procedure 
could be applied not only to bands 1 and 2, but also to the TSD1 and 
TSD2 sequences, since the latter deexcite into bands 1 and 2. In the 
process of selecting appropriate gating conditions, special care 
was taken to avoid numerous contaminant lines from either 
other $^{163}$Tm band structures or other reaction products, as well 
as some in-band doublet $\gamma$ rays. Actually, from the data analysis process 
it became clear that the cases of contamination can be grouped into two types: 
(1) the real $\gamma$ ray of interest can be seen free of contaminant 
lines in certain spectra with appropriate gating conditions; and (2) the real 
$\gamma$ ray of interest can not be distinguished from contaminant lines 
in any spectroscopic gating technique. For the latter case, $\it{e.g.}$, the 680-$keV$ 
doublet in band TSD1, which corresponds to the $51/2^{-}$$\rightarrow$$47/2^{-}$ transition 
and to the $43/2^{-}$$\rightarrow$$39/2^{-}$ transition (see Figure~\ref{fig:163Tm_level_scheme_PRC}), 
the impact from contaminant can be weakened mostly through a procedure applied in the 
process of a linear fit to obtain the corresponding $F(\tau)$ value. This method 
will be discussed later in this subsection. For the transitions affected by contaminants 
of the first case (distinguishable type), the impact from the contaminant lines can be 
eliminated and, hence, the real $\gamma$ rays of interest can be studied accurately 
in spectra generated with specific gating conditions, as illustrated in 
Figure~\ref{fig:sp_b1_b2_dblt_944_945} with the example of the 944-$keV$ 
line in band 1 and the 945-$keV$ transition in band 2, which affect each other. 

%Actually, from the data analysis process 
%it became clear that the impactive contaminations can be grouped 
%into two types, $\it{i.e.}$, the distinguishable ones with appropriate gating conditions 
%and the non-distinguishable ones with any available combination of gates. For the contaminant 
%lines of the latter type, $\it{e.g.}$, the 680-$keV$ doublet in band TSD1, which 
%corresponds to the $51/2^{-}$$\rightarrow$$47/2^{-}$ transition and to the 
%$43/2^{-}$$\rightarrow$$39/2^{-}$ transition (see Figure~\ref{fig:163Tm_level_scheme_PRC}), 
%their affections can be weakened mostly through a procedure applied in the 
%process of the corresponding linear fits, which will be discussed later in this section, 
%while it is illustrated in Figure~\ref{fig:sp_b1_b2_dblt_944_945} that the transitions accompanied by the 
%contaminations of the first type (with degenerate energies but distinguishable) can be 
%studied accurately in the spectra generated with some specific coincidence requirement, 
%using a example of the 944-$keV$ line in band 1 and the 945-$keV$ transition in band 2. 

\begin{figure}
\begin{center}
\includegraphics[angle=270,width=0.90\columnwidth]{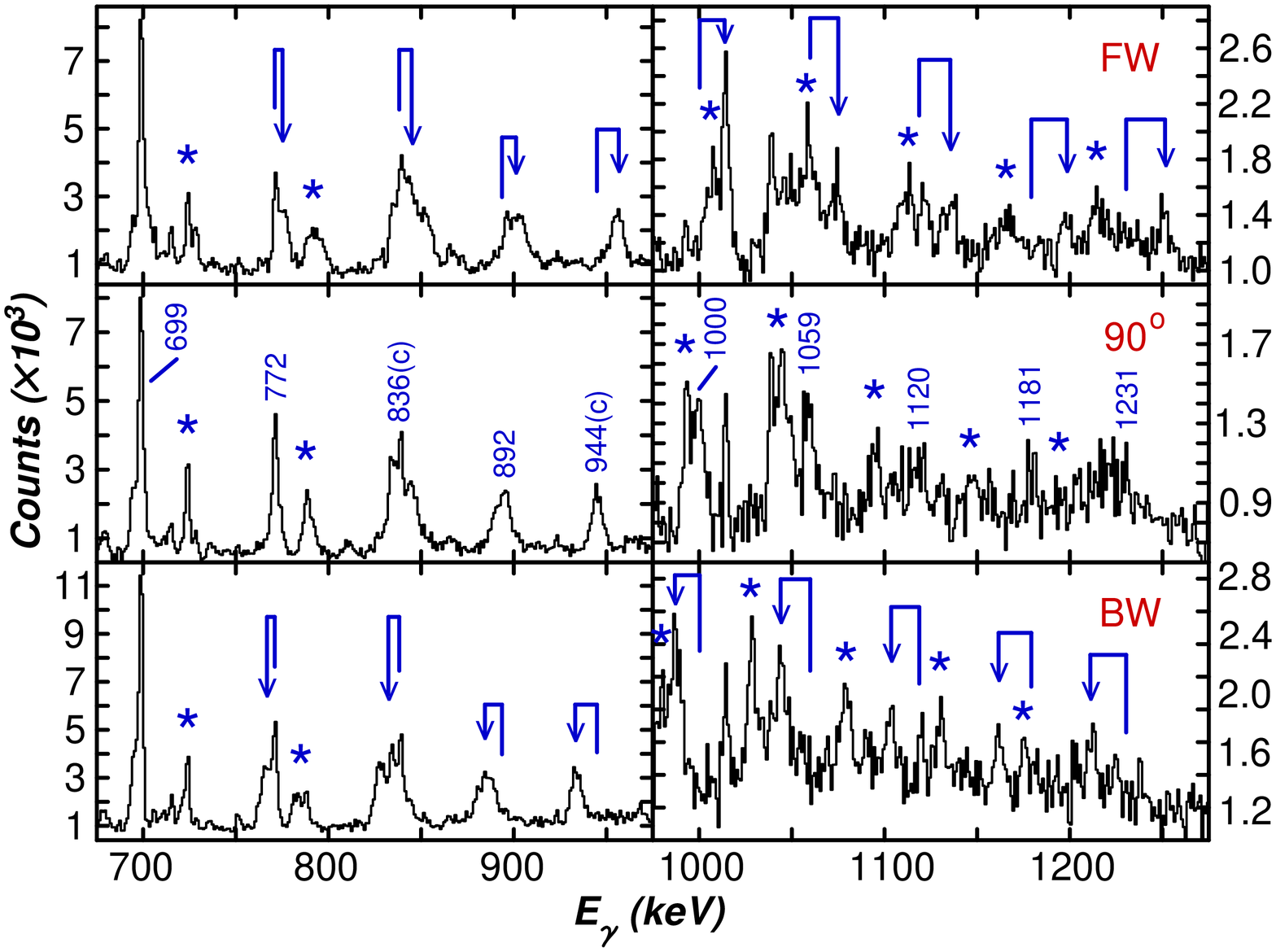}
\caption[Sum of spectra gated on in-band transitions for 
band 1.]{Sum of spectra gated on in-band transitions for 
band 1 at 3 detector angles: 35$^{\circ}$ (FW), 90$^{\circ}$, and 145$^{\circ}$ (BW). 
The positions of unshifted and shifted $\gamma$ rays are marked by energy values 
and arrows, respectively. Note that transitions from band 2 appear in these spectra 
(marked with $\star$ symbols) due to the fact that intense connecting transitions occur 
between the two structures. The transitions labeled by signs in the form of ``energy value(c)'' 
are the contaminated lines which need to be taken care of separately 
(see text for details).\label{fig:sp_band1}}
\end{center}
\end{figure}

\begin{figure}
\begin{center}
\includegraphics[angle=270,width=0.90\columnwidth]{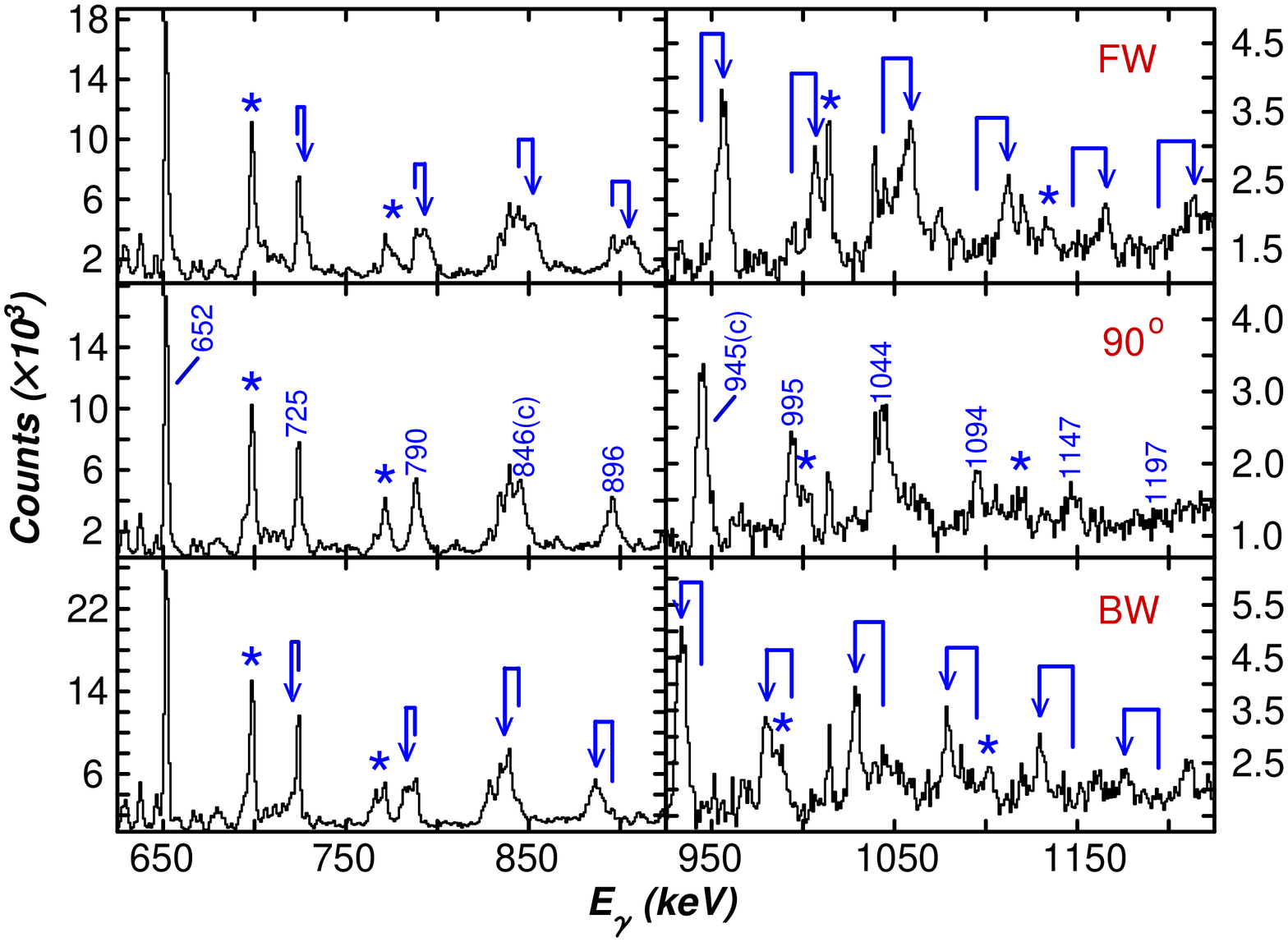}
\caption[Sum of spectra gated on in-band transitions for 
band 2.]{Sum of spectra gated on in-band transitions for 
band 2 at 3 detector angles: 35$^{\circ}$ (FW), 90$^{\circ}$, and 145$^{\circ}$ (BW). 
The positions of unshifted and shifted $\gamma$ rays are marked by energy values 
and arrows, respectively. Note that transitions from band 1 appear in these spectra 
(marked with $\star$ symbols) due to the fact that intense connecting transitions occur 
between the two structures. The transitions labeled by signs in the form of ``energy value(c)'' 
are the contaminated lines which need to be taken care of separately 
(see text for details).\label{fig:sp_band2}}
\end{center}
\end{figure}

\begin{figure}
\begin{center}
\includegraphics[angle=270,width=0.90\columnwidth]{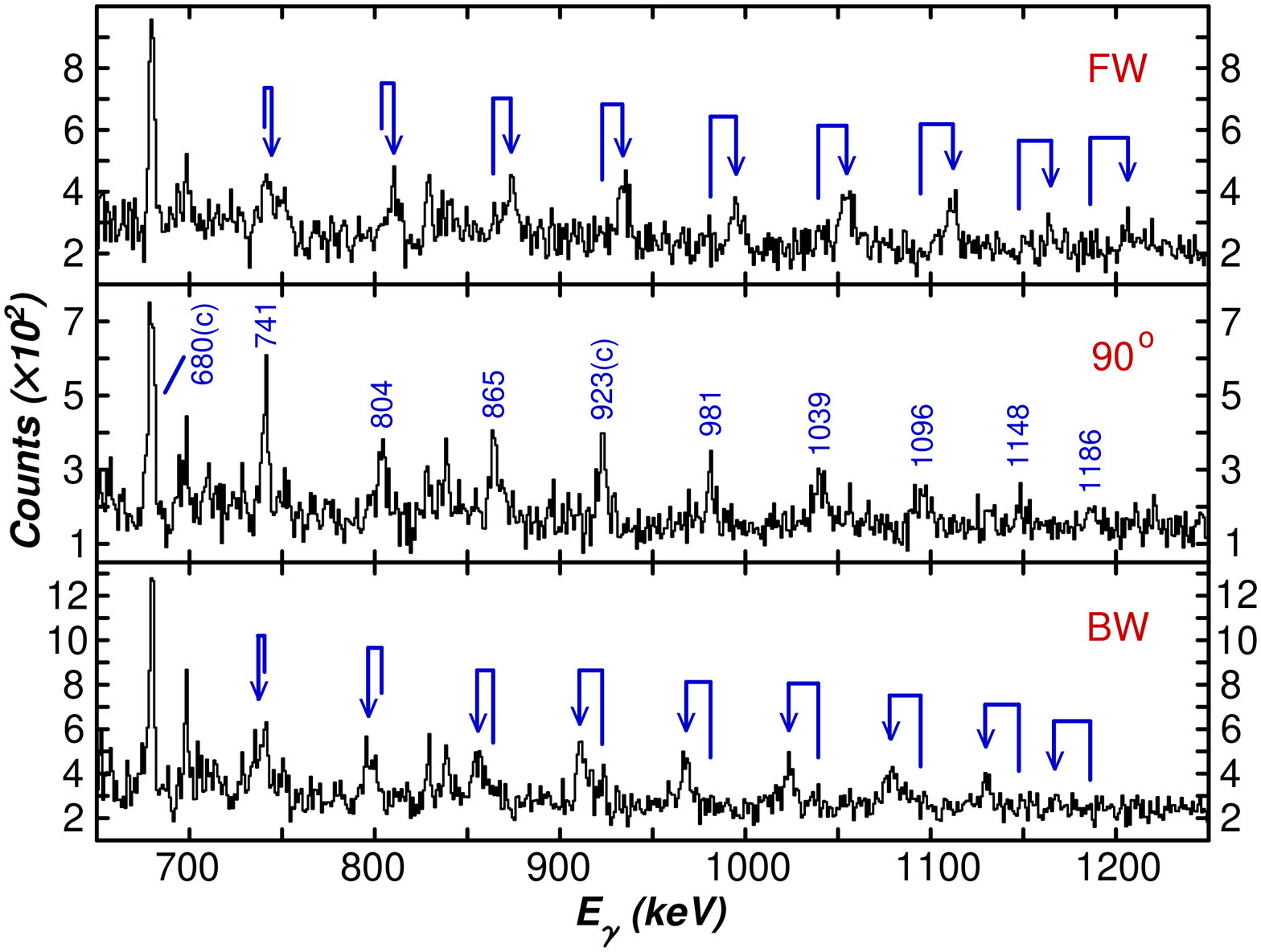}
\caption[Sum of spectra gated on in-band transitions for 
band TSD1.]{Sum of spectra gated on in-band transitions for 
band TSD1 at 3 detector angles: 35$^{\circ}$ (FW), 90$^{\circ}$, and 145$^{\circ}$ (BW). 
The positions of unshifted and shifted $\gamma$ rays are marked by energy values 
and arrows, respectively. Note that the transitions labeled by signs in the form of 
``energy value(c)'' are the contaminated lines which need to be taken care of separately 
(see text for details).\label{fig:sp_band3}}
\end{center}
\end{figure}

\begin{figure}
\begin{center}
\includegraphics[angle=270,width=0.90\columnwidth]{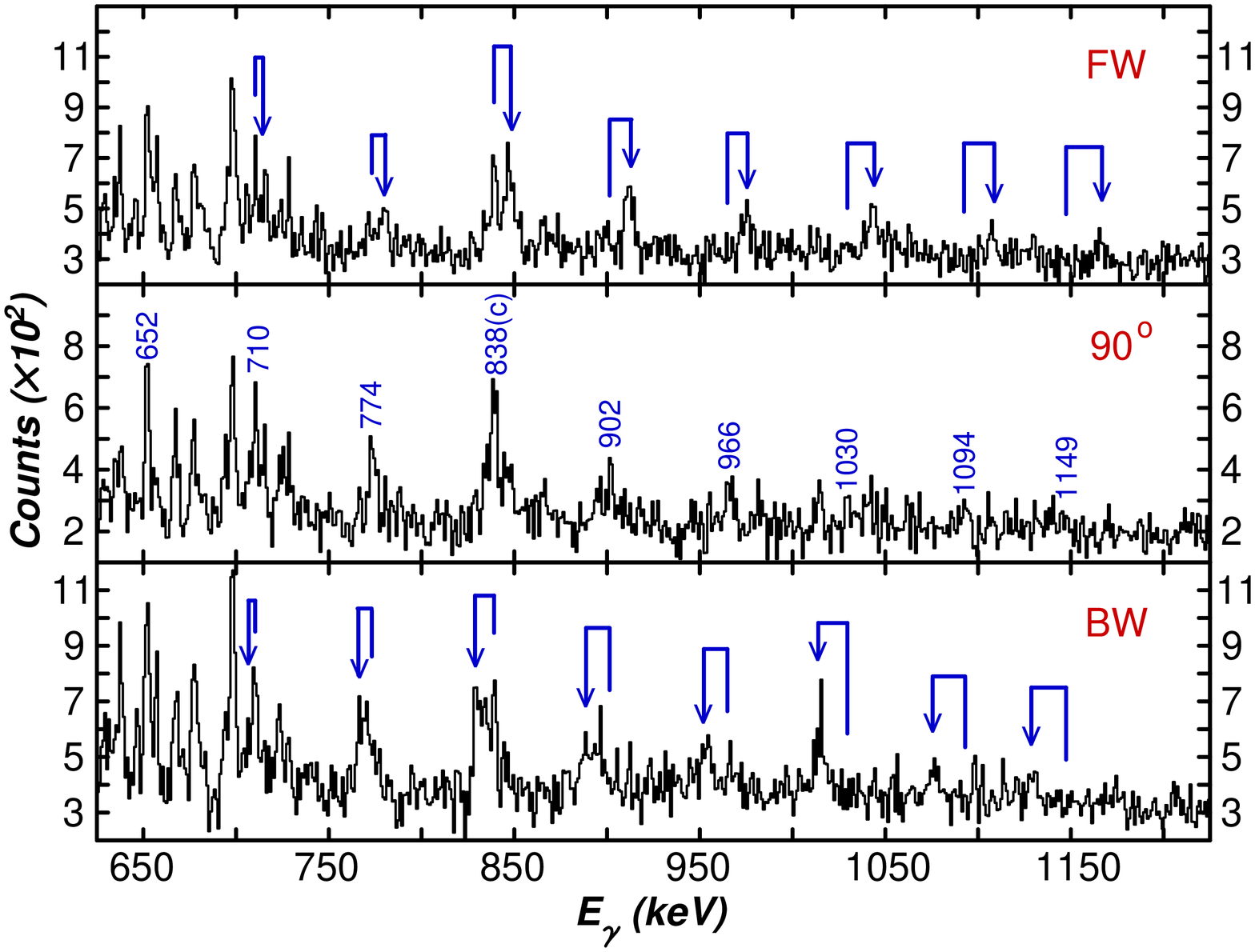}
\caption[Sum of spectra gated on in-band transitions for 
band TSD2.]{Sum of spectra gated on in-band transitions for 
band TSD2 at 3 detector angles: 35$^{\circ}$ (FW), 90$^{\circ}$, and 145$^{\circ}$ (BW). 
The positions of unshifted and shifted $\gamma$ rays are marked by energy values 
and arrows, respectively. Note that the transitions labeled by signs in the form of 
``energy value(c)'' are the contaminated lines which need to be taken care of separately 
(see text for details).\label{fig:sp_band4}}
\end{center}
\end{figure}

Proceeding in this careful manner, an optimized spectrum was obtained at each 
detector angle by summing up all clean double-gated coincidence spectra with 
the appropriate gating conditions. These optimized spectra for all four bands 
(presented in Figures~\ref{fig:sp_band1} -- \ref{fig:sp_band4}) were used 
to determine the centroid of $\gamma$-ray peaks of interest at each available 
detector angle, except for those corresponding to distinguishable contaminated 
lines. For such transitions, other specific spectra, described above, were used 
instead. In this context, the analysis of band TSD2 proved to be particularly 
challenging as it is the one most affected by the closeness in energy of many 
in-band transitions with either those in band 2 or other contaminant peaks. 
Hence, as can be seen in Figure~\ref{fig:sp_band4}, the peak positions of 
$\gamma$ rays in band TSD2 are the most difficult to determine, compared with 
the ones in the other three bands, because of the relatively poor quality of spectra. 

\begin{figure}
\begin{center}
\includegraphics[angle=270,width=0.95\columnwidth]{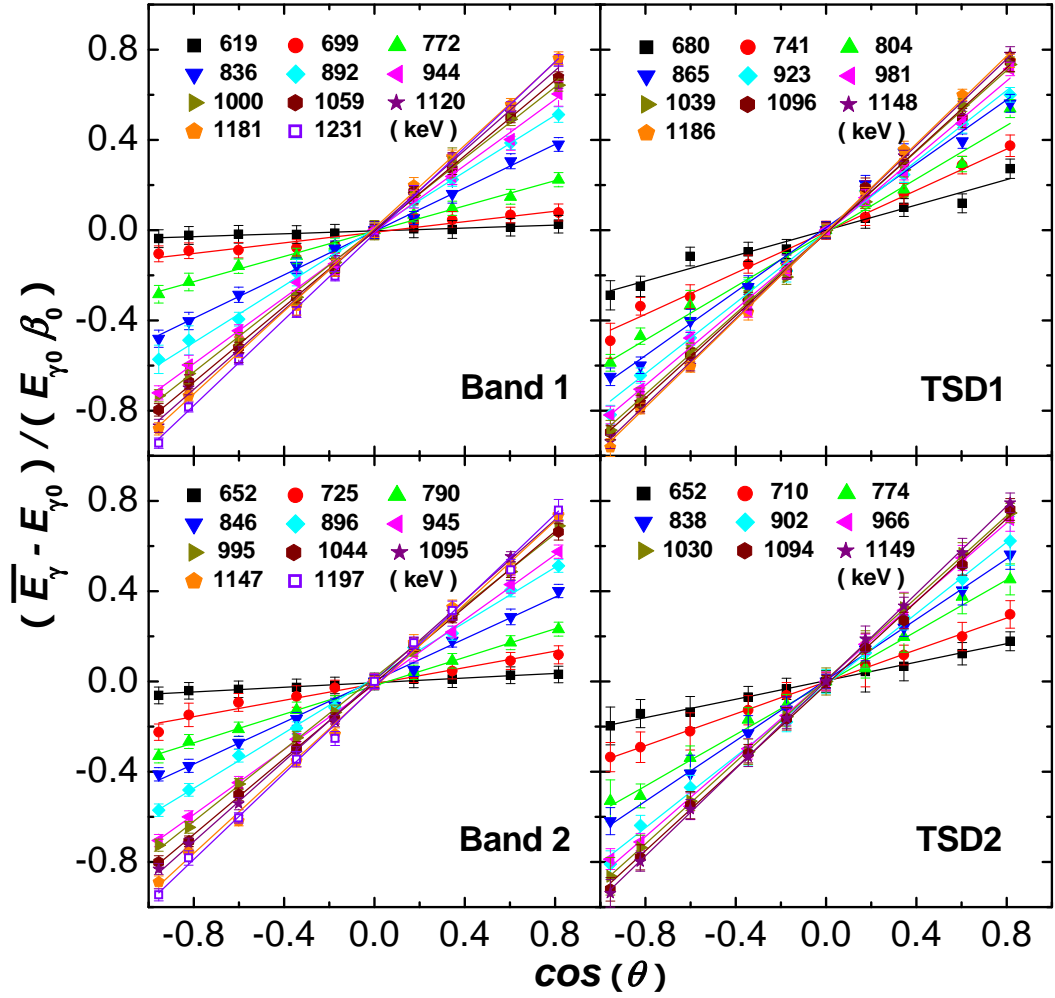}
\caption{Linear fits to the $\gamma$-ray energy shifts 
as a function of $cos(\theta)$ for bands 1, 2, TSD1 and TSD2.\label{fig:bands_1_2_3_4_full_DSfit}} 
\end{center}
\end{figure}

The fractions of full Doppler shift $F(\tau)$ and the associated errors 
were subsequently extracted for transitions in the four $^{163}$Tm bands 
of interest through linear fits of the shifts measured at 10 angles with 
the transformed expression of Eq.~\ref{eq:DopplerEffectGam1stOdApprx} in 
Sec.~\ref{subsec:DSAMtech} of Chapter~\ref{chap:exp_techs}:
\begin{equation}
F(\tau) = \frac{\overline{E_\gamma} - E_{\gamma 0}}{E_{\gamma 0}{\beta_0}\cos(\theta)}.\label{eq:DSftauexps}
\end{equation}
Here, for every transition $E_{\gamma 0}$ is the nominal $\gamma$-ray energy, 
$\overline{E_\gamma}$ is the measured energy at the angle $\theta$, and 
$\beta_0$ is the initial recoil velocity of the $^{163}$Tm residues formed 
in the center of the $^{130}$Te target layer. This quantity was calculated 
to be ${\beta_0}={v_0 / c}=0.021$ with the help of the stopping powers 
computed with the code SRIM 2003~\cite{Ziegler-85-book}. Illustration of the linear fits 
can be seen in Figure~\ref{fig:bands_1_2_3_4_full_DSfit} for bands 1, 2, TSD1 and TSD2. 

\begin{figure}
\begin{center}
\includegraphics[angle=270,width=0.80\columnwidth]{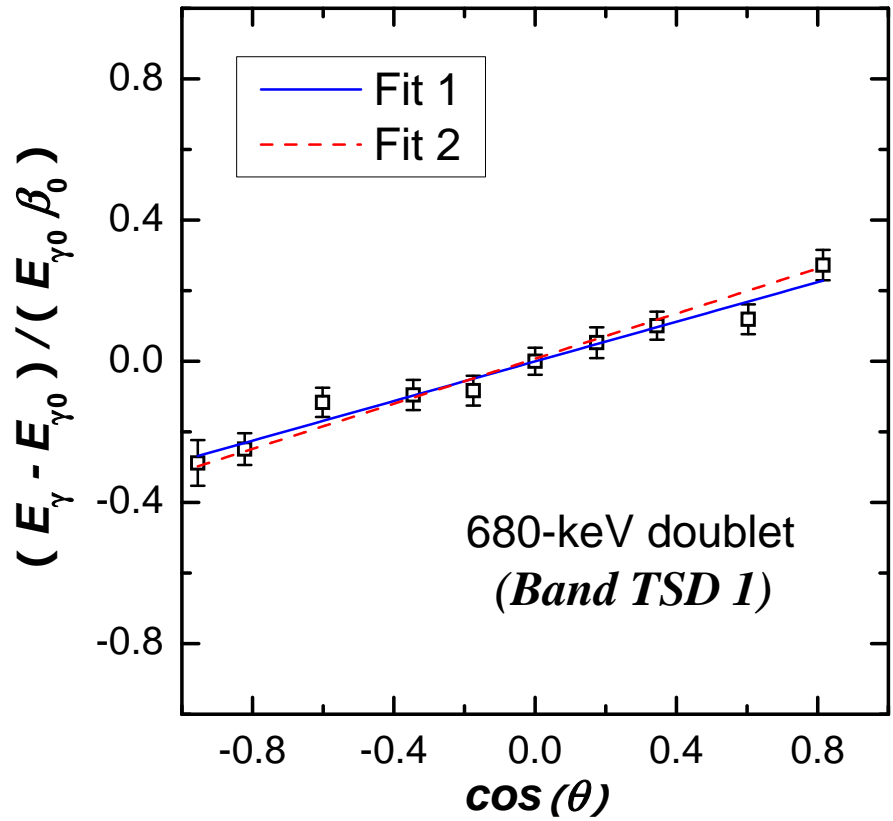}
\caption[Linear fits to the $\gamma$-ray energy shifts as a function 
of $cos(\theta)$ for the 680-$keV$ doublet in band TSD1.]{Linear 
fits to the $\gamma$-ray energy shifts as a function 
of $cos(\theta)$ for the 680-$keV$ doublet in band TSD1. Fit 1 covers all 
data points and Fit 2 considers data points only at three angles ($\theta$): 
35$^{\circ}$, 90$^{\circ}$, and 145$^{\circ}$. See text for details.\label{fig:band3_680keV_doublet}} 
\end{center}
\end{figure}

It has been described earlier in this section that for some of the transitions contaminated by the 
non-distinguishable lines, such as the 680-$keV$ doublet in band TSD1 and the 838-$keV$ $\gamma$ ray in band TSD2, 
the impact from contaminations can be weakened or eliminated through the process of a linear fit. The principle 
of this method is explained in the example of the 680-$keV$ doublet. As can be seen in the spectra for band 
TSD1 (figure~\ref{fig:sp_band3}), the 680-$keV$ transition is characterized by a marked Doppler 
shift clearly visible at the detector angles 
far away from 90$^{\circ}$, $\it{e.g.}$, 35$^{\circ}$ (FW) and 145$^{\circ}$ (BW). In this doublet, the lower-spin 
680-$keV$ line, located in the decay sequence of bands TSD1 and TSD2 towards bands 1 and 2, is a 
``stopped'' (no Doppler shift at every angle) transition, while the higher-spin 680-$keV$ line is an 
in-band transition of band TSD1. The latter is the only possible source of the observed Doppler shift in 
the spectra. Considering both reliability and accuracy, two fits were performed for this transition 
(shown in Figure~\ref{fig:band3_680keV_doublet}), $\it{i.e.}$, a fit (Fit 1) covering the data points at all 
angles and second fit (Fit 2) covering the data points only at three angles ($\theta$): 
35$^{\circ}$ (the largest Doppler shift), 90$^{\circ}$ (always zero Doppler shift for any transition), 
and 145$^{\circ}$ (the corresponding largest Doppler shift at backward angles). The adopted $F(\tau)$ 
value for the 680-$keV$ transition in band TSD1 is the average of the slopes of the two fits, 
while the corresponding error combines the errors 
obtained in both fits and the variation between the two extracted slopes. As a result, the error bars 
for the $F(\tau)$ values for such a contaminated transition are considerably larger than those for the 
neighboring, contaminant-free transitions. This effect is clearly indicated in Figure~\ref{fig:curve}, in which 
the measured $F(\tau)$ values are presented as a function of the $\gamma$-ray energy for 
all four bands. 

A cursory inspection of this figure 
indicates two families of $F(\tau)$ curves: for similar transition 
energies bands 1 and 2 have distinctly smaller $F(\tau)$ values than bands TSD1 and 
TSD2. It is also worth noting that the larger $F(\tau)$ uncertainties 
associated with band TSD2 relate to the difficulty of obtaining 
suitable spectra as discussed above. 

\begin{figure}
\begin{center}
\includegraphics[angle=0,width=0.75\columnwidth]{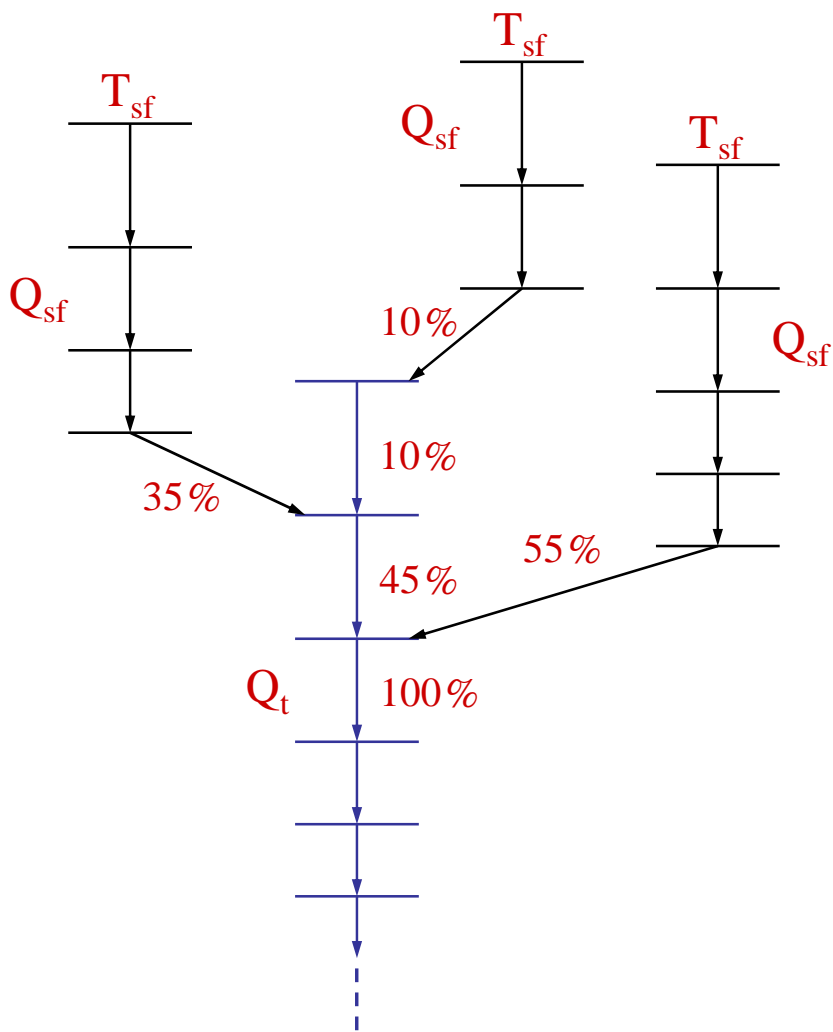}
\caption[A schematic illustration of the model of the cascade used in the 
analysis to determine the quadrupole moments $Q_t$.]{A schematic 
illustration of the model of the cascade used in the 
analysis to determine the quadrupole moments $Q_t$. The marked numbers (as examples) 
represent the relative intensities of associated transitions which 
affect the fit of the measured $F(\tau)$ values. 
Adapted from Ref.~\cite{Nisius-96-thesis}.\label{fig:FtauFitCascdModel}}
\end{center}
\end{figure}

The intrinsic transition quadrupole moments $Q_t$ of the four bands were 
extracted from the measured $F(\tau)$ values using the new Monte Carlo computer 
code MLIFETIME, developed by E. F. Moore at ANL. The model of 
the cascade used in fitting the measured data points 
is illustrated schematically in Figure~\ref{fig:FtauFitCascdModel} as a reference 
for the reader. The following assumptions have been made: (1) all levels in a given 
band have the same transition quadrupole moment $Q_t$; (2) the sidefeeding 
into each level in a band is modeled as a single cascade with a common, 
constant quadrupole moment $Q_{sf}$, and characterized by the same dynamic 
moment of inertia $\Im^{(2)}$ as the main band into which they feed; the 
number of transitions in each sidefeeding band is proportional to the number 
of transitions in the main band above the state of interest; (3) the sidefeeding 
intensity profile and the pattern of the decay out of the main band are 
determined directly from the measured $\gamma$-ray intensities 
within the bands; (4) the internal conversion coefficients ($\alpha$) are also taken into 
account, and (5) a one-step delay at the top of all feeder bands was 
parameterized by a single lifetime $T_{sf}$. 
The lifetimes of the individual states of interest depend on $E_{\gamma}$ (transition energy), 
$I_{\gamma}$ (intensity of the transition), $\alpha$, $Q_t$, $Q_{sf}$ and $T_{sf}$, in which 
$E_{\gamma}$ and $I_{\gamma}$ are the quantities measured directly from the 
spectra. In the simplest case, the 
mean lifetime $T_{\gamma}$ for a particular state, which deexcites only by a $E2$ 
$\gamma$ ray, is:
\begin{equation}
T_{\gamma}=\frac{8.210565{\times}10^6}{{Q_t^2}{E_{\gamma}^5}{{\langle}IK20|(I-2)K{\rangle}}^2},\label{eq:FtFitStatLifetime}
\end{equation}
where $T_{\gamma}$ is in femto-second $(fs)$, $E_{\gamma}$ in $MeV$ and $Q_t$ in 
electron-barn ($eb$). 

In order to determine the average recoil velocity at which the decay from a particular 
state occurs, the detailed slowing-down histories of the recoiling $^{163}$Tm ions in both 
the target and the Au backing were calculated using the SRIM 2003~\cite{Ziegler-85-book} 
Monte Carlo code. The initial positions and velocity vectors for each of 10,000 
starting ions were calculated in a Monte Carlo fashion which included the 
broadening of the recoil cone due to the evaporation of neutrons from the 
$^{167}$Tm compound nucleus. The production cross section was assumed to be 
constant over the range of energies due to the beam slowing down in the target. 
This resulted in an even distribution for the starting positions of the 
$^{163}$Tm ions throughout the target thickness. The initial $^{163}$Tm ion 
positions in the target, ion energies, and recoil direction were supplied as 
input to the SRIM 2003 code, which then transported each ion through the 
target and the backing. The detailed recoil history for each ion was 
written out to a computer file which listed the energy, direction, and position 
at which each collision of the recoiling $^{163}$Tm ions with the target and 
backing atoms occurred. The lifetime code then read in this file and tracked 
each ion history in one femto-second (1 $fs$) time steps from the initial 
formation until the ion came to rest. 

\begin{figure}
\begin{center}
\includegraphics[angle=270,width=0.75\columnwidth]{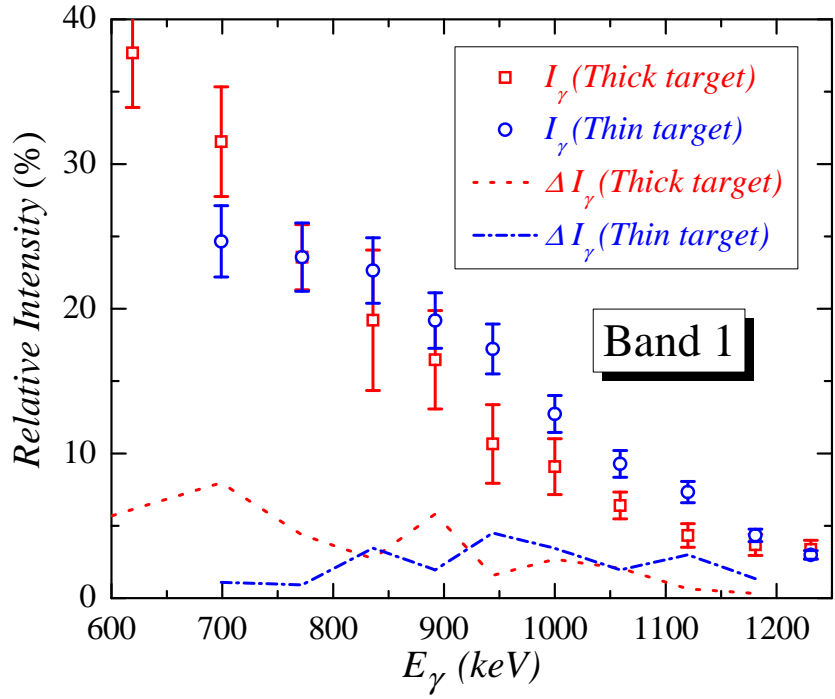}
\caption[The intensity distribution for inband transitions of band 1.]{The 
intensity distribution ($I_{\gamma}$) for inband transitions of band 1 
obtained in our thick-target experiment, compared with the one from the thin-target 
data. The parameter, $\Delta{I}_{\gamma}=I_{\gamma}(J{\rightarrow}J-2)-I_{\gamma}(J+2{\rightarrow}J)$ 
($J$ is the spin of level in band 1), represents the strength of the feeding for this 
band.\label{fig:band1_thin_thick_intens}}
\end{center}
\end{figure}

\begin{figure}
\begin{center}
\includegraphics[angle=270,width=0.75\columnwidth]{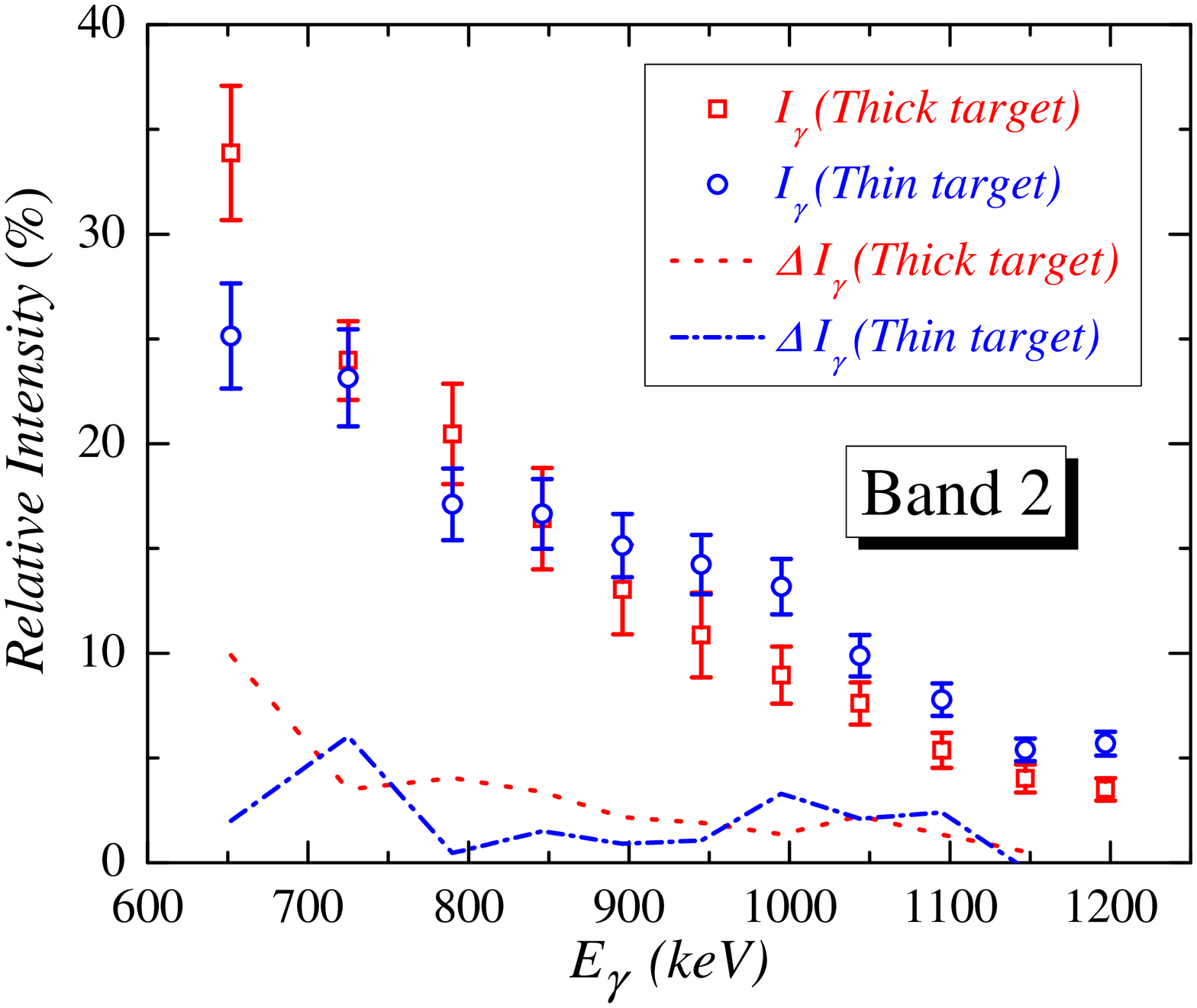}
\caption[The intensity distribution for inband transitions of band 2.]{The 
intensity distribution ($I_{\gamma}$) for inband transitions of band 2 
obtained in our thick-target experiment, compared with the one from the thin-target 
data. The parameter, $\Delta{I}_{\gamma}=I_{\gamma}(J{\rightarrow}J-2)-I_{\gamma}(J+2{\rightarrow}J)$ 
($J$ is the spin of level in band 2), represents the strength of the feeding for this 
band.\label{fig:band2_thin_thick_intens}}
\end{center}
\end{figure}

\begin{figure}
\begin{center}
\includegraphics[angle=270,width=0.75\columnwidth]{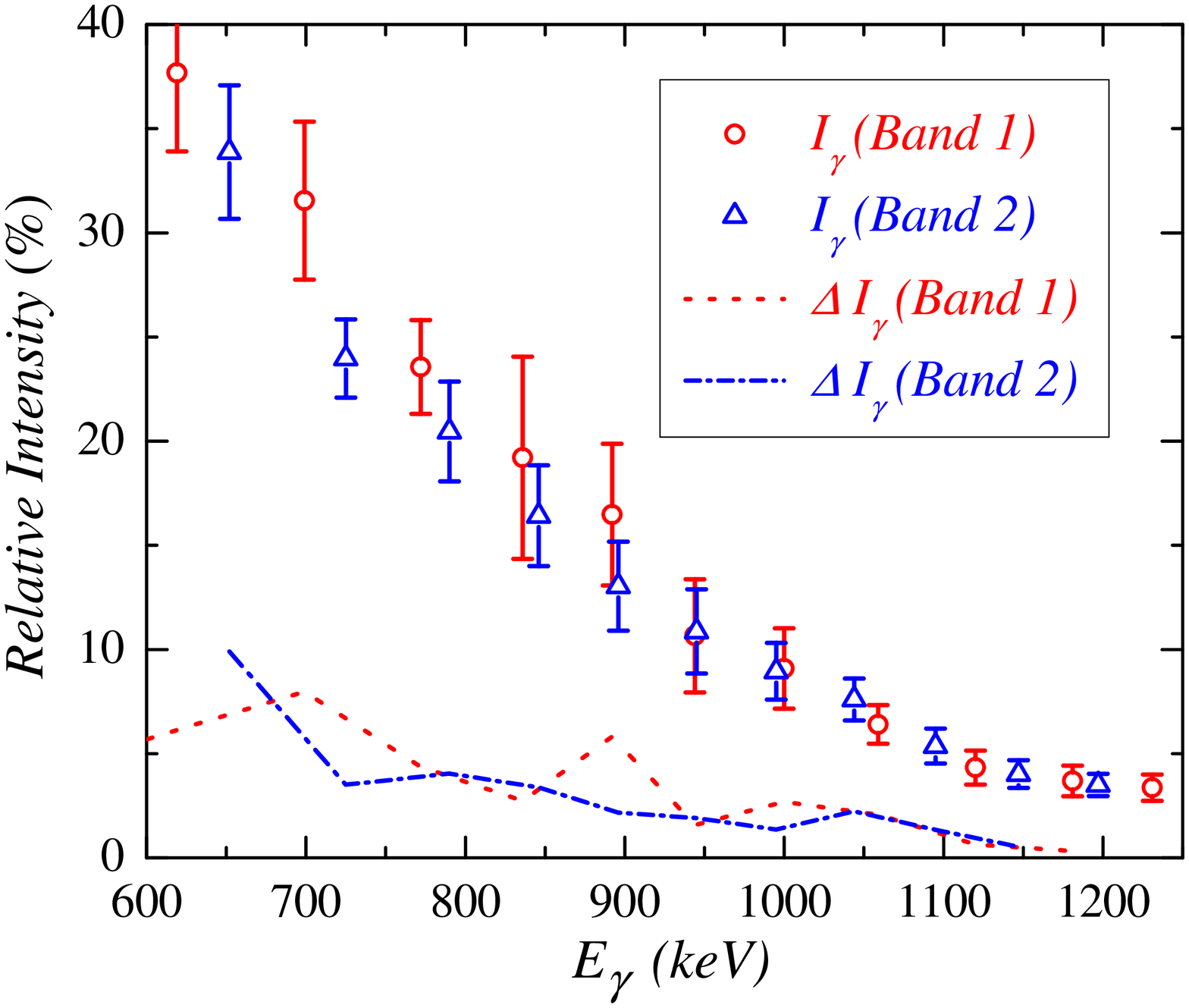}
\caption[The intensity distribution for inband transitions of bands 1 
and 2 obtained in our DSAM experiment.]{The intensity 
distribution ($I_{\gamma}$) for inband transitions of bands 1 
and 2 obtained in our DSAM experiment. The parameter, 
$\Delta{I}_{\gamma}=I_{\gamma}(J{\rightarrow}J-2)-I_{\gamma}(J+2{\rightarrow}J)$ 
($J$ is the spin of level in band 1 or 2), represents the strength of the feeding for 
each band.\label{fig:band1_2_thick_intens}}
\end{center}
\end{figure}

\begin{figure}
\begin{center}
\includegraphics[angle=270,width=0.75\columnwidth]{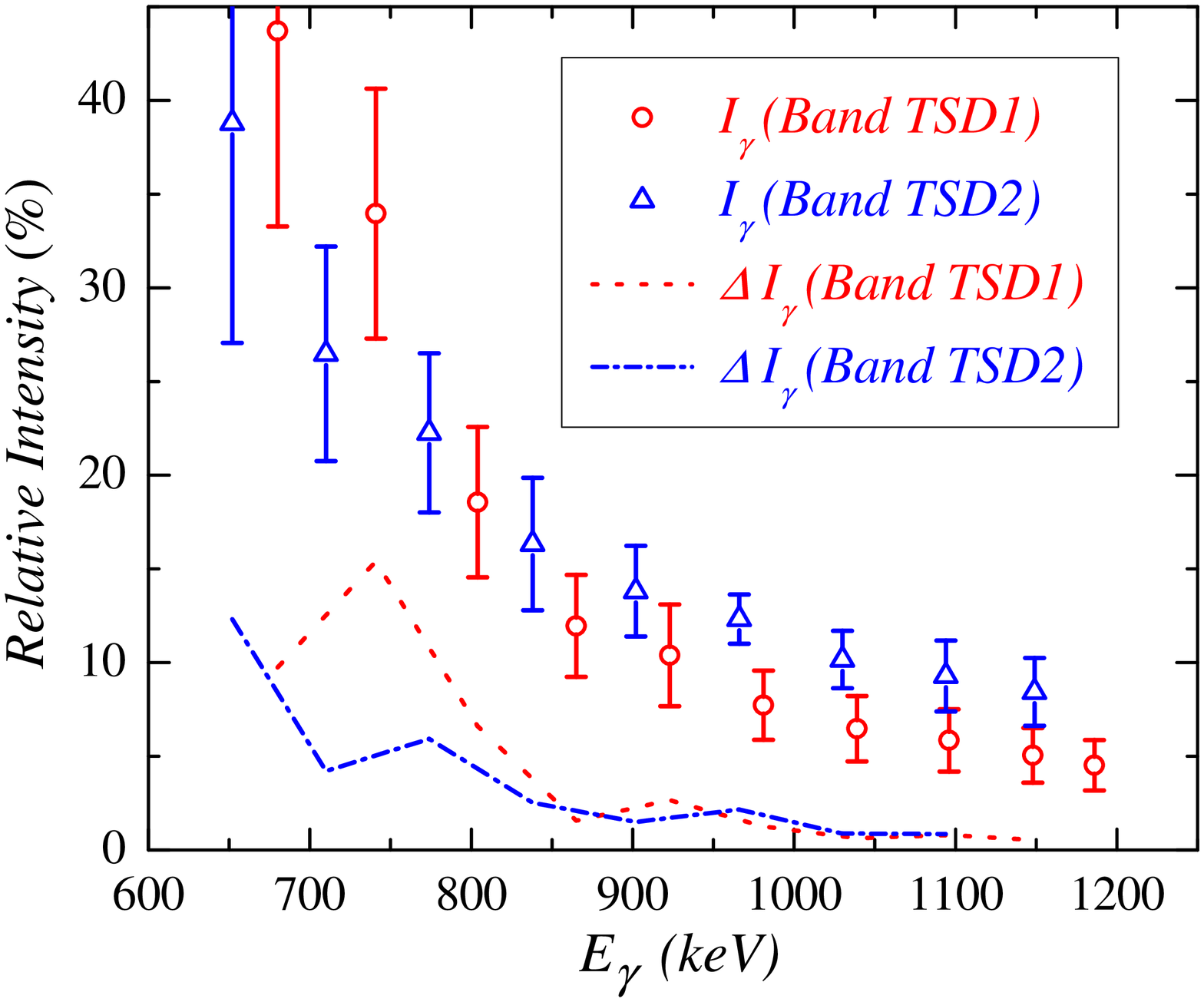}
\caption[The intensity distribution for inband transitions of bands TSD1 
and TSD2 obtained in our DSAM experiment.]{The intensity 
distribution ($I_{\gamma}$) for inband transitions of bands TSD1 
and TSD2 obtained in our DSAM experiment. The parameter, 
$\Delta{I}_{\gamma}=I_{\gamma}(J{\rightarrow}J-2)-I_{\gamma}(J+2{\rightarrow}J)$ 
($J$ is the spin of level in band TSD1 or TSD2), represents the strength of the feeding for 
each band.\label{fig:band3_4_thick_intens}}
\end{center}
\end{figure}

In order to compute the Doppler shifted energies of each $\gamma$ ray emitted by 
the recoiling $^{163}$Tm ions in a Monte-Carlo fashion, feeder bands into each 
state in the main band were randomly populated according to the measured 
intensity distribution, as shown in Tables~\ref{tab:163Tm_band1_intense}, 
\ref{tab:163Tm_band2_intense}, \ref{tab:163Tm_band3_intense} and \ref{tab:163Tm_band4_intense} 
(refer to Figure~\ref{fig:FtauFitCascdModel} for obtaining the relative 
intensity of the sidefeeding). 
%The mean lifetime for each decay step is known, but this is 
%still a radioactive decay process involving several parent-daughter decay steps. 
%The number of nuclei in a state $k$ at time $t$ in the process can be calculated 
%by Eq.~\ref{eq:StateKpopult} in Chapter~\ref{chap:exp_techs}, $\it{i.e.}$, the Bateman equation. 
It is necessary to point out here that the energy of the beam for the thick-target experiment 
(165 $MeV$) is slightly different from that for the thin-target run (170 $MeV$). Hence, in the 
process of fitting $F(\tau)$ values, we adopted the intensity values measured in our DSAM experiment 
instead of the values obtained in the earlier thin-target measurement. In order to validate 
this approach, the intensity distributions for bands 1 and 2 obtained in the thick-target 
experiment were compared with those from the thin-target data, as illustrated in 
Figures~\ref{fig:band1_thin_thick_intens} and \ref{fig:band2_thin_thick_intens}. 
This comparison indicates that the intensity distribution varies slightly between the 
two experiments. It appears that there is slightly more $\gamma$-ray intensity at higher 
angular momentum in the thin-target data, as would be expected since the higher beam 
energy translates in an higher angular momentum in the compound nucleus. Further, the intensity 
distribution for bands 1 and 2, and for bands TSD1 TSD2 are compared in 
Figures~\ref{fig:band1_2_thick_intens} and \ref{fig:band3_4_thick_intens}. 
The populated intensity of bands TSD1 and TSD2 is approximately $5\%$ relative to the 
one of bands 1 and 2 in the DSAM experiment. Since these bands are either signature partners 
(bands 1 and 2) or members of a single TSD family (bands TSD1 and TSD2), the populated intensity 
for one band should be close to that of the partner band, which is a feature exhibited by the 
data as seen in Figures~\ref{fig:band1_2_thick_intens} and \ref{fig:band3_4_thick_intens}. This 
observation provides further support for the validity of the data analysis. 

The subsequent decay profile through the feeder and main band was tracked in 1 $fs$ steps, 
with the decay probability given by the radioactive decay law using the 
$T_{sf}$ parameter and lifetimes generated from each $Q_{sf}$, $Q_t$ parameter 
set (see Eq.~\ref{eq:FtFitStatLifetime}). The velocity vector 
of the $\gamma$ emitting ion was recorded at the 
time of decay of each state of interest. The calculated average fraction of 
the full Doppler shift was generated by accumulating a large number of histories. 
In the present analysis, each of the 10,000 ion histories was used 10 times, 
resulting in better than $1\%$ statistical uncertainty in the 
calculated $F(\tau)$ values. 

\begin{figure}
\begin{center}
\includegraphics[angle=270,width=0.75\columnwidth]{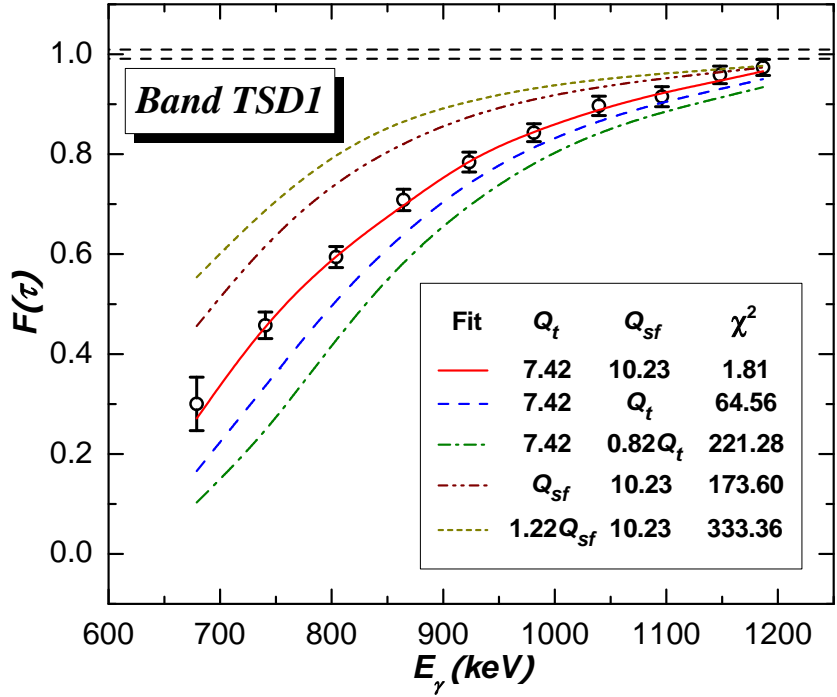}
\caption[The $f(\tau)$ data points for band TSD1 as well as the fit line with the best-fit 
parameters and four other fit curves.]{The $f(\tau)$ 
data points for band TSD1 as well as the fit line with the best-fit 
parameters which resulted from the running of the computer code MINUIT and four fit curves 
with some other values of $Q_t$ and $Q_{sf}$ (either the best-fit $Q_t$ value and various 
$Q_{sf}$ values or various $Q_t$ parameters and the best-fit $Q_{sf}$ value). See 
text for explanation. The two horizontal dashed lines show the range of initial recoil 
velocities within the $^{130}$Te target layer.\label{fig:ftau_b3_fit_argument}}
\end{center}
\end{figure}

\begin{figure}
\begin{center}
\includegraphics[angle=270,width=0.75\columnwidth]{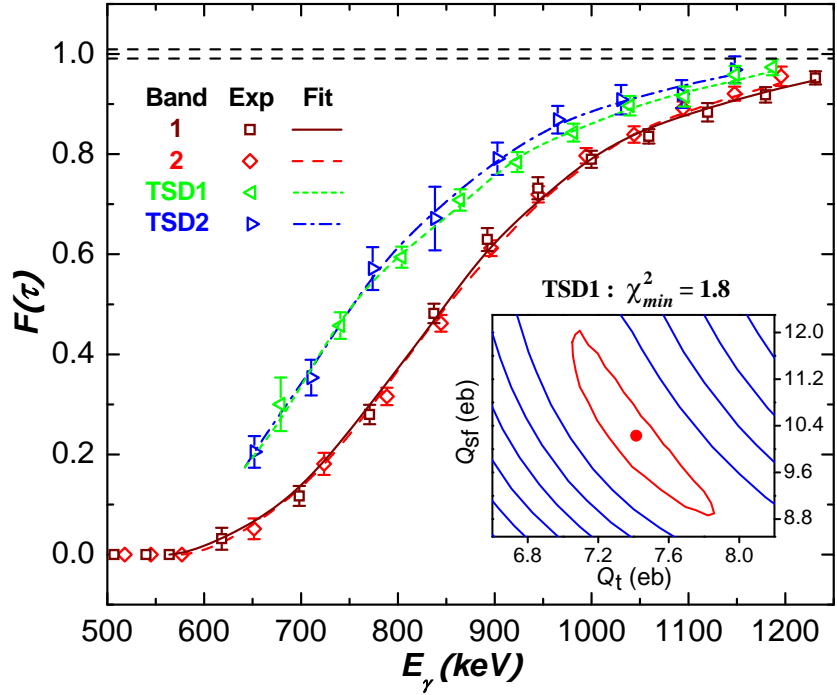}
\caption[Measured $F(\tau)$ values with best-fit curves 
for the bands in $^{163}$Tm.]{Measured $F(\tau)$ values with best-fit curves 
as described in the text for the four bands in $^{163}$Tm. The two 
horizontal dashed lines show the range of initial recoil velocities within 
the $^{130}$Te target layer. Insert: plot of the ${\chi}^2(Q_t,Q_{sf})$ 
surface for band TSD1. The central dot indicates the location of the minimum 
$({\chi}^{2}_{min}=1.8)$, with the first contour plotted in an increment 
of one.\label{fig:curve}}
\end{center}
\end{figure}

\begin{figure}
\begin{center}
\includegraphics[angle=270,width=0.75\columnwidth]{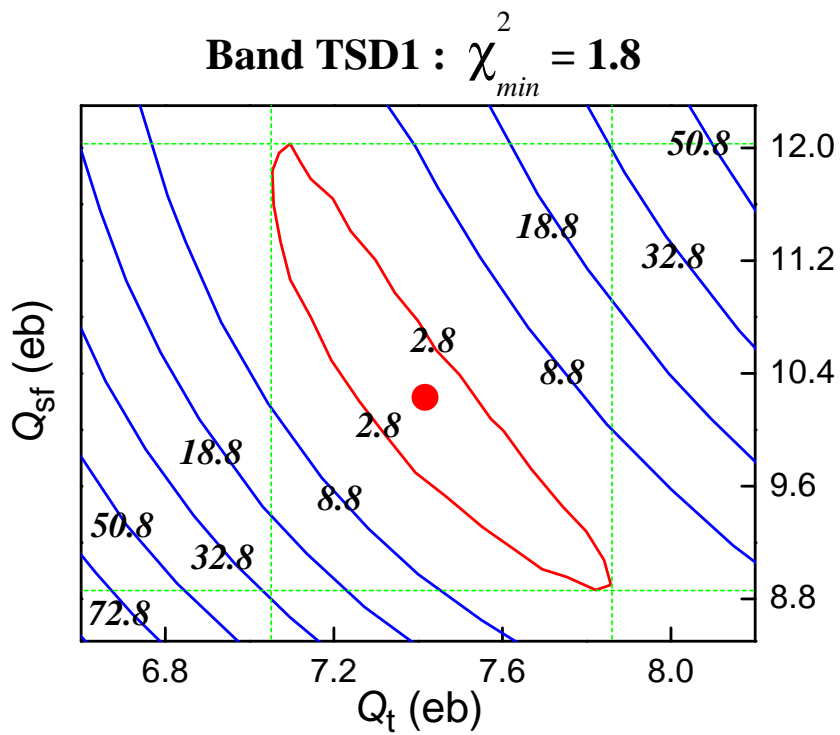}
\caption[An illustration of the method to determine the error bars for the parameters $Q_t$.]{An 
illustration (using the example of band TSD1) of the method 
used in present work to determine the error bars for the parameters $Q_t$ 
and $Q_{sf}$. See text for details.\label{fig:band3_error_determn}}
\end{center}
\end{figure}

A ${\chi}^2$ minimization using the computer code MINUIT with the fit parameters 
$Q_t$, $Q_{sf}$ and $T_{sf}$ was performed to the measured $F(\tau)$ values 
for the four bands. The results of the fitting process are summarized 
in Table~\ref{tab:fitmoments}, where the quoted errors include the 
covariance between the fit parameters. In order to illustrate the sensitivity 
of the $F(\tau)$ values to the values of the parameters $Q_t$ and $Q_{sf}$, 
the data points of $F(\tau)$ for band TSD1 are compared to the calculated values 
with the best-fit parameters as well as to those with other values of $Q_t$ and 
$Q_{sf}$ in Figure~\ref{fig:ftau_b3_fit_argument}. The four fit lines with 
either the best-fit $Q_t$ value and various $Q_{sf}$ values or with various 
$Q_t$ parameters and the best-fit $Q_{sf}$ value clearly lie away from data points, and the 
correponding $\chi^{2}$ values are large. In contrast, the curve with the best-fit parameters 
agrees well with almost all data points and the corresponding $\chi^{2}$ value reaches its 
minimum. As can be seen from Figure~\ref{fig:curve}, the fit of the $F(\tau)$ data is satisfactory 
in all cases. This is illustrated further in the case of band TSD1 in the insert to Figure~\ref{fig:curve}, 
where contours of ${\chi}^2$ values are presented in a $(Q_t,Q_{sf})$ plane and a clear minimum 
can be seen. The corresponding error bars are determined by allowing the total $\chi^{2}$ of the 
fit to increase by 1 relative to its minimum ($\chi_{min}^2$) and projecting the contour of 
($\chi_{min}^2+1$) onto the $Q_{t}$ and $Q_{sf}$ axes~\cite{Bevington-92-book}. This determination 
of the error bars is illustrated in Figure~\ref{fig:band3_error_determn} for the case of 
band TSD1. 

\begin{table}
\begin{center}
\caption[THE RELATIVE TOTAL INTENSITY AND ERROR OF EACH TRANSITION OF INTEREST IN 
BAND 1 OF $^{163}$Tm]{THE RELATIVE TOTAL INTENSITY ($\it{i.e.}$, CORRECTED FOR EFFICIENCY AND INTERNAL CONVERSION) AND 
ERROR OF EACH TRANSITION OF INTEREST IN BAND 1 OF $^{163}$Tm, MEASURED IN THE DSAM 
EXPERIMENT\label{tab:163Tm_band1_intense}}
\begin{tabular}{ccc}
\hline \hline 
$E_{\gamma}$~($keV$) & Intensity~($\%$) & Error~($\%$)\\ \hline
619 & 38 & 4\\
699 & 32 & 4\\
772 & 24 & 2\\
836 & 19 & 5\\
892 & 16 & 3\\
944 & 11 & 3\\
1000 & 9 & 2\\
1059 & 6.4 & 0.9\\
1120 & 4.3 & 0.8\\
1181 & 3.7 & 0.7\\
1231 & 3.4 & 0.6\\
\hline 
\end{tabular}
\end{center}
\end{table}

\begin{table}
\begin{center}
\caption[THE RELATIVE TOTAL INTENSITY AND ERROR OF EACH TRANSITION OF INTEREST IN 
BAND 2 OF $^{163}$Tm]{THE RELATIVE TOTAL INTENSITY ($\it{i.e.}$, CORRECTED FOR EFFICIENCY AND INTERNAL CONVERSION) AND 
ERROR OF EACH TRANSITION OF INTEREST IN BAND 2 OF $^{163}$Tm, MEASURED IN THE DSAM 
EXPERIMENT\label{tab:163Tm_band2_intense}}
\begin{tabular}{ccc}
\hline \hline 
$E_{\gamma}$~($keV$) & Intensity~($\%$) & Error~($\%$)\\ \hline
652 & 34 & 3\\
725 & 24 & 2\\
790 & 20 & 2\\
846 & 16 & 2\\
896 & 13 & 2\\
945 & 11 & 2\\
995 & 9 & 1\\
1044 & 7.6 & 1.0\\
1095 & 5.4 & 0.8\\
1147 & 4.0 & 0.7\\
1197 & 3.5 & 0.5\\
\hline 
\end{tabular}
\end{center}
\end{table}

\begin{table}
\begin{center}
\caption[THE RELATIVE TOTAL INTENSITY AND ERROR OF EACH TRANSITION OF INTEREST IN 
BAND TSD1 OF $^{163}$Tm]{THE RELATIVE TOTAL INTENSITY ($\it{i.e.}$, CORRECTED FOR EFFICIENCY AND INTERNAL CONVERSION) AND 
ERROR OF EACH TRANSITION OF INTEREST IN BAND TSD1 OF $^{163}$Tm, MEASURED IN THE DSAM 
EXPERIMENT\label{tab:163Tm_band3_intense}}
\begin{tabular}{ccc}
\hline \hline 
$E_{\gamma}$~($keV$) & Intensity~($\%$) & Error~($\%$)\\ \hline
680 & 44 & 10\\
741 & 34 & 7\\
804 & 19 & 4\\
865 & 12 & 3\\
923 & 10 & 3\\
981 & 8 & 2\\
1039 & 7 & 2\\
1096 & 6 & 2\\
1148 & 5 & 2\\
1186 & 5 & 1\\
\hline 
\end{tabular}
\end{center}
\end{table}

\begin{table}
\begin{center}
\caption[THE RELATIVE TOTAL INTENSITY AND ERROR OF EACH TRANSITION OF INTEREST IN 
BAND TSD2 OF $^{163}$Tm]{THE RELATIVE TOTAL INTENSITY ($\it{i.e.}$, CORRECTED FOR EFFICIENCY AND INTERNAL CONVERSION) AND 
ERROR OF EACH TRANSITION OF INTEREST IN BAND TSD2 OF $^{163}$Tm, MEASURED IN THE DSAM 
EXPERIMENT\label{tab:163Tm_band4_intense}}
\begin{tabular}{ccc}
\hline \hline 
$E_{\gamma}$~($keV$) & Intensity~($\%$) & Error~($\%$)\\ \hline
652 & 39 & 12\\
710 & 26 & 6\\
774 & 22 & 4\\
838 & 16 & 4\\
902 & 14 & 2\\
966 & 12 & 1\\
1030 & 10 & 2\\
1094 & 9 & 2\\
1149 & 8 & 2\\
\hline 
\end{tabular}
\end{center}
\end{table}

\begin{table}
\begin{center}
\caption[SUMMARY OF QUADRUPOLE MOMENTS RESULTING FROM DSAM CENTROID 
SHIFT ANALYSIS FOR THE 4 BANDS IN $^{163}$Tm]{SUMMARY 
OF QUADRUPOLE MOMENTS RESULTING FROM DSAM CENTROID 
SHIFT ANALYSIS FOR THE 4 BANDS IN $^{163}$Tm. IN ALL CASES THE VALUE 
OF $T_{sf}$ IS VERY SMALL, $\it{i.e.}$, $T_{sf}~{\sim}~1~fs$. THE ERROR BARS ARE 
STATISTICAL ONLY, $\it{i.e.}$, THEY DO NOT INCLUDE THE $\sim$ $15\%$ ERROR ASSOCIATED WITH 
THE SYSTEMATIC UNCERTAINTY IN THE STOPPING POWERS (SEE TEXT FOR DETAILS)\label{tab:fitmoments}}
\begin{tabular}{cccc}
\hline \hline 
Band & $Q_t~(eb)$ & $Q_{sf}~(eb)$ & ${\chi}^{2}_{min}$\\ \hline
1 & $6.4^{+0.6}_{-0.3}$ & $6.7^{+0.7}_{-0.8}$ & 6.51\\
2 & $6.4^{+0.3}_{-0.3}$ & $7.0^{+0.9}_{-0.6}$ & 8.01\\
TSD1 & $7.4^{+0.4}_{-0.4}$ & $10.2^{+1.8}_{-1.3}$ & 1.81\\
TSD2 & $7.7^{+1.0}_{-0.6}$ & $9.7^{+2.9}_{-2.3}$ & 1.15\\
\hline 
\end{tabular}
\end{center}
\end{table}

\subsection{\label{subsec:163Tm_life_discuss}Interpretation and discussion}

Before discussing the significance of the difference in the measured $Q_t$ values for 
bands 1 and 2, on the one hand, and the TSD1 and TSD2 sequences on the other, it is worth 
to examine the relevance of the results through a comparison with other nuclei in the 
region. Since bands 1 and 2 are based on the $[523]7/2^{-}$ configuration, a search of 
the literature was undertaken for quadrupole moment measurements of this configuration 
in neighboring nuclei. The results are given in Table~\ref{tab:NDcompare}. The 
$[523]7/2^{-}$ configuration is yrast in $^{163,165}$Ho, and, perhaps more importantly, 
in $^{163}$Lu, one of the isotopes where TSD and wobbling bands are known as well. It 
should be noted that the $^{163}$Lu yrast sequence had first been associated with the 
$[514]9/2^{-}$ configuration~\cite{Schmitz-NPA-539-112-92}. However, following the work of 
Ref.~\cite{Schmitz-PLB-303-230-93}, the $[523]7/2^{-}$ configuration was adopted on the basis of 
the $B(E2)$ and $B(M1)$ transition probabilities deduced from the measured lifetimes 
and branching ratios. 

\begin{table}
\begin{center}
\caption[QUADRUPOLE MOMENTS OF ND BANDS IN Tm, Ho, AND Lu NUCLEI]{QUADRUPOLE 
MOMENTS OF ND BANDS BASED ON THE $[523]7/2^{-}$ CONFIGURATION 
IN Tm, Ho, AND Lu NUCLEI. THE LAST COLUMN PROVIDES THE REFERENCE AND IDENTIFIES THE 
METHOD USED TO MEASURE THE MOMENTS BY THE FOLLOWING SYMBOLS: FT - DSAM $F(\tau)$; 
LS - DSAM LINE SHAPE; RD - RECOIL DISTANCE; LRIMS - LASER RESONANCE IONIZATION; 
KaX - KAONIC X-RAY; PiX - PIONIC X-RAY; MuX - MUONIC X-RAY. THE ERROR BARS 
ARE STATISTICAL ONLY AND DO NOT INCLUDE THE SYSTEMATIC UNCERTAINTY IN THE 
STOPPING POWERS. NOTE THAT FOR SOME ENTRIES IN THE TABLE, A RANGE OF VALUES IS GIVEN. 
THE READER IS REFERRED TO THE CITED REFERENCE FOR FURTHER DETAILS\label{tab:NDcompare}}
\begin{tabular}{cccc}
\hline \hline 
Nuclide & Band & $Q_t~(eb)$ & Method~[REF]\\ \hline
$^{163}$Tm & 1 & $6.4^{+0.6}_{-0.3}$ & FT~[present work]\\
$^{163}$Tm & 2 & $6.4^{+0.3}_{-0.3}$ & FT~[present work]\\
$^{163}$Ho & ND & $6.78\pm1.13$ & LRIMS~\cite{Alkh-NPA-504-549-89}\\
$^{165}$Ho & ND1 & $6.42\pm0.15$, $6.78\pm0.04$ & KaX, PiX~\cite{Batty-NPA-355-383-81}\\
 & & $6.74\pm0.04$ & PiX~\cite{Olani-NPA-403-572-83}\\
 & & $6.57\pm0.06$ & MuX~\cite{Powers-NPA-262-493-76}\\
$^{165}$Ho & ND2 & $5.76\pm0.07$ & MuX~\cite{Powers-NPA-262-493-76}\\
$^{163}$Lu & ND1 & $4.88^{+1.36}_{-0.68}$ -- $6.78^{+2.66}_{-1.39}$ & LS + RD~\cite{Schmitz-PLB-303-230-93}\\
$^{163}$Lu & ND2 & $2.13^{+0.62}_{-0.43}$ -- $6.72^{+0.77}_{-0.40}$ & LS + RD~\cite{Schmitz-PLB-303-230-93}\\
\hline 
\end{tabular}
\end{center}
\end{table}

As can be seen from Table~\ref{tab:NDcompare}, the $Q_t$ moments have been obtained 
using a number of techniques ranging from the analysis of $F(\tau)$ values, such as 
those in the present work, and full line shape analyses of data taken using the 
DSAM technique, to measurements with the recoil distance method ($\it{e.g.}$, so-called 
plunger data), and even to laser resonance ionization as well as detection of the 
characteristic $X$ rays of kaonic, pionic or muonic atoms. It can be concluded from 
Table~\ref{tab:NDcompare} that the moments measured in the present work for bands 
1 and 2 ($Q_t$ $\sim$ $6.4~eb$) are in good agreement with those reported for the same 
configuration in the literature. This observation provides further confidence in the 
analysis presented above. 

The $Q_t$ moments of bands 1 and 2 can then also be compared with the calculations first 
outlined in the earlier $^{163}$Tm spectroscopy work~\cite{Pattabi-PLB-163Tm-07}. These predict the value 
to be $Q_t$ = $5.8~eb$ at spin $I$ = 30, with an associated axial quadrupole deformation 
of ${\epsilon}_2$ = 0.21. Considering the fact that the errors quoted for the 
$Q_t$ moments in Table~\ref{tab:fitmoments} are statistical only and do not include 
the additional systematic error of $\sim$ $15\%$ due to the uncertainties in the stopping 
powers, the agreement between experiment and theory can be considered as satisfactory. 
Nevertheless, the fact remains that deformations calculated with the Cranked Nilsson-Strutinsky 
(CNS) model~\cite{Bengtsson-NPA-436-14-85}, the Tilted-Axis Cranking (TAC) model~\cite{Frauendorf-NPA-677-115-00} 
or the Ultimate Cranker (UC) code~\cite{Bengtsson-UC-code-URL}, all using the same Nilsson potential, 
tend to be systematically somewhat smaller than the values derived from experiment, 
an observation that warrants further theoretical investigation. 

The present data clearly indicate that the deformation 
associated with bands TSD1 and TSD2 is larger 
than that of the yrast structure: as can be seen from Table~\ref{tab:fitmoments}, 
the $Q_t$ moments of bands TSD1 and TSD2 ($\sim$ $7.5~eb$) exceed those for bands 
1 and 2 by $\sim$ $1~eb$. The larger deformation agrees with the interpretation 
proposed in the earlier work~\cite{Pattabi-PLB-163Tm-07}. However, the magnitude of the increase in $Q_t$ 
moments is not reproduced as the TAC calculations indicate a transitional 
quadrupole moment increasing slightly from $8.7~eb$ at spin $I$ = 24 to $9.6~eb$ 
for 34 $<$ $I$ $<$ 50. At present, this discrepancy between data and calculations 
is not understood. It is, however, not unique to $^{163}$Tm. Table~\ref{tab:TSDcompare} 
compares $Q_t$ moments for TSD bands in all nuclei of the region where this information 
is available. Just as was the case above, the systematic uncertainty associated with 
the stopping powers has been ignored. Nevertheless, three rather striking observations 
can be made from Table~\ref{tab:TSDcompare}: (1) the $Q_t$ values for the TSD bands in 
$^{163}$Lu and $^{163}$Tm are essentially the same, (2) the $Q_t$ moments of the TSD 
bands decrease from $^{163}$Lu and $^{163}$Tm to $^{165}$Lu, an observation already 
made for Lu isotopes in Refs.~\cite{Schonwa-EPJA-13-291-02,Schonwa-EPJA-15-435-02}, and (3) all the TSD bands in 
Hf nuclei are characterized by $Q_t$ moments that are larger than those in Lu and Tm 
by $\sim$ 4 -- $6~eb$, possibly pointing to a rather different nature for these bands. 
Just as in the present $^{163}$Tm case, a discrepancy between the measured and 
calculated $Q_t$ moments was found for the Lu isotopes: the UC calculations predicted 
values of $Q_t$ $\sim$ $9.2~eb$ and $11.5~eb$ for positive and negative values of the 
deformation parameter $\gamma$, and these values were computed to be essentially 
the same for the three Lu isotopes ($A$ = 163, 164, 165)~\cite{Schonwa-EPJA-13-291-02,Schonwa-EPJA-15-435-02}, 
but with the configuration associated with a rotation about the short axis 
($\gamma$ $>$ 0) being lower in energy; while the calculations in the TAC model gave the similar $Q_t$ 
moments ($\sim$ 9.7 -- $10.3~eb$) for the TSD bands in $^{163}$Lu. As stated above, the physical 
origin of the discrepancy between theory and experiment is at present 
unclear, although it was pointed out in Refs.~\cite{Schonwa-EPJA-13-291-02,Schonwa-EPJA-15-435-02} that the 
exact location in energy of the $i_{13/2}$ and $h_{9/2}$ proton- and $i_{11/2}$ 
neutron-intruder orbitals is crucial for the deformation. These orbitals are 
deformation driving and, hence, might have a considerable impact on the $Q_t$ 
moments. It is possible that the use of the standard Nilsson potential parameters, 
questioned above for normal deformed configurations, needs also to be reconsidered 
for the precise description of TSD bands. 

\begin{table}
\begin{center}
\caption[QUADRUPOLE MOMENTS OF TSD BANDS IN Tm, Lu, AND Hf NUCLEI]{QUADRUPOLE 
MOMENTS OF TSD BANDS IN Tm, Lu, AND Hf NUCLEI. 
THE LAST COLUMN PROVIDES THE REFERENCE AND IDENTIFIES THE METHOD 
USED TO MEASURE THE MOMENTS BY THE FOLLOWING SYMBOLS: 
FT - DSAM $F(\tau)$; LS - DSAM LINE SHAPE. THE ERROR BARS 
ARE STATISTICAL ONLY AND DO NOT INCLUDE THE SYSTEMATIC UNCERTAINTY IN THE 
STOPPING POWERS. NOTE THAT FOR SOME ENTRIES IN THE TABLE, A RANGE OF 
VALUES IS GIVEN. THE READER IS REFERRED TO THE CITED REFERENCE FOR 
FURTHER DETAILS\label{tab:TSDcompare}}
\begin{tabular}{ccccc}
\hline \hline 
Nuclide & Band & $Q_t~(eb)$ & $Q_{sf}~(eb)$ & Method~[REF]\\ \hline
$^{163}$Tm & TSD1 & $7.4^{+0.4}_{-0.4}$ & $10.2^{+1.8}_{-1.3}$ & FT~[present work]\\
$^{163}$Tm & TSD2 & $7.7^{+1.0}_{-0.6}$ & $9.7^{+2.9}_{-2.3}$ & FT~[present work]\\
$^{163}$Lu & TSD1 & $7.4^{+0.7}_{-0.4}$, $7.7^{+2.3}_{-1.3}$ 
& $6.7^{+0.7}_{-0.7}$, $7.0^{+0.7}_{-0.7}$ & FT~\cite{Schonwa-EPJA-15-435-02}\\
 & & $7.63^{+1.46}_{-0.88}$ -- $9.93^{+1.14}_{-0.99}$ & & LS~\cite{Gorgen-PRC-69-031301-04}\\
$^{163}$Lu & TSD2 & $6.68^{+1.70}_{-1.02}$ -- $8.51^{+0.95}_{-0.73}$ & & LS~\cite{Gorgen-PRC-69-031301-04}\\
$^{164}$Lu & TSD1 & $7.4^{+2.5}_{-1.3}$ & $6.7^{+0.7}_{-0.7}$ & FT~\cite{Schonwa-EPJA-15-435-02}\\
$^{165}$Lu & TSD1 & $6.0^{+1.2}_{-0.2}$, $6.4^{+1.9}_{-0.7}$ 
& $5.4^{+0.5}_{-0.5}$, $5.8^{+0.6}_{-0.6}$ & FT~\cite{Schonwa-EPJA-15-435-02}\\
$^{167}$Lu & TSD1 & $6.9^{+0.3}_{-0.3}$ & $4.4^{+0.4}_{-0.2}$ & FT~\cite{Gurdal-JPG-31-S1873-05}\\
$^{168}$Hf & TSD1 & $11.4^{+1.1}_{-1.2}$ & $10.5^{+1.7}_{-1.6}$ & FT~\cite{Amro-PLB-506-39-01}\\
$^{174}$Hf & TSD1 & $13.8^{+0.3}_{-0.4}$ & $8.4^{+0.3}_{-0.3}$ & FT~\cite{Hartley-PLB-608-31-05}\\
$^{174}$Hf & TSD2 & $13.5^{+0.2}_{-0.3}$ & $8.0^{+0.3}_{-0.2}$ & FT~\cite{Hartley-PLB-608-31-05}\\
$^{174}$Hf & TSD3 & $13.0^{+0.8}_{-0.4}$ & $10.3^{+0.6}_{-0.8}$ & FT~\cite{Hartley-PLB-608-31-05}\\
$^{174}$Hf & TSD4 & $12.6^{+0.8}_{-0.8}$ & $10.2^{+1.6}_{-1.3}$ & FT~\cite{Hartley-PLB-608-31-05}\\
\hline 
\end{tabular}
\end{center}
\end{table}

In Ref.~\cite{Schonwa-EPJA-13-291-02} it was argued that the fact that the measured $Q_t$ moments 
in the $^{163}$Lu TSD bands are smaller than the calculated values points towards 
a positive $\gamma$ deformation because the latter is associated with the smaller 
computed moments. As already discussed in the earlier $^{163}$Tm paper~\cite{Pattabi-PLB-163Tm-07} 
as well as in the discussion above, the same conclusion 
can not be drawn in the case of $^{163}$Tm. Indeed, TAC calculations, which 
do not restrict the orientation of rotational axis to one of the principal axes, 
point to a tilted solution that smoothly connects two minima of opposite sign in 
$\gamma$ deformation. The average deformation parameters are ${\epsilon}_2$ = 0.39, 
$|\gamma|$ = 17$^{\circ}$. For $I$ $>$ 23 the angular 
momentum vector gradually moves away from the intermediate axis ($\gamma$ $<$ 0) 
toward the short one ($\gamma$ $>$ 0), without reaching the latter by $I$ = 50. 

The TAC calculations of Ref.~\cite{Pattabi-PLB-163Tm-07} have also 
been extended to the case of $^{163}$Lu and the computed $Q_t$ moments for the TSD 
bands are larger than the measured ones, in agreement with the general findings 
discussed above. These $Q_t$ moments in $^{163}$Lu were also found to decrease slightly 
from $10.3~eb$ at $I$ = 20 to $9.7~eb$ at $I$ = 40 just as in $^{163}$Tm. Moreover, 
the $^{163}$Lu values are also somewhat larger than the corresponding ones in $^{163}$Tm, 
reflecting the additional drive towards larger deformation brought about 
by the $i_{13/2}$ proton orbital which is occupied in this case. However, it 
should be pointed out that within the framework of these 
calculations~\cite{Pattabi-PLB-163Tm-07}, the occupation of the $i_{13/2}$ 
proton orbital is not a necessary condition to achieve a TSD minimum. Rather, 
the deformation is driven mainly by the $N$ = 94 neutron gap (the same point 
has been indicated in the earlier work~\cite{Pattabi-PLB-163Tm-07}), as is 
illustrated in Figure~\ref{fig:163Tm_neutron_routhian}, where the single-neutron 
routhians are presented and the large $N=94$ gap associated with the TSD 
shapes at positive and negative $\gamma$ values is clearly visible. The 
corresponding single-proton routhians can be seen in 
Figure~\ref{fig:163Tm_proton_routhian}. The occupation of the $i_{13/2}$ proton 
level in the Lu isotopes adds an additional degree of shape driving towards 
larger deformation. However, as stated above, the data indicate that its 
impact is rather minor. This is borne out by the calculations where average 
deformations of ${\epsilon}_2=0.39$, $|\gamma|=17^{\circ}$ for $^{163}$Tm should 
be compared with computed values of ${\epsilon}_2=0.41$, ${\gamma}=+19^{\circ}$ 
for $^{163}$Lu. The nearly equal deformations find their origin in the 
following physical effect: $^{163}$Lu does not make full use of the $N=94$ 
gap, because it has two fewer neutrons, but this absence is compensated by 
the additional drive provided by the $i_{13/2}$ proton. As argued in the earlier 
paper~\cite{Pattabi-PLB-163Tm-07}, the large $N=94$ gap makes it unlikely 
that the $^{163}$Tm TSD bands involve a three-quasiparticle structure with 
a proton coupled to a neutron particle-hole excitation. 
The possibility that these bands correspond to configurations with the 
odd proton occupying the $[541]1/2^{-}$ level (labeled as $h_{9/2}$ in 
Figure~\ref{fig:163Tm_proton_routhian}) is also unlikely. This 
orbital is characterized by a large signature splitting and small 
$B(M1)$ values for inter-band transitions, in clear contradiction 
with the $^{163}$Tm data~\cite{Pattabi-PLB-163Tm-07}. While it is possible that 
combining the occupation of the $[541]1/2^{-}$ level with a neutron particle-hole 
excitation would alter these observables, it would result in an excitation 
energy much larger than seen experimentally because of the large $N=94$ gap. 
Furthermore, as can be seen in Figure~\ref{fig:163Tm_proton_routhian}, 
there are no other low-lying proton excitations that lead to small signature splitting. 

Finally, it is worth noting that the values of the $Q_{sf}$ moments associated 
with the sidefeeding differ significantly between bands 1 and 2, $Q_{sf}$ $\sim$ 
$6.8~eb$, and bands TSD1 and TSD2, $Q_{sf}$ $\sim$ $10~eb$ 
(see Table~\ref{tab:fitmoments}). This change in $Q_{sf}$ values is in part 
responsible for the large difference in the $F(\tau)$ curves as a function of 
energy seen in Figure~\ref{fig:curve}. It implies that the $\gamma$-ray intensity 
responsible for the feeding of the bands originates from states associated with 
different intrinsic structures. The calculations presented in Ref.~\cite{Pattabi-PLB-163Tm-07} 
suggested that several other TSD bands, corresponding to various p-h excitations, 
should be present in $^{163}$Tm at excitation energies comparable 
to those of bands TSD1 and TSD2. It is plausible that these other TSD bands are 
part of the final stages in the deexcitation process towards the yrast TSD 
bands. If this is the case, the present observations suggest 
that the average deformation associated with the feeding TSD bands is larger 
than that of their yrast counter parts. Conversely, the feeding of bands 1 and 
2 then appears to occur from levels associated with a smaller deformation, 
similar to that of the bands themselves. 

\section{\label{sec:163Tm_summary}Conclusions and outlook}

In summary, two new bands observed in the $^{163}$Tm nucleus have been identified to 
be the triaxial strongly deformed (TSD) bands built on a minimum with calculated 
deformation parameters: $\varepsilon_2{\sim}0.39$, $|\gamma|{\sim}17^{\circ}$. 
The measured intrinsic properties, $\it{e.g.}$, alignments and moments of inertia, of these two 
TSD bands appear to be similar to the ones of other TSD bands observed in the region 
and have been reproduced by TAC calculations with parameters representative of a triaxial 
shape. It has been confirmed that these two TSD bands are associated with a larger 
deformation than the yrast (ND) bands, $\it{i.e.}$, $Q_t$(TSD) ${\sim}$ $7.5~eb$ and $Q_t$(ND) ${\sim}$ $6.4~eb$. 
Within the framework of present calculations, the deformation of the TSD bands is 
driven mainly by a large neutron gap at $N=94$. The data also indicates that the feeding 
of these two TSD bands is associated with states of larger deformation ($Q_{sf}{\sim}10~eb$). 
A surprising discrepancy ($\sim$ 20 -- 30$\%$) between data and calculations for the 
quadrupole moments of the TSD bands, which seems to be a general feature in the region, 
requires further investigation. The TSD bands in $^{163}$Tm, which are distinctly 
different from the wobbling bands, have been interpreted by the TAC calculations 
to be structures associated with particle-hole excitations in the TSD well. Perhaps more 
importantly, the TAC calculations performed in present work provide a natural explanation 
for the presence of wobbling bands in the $_{71}$Lu isotopes and the absence of such bands 
in all neighboring $_{69}$Tm, $_{72}$Hf and $_{73}$Ta nuclei. The explanation is related to: 
(I) the level density around the Fermi surface; (II) the presence of a strong 
shell gap at $N=94$; and (III) the shape driving effects of the $i_{13/2}$ proton orbital.

%
% Chapter 4
%

%
% Chapter 4
%

%\caption[short_title]{caption_text}

\chapter{\label{chap:PuIsotopes}OCTUPOLE CORRELATIONS IN $^{238,240,242}$Pu}

\section{Reflection asymmetry in nuclei}

\subsection{Introduction}
\begin{figure}
\begin{center}
\includegraphics[angle=0,width=0.90\columnwidth]{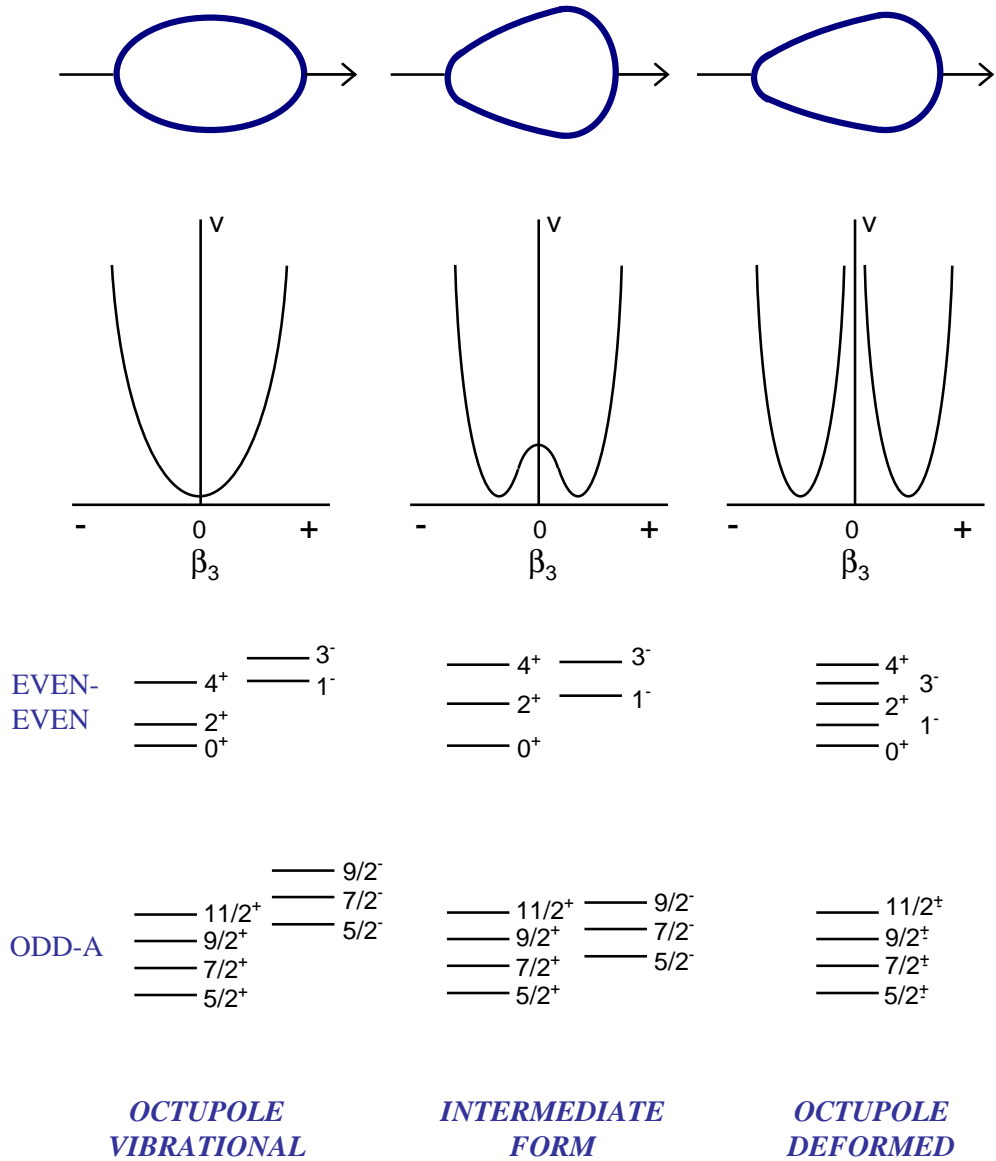}
\caption[Energy level diagram and potential energy as a function of $\beta_3$ deformation for 
different axially symmetric $(K=0)$ shapes.]{Energy level 
diagram and potential energy as a function of $\beta_3$ deformation for 
different axially symmetric $(K=0)$ shapes. The left panel (octupole vibration) represents a nucleus 
which has a spheroidal equilibrium shape in its ground state, which is relection symmetric as 
well. The right panel (octupole deformed) shows a nucleus with a rigid octupole shape (pear shape 
used as an example here). The middle panel (intermediate form) represents a soft pear-shaped nucleus, 
with deformation parameters: $\beta_2{\sim}0.15$ and $\beta_3{\sim}0.09$. Taken from 
Ref.~\cite{Ahmad-ARNPS-43-71-93}.\label{fig:OctuExtms}}
\end{center}
\end{figure}

For most deformed nuclei, a description in terms of an axial- and reflection-symmetric shape is 
adequate to interpret the observed band structures. Since such a shape is symmetric under space 
inversion, all levels in the associated rotational bands should have the same parity. However, the 
observation of negative-parity states near the ground level in even-even Ra and Th nuclei, 
which was first made by a Berkeley group in the 
1950s~\cite{Asaro-PR-92-1495-53,Stephens-PR-96-1568-54,Stephens-PR-100-1543-55}, indicated that 
some nuclei might have a shape asymmetric under reflection, such as a pear shape, for example. 
Further measurements showed that these negative-parity states form bands with a spin-parity sequence 
$1^{-}$, $3^{-}$, $5^{-}$, .... They are also characterized by a principal quantum number $K=0$. 
These observations were interpreted as signatures for octupole 
vibrations about a spheroidal equilibrium shape~\cite{Stephens-PR-96-1568-54,Stephens-PR-100-1543-55} 
(illustrated in Figure~\ref{fig:OctuExtms}). 
Soon after the discovery of octupole vibrations, suggestions were made regarding the possibility of 
the onset of stable octupole deformation in nuclei~\cite{Alder-RMP-28-432-56,Lee-PR-108-774-57} 
(illustrated in Figure~\ref{fig:OctuExtms}). 

Interest in octupole correlations and octupole deformation was revived in the early 1980s by two theoretical 
developments. First, microscopic many-body calculations by Chasman~\cite{Chasman-PRL-42-630-79,Chasman-PLB-96-7-80} 
predicted the presence of parity doublets in several odd-mass Th, Ac and Pa nuclei as a sign of strong octupole 
deformation. Second, calculation of atomic masses with the Strutinsky method indicated that, in some nuclei, the 
addition of octupole deformation in the parameterization of the potential resulted in extra binding, typically of 
the order of $\sim$ 1.5 $MeV$ in the $A{\sim}224$ region~\cite{Moller-NPA-361-117-81}. 

\subsection{Reflection-asymmetric shape}
As described in Sec.~\ref{subsec:Deformation} of Chapter~\ref{chap:theoriBkgd}, the nuclear shape 
is often parameterized in terms of a spherical harmonic expansion like Eq.~\ref{eq:NuclRadius}:
\[
R(\theta,\phi) = {R_0}\left(1+{\alpha_{00}}+\sum_{\lambda=1}^{\infty}
\sum_{\mu=-\lambda}^{\lambda}{\alpha_{\lambda\mu}}
{Y_{\lambda\mu}(\theta,\phi)}\right).
\]
The requirement that the radius is real imposes the condition: 
$\alpha_{{\lambda}{-\mu}}=(-1)^{\mu}{\alpha}_{{\lambda}{\mu}}$. 
For axially symmetric shapes, all deformation parameters ($\alpha_{{\lambda}{\mu}}$) with 
$\mu{\neq}0$ vanish. The remaining shape parameters $\alpha_{{\lambda}0}$ are usually called 
$\beta_{\lambda}$, $\it{i.e.}$, $\beta_{\lambda}\equiv{\alpha}_{{\lambda}0}$. For nuclei with reflection 
symmetry, only $\beta_{\lambda}$ terms with even $\lambda$ have nonzero values; on the other hand 
$\beta_3$ ($\lambda=3$) represents the magnitude of octupole deformation of an axially reflection-asymmetric 
nucleus. It is worth noting that the terms octupole shape and reflection-asymmetric shape are used 
interchangeably in this thesis. 

\begin{figure}
\begin{center}
\includegraphics[angle=0,width=0.85\columnwidth]{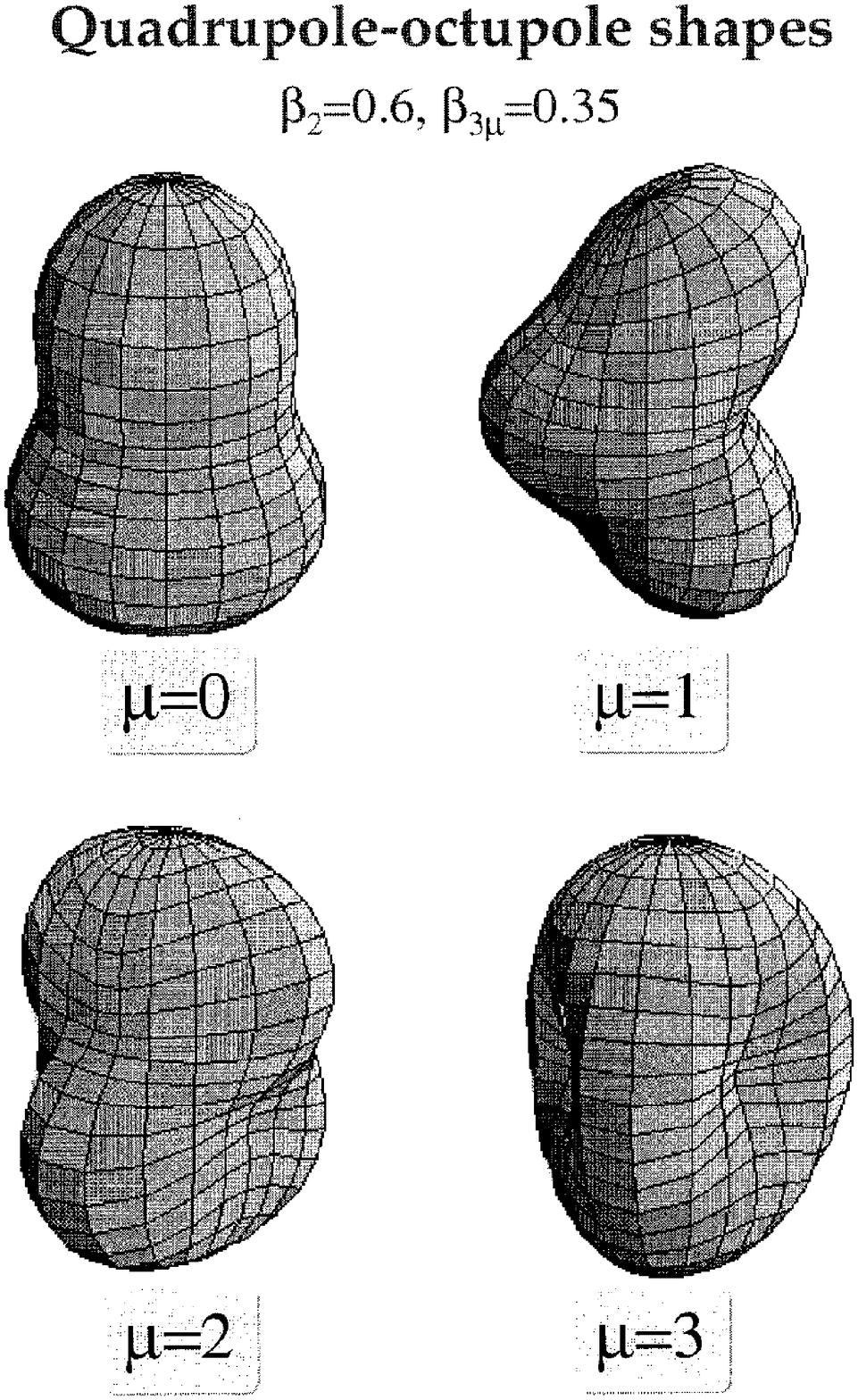}
\caption[Four quadrupole-octupole shapes correspond to octupole deformation with $\mu=0$, 1, 2, 3.]{Four 
quadrupole-octupole shapes correspond to octupole deformation with $\mu=0$, 1, 2, 3. 
In all cases, the same axial quadrupole $\alpha_{20}=\beta_{2}=0.6$ is assumed for illustration purposes. 
Taken from Ref.~\cite{Butler-RMP-68-349-96}.\label{fig:Quad-Octu-Shape}}
\end{center}
\end{figure}

Attempts to find a unique parameterization of a pure-octupole shape ($\it{i.e.}$, without involving the 
quadrupole deformation) turned out to be less successful~\cite{Butler-RMP-68-349-96}. The octupole 
deformation is often seen in quadrupole deformed nuclei. The general quadrupole-octupole shape is 
described by two quadrupole shape parameters. $\it{i.e.}$, $\alpha_{20}$ and $\alpha_{22}$, or $\beta_2$ 
and $\gamma$, and seven independent octupole parameters $\alpha_{3{\mu}}$ ($\mu=0,{\pm}1,{\pm}2,{\pm}3$). 
Figure~\ref{fig:Quad-Octu-Shape} displays four shapes resulting from the superposition of axial 
quadrupole and octupole deformation with $\mu=0$, 1, 2, and 3. A general parameterization of 
the combined quadrupole-octupole field, covering all possible shapes without double counting, 
was proposed by Rohozi{\'n}ski~\cite{Rohoz-JPG-16-L173-90}. 

\begin{figure}
\begin{center}
\includegraphics[angle=270,width=0.60\columnwidth]{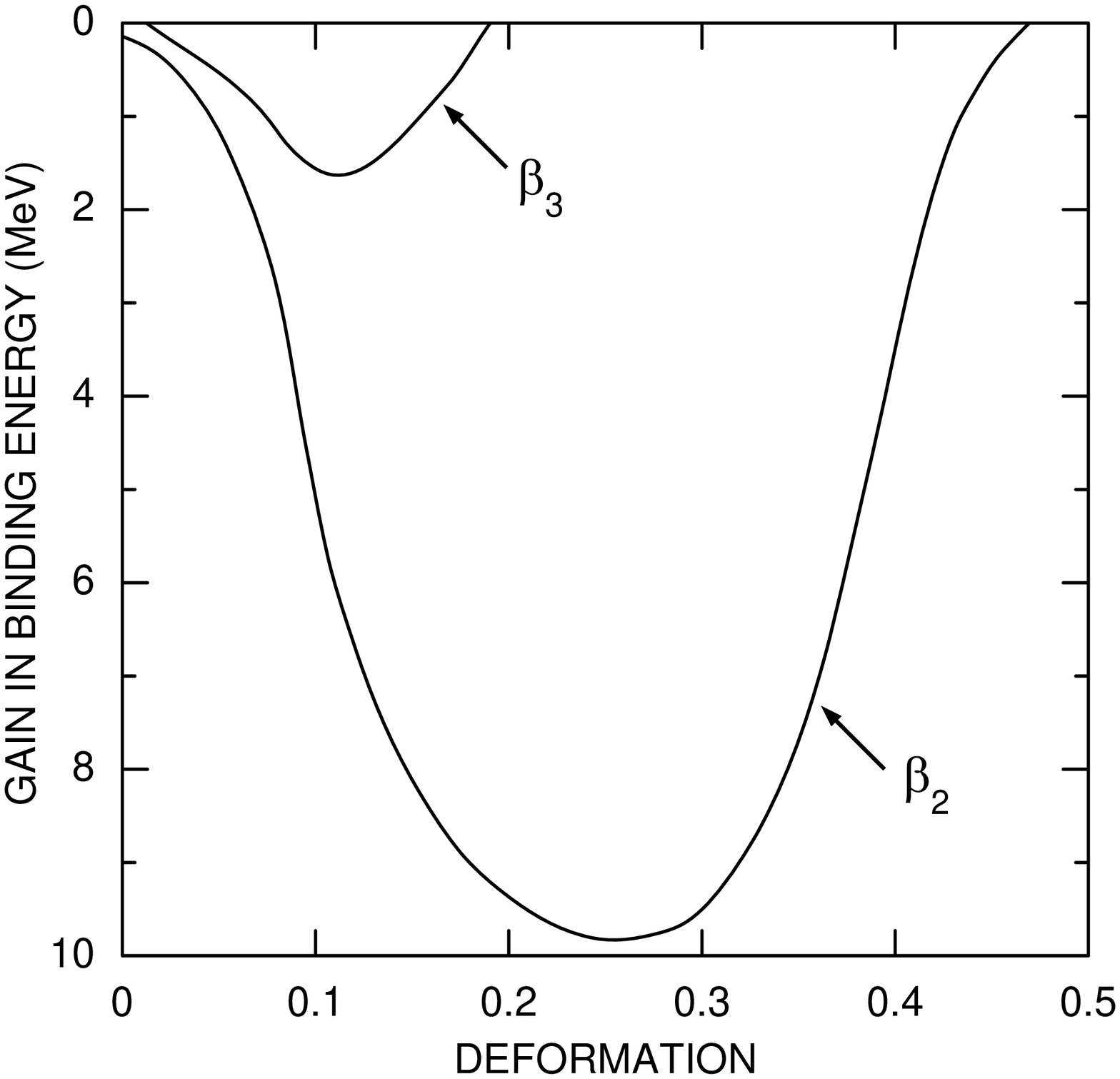}
\caption[Calculated gain in binding energy as a function of deformation for an actinide nucleus.]{Calculated 
gain in binding energy as a function of deformation for an actinide nucleus. 
The bottom curve shows the gain in binding energy as a spherical nucleus develops $\beta_2$ deformation. 
The upper curve shows the gain with increase in $\beta_3$ deformation in an actinide nucleus with some 
$\beta_2$ deformation. Taken from Ref.~\cite{Ahmad-ARNPS-43-71-93}.\label{fig:GainBnEnerDefPara}}
\end{center}
\end{figure}

It was found that the contribution of the $\beta_{6}$ degree of freedom on binding energies 
is comparable to the contribution from the $\beta_{3}$ deformation~\cite{Chasman-PLB-175-254-86}. 
Calculations with average potentials that included $\beta_{2}$--$\beta_{7}$ deformation parameters 
found that the $\beta_{5}$ mode helps stabilize the reflection-asymmetric shape, and the well depth 
was determined to be $\sim$ 1 $MeV$~\cite{Sobiczewski-NPA-485-16-88}. This is small compared to the 
gain in the binding energy of 10 $MeV$ due to the quadrupole deformation~\cite{Chasman-book-WS-86} 
(see Figure~\ref{fig:GainBnEnerDefPara}).

Unlike for the quadrupole deformation, the magnitude of octupole deformation is difficult to determine 
from direct measurements. Several experimental signatures associated with octupole correlations 
will be discussed later in Sec.~\ref{subsec:OctuExpEvd}. The three shapes and the potential 
energy plots for nuclei associated with octupole correlations are illustrated in Figure~\ref{fig:OctuExtms}. 
The situation seen in the left panel occurs for a nucleus with a spheroidal equilibrium shape in its 
ground state for which a vibrational band develops at an the excitation of $\sim$ 1 $MeV$ in the corresponding level 
diagram; this corresponds to the so-called ``octupole vibrator''. The other limit (in the right panel) occurs when 
the nucleus has a sizable $\beta_3$ deformation in its ground state and there is an infinite barrier between the 
reflection-asymmetric shape and its mirror image; this is the so-called ``octupole rotor'' and is associated with ``stable 
octupole deformation''. The level diagram characteristic of a nucleus with such a shape, shown in the bottom of 
Figure~\ref{fig:OctuExtms}, will be discussed further in Sec.~\ref{subsec:OctuExpEvd}. The third possible 
shape (in the middle panel) is intermediate between the two limits. Here, a finite barrier ($<~0.5~MeV$) exists 
between the reflection-asymmetric shape and its mirror image and tunneling motion is possible between the 
two mirror shapes~\cite{Ahmad-ARNPS-43-71-93}. 

One of the most distinct differences in the properties of reflection-symmetric and 
reflection-asymmetric nuclei is the value of the electric dipole moment. The nuclear electric 
dipole moment ($\vec{D}$) is a measure of the shift between the center of charge and the 
center of mass of the nucleus. It can be written as:
\begin{equation}
\vec{D}=e\frac{ZN}{A}(\vec{r}_{p,c.m.}-\vec{r}_{n,c.m.})\label{eq:DiplMntDef}
\end{equation}
($\it{i.e.}$, Eq. 12 in Ref.~\cite{Butler-RMP-68-349-96}), where $e$ is the charge of electron, $\vec{r}_{p,c.m.}=\vec{r}_{p}/Z$ 
and $\vec{r}_{n,c.m.}=\vec{r}_{n}/N$ are the center-of-mass coordinates for protons and neutrons, respectively. 
For reflection-symmetric nuclei, the nucleonic (proton and neutron) densities have three symmetry planes, so that 
$|\vec{r}_{n}|=|\vec{r}_{p}|=0$, and, hence, $|\vec{D}|=0$. However, $\vec{r}_{p,c.m.}~{\neq}~\vec{r}_{n,c.m.}$ 
generally if the density distributions are reflection asymmetric. Thus, a large static $E1$ moment may arise. 
For an axially deformed system in the reflection-asymmetric case, the intrinsic dipole moment is aligned along 
the symmetry axis ($z$ axis), and its value $D_{0}$ can be calculated directly from Eq.~\ref{eq:DiplMntDef}. 
In the most general case of triaxial and reflection-asymmetric density distributions, the intrinsic dipole moment 
is characterized by three spherical components, $D_{\pm{1}}$ and $D_{0}$. 

\subsection{\label{subsec:Reft_asym_theory}Theoretical description}
\begin{figure}
\begin{center}
\includegraphics[angle=0,width=0.85\columnwidth]{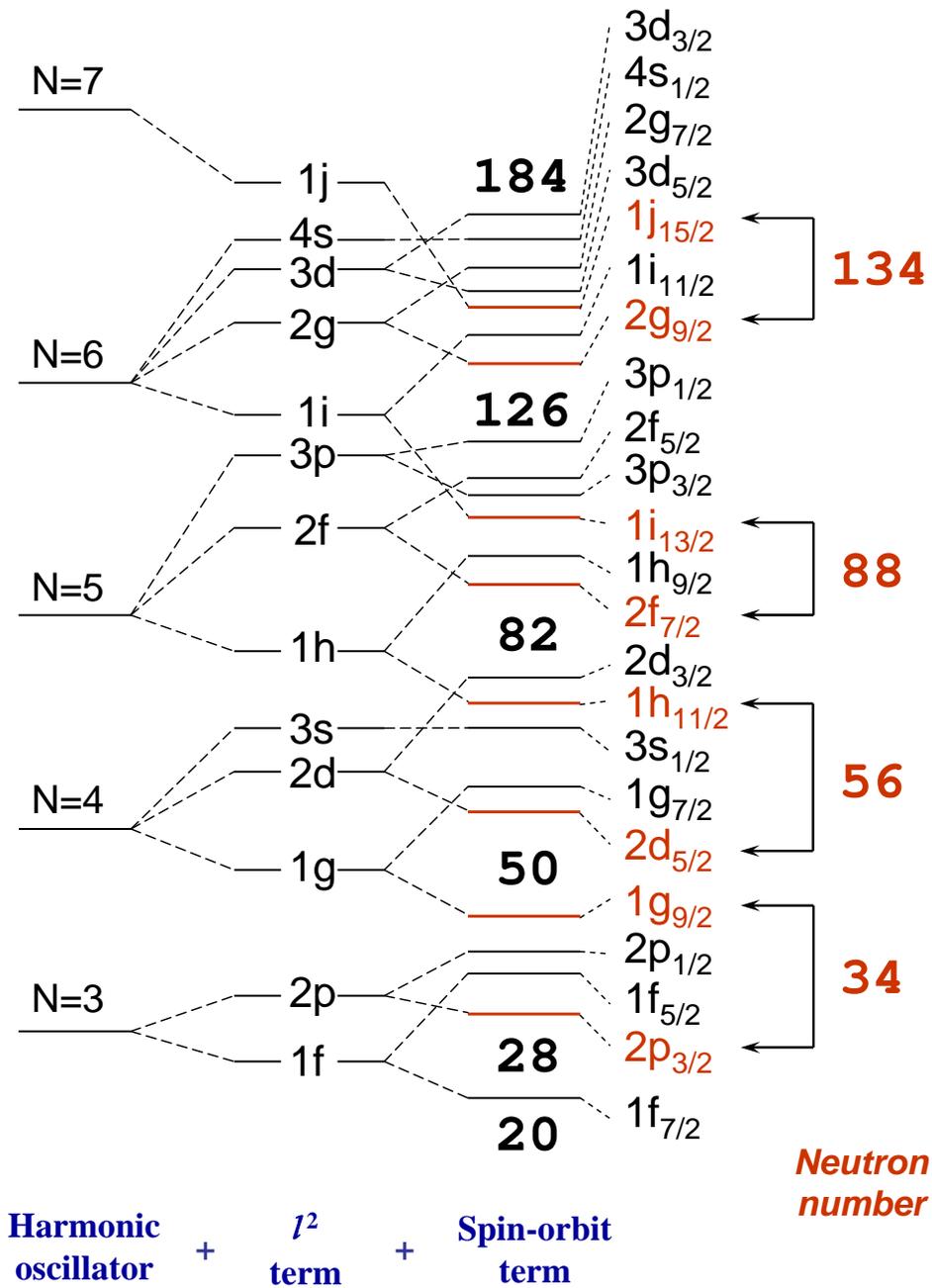}
\caption[Partial energy scheme of single-particle levels of the shell model for neutrons showing the locations of 
the octupole-driving orbits.]{Partial energy scheme of single-particle 
levels of the shell model for neutrons showing the locations of 
the octupole-driving $\Delta{j}=\Delta{l}=3$ coupled orbits, highlighted in red color. The numbers 
on the right correspond to the neutron numbers for which octupole correlations are strongest. The proton 
single-particle scheme is quite similar. Adapted from Figure~\ref{fig:ShellScheme} in 
Chapter~\ref{chap:theoriBkgd}.\label{fig:OctupoleShellScheme}}
\end{center}
\end{figure}

Octupole correlations in nuclei are produced by the long-range, 
octupole-octupole interaction between nucleons. These correlations depend on the matrix 
elements of $Y_{0}^{3}$ between single-particle states with $\Delta{j}=\Delta{l}=3$ and the energy spacings 
between them. As can be seen in Figure~\ref{fig:OctupoleShellScheme}, certain intruder states 
(defined in Sec.~\ref{subsec:NUSMD} of Chapter~\ref{chap:theoriBkgd}) with $l$, $j$ quantum numbers 
lie close in energy to other states with $(l-3)$, $(j-3)$ quantum numbers. These pairs of 
orbits can be stongly coupled by the octupole correlations. The most important 
couplings are $(1g_{9/2}{\rightarrow}2p_{3/2})$, $(1h_{11/2}{\rightarrow}2d_{5/2})$, 
$(1i_{13/2}{\rightarrow}2f_{7/2})$ and $(1j_{15/2}{\rightarrow}2g_{9/2})$, highlighted in red color 
in Figure~\ref{fig:OctupoleShellScheme}. Nuclei with a Fermi surface close to these pairs of 
states are particularly affected by strong octupole correlations. For spherical or near 
spherical nuclei, the corresponding nucleon numbers for these coupled orbits are 34, 56, 88 and 134, 
and these numbers have been labeled in Figure~\ref{fig:OctupoleShellScheme}. For 
nuclei with large quadrupole deformation, the strong octupole correlations resulting from the coupled 
orbits generally have an impact over a range of several nucleon numbers. It can also be seen in 
Figure~\ref{fig:OctupoleShellScheme} that the octupole-driving $\Delta{j}=\Delta{l}=3$ orbits are 
close together for larger particle numbers. Thus, nuclei with $Z{\sim}88$ and $N{\sim}134$ 
(the light actinides) are expected to possess the largest octupole correlations. 

In order to reproduce and predict octupole correlations in nuclei, a number of theoretical approaches, microscopic as 
well as macroscopic, have been developed since the $K^{\pi}=0^{-}$ octupole vibrational bands in the actinides 
were discovered in early 1950s. One model that explained the octupole vibrational bands in heavier 
actinides satisfactorily is the Random Phase Approximation (RPA) model by Neerg{\r{a}}rd and 
Vogel~\cite{Neergard-NPA-149-209-70,Neergard-NPA-149-217-70}. This model was later extended to the so-called 
``the Cranking plus RPA'' method~\cite{Marshalek-NPA-266-317-76} through combining the approach with the cranked shell model, 
introduced in Sec.~\ref{subsec:CSMintro} of Chapter~\ref{chap:theoriBkgd}. In the RPA model based on the Cranked 
Nilsson potential, the Hamiltonian can be written as~\cite{Mizutori-PTP-86-131-91}:
\begin{equation}
H=H_{Nilsson}-\omega{J}_{x}-\frac{1}{2}\sum_{K}{\chi}_{3K}Q_{3K}^{''{\dagger}}Q_{3K}^{''},\label{eq:RPAhamil}
\end{equation}
where $H_{Nilsson}$ is a Nilsson Hamiltonian, defined in Eq.~\ref{eq:NilsnHamil} of Chapter~\ref{chap:theoriBkgd}, 
and $Q_{3K}^{''}$ are the doubly stretched octupole operators defined by coordinates $x_{i}^{''}=({\omega}_i/{\omega}_0)x_i$ 
with $i=1$, 2 and 3~\cite{Sakamoto-NPA-501-205-89}. Here $\omega_{i}/\omega_{0}$ denotes the ratios of 
the frequencies of the deformed harmonic-oscillator potential to that of the spherical one, while $K$ represents 
the components of angular momentum on the symmetry axis of the potential. The force-strengths $\chi_{3K}$ can be 
determined from the self-consistency conditions between the potential and the density, once the single-particle 
potential at the equilibrium (for example, Nilsson potential, described in Sec.~\ref{subsec:DefmShelModl} 
of chapter~\ref{chap:theoriBkgd}) is given~\cite{Sakamoto-NPA-501-205-89}. 
For the cases in which the pairing interactions are not negligible, another term, the pairing Hamiltonian 
(see Eq.~\ref{eq:PairHamilt} in Chapter~\ref{chap:theoriBkgd}), needs to be added in Eq.~\ref{eq:RPAhamil}. 
With some appropriate assumptions and deductions~\cite{Marshalek-NPA-266-317-76,Mizutori-PTP-86-131-91}, 
the RPA energy eigenvalues, the so-called ``Octupole strength'', $S(Q_{3K}^{''},\omega)$~\cite{Mizutori-PTP-86-131-91}, and, 
hence, the neutron and proton single-particle routhians can be calculated as functions of the rotational frequency $\omega$. 
The octupole vibrational states are then described in terms of a superposition of many particle-hole excitations. 
Thus, individual vibrations can be classified according to the different degrees of freedom of the particle-hole pairs. 
The detailed description of the RPA method is beyond the scope of this thesis work, but can be found in 
Refs.~\cite{Marshalek-NPA-266-317-76,Ring-book-80}. 
The present work focuses mostly on the observation for octupole bands in Pu isotopes. 
Therefore, RPA calculations based on the cranked shell model, which can account well for the interplay between 
octupole vibrations and collective quasi-particle excitations under the stress of rotation, were carried out 
in Japan by T. Nakatsukasa of RIKEN. 

Other models, such as microscopic many-body calculations, the particle-plus-rotor model, shell correction calculations, 
algebraic models, $\it{etc.}$, are also available and may perhaps interpret the data better for nuclei with more intense octupole 
correlations ($\it{i.e.}$, nuclei that may be octupole deformed rather than octupole vibrational). 
Readers with further interest are referred to Refs.~\cite{Ahmad-ARNPS-43-71-93,Butler-RMP-68-349-96} for details. 

\subsection{\label{subsec:OctuExpEvd}Experimental evidence}
Though the existence of reflection asymmetry in nuclei is hard to prove conclusively, some of the experimental 
evidence, discussed below, can be reproduced well by theoretical calculations including octupole 
correlations. 

\begin{figure}[h]
\begin{center}
\includegraphics[angle=270,width=\columnwidth]{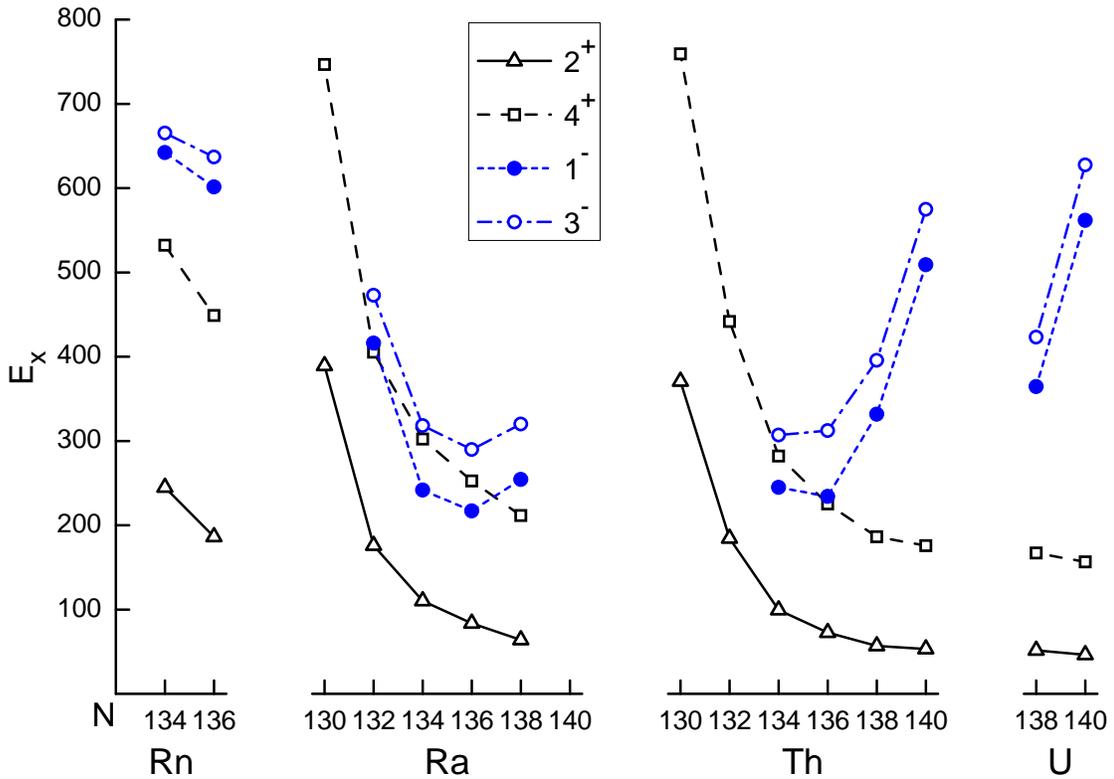}
\caption[Excitation energies ($keV$) of the yrast $2^{+}$, $4^{+}$, $1^{-}$, and $3^{-}$ states in the 
$Z=$ 86--92 region.]{Excitation energies ($keV$) of the yrast $2^{+}$, $4^{+}$, $1^{-}$, and $3^{-}$ states in the 
$Z=$ 86--92 region. Adapted from Fig. 14 in Ref.~\cite{Butler-RMP-68-349-96}.\label{fig:Acti_nucl_1_3_minus_ener}}
\end{center}
\end{figure}

As discussed at the beginning of this chapter, the observation of low-lying $1^{-}$ and $3^{-}$ states in the 
even-even Ra and Th nuclei provided the first experimental evidence that some nuclei may possess 
reflection-asymmetric shapes. In this mass region, the observed $1^{-}$ and $3^{-}$ states remain higher in energy 
than the $2^{+}$ and $4^{+}$ levels, respectively (as can be seen in Figure~\ref{fig:Acti_nucl_1_3_minus_ener}), 
an observation which conflicts with a well-known property of octupole deformed nuclei, $\it{i.e.}$, the positive- and negative-parity 
states are perfectly interleaved in a single rotational band. Hence, this observation leads to the hypothesis that 
these nuclei are not rigidly octupole deformed. Rather, they may fluctuate back to a reflection-symmetric shape. 
Figure~\ref{fig:Acti_nucl_1_3_minus_ener} indicates that the energies of negative-parity states are the lowest 
in the even-even isotopes of Rn, Ra, Th and U with $N{\sim}134$, where a reflection-asymmetric shape is predicted 
to occur due to strong octupole correlations. It can be seen in this figure that the lowest-lying states 
are also very localized in $N$. As yet, in none of these nuclei has the lowest $1^{-}$ state been observed to be 
lower in energy than the lowest $2^{+}$ state. Phenomena similar to those described above were also observed in the region 
of the lanthanide nuclei (Xe, Ba, Ce, Nd, Sm and Gd) with $N{\sim}88$ (see Fig. 13 in Ref.\cite{Butler-RMP-68-349-96}), 
even though the octupole correlations appear to be somewhat less intense. The systematic behavior of excited negative-parity states has 
been discussed by several authors. It has been concluded that a vibrational interpretation in terms of RPA 
calculations is most often appropriate. Further discussion of the theoretical attempts to describe the properties of the 
low-lying $1^{-}$ and $3^{-}$ states can be found, along with references, in the review article of Butler and 
Nazarewicz~\cite{Butler-RMP-68-349-96}. 

\begin{figure}
\begin{center}
\includegraphics[angle=0,width=0.50\columnwidth]{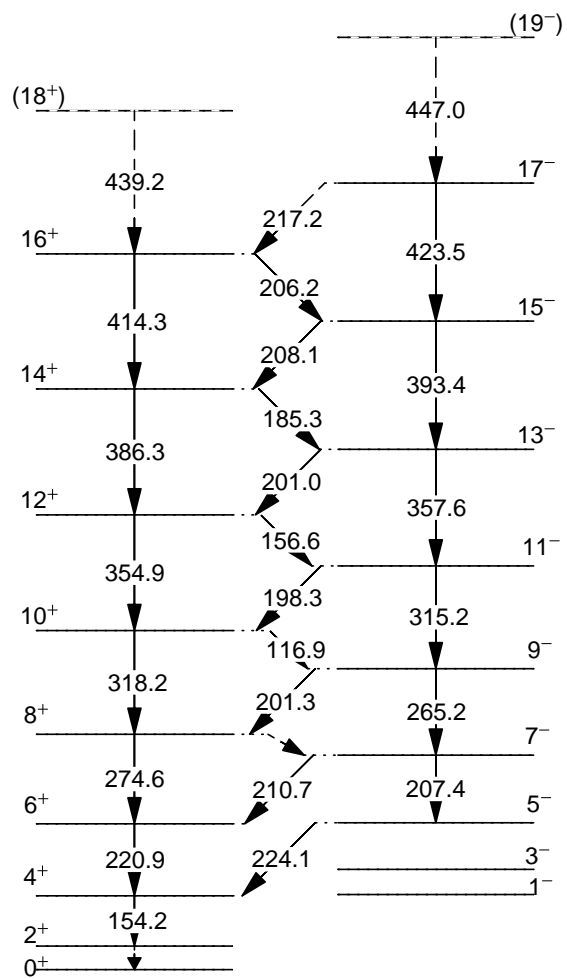}
\caption[Partial level scheme of $^{226}$Th displaying a characteristic sequence of states with alternating 
parity.]{Partial level scheme of $^{226}$Th displaying a characteristic sequence of states with alternating 
parity. Taken from Ref.~\cite{Schuler-PLB-174-241-86}.\label{fig:226Th_lev_schm}}
\end{center}
\end{figure}

The presence of a rotational band consisting of states with alternating parity, 
$\it{e.g.}$, $I^{+}$, $(I+1)^{-}$, $(I+2)^{+}$,..., is widely agreed 
to be one of the signatures for octupole deformed nuclei. The first observations of this striking 
feature in heavy, even-even nuclei were reported in $^{218}$Ra~\cite{Fernandez-NPA-391-221-82} and 
$^{222}$Th~\cite{Ward-NPA-406-591-83,Bonin-ZPA-310-249-83}. In the medium-mass region, band structures with 
similar features were seen much earlier, for example, in $^{152}$Gd~\cite{Zolnowski-PLB-55-453-75} and in 
$^{150}$Sm~\cite{Sujkowski-NPA-291-365-77}. Figure~\ref{fig:226Th_lev_schm} shows a typical example, $^{226}$Th, 
in which the sequence has been observed up to spin 20$\hbar$~\cite{Schuler-PLB-174-241-86,Ackermann-NPA-559-61-93}. 
Structures similar to those of even-even octupole nuclei are observed in transitional odd-mass and odd-odd 
nuclei in which the odd particles are weakly coupled to the core. This situation exists, for example, in $^{219}$Ra~\cite{Cottle-PRC-33-1855-86} 
and $^{216}$Fr~\cite{Debray-PRC-41-R1895-90}. Bands of this type have been observed in over 50 nuclei. 

In reflection-asymmetric nuclei, the $E1$ transitions observed between the yrast positive- and negative-parity bands, 
for example, $(I+1)^{-}{\rightarrow}I^{+}$ and $I^{+}{\rightarrow}(I-1)^{-}$, compete favorably with the $E2$ in-band 
$\gamma$ rays, since they exhibit relatively large transition probabilities, $B(E1)$. Typical $B(E1)$ values 
are less than $10^{-4}$ Weisskopf units (W.u.), defined in Sec.~\ref{subsec:Gamma-rayProperty} of 
Chapter~\ref{chap:theoriBkgd}, but the $B(E1)$ values in the mass regions where octupole correlations 
are strong, for example, in the actinides, range from $10^{-3}$ to $10^{-2}$ W.u. This phenomenon was first 
described by Bohr and Mottelson~\cite{Bohr-NP-4-529-57,Bohr-NP-9-687-58}, and Strutinsky~\cite{Strutinsky-AE-4-150-56} 
in terms of the macroscopic liquid-drop model. The related models are reviewed in Ref.~\cite{Butler-RMP-68-349-96} and 
the calculations using them reproduced the data of dipole moments ($D_0$) in both actinide and lanthanide nuclei. The relations 
between electric moments ($D_0$ and $Q_0$) and transition rates for $E1$ and $E2$ transitions have been described by 
Eqs.~\ref{eq:E1ReltQ0B} and \ref{eq:E2ReltQ0B} in Chapter~\ref{chap:theoriBkgd}. In most cases, absolute values of 
$B(E1)$ probabilities are not available, and the $D_0$ moments have to be extracted from the branching ratios 
$I_{\gamma}(J{\rightarrow}J-1)/I_{\gamma}(J{\rightarrow}J-2)$ ($I_{\gamma}$ is the $\gamma$-ray intensity), 
which are related to $B(E1)/B(E2)$ ratios in the form of Eq.~\ref{eq:BrRtoE1E2}. It is worth noting that many nuclei 
in the mass regions of interest are not good rotors, hence, the use of formulae for axial-symmetric nuclei, 
Eqs.~\ref{eq:E1ReltQ0B} and \ref{eq:E2ReltQ0B}, is questionable~\cite{Butler-RMP-68-349-96}. Nevertheless, 
they provide a consistent way to extract the $D_0$ moments from the data. 

In nuclei with one of the other two shapes shown in Figure~\ref{fig:OctuExtms}, $\it{i.e.}$, with the octupole vibrational 
shape or the intermediate form, the states in the two bands with opposite parity are not interleaved. 
However, the displacement in energy of a state, $I^{-}$, from the middle point between two adjacent states with opposite 
parity, $(I-1)^{+}$ and $(I+1)^{+}$, often decreases with an increase in spin. Actually, even in the well-defined reflection-asymmetric 
nuclei with strong octupole correlations, $\it{e.g.}$, $^{226}$Th, the states are not interleaved at low spin ($I<5$). This 
observation suggests that, in reality, the extreme of stable octupole deformation is never reached at the ground state. 
The barriers between two degenerate octupole minima are finite, like depicted for nuclei with intermediate shapes in 
Figure~\ref{fig:OctuExtms}. An important parameter, $S(I)$, $\it{i.e.}$, the energy staggering factor, is defined as:
\begin{equation}
S(I)=E_{I}-\frac{(I+1)E_{I-1}+{I}E_{I+1}}{2I+1},\label{eq:EnerStag}
\end{equation}
in Ref.~\cite{Ahmad-ARNPS-43-71-93}, where $E_{I}$, $E_{I-1}$ and $E_{I+1}$ are the energy of states $I^{-}$, $(I-1)^{+}$ and 
$(I+1)^{+}$, respectively. This quantity is a measure of the extent of that the two sequences of opposite parity are 
interleaved in spin and can be regarded as a single octupole rotational band. The values of the energy staggering for the 
yrast cascade and the lowest-lying band with opposite parity are observed to be zero at spins around $10{\hbar}$ in octupole 
deformed nuclei, for example in $^{223}$Th and $^{220}$Ra, while they tend to reach zero at high spin ($I{\sim}30$) in some of 
nuclei that are characteristic of octupole vibrations at low spin, such as $^{239,240}$Pu 
(see Fig. 4b in Ref.~\cite{Zhu-PLB-618-51-05} or Figure~\ref{fig:ener_stagr_Pu_Ra_Th} in Sec.~\ref{subsec:DirtMotvPresWk}). 
In both cases, it seems that rotation acts to enhance the strength of the octupole correlations. In other words, 
rotation appears to stabilize octupole deformation. 
An explanation for the occurrence of this phenomenon has been presented in Ref.~\cite{Butler-RMP-68-349-96}: 
(a) the octupole shape has weaker pairing correlations, which increases the moments of inertia; and (b) the rotational 
motion perturbs the single-particle states of opposite parity, which makes the octupole driving ($\Delta{l}=\Delta{j}=3$) 
orbitals approach each other with increasing frequency of rotation. 

\begin{figure}
\begin{center}
\includegraphics[angle=270,width=0.90\columnwidth]{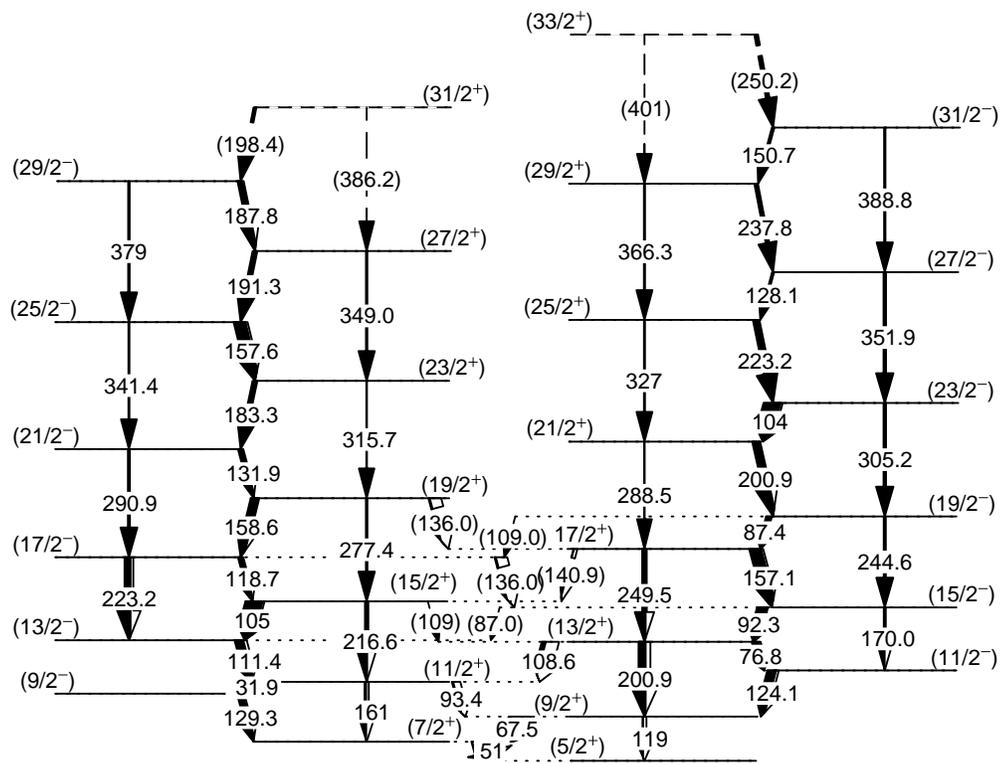}
\caption[Partial level scheme of $^{223}$Th.]{Partial 
level scheme of $^{223}$Th. Taken from Ref.~\cite{Dahlinger-NPA-484-337-88}.\label{fig:223Th_lev_schm}}
\end{center}
\end{figure}

It is important to point out another experimental fingerprint for octupole deformation in odd-$A$ and odd-odd nuclei: 
the so-called parity doublets. These were first discussed in Ref.~\cite{Bohr-98-book} and re-emphasized later by 
Chasman~\cite{Chasman-PLB-96-7-80}. The definition of parity 
doublets is that bands come in pairs close in energy with states of the same spin, but opposite parity. In other words, 
there is always another band close in energy with the same value of $K$ and opposite parity for each 
band, and the bands of different parity and signature are connected by strong $E1$ and $M1$ transitions. A good 
example is the level structures of the odd-even $^{223}$Th nucleus~\cite{Dahlinger-NPA-484-337-88}, shown in 
Figure~\ref{fig:223Th_lev_schm}, and the odd-odd nucleus, $^{224}$Ac~\cite{Sheline-PRC-44-R1732-91}, for example. 

In a nucleus with a stable reflection-asymmetric shape, the signature ($\alpha$) and parity ($p$), 
defined in Chapter~\ref{chap:theoriBkgd}, are no longer good quantum numbers. The only remaining symmetry is a combination of them, 
known as simplex ($S$)~\cite{Goodman-NPA-230-466-74,Frauendorf-PLB-141-23-84}, which has properties similar to those of the signature quantum 
numbers in the absence of reflection symmetry~\cite{Nazarewicz-PRL-52-1272-84}. Simplex is equivalent to reflection in 
a plane containing the symmetry axis (more generally, $S$ represents the symmetry with respect to a plane perpendicular to 
the rotational axis), and is defined as the eigenvalue of the $S$ operator, $S={\mathcal{P}}{\mathcal{R}}^{-1}$, 
where ${\mathcal{P}}$ is the space-reversing operator (its eigenvalue $p$ is parity) and ${\mathcal{R}}$ is the 
rotation operator. The rotational band with simplex $s$ is characterized by states with spin $I$ of alternating parity 
$p$~\cite{Bohr-98-book}, 
\begin{equation}
p=se^{-i{\pi}I}.\label{eq:SimplxDef}
\end{equation}
Thus, for a reflection-asymmetric nucleus with even mass number, the sequences are: 
\begin{eqnarray}
s=+1,~~I^{p}=0^{+},1^{-},2^{+},3^{-},...,\label{eq:EvenOctSimpP} \\
s=-1,~~I^{p}=0^{-},1^{+},2^{-},3^{+},...,\label{eq:EvenOctSimpM}
\end{eqnarray}
while in the case of an odd-$A$ nucleus, they are:
\begin{eqnarray}
s=+i,~~I^{p}={1/2}^{+},{3/2}^{-},{5/2}^{+},{7/2}^{-},...,\label{eq:OddOctSimpP} \\
s=-i,~~I^{p}={1/2}^{-},{3/2}^{+},{5/2}^{-},{7/2}^{+},....\label{eq:OddOctSimpM}
\end{eqnarray}

As discussed in Chapter~\ref{chap:theoriBkgd}, the experimental aligned angular momentum can be transfered 
to the intrinsic frame of the nucleus and then be compared with calculations. In an investigation of alignments 
up to high spin in several isotopes of Rn, Ra and Th~\cite{Cocks-PRL-78-2920-97,Cocks-NPA-645-61-99}, 
some features of the behavior of alignment as a function of rotational frequency ($\hbar\omega$) for structures associated 
with octupole correlations have been summarized. The difference of aligned angular momentum, 
$\Delta{i}_{x}=i_{x}^{-}-i_{x}^{+}$, between the positive- and negative-parity bands was extracted by subtracting 
a smoothed, interpolated value of $i_x$ for positive-parity states from the $i_x$ value for each negative-parity 
state at the same value of $\hbar\omega$. In an octupole-vibrational nucleus, where the negative-parity states 
are described in terms of an octupole phonon coupled to the positive-parity band, the value of $\Delta{i}_x$ 
is approximately 3$\hbar$, as the octupole phonon quickly aligns with the rotation axis as the rotational 
frequency increases. For a nucleus with stable octupole deformation, the value of $\Delta{i}_x$ should be 
zero because the positive- and negative-parity states are perfectly interleaved of energy in an octupole 
rotational sequence. 

Some other experimental observations, which can be associated with reflection asymmetry in nuclei, for example, 
relatively large $E3$ transition rates, $B(E3)$, or enhanced $\alpha$-decay probabilities to low-lying ($1^{-}$) states, 
are not discussed here. The interested reader is referred to dedicated review 
articles~\cite{Ahmad-ARNPS-43-71-93,Butler-RMP-68-349-96} and the references therein. 

\section{Motivations of present work}

\subsection{\label{subsec:Motiv_regn_octu}Regions of strong octupole correlations}
As pointed out earlier in this chapter, nuclei with proton and neutron numbers close to 
34, 56, 88 and 134 are predicted to possess strong octupole-octupole interactions. The heavy nuclei 
in the lanthanide ($Z{\sim}60$, $N{\sim}90$) and actinide ($Z{\sim}90$, $N{\sim}140$) regions 
attracted much attention since they have shown evidence for the predicted strong 
octupole correlations. Nuclei in these regions offer the golden possibility to investigate the interplay 
between collective rotation and octupole degrees of freedom, because they are also characterized, 
at the minimum, by a modest quadrupole deformation. 

\begin{figure}
\begin{center}
\includegraphics[angle=0,width=0.85\columnwidth]{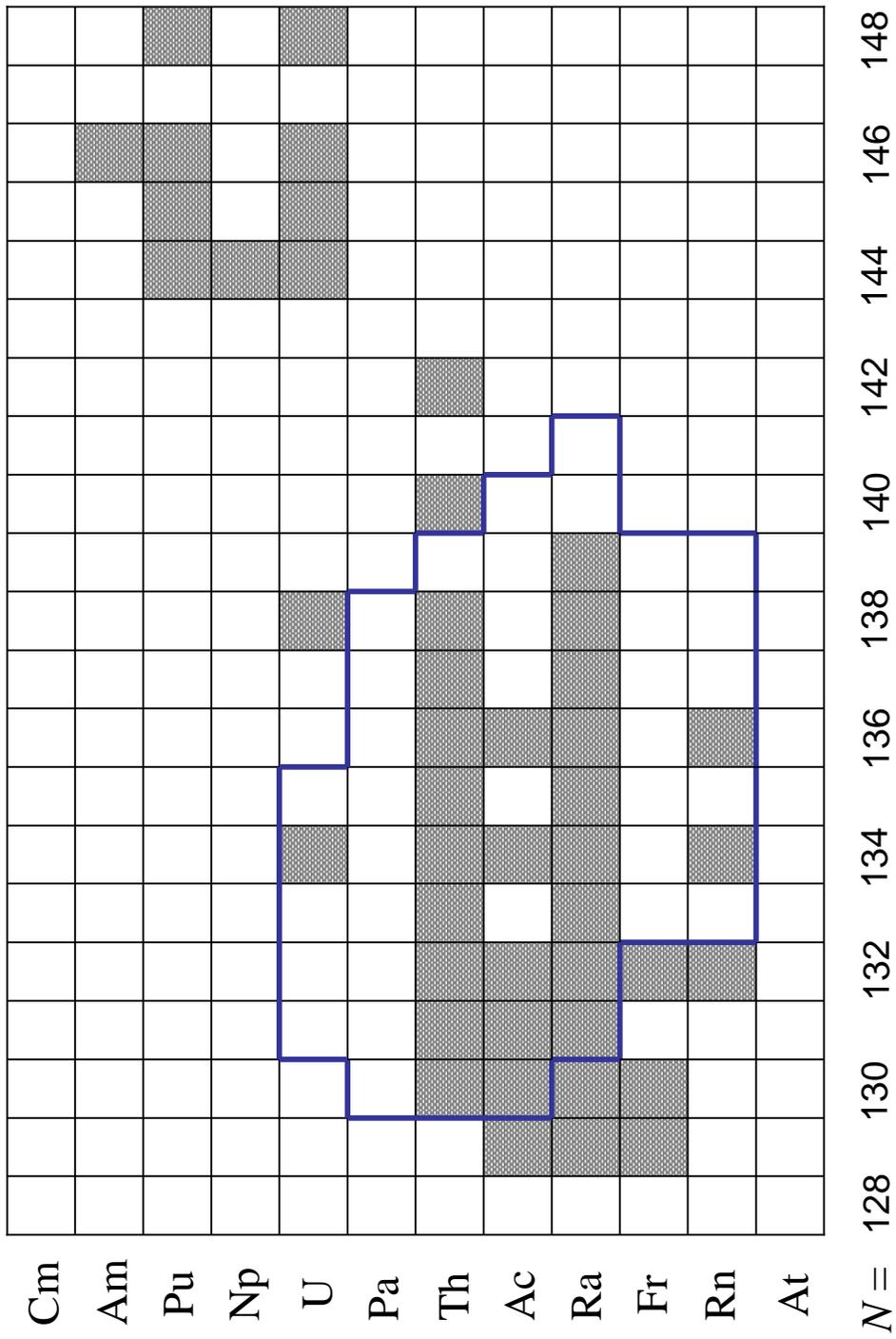}
\caption[The partial actinide region of the nuclear chart with the purpose to illustrate the impact of octupole 
correlations.]{The partial actinide region of the nuclear chart with the purpose to illustrate the impact of octupole 
correlations. The thick line represents the predicted boundary of octupole 
deformation~\cite{Nazarewicz-NPA-429-269-84,Sheline-PLB-197-500-87}, while the shaded squares are located at 
the nuclei in which octupole correlations have been investigated experimentally.\label{fig:Actinide_chart}}
\end{center}
\end{figure}

The present work concentrates on the actinide region. In this region, nuclei are difficult to populate 
experimentally, but, many of them exhibit properties associated with more or less intense octupole 
correlations. As a result, some observed bands are interpreted as structures built 
on an octupole vibration (vibrator) or an octupole deformation (rotor). As can be seen in Figure~\ref{fig:Actinide_chart}, 
among all these nuclei, the Ra and Th isotopes are the ones that have been studied most thoroughly, 
since they exhibit the strongest octupole correlations based on both the experimental evidence and 
theoretical calculations. In Figure~\ref{fig:Actinide_chart}, the thick line represents the predicted 
boundary of octupole deformation~\cite{Nazarewicz-NPA-429-269-84,Sheline-PLB-197-500-87}, while the 
shaded squares highlight the nuclei in which octupole correlations have been investigated experimentally. 
In the area within this boundary and nearby, for even-even nuclei, $^{220}$Ra~\cite{Burrows-JPG-10-1449-84,Cottle-PRC-30-1768-84} 
and $^{222}$Th~\cite{Ward-NPA-406-591-83} are two of the best examples of octupole rotors, and, 
some negative-parity cascades observed, for example, the ones found in $^{228}$Th~\cite{Schuler-PLB-174-241-86} 
and $^{230,232}$Th~\cite{Cocks-NPA-645-61-99}, have 
been associated with the octupole vibration; while, the alternating-parity bands observed in $^{216}$Fr~\cite{Debray-PRC-41-R1895-90} 
and $^{220}$Ac~\cite{Schulz-ZPA-335-107-90,Schulz-ZPA-339-325-91} and the parity doublets established in 
$^{221}$Ra~\cite{Fernandez-NPA-531-164-91} and $^{223}$Th~\cite{Dahlinger-NPA-484-337-88} represent the occurrence 
of reflection asymmetry in odd-odd nuclei and in odd-$A$ ones, respectively. Further, it was found that octupole 
correlations also have a significant impact on the properties of some nuclei away from the region defined above. 
For example, the low lying negative-parity bands observed in $^{238}$U~\cite{Ward-NPA-600-88-96} and 
$^{240}$U~\cite{Ishii-PRC-72-R021301-05} have been interpreted as structures built on an octupole vibration. 
In these heavier actinide nuclei (U, Np, Pu, Am,... isotopes with $A>236$), octupole correlations may still play an important 
role in explaining many observations. However, thus far, there is little information regarding octupole correlations in these 
nuclei, compared with the nuclei within or close to the bounded area illustrated in Figure~\ref{fig:Actinide_chart}. 
From this point of view, the present work, which is centered on the Pu isotopes ($Z=94$) with $A{\sim}240$, expands the limited 
knowledge of octupole correlations in the actinides, and, some characters of octupole degrees of freedom in the heavier actinide 
nuclei different from those in the lighter ones might be revealed as well. 

\subsection{\label{subsec:DirtMotvPresWk}Direct motivations}

\begin{figure}
\begin{center}
\includegraphics[angle=270,width=0.80\columnwidth]{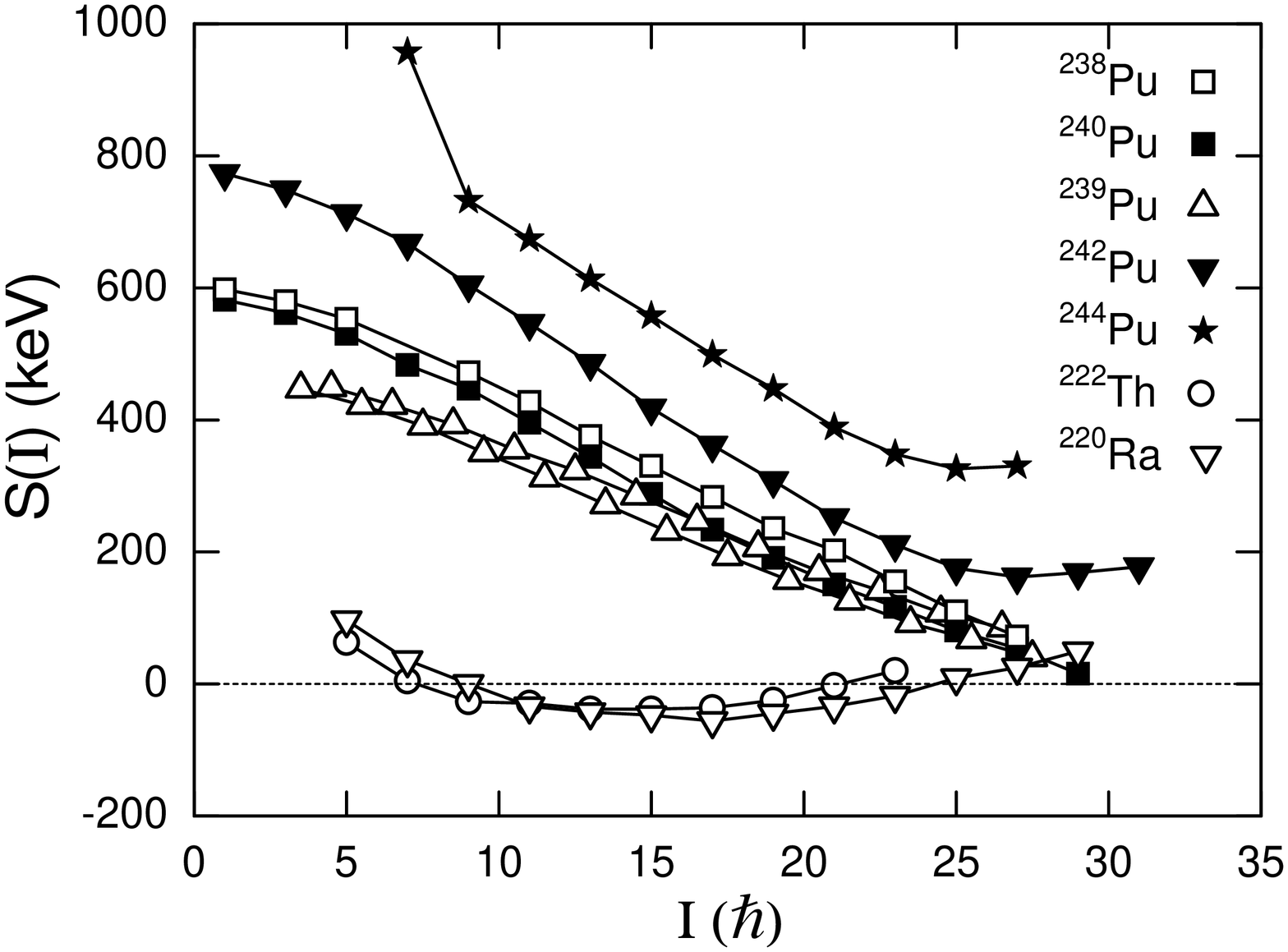}
\caption[Comparison of the energy staggering $S(I)$ as a function of spin $I$ in the Pu isotopes 
and in $^{220}$Ra and $^{222}$Th.]{Comparison of the energy staggering $S(I)$ as a function of spin $I$ in the Pu isotopes 
and in $^{220}$Ra and $^{222}$Th. Taken from Ref.~\cite{Wiedenhover-PRL-83-2143-99}.\label{fig:ener_stagr_Pu_Ra_Th}}
\end{center}
\end{figure}

A few years ago, a study of the properties of the yrast and lowest negative-parity bands in a number 
of Pu isotopes ($A$ $\sim$ 238--244) was performed by I. Wiedenh{\"o}ver, $\it{et~al.}$~\cite{Wiedenhover-PRL-83-2143-99}. 
It was suggested in this work that strong octupole correlations lead to the absence (in $A=$239, 240) or delay in 
frequency (in $A=$ 238) of the strong proton alignment observed in the heavier ($A=$ 241, 242, 243 and 244) 
Pu isotopes. Further, the $^{240}$Pu nucleus was found to possibly evolve from an octupole vibrator at low spin to 
an octupole rotor at high spin, in agreement with theoretical predictions by Jolos 
and von Brentano~\cite{Jolos-PRC-49-R2301-94,Jolos-NPA-587-377-95}. The evidence was based mostly on (but not limited to) (a) the 
fact that, at the highest spins in $^{238,239,240}$Pu, the energy staggering values become very small and 
comparable to the ones in $^{220}$Ra and $^{222}$Th, two of the best examples of static octupole rotors, 
as shown in Figure~\ref{fig:ener_stagr_Pu_Ra_Th}, $\it{i.e.}$, the states in the yrast band with positive parity 
become interleaved with the states in the octupole band with negative parity; (b) the comparison of aligned spins as 
a function of angular frequency ($\hbar{\omega}$) of the yrast bands in several Pu isotopes and other neighboring nuclei; 
and (c) the strength of the connecting transitions between the octupole and yrast bands. 

The above suggestion received further support from a later work by Sheline and Riley~\cite{Sheline-PRC-61-057301-00}. 
It was pointed out in their paper that a smaller depression of the excitation energies of the lowest 
$1^{-}$ states was found at neutron numbers of 144 and 146 for Pu isotopes, $\it{i.e.}$, $^{238,240}$Pu, which is similar with 
the deep $1^{-}$-state depression observed in the region of neutron numbers from 132 to 140 for Ra and Th nuclei 
(the region where strong octupole correlations lead to stable octupole deformation), as can be seen in 
Fig. 1(a) of Ref.~\cite{Sheline-PRC-61-057301-00}. 

\begin{figure}
\begin{center}
\includegraphics[angle=0,width=0.90\columnwidth]{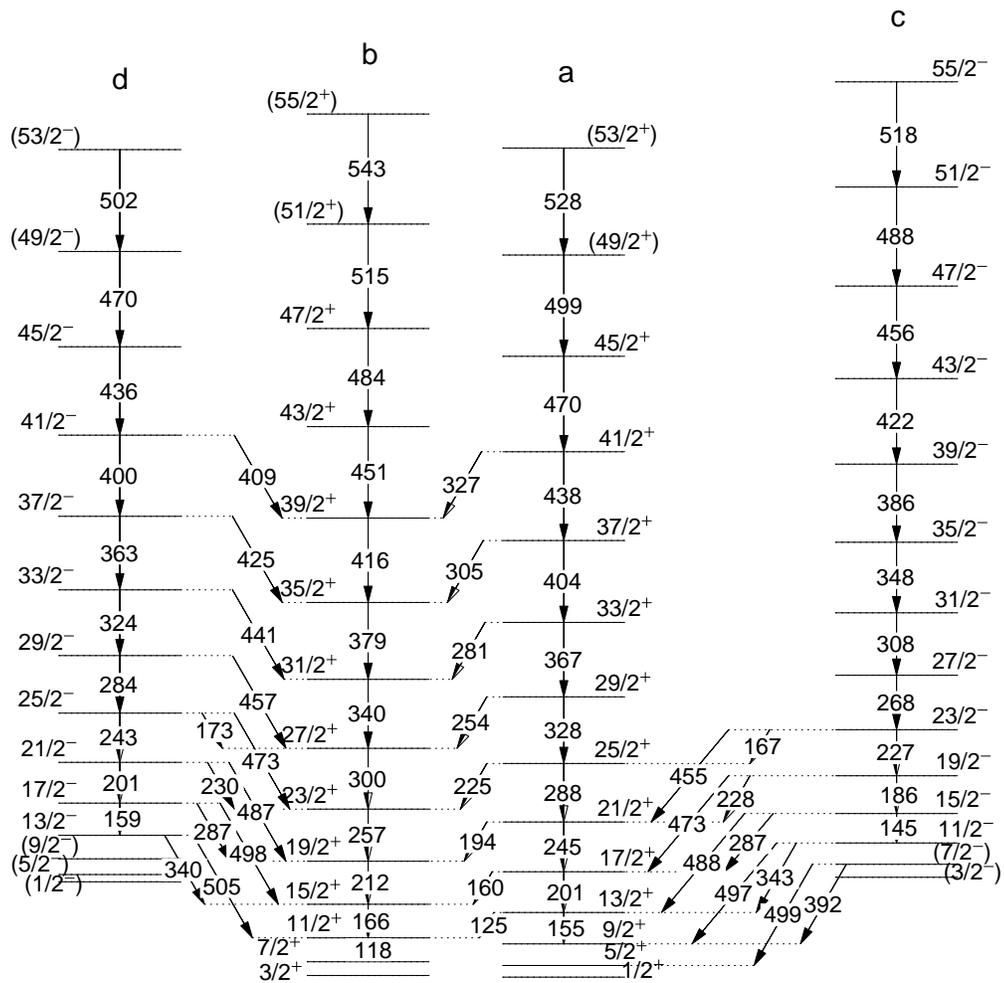}
\caption[Partial level scheme of $^{239}$Pu.]{Partial level scheme of $^{239}$Pu. 
Bands $c$ and $d$ are the octupole bands of interest. 
Taken from Ref.~\cite{Zhu-PLB-618-51-05}.\label{fig:239Pu_lev_schm}}
\end{center}
\end{figure}

Very recently, the properties of the odd-$A$ $^{239}$Pu nucleus, which was also proposed to have strong octupole 
correlations, was investigated and compared with its isotone $^{237}$U and other associated neighboring nuclei 
by Zhu $\it{et~al.}$~\cite{Zhu-PLB-618-51-05}. The resulting level scheme of $^{239}$Pu is shown in 
Figure~\ref{fig:239Pu_lev_schm}. This is the first observation of negative-parity bands in an odd-$A$ nucleus 
that would be associated with an octupole vibration. In contrast to its isotone $^{237}$U, 
the energy differences in $^{239}$Pu between states of same spin and opposite parity are getting smaller at high spin. 
Hence, the observed positive- and negative-parity bands, connected by strong $E1$ transitions 
at lower spins, tend to form parity doublets (defined in Sec.~\ref{subsec:OctuExpEvd}), one of the most important 
signatures for octupole deformation in odd-$A$ nuclei, at the highest spins ($\frac{53}{2}{\hbar}$). More 
experimental evidence supporting the occurrence of strong octupole correlations at high spin in $^{239}$Pu, for 
example, the aligned spins of the yrast and octupole bands in $^{239}$Pu compared with those in several neighboring 
nuclei, were also presented in the paper~\cite{Zhu-PLB-618-51-05}. 

The present work focuses on the octupole correlations in three even-even Pu nuclei ($A=$ 238, 240, 242). We 
are especially interested in the strong octupole correlations in $^{240}$Pu, which are proposed to be enhanced 
with the increase in angular momentum and be sufficiently strong to result in a transition from an octupole vibration 
at low spin to octupole deformation at high spin. Hence, the purpose of this work includes: (a) establishing 
the possible existence of an octupole rotational band consisting of states with alternating parity and spin, 
$I^{+}$, $(I+1)^{-}$, $(I+2)^{+}$,..., which are connected by strong $E1$ transitions, at high spin in $^{240}$Pu 
as this is one of the best fingerprints for stable octupole deformation in even-even nuclei, and it could not be 
established in the earlier work~\cite{Wiedenhover-PRL-83-2143-99} due to the lack of statistics; 
(b) exploring the magnitude of octupole correlations and its impact on the intrinsic structures of three 
Pu nuclei by extracting the relevant properties of the observed bands, such as alignments, routhians, $\it{etc.}$, 
and compairng them with results from RPA calculations including octupole interactions; and (c) searching for new 
levels and new band structures in the three Pu isotopes in order to get a more complete understanding of their 
intrinsic structures. 

\section{\label{sec:Pu_exp_datana}Experiment and data analysis}
A series of "unsafe" Coulomb excitation experiments (for $^{240,242}$Pu) as well as a single-neutron transfer 
measurement (for $^{238}$Pu) have been carried out at ATLAS with Gammasphere (GS). The number of Compton-suppressed 
HPGe detectors in working condition was 99 in the $^{240}$Pu experiment and 101 for the two other measurements. 
The so-called ``unsafe'' Coulomb excitation (Coulex) technique, 
which has been discussed in detail in Sec.~\ref{subsec:UnsafeCulxTech} of chapter~\ref{chap:exp_techs}, was 
used in order to enhance the feeding of the highest-spin states. In the cases of $^{240}$Pu and $^{242}$Pu, 
beams of $^{208}$Pb ions at an energy of 1300 $MeV$ bombarded targets consisting of 0.35-$mg/{cm^2}$ layers of $^{240}$Pu 
or $^{242}$Pu (98$\%$ enriched), electroplated onto 50-$mg/{cm^2}$ Au backing foils. In addition, each target also 
had a thin (0.05-$mg/{cm^2}$) Au layer in front of the Pu material in order to avoid sputtering of the activity into 
the scattering chamber. For the single-neutron transfer reaction, an odd-neutron $^{207}$Pb beam at 1300 $MeV$ 
was used in conjunction with a $^{239}$Pu target with characteristics similar to those just described. 

About $1.2{\times}10^9$, $3{\times}10^9$ and $0.4{\times}10^9$ events with fold three or higher were collected for 
$^{239}$Pu, $^{240}$Pu and $^{242}$Pu targets, respectively. In the subsequent data analysis, 
the raw data were converted into both the traditional Radware format (Cube and Hypercube) and the latest 
Blue database format. The Cube and Hypercube were used to search for new band structures, while the Blue database 
made an important contribution in extending the states in the observed bands up to the highest spins. 
The application of the Blue database technique in the present work also made it convenient to produce spectra at 
different detector angles for the important angular distribution analysis, which will be discussed later in this section. 
As discussed in Sec.~\ref{subsec:UnsafeCulxTech} of chapter~\ref{chap:exp_techs}, one of the drawbacks of the unsafe Coulex 
technique is that the transitions at high spin are usually emitted when the recoiling nuclei are still moving 
in the thick backing (Au) or in the target (Pu), and, as a result, they can be hard to resolve because of Doppler shifts 
and/or broadening. On the other hand, stable octupole deformation, experimentally exhibited in the form of the 
presence of a single band structure of states with alternating spins (even and odd) and 
parities (``$+$'' and ``$-$'') in even-even nuclei, appears to occur at high spin in $^{240}$Pu based on 
previous experimental observations~\cite{Wiedenhover-PRL-83-2143-99}. 
Hence, these in-band and inter-band transitions at high spin, affected 
by Doppler broadening and shift, are very critical for the success of the present work. In other words, the higher in 
spin both the yrast and octupole vibrational bands are extended, the larger the probability that evidence for an 
octupole rotor will be found is. The best solution so far to overcome the difficulty of observing high-spin transitions 
is to make use of the Blue database analysis technique, so that the inspection of data angle by angle becomes possible. 
Because of its unique data structure, the Blue database is very useful and powerful 
for dealing with the thick-target data of $\gamma$ rays with Doppler-shift and/or -broadening. This has been proven 
above in the $^{163}$Tm DSAM measurement (see Chapter~\ref{chap:163Tm-lifetime}). With this technique, individual 
gating conditions can be set conveniently on $\gamma$ rays at each detector angle during the process 
of generating $\gamma$-ray histograms from the data. Then, the resulting spectra with appropriate 
background subtraction would have excellent quality providing, perhaps, the opportunity to observe the transitions 
of interest at high spin. Moreover, as discussed in 
Sec.~\ref{subsec:GenSpecBKsub} of Chapter~\ref{chap:exp_techs}, a coincidence requirement with additional 
conditions on $\gamma$-ray time, total multiplicity and sum-energy in the event and an appropriate subtraction 
of random signals in the data analysis are also important for achieving reliable coincidence spectra. 
This is essential because of the negative influence of the strong contaminations from Coulomb excitation of the 
Au backings. This background will be effectively eliminated or at least strongly suppressed. 
Therefore, any spectrum used in the 
following analysis of the Pu data was generated with these additional gating conditions. The method of 
extending the sequence of states up to higher spins in the Blue database as well as the way of searching for 
new band structures in Radware have been described in Sec.~\ref{sec:DataAnytech} of Chapter~\ref{chap:exp_techs} 
and in Sec.~\ref{subsec:163Tm_life_expe_result} of Chapter~\ref{chap:163Tm-lifetime}. It is also worth noting 
here that only the Radware Hypercube ($\gamma$-$\gamma$-$\gamma$-$\gamma$) was used for developing the level 
scheme of $^{238}$Pu, while the $\gamma$-$\gamma$-$\gamma$ Cube was applied mostly in the analysis of band 
structures in $^{240}$Pu and $^{242}$Pu. Indeed, since the $^{238}$Pu data was obtained 
from the weak, single-neutron transfer channel in the reaction with a $^{207}$Pb beam bombarding a $^{239}$Pu target, 
another coincidence gate was set on one of the strongest transitions in $^{208}$Pb, the reaction partner of 
$^{238}$Pu, to enhance the channel of interest. 

\begin{figure}
\begin{center}
\includegraphics[angle=270,width=0.60\columnwidth]{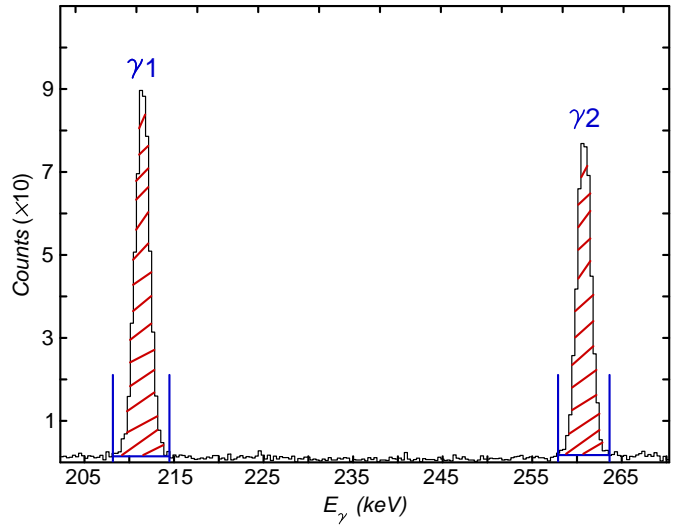}
\caption[Coincidence spectrum double gated on the 102.8-$keV$ and 158.5-$keV$ transitions in band 1 of $^{242}$Pu. ]{Coincidence 
spectrum double gated on the 102.8-$keV$ and 158.5-$keV$ transitions in band 1 of $^{242}$Pu. The 211.3-$keV$ and 
260.7-$keV$ $\gamma$ rays above the gate lines are the transitions of interest, labeled as $\gamma1$ and $\gamma2$, respectively. 
See text for details.\label{fig:Pu242_sp_4_intens_smp}}
\end{center}
\end{figure}

The example of the 260.7-$keV$ ($10^{+}{\rightarrow}8^{+}$) transition in band 1 of $^{242}$Pu (see Figure~\ref{fig:Pu242_levl_sche}) 
is used here to introduce the manner in which the relative intensities of transitions of interest (shown in the column ``$I_{\gamma}$'' 
of Tables~\ref{tab:240Pu-B1-gintns-agdscff}--\ref{tab:238Pu-B2-gintns-agdscff}) were extracted from the data. Given that the relative intensity of 
the 211.3-$keV$ ($8^{+}{\rightarrow}6^{+}$) transition in band 1 ($I_{\gamma1}{\pm}{\delta}{I}_{\gamma1}$) is known from the total 
projection spectrum, a spectrum double gated on the two lines below the 211.3-$keV$ transition (the 102.8-$keV$ and 158.5-$keV$ lines) 
was calibrated with the efficiency curve (see Sec.~\ref{subsec:calibration} of 
Chapter~\ref{chap:exp_techs}) and, then, adopted after inspecting the level scheme (Figure~\ref{fig:Pu242_levl_sche}). 
As can be seen in Figure~\ref{fig:Pu242_sp_4_intens_smp}, the peak areas (shaded areas) for the 211.3-$keV$ ($\gamma1$) and 
the 260.7-$keV$ lines ($\gamma2$) obtained from the analysis of this spectrum are $S_{\gamma1}$ and $S_{\gamma2}$, respectively. 
Thus, the relative intensity of the 260.7-$keV$ transition ($I_{\gamma2}$) was calculated by: $I_{\gamma2}=\frac{S_{\gamma2}}{S_{\gamma1}}I_{\gamma1}$. 
The errors of $S_{\gamma1}$ and $S_{\gamma2}$ ($\delta{S}_{\gamma1}$ and $\delta{S}_{\gamma2}$) reflect the statistical 
fluctuations only, while the determination of the error of $I_{\gamma2}$ (${\delta}I_{\gamma2}$), which is a function 
of $I_{\gamma1}$, $I_{\gamma2}$, $S_{\gamma1}$, $S_{\gamma2}$, ${\delta}{I}_{\gamma1}$, $\delta{S}_{\gamma1}$ and 
$\delta{S}_{\gamma2}$, obeys the general rules of propagation of errors~\cite{Bevington-92-book}. The coincidence gates needed 
to generate the appropriate spectra were set either on the lines above or on the lines below the transition of interest depending 
on whether this transition shares the same initial or final state with the reference transition. The real gates used in the data 
analysis for getting the intensity values will be described in the following sections. 

\begin{figure}
\begin{center}
\includegraphics[angle=0,width=0.80\columnwidth]{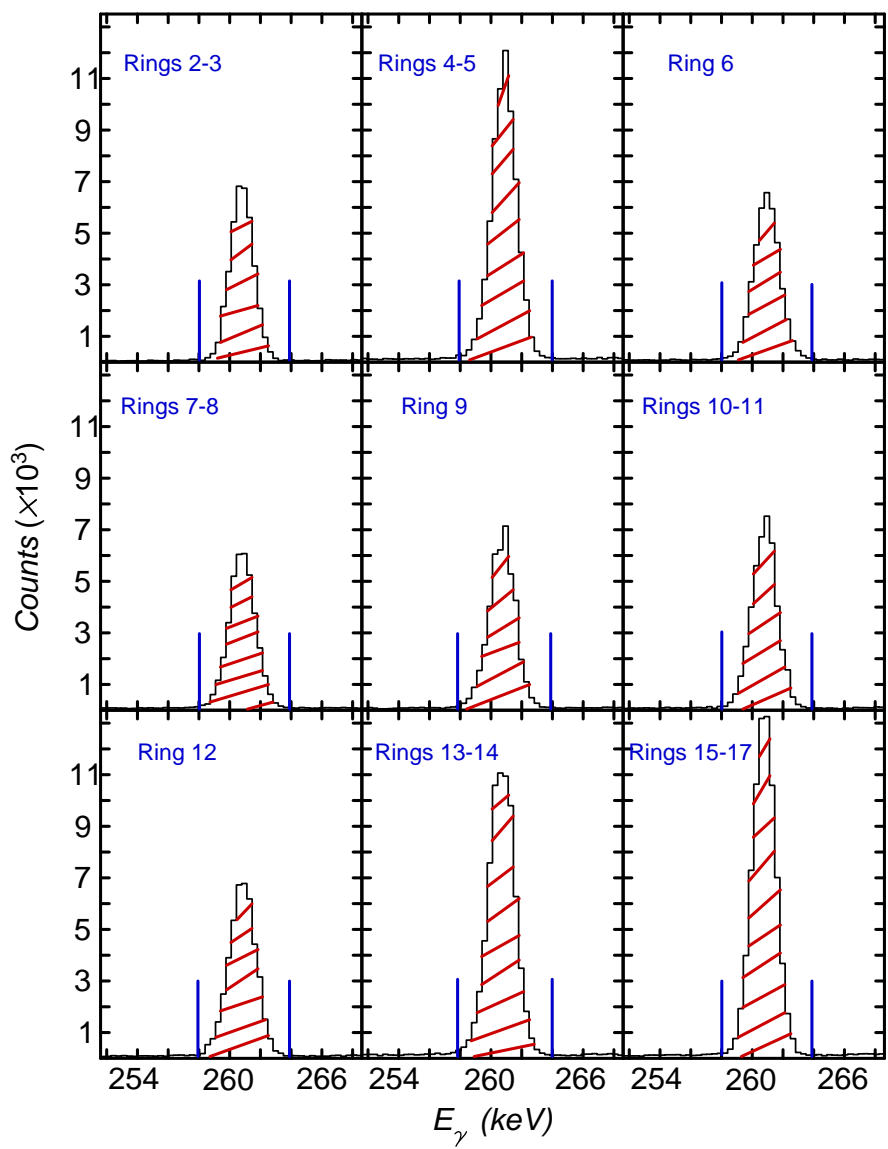}
\caption[The summed spectra gated on two of 11 transitions in band 1 of $^{242}$Pu at nine detector angles.]{The 
summed spectra gated on two of 11 transitions in band 1 of $^{242}$Pu at nine detector angles. The peak area 
(shaded area) of the 260.7-$keV$ transition in each spectrum was obtained for the subsequent angular distribution analysis. See 
text for details.\label{fig:Pu242_sp_4_angu_dist_smp}}
\end{center}
\end{figure}

In order to assign or confirm the spins and parities of the observed states, the multipolarities of related transitions were 
studied by obtaining their angular distribution coefficients (shown in the columns ``$A_2$'' and ``$A_4$'' 
of Tables~\ref{tab:240Pu-B1-gintns-agdscff}--\ref{tab:238Pu-B2-gintns-agdscff}). The general process of extracting 
the $A_2$ and $A_4$ coefficients for the transitions of interest from the data is described using the example of the 260.7-$keV$ 
($10^{+}{\rightarrow}8^{+}$) transition in band 1 of $^{242}$Pu. The nine summed spectra, shown in 
Figure~\ref{fig:Pu242_sp_4_angu_dist_smp}, represent the $\gamma$-ray signals collected by the detectors at nine angles 
with the gating conditions set on two of 11 transitions in band 1 of $^{242}$Pu (102.8-$keV$ -- 511.0-$keV$ lines, excluding 
the 260.7-$keV$ line). The detectors in 16 rings of GS (no detector installed in ring 1) 
were arranged into nine groups (rings 2--3, rings 4--5, ring 6, rings 7--8, ring 9, rings 10--11, ring 12, rings 13-14, 
rings 15--17) in the data analysis. This implies that detectors in some neighboring rings were placed in one group in order to 
achieve sufficient statistics for each spectrum. Hence, nine values of peak areas (shaded areas) for the 260.7-$keV$ transition 
were obtained. As seen in Table~\ref{tab:GSdetRings} of Chapter~\ref{chap:exp_techs}, each detector ring has its own angle 
($\theta_{i}$ for ring $i$). From the calibration (see Sec.~\ref{subsec:calibration} of Chapter~\ref{chap:exp_techs}) done earlier, 
the relative efficiency for detecting $\gamma$ rays with energies in the range $\sim$ 260 $keV$ for each detector 
ring has been obtained (the efficiency for ring $i$ is $Eff_{i}$). The effective efficiency for a group consisting of detectors 
in more than one ring, for example, rings 15--17, is the sum of the efficiencies for all involved ring ($Eff_{15}+Eff_{16}+Eff_{17}$). 
And, the effective angle for such a group is the weighted mean value of the angles of all involved rings with the efficiency of 
the individual ring as the weighting factor. For example, 
\[
\cos(\theta_{eff})=\frac{Eff_{15}\cos{\theta_{15}}+Eff_{16}\cos{\theta_{16}}+Eff_{17}\cos{\theta_{17}}}{Eff_{15}+Eff_{16}+Eff_{17}}
\]
for rings 15--17. Then, the process of fitting the measured intensity (peak area) as a function of $\cos(\theta)$ ($\theta$ is the detector 
angle or effective angle for each group) in the formula of Eq.~\ref{eq:angu_dist}, introduced in Sec.~\ref{subsec:LS-AngDist} of 
Chapter~\ref{chap:exp_techs}, was fulfilled by running the codes ``Legft'' of the Radware software package. An input file for ``Legft'' 
is composed of the values of the peak areas, effective angle ($\cos(\theta)$) and effective efficiency obtained above, while the output 
of ``Legft'' gives the $A_0$ (amplitude), $A_2$ and $A_4$ coefficients with the errors for the best fit. Examples of angular distribution 
curves can be seen in Figures~\ref{fig:240Pu_angu_dist_smp}, \ref{fig:242Pu_angu_dist_smp} and \ref{fig:238Pu_angu_dist_smp}. 
The gating conditions to generate the appropriate spectra for the angular distribution analysis were 
either both placed on transitions in the same band (for studying in-band lines) or one set on the in-band transitions in one band 
and the other placed on the lines in the other band (for studying inter-band lines). The actual gates used in the present work 
will be described in the following sections. 

The errors on the $\gamma$-ray energies given in the column ``$E_{\gamma}$'' of 
Tables~\ref{tab:240Pu-B1-gintns-agdscff}--\ref{tab:238Pu-B2-gintns-agdscff} include the statistical errors associated with 
the peak fitting and the systematic ones associated with the calibration of the $\gamma$-ray energy. 
The error arising from the energy calibration (see Sec.~\ref{subsec:calibration} of Chapter~\ref{chap:exp_techs}) 
can be studied through comparing the energy values translated from the measured channel numbers with the adopted energy values 
for the $\gamma$ rays from the standard radioactive sources. These errors are usually small, 
typically less than 0.1 $keV$ over a wide range of $\gamma$-ray energies (10 $keV$ -- 2 $MeV$). 

It should be noted that, due to the weak $\gamma$-ray intensity, the strong contamination, $\it{etc.}$, 
the values of the $\gamma$-ray intensity ($I_{\gamma}$) and angular-distribution coefficients ($A_2$, $A_4$) for some 
transitions are missing in Tables~\ref{tab:240Pu-B1-gintns-agdscff}--\ref{tab:238Pu-B2-gintns-agdscff}. 

\section{$^{240}$Pu data}

\begin{figure}
\begin{center}
\includegraphics[angle=0,width=0.70\columnwidth]{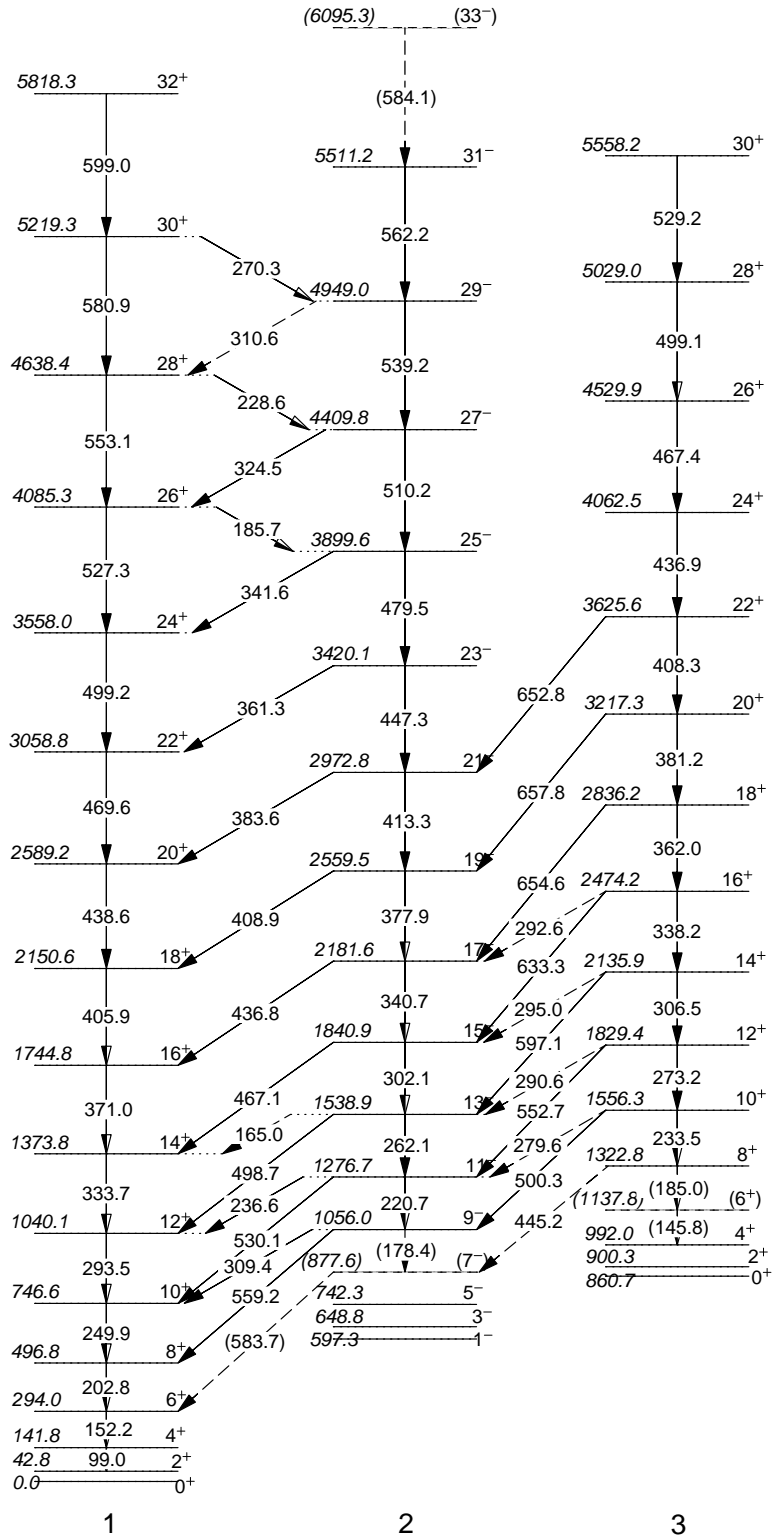}
\caption[Partial level scheme of $^{240}$Pu resulting from present work. ]{Partial level scheme of $^{240}$Pu 
resulting from present work. See text for details.\label{fig:Pu240_levl_sche}}
\end{center}
\end{figure}

The level scheme of $^{240}$Pu resulting from the above data analysis is presented in Figure~\ref{fig:Pu240_levl_sche}. 
The three bands, observed in our experiment, are labeled as bands 1, 2 and 3, respectively, and will be 
discussed in detail one by one below. In the level scheme, a state is labeled by its spin, parity and excitation energy 
relative to the ground state (unit: $keV$), while a transition is labeled by its energy (unit: $keV$) only. The states 
and transitions drawn as dashed lines or labeled by the energy, spin-parity symbols in parentheses are tentative. This means of 
labeling states and transitions in the level scheme was also adopted in the $^{242}$Pu and $^{238}$Pu cases below. 

\begin{figure}[h]
\begin{center}
\includegraphics[angle=270,width=\columnwidth]{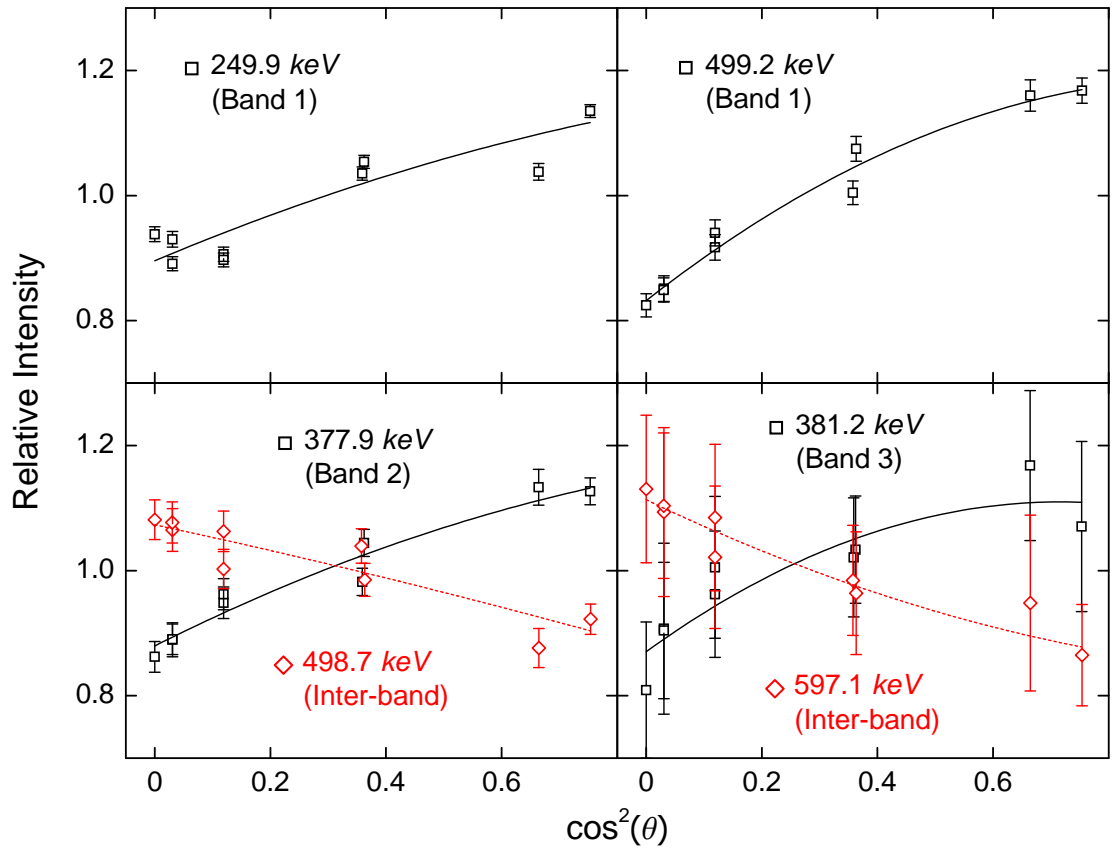}
\caption[Samples of angular distributions for transitions in $^{240}$Pu.]{Samples 
of angular distributions for transitions in $^{240}$Pu. The drawn curves (solid or dashed) represent 
the best fit of the data points. See text for details.\label{fig:240Pu_angu_dist_smp}}
\end{center}
\end{figure}

For calculating the values of relative intensity in the column ``$I_{\gamma}$'' of Tables~\ref{tab:240Pu-B1-gintns-agdscff}, 
\ref{tab:240Pu-B2-gintns-agdscff} and \ref{tab:240Pu-B3-gintns-agdscff}, the 249.9-$keV$ ($10^{+}{\rightarrow}8^{+}$) transition 
in band 1 was taken as the ``standard'', and its intensity has been normalized to ``1000'' for convenience. Based on the 
relative $\gamma$-ray intensities obtained, the population of bands 2 and 3 relative to band 1 in this experiment 
was estimated to be $10\%$ and $2\%$, respectively. Representative angular distributions for in-band and inter-band transitions 
associated with each band in $^{240}$Pu are compared in Figure~\ref{fig:240Pu_angu_dist_smp}, $\it{i.e.}$, the examples of the 249.9-$keV$ 
($10^{+}{\rightarrow}8^{+}$) and 499.2-$keV$ ($24^{+}{\rightarrow}22^{+}$) lines in band 1, the 377.9-$keV$ ($19^{-}{\rightarrow}17^{-}$) 
line in band 2, the 498.7-$keV$ ($13^{-}{\rightarrow}12^{+}$) transition linking bands 1 and 2, the 381.2-$keV$ ($20^{+}{\rightarrow}18^{+}$) 
line in band 3 and the 597.1-$keV$ ($14^{+}{\rightarrow}13^{-}$) transition linking bands 2 and 3. It was found in the analysis that the measured 
angular distribution coefficients ($A_2$ and $A_4$) for in-band and inter-band transitions in $^{240}$Pu (see 
Tables~\ref{tab:240Pu-B1-gintns-agdscff}, \ref{tab:240Pu-B2-gintns-agdscff} and \ref{tab:240Pu-B3-gintns-agdscff}) are very close to 
the typical values expected for quadrupole and dipole $\gamma$ rays, respectively (see Table~\ref{tab:angu_dist_coeff_values} 
in Sec.~\ref{subsec:LS-AngDist} of Chapter~\ref{chap:exp_techs}). 

\subsection{\label{subsec:240Pu_band1}$^{240}$Pu band 1}
\begin{figure}
\begin{center}
\includegraphics[angle=270,width=\columnwidth]{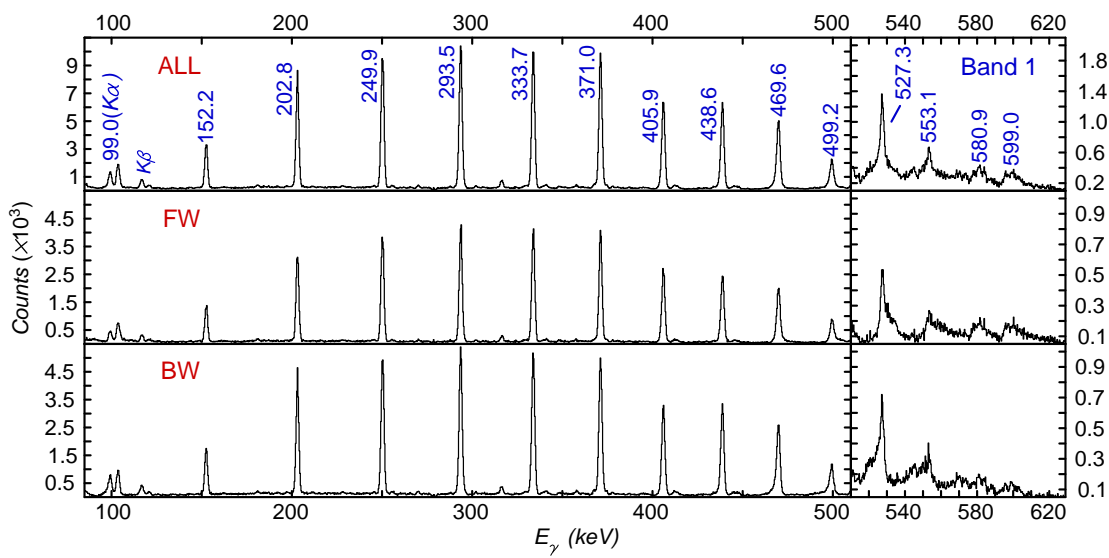}
\caption[Three spectra representative of band 1 in $^{240}$Pu.]{Three 
spectra representative of band 1 in $^{240}$Pu. They are the sum of coincidence spectra, double gated on any two of 
11 in-band $\gamma$ rays (from the 152.2-$keV$ line to the 527.3-$keV$ transition) of band 1, at all detector rings (ALL), at 
rings 2--8 (FW) and at rings 10--17 (BW), respectively. $K{\alpha}$ and $K{\beta}$ denote the 
characteristic $X$ rays of Pu.\label{fig:Pu240_sp_4_band1}}
\end{center}
\end{figure}

Band 1 in Figure~\ref{fig:Pu240_levl_sche} is the yrast band, which has been identified in several previous 
measurements~\cite{Hseuh-PRC-23-1217-81,Parekh-PRC-26-2178-82,Hardt-NPA-407-127-83,Hackman-PRC-57-R1056-98}. It consists 
of 16 transitions. The present work extends this band by three additional transitions with respect to the work of 
Ref.~\cite{Hackman-PRC-57-R1056-98}. All in-band transitions of band 1 can be seen in Figure~\ref{fig:Pu240_sp_4_band1}, 
with the exception of the 42.8-$keV$ ($2^{+}{\rightarrow}0^{+}$) transition. The latter was not observed due to the high internal 
electron conversion probability in this low-energy range. The sum of spectra double gated on any two of the 11 in-band $\gamma$ rays 
(from 152.2-$keV$ to 527.3-$keV$) at all detector angles (rings 2--17), at forward angles (rings 2--8) and at backward angles 
(rings 10--17) are given in the top, middle and bottom panel, respectively. It is not surprising to see in the spectra 
that the 99.0-$keV$ ($4^{+}{\rightarrow}2^{+}$) $\gamma$ ray is contaminated by the $K{\alpha}$ characteristic $X$ rays of Pu. 
From the comparison of these three spectra, it can be seen clearly that the transitions at high spin, $\it{i.e.}$, the 527.3-$keV$ $\gamma$ ray 
and all other lines above it, are associated with Doppler shifts and broadenings. This observation provides a direct indication that making use of the 
angular gating technique ($\it{i.e.}$, using individual gating conditions for each detector angle when generating the spectra with certain 
coincidence requirements) in the Blue database is absolutely necessary in order to see the transitions at the highest 
spins (as discussed in Sec.~\ref{sec:Pu_exp_datana}). Conversely, it is just the application of these proper data analysis 
techniques that makes the observation of such Doppler-shifted and -broadened transitions possible. 

\begin{table}
\begin{center}
\caption{THE EXCITATION ENERGIES ($E_x$) OF INITIAL STATES, ASSIGNED SPINS, $\gamma$-RAY ENERGIES ($E_{\gamma}$), RELATIVE 
$\gamma$-RAY INTENSITIES ($I_{\gamma}$) AND ANGULAR DISTRIBUTION COEFFICIENTS ($A_2$ AND $A_4$) FOR THE TRANSITIONS ASSOCIATED 
WITH BAND 1 IN $^{240}$Pu\label{tab:240Pu-B1-gintns-agdscff}}
\begin{tabular}{cccccc}
\hline \hline 
\multicolumn{6}{c}{Band 1 in $^{240}$Pu}\\
\hline 
$E_x$ ($keV$) & Assigned spin ($\hbar$) & $E_{\gamma}$ ($keV$) & $I_{\gamma}$ (rel.) & $A_{2}$ & $A_{4}$\\ 
\hline
141.8 & $4^{+}{\rightarrow}2^{+}$ & 99.0(3) & & & \\
294.0 & $6^{+}{\rightarrow}4^{+}$ & 152.2(3) & 370(10) & 0.29(5) & -0.04(5)\\
496.8 & $8^{+}{\rightarrow}6^{+}$ & 202.8(3) & 844(21) & 0.29(4) & -0.04(4)\\
746.6 & $10^{+}{\rightarrow}8^{+}$ & 249.9(3) & 1000(14) & 0.19(4) & -0.03(5)\\
1040.1 & $12^{+}{\rightarrow}10^{+}$ & 293.5(3) & 966(23) & 0.21(3) & -0.02(5)\\
1373.8 & $14^{+}{\rightarrow}12^{+}$ & 333.7(3) & 841(23) & 0.21(2) & -0.04(3)\\
1744.8 & $16^{+}{\rightarrow}14^{+}$ & 371.0(3) & 705(20) & 0.19(3) & -0.01(5)\\
2150.6 & $18^{+}{\rightarrow}16^{+}$ & 405.9(3) & 529(25) & 0.19(4) & -0.05(5)\\
2589.2 & $20^{+}{\rightarrow}18^{+}$ & 438.6(3) & 347(29) & 0.21(4) & -0.02(6)\\
3058.8 & $22^{+}{\rightarrow}20^{+}$ & 469.6(3) & 210(14) & 0.19(5) & -0.02(6)\\
3558.0 & $24^{+}{\rightarrow}22^{+}$ & 499.2(3) & 125(13) & 0.27(3) & -0.08(3)\\
4085.3 & $26^{+}{\rightarrow}24^{+}$ & 527.3(3) & 54(9) & 0.33(5) & -0.05(5)\\
 & $26^{+}{\rightarrow}25^{-}$ & 185.7(3) & 6(3) & & \\
4638.4 & $28^{+}{\rightarrow}26^{+}$ & 553.1(3) & 26(9) & & \\
 & $28^{+}{\rightarrow}27^{-}$ & 228.6(4) & 7(4) & & \\
\hline 
\end{tabular}
\end{center}
\end{table}

\begin{table}
\begin{center}
\centerline{TABLE~\ref{tab:240Pu-B1-gintns-agdscff} (contd.)}
\begin{tabular}{cccccc}
\hline \hline 
$E_x$ ($keV$) & Assigned spin ($\hbar$) & $E_{\gamma}$ ($keV$) & $I_{\gamma}$ (rel.) & $A_{2}$ & $A_{4}$\\ 
\hline
5219.3 & $30^{+}{\rightarrow}28^{+}$ & 580.9(3) & 19(7) & & \\
 & $30^{+}{\rightarrow}29^{-}$ & 270.3(4) & & & \\
5818.3 & $32^{+}{\rightarrow}30^{+}$ & 599.0(3) & 19(9) & & \\
\hline 
\end{tabular}
\end{center}
\end{table}

The intensities of the three low-lying transitions, $\it{i.e.}$, the 152.2-$keV$ ($6^{+}{\rightarrow}4^{+}$), the 202.8-$keV$ 
($8^{+}{\rightarrow}6^{+}$) and the 249.9-$keV$ ($10^{+}{\rightarrow}8^{+}$) transitions, were obtained from the total 
projection spectrum of the data. For the 293.5-$keV$ ($12^{+}{\rightarrow}10^{+}$) transition, the coincidence spectrum double gated on 
the 152.2-$keV$ line and the 202.8-$keV$ $\gamma$ ray 
was analyzed to achieve the ratio of intensity between the 249.9-$keV$ and 293.5-$keV$ transitions. The relative intensity 
of the 293.5-$keV$ line was then obtained by the means described earlier in Sec.~\ref{sec:Pu_exp_datana}. Similarly, the intensities 
for the 333.7-$keV$ and higher transitions were acquired from the appropriate spectra. For analyzing the angular distributions of the 
transitions in this band, the summed spectra with all available double gates were used. Nevertheless, when studying the intensity and 
angular distribution of transitions impacted by contamination, for example, the 527.3-$keV$ ($26^{+}{\rightarrow}24^{+}$) transition that is 
contaminated by the 530.1-$keV$ ($11^{-}{\rightarrow}10^{+}$) line, the use of gates that can lead to the observation of 
contamination, $\it{i.e.}$, any two of the 249.9-$keV$ line and transitions located below, was carefully avoided. 

The spin and parity of each state below the $26^{+}$ level in this band, established in previous 
measurements~\cite{Parekh-PRC-26-2178-82,Hardt-NPA-407-127-83,Hackman-PRC-57-R1056-98}, was confirmed here 
through the measured angular distributions. 
As the transitions above the $26^{+}$ state form a natural extension of the lower-spin sequence in this band, the spin 
and parity for states above $26^{+}$ was assigned based on an $E2$ multipolarity for these in-band 
transitions, although the information of $A_2$ and $A_4$ coefficients for them is not available in the present work. 
The fact that all these transitions below the $26^{+}$ level in band 1 exhibit the expected angular distribution patterns 
can be viewed as support for the analysis techniques. 

%The measured angular distribution coefficients for transitions 
%in band 1 in the present experiment agree with the expected values of such coefficients for $E2$ $\gamma$ rays according to the 
%rule of $\gamma$-ray angular disrtibution discussed in Sec.~\ref{subsec:LS-AngDist} of Chapter~\ref{chap:exp_techs}. 
%This agreement provides confidence for the validity of the approach in the present work. 

\subsection{\label{subsec:240Pu_band2}$^{240}$Pu band 2}
\begin{figure}
\begin{center}
\includegraphics[angle=270,width=\columnwidth]{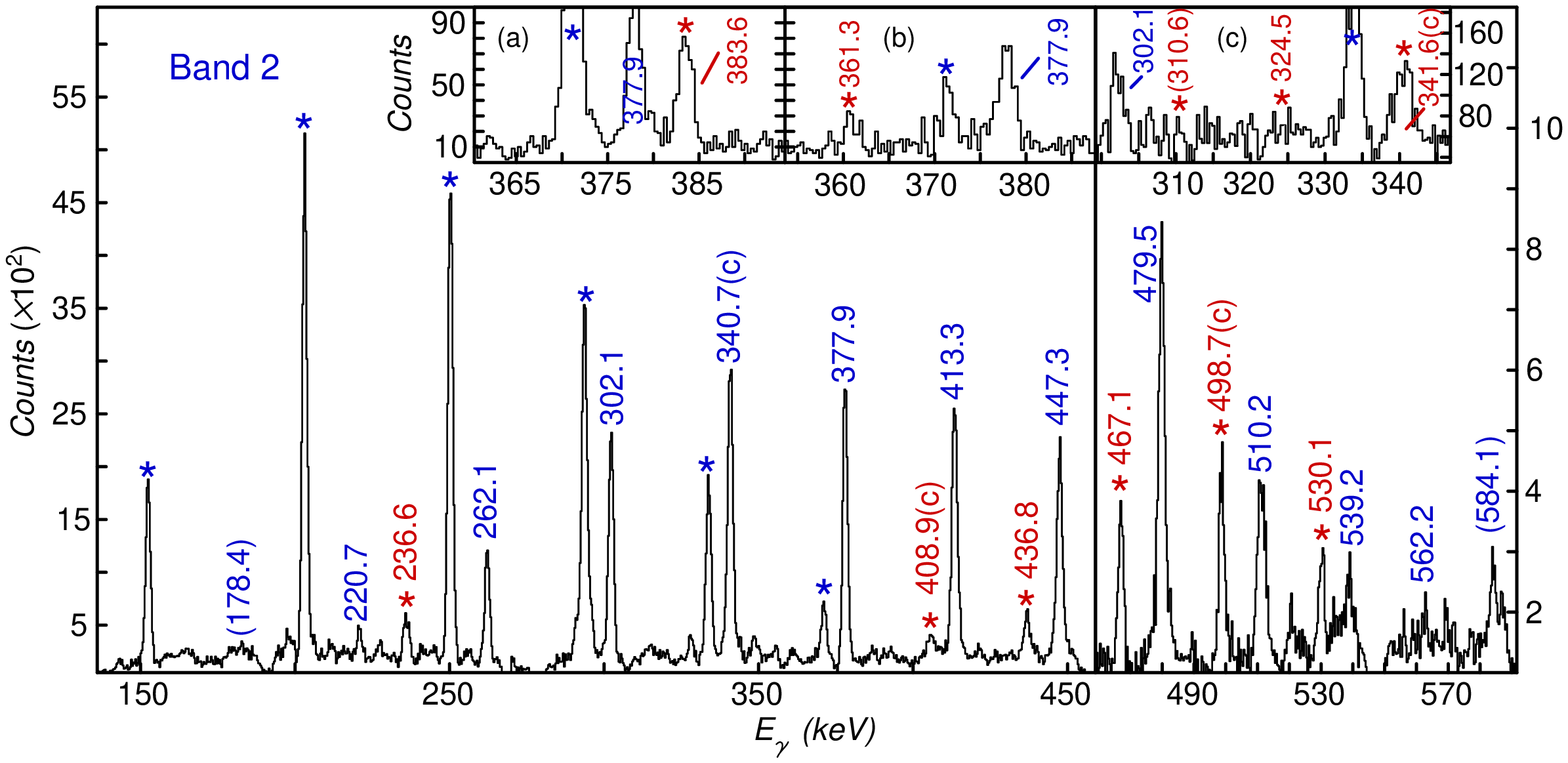}
\caption[Spectra representative of band 2 in $^{240}$Pu.]{Spectra 
representative of band 2 in $^{240}$Pu. The main spectrum is the sum of spectra double gated on any two of 
the 10 in-band $\gamma$ rays (from 220.7-$keV$ to 539.2-$keV$) of band 2. Insert (a, left): the spectrum gated on the 438.6-$keV$ 
line in band 1 and the 447.3-$keV$ line in band 2; insert (b, middle): the spectrum gated on the 469.6-$keV$ line in band 1 and 
the 479.5-$keV$ line in band 2; insert (c, right): the sum of spectra gated on the 499.2-$keV$ line (band 1) / the 510.2-$keV$ line (band 2), 
the 527.3-$keV$ line (band 1) / the 539.2-$keV$ line (band 2), the 527.3-$keV$ line (band 1) / the 562.2-$keV$ line (band 2) and 
the 553.1-$keV$ line (band 1) / the 562.2-$keV$ line (band 2). The in-band transitions in band 2 are labeled by their energy values. 
The in-band transitions in other bands ($\it{e.g.}$, band 1) of $^{240}$Pu are highlighted by ``$\star$'' symbols, while the inter-band 
transitions are marked with ``$\star$'' symbols plus their energy values. The energy values followed by ``(c)'' symbols and 
such values placed in parentheses denote contaminated transitions and tentative ones, respectively.\label{fig:Pu240_sp_4_band2}}
\end{center}
\end{figure}

Band 2 in Figure~\ref{fig:Pu240_levl_sche}, previously observed by Hackman $\it{et~al.}$~\cite{Hackman-PRC-57-R1056-98}, 
has been interpreted as the $K^{\pi}=0^{-}$ octupole vibrational band. It consists of 16 transitions. All 
transitions above the $7^{-}$ level in this band can be seen in Figure~\ref{fig:Pu240_sp_4_band2}. In this spectrum, 
the 178.4-$keV$ ($9^{-}{\rightarrow}7^{-}$) $\gamma$ ray is too weak to be 
observed clearly, and, the 584.1-$keV$ ($33^{-}{\rightarrow}31^{-}$) transition is hard to establish due to the impact 
of contamination and Doppler shift and/or broadening. As a result, these two transitions as 
well as the associated $7^{-}$ and $33^{-}$ states were assigned as tentative in the level scheme 
(Figure~\ref{fig:Pu240_levl_sche}). The three bottom (below the $7^{-}$ level) transitions were not observed here 
due to the high internal electron conversion probability, but the associated energy levels, $\it{i.e.}$, the $1^{-}$, $3^{-}$ 
and $5^{-}$ states, have been established in previous decay studies~\cite{Hseuh-PRC-23-1217-81,Parekh-PRC-26-2178-82}. 
Compared with the level scheme in Ref.~\cite{Hackman-PRC-57-R1056-98}, three new transitions at the highest spins (above $27^{-}$) 
have been added, though the top transition at 584.1-$keV$ ($33^{-}{\rightarrow}31^{-}$) remains tentative. All of 
the transitions connecting bands 1 and 2, observed in previous experiment~\cite{Hackman-PRC-57-R1056-98}, 
are $\gamma$ rays associated with the deexcitations from band 2 (octupole band) to band 1 (yrast band). 
They can be grouped into two types: $J^{-}{\rightarrow}(J-1)^{+}$ 
and $J^{-}{\rightarrow}(J+1)^{+}$, with the latter being much weaker in intensity (see Table~\ref{tab:240Pu-B2-gintns-agdscff}). 
These inter-band transitions were also observed in the present work, as indicated in the main spectrum of 
Figure~\ref{fig:Pu240_sp_4_band2}. 

\begin{table}
\begin{center}
\caption{THE EXCITATION ENERGIES ($E_x$) OF INITIAL STATES, ASSIGNED SPINS, $\gamma$-RAY ENERGIES ($E_{\gamma}$), RELATIVE 
$\gamma$-RAY INTENSITIES ($I_{\gamma}$) AND ANGULAR DISTRIBUTION COEFFICIENTS ($A_2$ AND $A_4$) FOR THE TRANSITIONS ASSOCIATED 
WITH BAND 2 IN $^{240}$Pu\label{tab:240Pu-B2-gintns-agdscff}}
\begin{tabular}{cccccc}
\hline \hline 
\multicolumn{6}{c}{Band 2 in $^{240}$Pu}\\
\hline 
$E_x$ ($keV$) & Assigned spin ($\hbar$) & $E_{\gamma}$ ($keV$) & $I_{\gamma}$ (rel.) & $A_{2}$ & $A_{4}$\\ 
\hline
877.6 & $7^{-}{\rightarrow}6^{+}$ & 583.7(4) & & & \\
1056.0 & $9^{-}{\rightarrow}7^{-}$ & 178.4(4) & & & \\
 & $9^{-}{\rightarrow}8^{+}$ & 559.2(3) & 14(10) & -0.16(5) & -0.02(7)\\
 & $9^{-}{\rightarrow}10^{+}$ & 309.4(3) & 5(5) & -0.20(5) & 0.04(8)\\
1276.7 & $11^{-}{\rightarrow}9^{-}$ & 220.7(3) & 4(2) & 0.22(4) & -0.06(6)\\
 & $11^{-}{\rightarrow}10^{+}$ & 530.1(3) & 18(9) & -0.14(5) & 0.04(4)\\
 & $11^{-}{\rightarrow}12^{+}$ & 236.6(3) & 7(4) & & \\
1538.9 & $13^{-}{\rightarrow}11^{-}$ & 262.1(3) & 21(10) & 0.18(4) & -0.03(5)\\
 & $13^{-}{\rightarrow}12^{+}$ & 498.7(3) & 28(18) & -0.15(3) & -0.01(5)\\
 & $13^{-}{\rightarrow}14^{+}$ & 165.0(3) & 2(2) & & \\
1840.9 & $15^{-}{\rightarrow}13^{-}$ & 302.1(3) & 57(25) & 0.23(3) & -0.05(4)\\
 & $15^{-}{\rightarrow}14^{+}$ & 467.1(3) & 37(21) & & \\
2181.6 & $17^{-}{\rightarrow}15^{-}$ & 340.7(3) & 90(84) & 0.17(4) & -0.03(4)\\
 & $17^{-}{\rightarrow}16^{+}$ & 436.8(3) & 31(38) & & \\
2559.5 & $19^{-}{\rightarrow}17^{-}$ & 377.9(3) & 68(52) & 0.21(3) & -0.04(4)\\
 & $19^{-}{\rightarrow}18^{+}$ & 408.9(3) & 14(3) & & \\
\hline 
\end{tabular}
\end{center}
\end{table}

\begin{table}
\begin{center}
\centerline{TABLE~\ref{tab:240Pu-B2-gintns-agdscff} (contd.)}
\begin{tabular}{cccccc}
\hline \hline 
$E_x$ ($keV$) & Assigned spin ($\hbar$) & $E_{\gamma}$ ($keV$) & $I_{\gamma}$ (rel.) & $A_{2}$ & $A_{4}$\\ 
\hline
2972.8 & $21^{-}{\rightarrow}19^{-}$ & 413.3(3) & 42(26) & 0.16(5) & -0.04(5)\\
 & $21^{-}{\rightarrow}20^{+}$ & 383.6(3) & 9(2) & & \\
3420.1 & $23^{-}{\rightarrow}21^{-}$ & 447.3(3) & 29(11) & 0.19(3) & -0.04(5)\\
 & $23^{-}{\rightarrow}22^{+}$ & 361.3(3) & 6(2) & & \\
3899.6 & $25^{-}{\rightarrow}23^{-}$ & 479.5(3) & 22(15) & 0.19(4) & -0.07(3)\\
 & $25^{-}{\rightarrow}24^{+}$ & 341.6(3) & 7(3) & & \\
4409.8 & $27^{-}{\rightarrow}25^{-}$ & 510.2(3) & 19(14) & 0.24(5) & -0.12(4)\\
 & $27^{-}{\rightarrow}26^{+}$ & 324.5(3) & & & \\
4949.0 & $29^{-}{\rightarrow}27^{-}$ & 539.2(3) & 13(9) & & \\
 & $29^{-}{\rightarrow}28^{+}$ & 310.6(4) & & & \\
5511.2 & $31^{-}{\rightarrow}29^{-}$ & 562.2(3) & 10(7) & & \\
6095.3 & $33^{-}{\rightarrow}31^{-}$ & 584.1(4) & & & \\
\hline 
\end{tabular}
\end{center}
\end{table}

\begin{figure}
\begin{center}
\includegraphics[angle=270,width=\columnwidth]{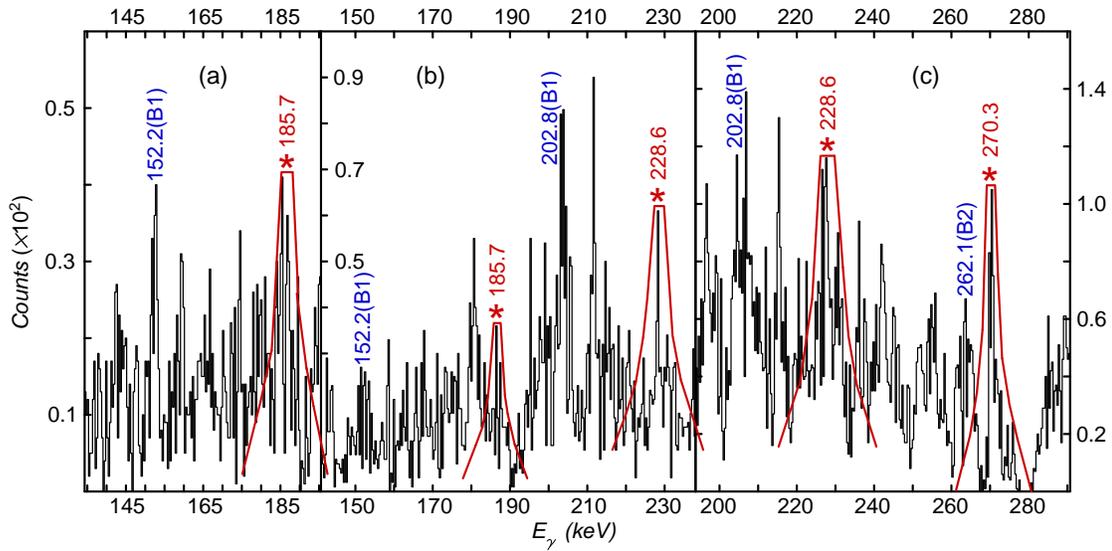}
\caption[Coincidence spectra supporting the observation of three important $(J+1)^{+}{\rightarrow}J^{-}$ inter-band 
transitions.]{Coincidence spectra supporting the observation of three important $(J+1)^{+}{\rightarrow}J^{-}$ inter-band transitions. 
Gates for the panel (a) are set on the 553.1-$keV$ transition in band 1 and one of the four lines (from 377.9-$keV$ to 479.5-$keV$) 
in band 2; gates for the panel (b) are set on the 580.9-$keV$ transition in band 1 and one of the five lines (from 377.9-$keV$ 
to 510.2-$keV$) in band 2; gates for the panel (c) are set on the 599.0-$keV$ transition in band 1 and one of the six lines 
(from 377.9-$keV$ to 539.2-$keV$) in band 2. The energy values followed by ``(B1)'' and ``(B2)'' symbols represent the transitions 
in band 1 and in band 2, respectively. The red outlines indicate the estimated broadened shapes of these $\gamma$-ray peaks. See 
text for details.\label{fig:Pu240_sp_4_3_linklines}}
\end{center}
\end{figure}

As described above, with the application of the angular gating technique in the Blue database, both the yrast band and the octupole 
sequence were extended towards higher spin. As a result, five additional $J^{-}{\rightarrow}(J-1)^{+}$ transitions between these 
two bands at higher spin, $\it{i.e.}$, the 383.6-, 361.3-, 341.6-, 324.5- and 310.6-$keV$ lines, were found in the 
present work. All of these five $\gamma$ rays are indicated in the inserted spectra of Figure~\ref{fig:Pu240_sp_4_band2}, but, the 
highest lying one, $\it{i.e.}$, the 310.6-$keV$ ($29^{-}{\rightarrow}28^{+}$) transition, was taken as tentative because of the 
poor associated statistics. More importantly, the linking transitions between bands 1 and 2 in an opposite direction, 
$\it{i.e.}$, $(J+1)^{+}{\rightarrow}J^{-}$, were revealed for the first time in the present work. 
Three such transitions above spin $25\hbar$, $\it{i.e.}$, the 185.7-$keV$ ($26^{+}{\rightarrow}25^{-}$), 228.6-$keV$ ($28^{+}{\rightarrow}27^{-}$) 
and 270.3-$keV$ ($30^{+}{\rightarrow}29^{-}$) lines, are indicated in spectra (see Figure~\ref{fig:Pu240_sp_4_3_linklines}) with 
appropriate gating conditions, $\it{i.e.}$, one gate set on one of transitions in one band and the other gate placed on one of the lines 
in the other band. As can be seen in Figure~\ref{fig:Pu240_sp_4_3_linklines}, due to the range of spin in which they are 
located, these transitions are weak and have significant Doppler shifts and/or broadenings. As a result, they are hard to 
identify experimentally. However, by using the angular gating technique on the data with the Blue database format, 
the difficulty of observing those inter-band transitions was overcome 
in the present work. As described in Sec.~\ref{subsec:DirtMotvPresWk}, one of the purposes of present work is to establish 
the possible existence of an octupole rotational band consisting of states with alternating parity, connected 
by $E1$ transitions at high spin in $^{240}$Pu. Therefore, the observation of these $J^{-}{\rightarrow}(J-1)^{+}$ and 
$(J+1)^{+}{\rightarrow}J^{-}$ inter-band transitions, $\it{i.e.}$, the so-called ``zig-zag'' structure, at the $24\hbar$ and 
higher spins in the present work constitutes one of the 
most important evidence for fulfilling the above purpose. Its importance will be discussed further in Sec.~\ref{subsec:Pu_octupl_bds}. 

In Ref.~\cite{Hackman-PRC-57-R1056-98}, the spins and parity of levels $9^{-}$ -- $27^{-}$ were assigned taking advantage of the fact that 
these levels form a natural extension of a $K^{\pi}=0^{-}$ level sequence ($1^{-}$, $3^{-}$, $5^{-}$) previously identified in 
Refs.~\cite{Hseuh-PRC-23-1217-81,Parekh-PRC-26-2178-82}. Here, this assignment has been confirmed by the measured angular 
distribution coefficients for the transitions connecting bands 1 and 2 (indicative of dipole multipolarity) and the 
ones in band 2 (indicative of quadrupole multipolarity), see Table~\ref{tab:240Pu-B2-gintns-agdscff}. The spins and parity 
for states above $27^{-}$ were assigned based on the fact that all transitions in a rotational band are 
associated with an $E2$ multipolarity, $\it{i.e.}$, the argument is similar to that used for the highest-spin states in band 1. 
The ratios of intensity between the transition in band 1 and the line connecting bands 1 and 2, which have a same final state, 
for example the 469.6-$keV$ ($22^{+}{\rightarrow}20^{+}$) line and the 383.6-$keV$ ($21^{-}{\rightarrow}20^{+}$) transition, can be 
determined in the summed spectra double gated on the in-band transitions below the associated state of band 1 (the $20^{+}$ level 
for the above example). Then, the relative intensities of these inter-band transitions can be calculated 
(see Table~\ref{tab:240Pu-B2-gintns-agdscff}) since the ones of in-band lines of band 1 have been obtained earlier. The ratios of 
intensity between a transition in band 2 and the line connecting bands 1 and 2 which have a same initial state, for example the 
413.3-$keV$ ($21^{-}{\rightarrow}19^{-}$) line and the 383.6-$keV$ ($21^{-}{\rightarrow}20^{+}$) $\gamma$ ray, can also be 
extracted in the summed coincidence spectra double gated on the in-band transitions above the associated 
state of band 2 (the $21^{-}$ level for the above example). Similarly, the relative intensities of the transitions in band 2 
can be obtained (see Table~\ref{tab:240Pu-B2-gintns-agdscff}). The difference in the data analysis of bands 1 and 2 is that 
it is harder to pick proper gating transitions to generate coincidence spectra for band 2 because of the large number of 
contaminants and doublet transitions. This observation, together with the weaker intensity of band 2, explains why the 
quoted uncertainties in the relative intensities are often large. 

\subsection{\label{subsec:240Pu_band3}$^{240}$Pu band 3}
\begin{figure}
\begin{center}
\includegraphics[angle=270,width=\columnwidth]{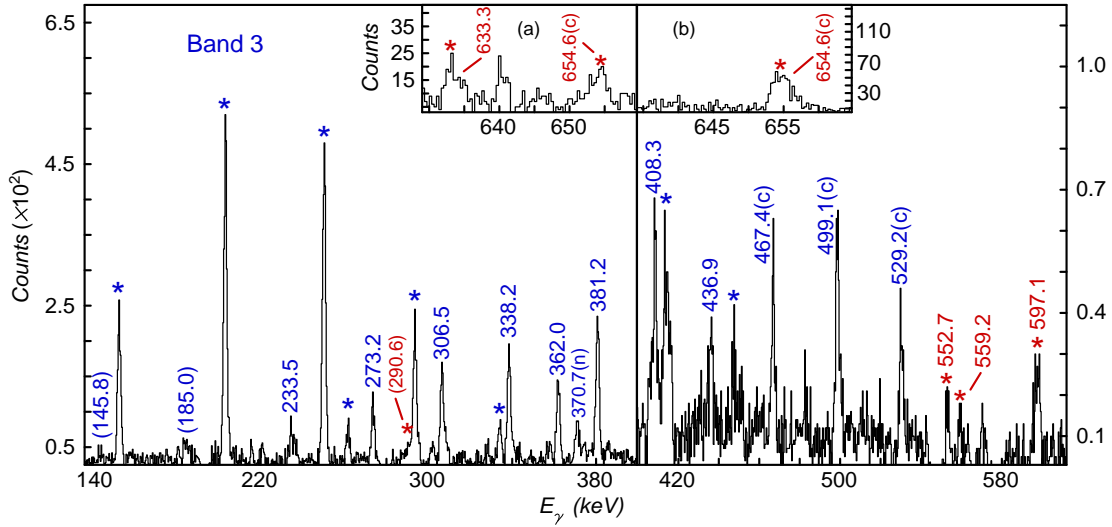}
\caption[Spectra representative of band 3 in $^{240}$Pu.]{Spectra 
representative of band 3 in $^{240}$Pu. The left panel of the main spectrum is the sum of several double-gated 
spectra. The gates were set on one of the two transitions with 273.2-$keV$ and 306.5-$keV$ energies, respectively, in band 3 
and one of another three lines in the same band (306.5-$keV$, 338.2-$keV$ and 362.0-$keV$), but, the two gates were not set on the 
306.5-$keV$ line at the same time. The right panel of the main spectrum is the spectrum double gated on the 338.2-$keV$ and the 
381.2-$keV$ $\gamma$ rays in band 3. Insert (a, left): the sum of spectra double gated on the 302.1-$keV$ line in 
band 2 and one of the transitions above the $16^{+}$ level in band 3; insert (b, right): the sum of spectra gated on the 340.7-$keV$ 
line in band 2 and one of the $\gamma$ rays above the $18^{+}$ level in band 3. The related transitions shown in the spectra 
are labeled in the manner used in Figure~\ref{fig:Pu240_sp_4_band2}. In addition, the energy value followed by the ``(n)'' symbol 
denotes a transition in the new band described in Sec.~\ref{subsec:240Pu_band4} and illustrated in 
Figure~\ref{fig:Pu240_unasgnd_band}.\label{fig:Pu240_sp_4_band3}}
\end{center}
\end{figure}

The second positive-parity band, band 3, in Figure~\ref{fig:Pu240_levl_sche} consists of 15 transitions. Starting from 
the $6^{+}$ level, none of the higher states of this band have been reported in the literature. The three bottom levels, on the other hand, were 
established previously by a decay study of $^{240}$Np~\cite{Parekh-PRC-26-2178-82}. All of the in-band transitions 
up to spin $30\hbar$ as well as the transitions connecting bands 2 and 3 are indicated in Figure~\ref{fig:Pu240_sp_4_band3}, 
except for the two $\gamma$ rays at the lowest spins (below the $4^{+}$ level). These were not observed in the present work due 
to the high internal conversion probability. Since the transitions associated with this band encounter intense 
contamination, such as the 408.3-$keV$ ($22^{+}{\rightarrow}20^{+}$) transition being affected by the 405.9-$keV$ 
($18^{+}{\rightarrow}16^{+}$) line in band 1 and the 408.9-$keV$ ($19^{-}{\rightarrow}18^{+}$) $\gamma$ ray between 
bands 1 and 2, for example, the number of transitions that can be used as gates for generating proper coincidence spectra is 
very limited. Hence, the spectra in Figure~\ref{fig:Pu240_sp_4_band3} were obtained with appropriate gating conditions 
following careful and thorough checking of the gates. As can be seen in the main spectrum of Figure~\ref{fig:Pu240_sp_4_band3}, 
the 145.8-$keV$ ($6^{+}{\rightarrow}4^{+}$) and the 185.0-$keV$ ($8^{+}{\rightarrow}6^{+}$) transitions are too weak to be 
identified firmly. Hence, these two transitions as well as the associated state, $\it{i.e.}$, the $6^{+}$ level, were assigned as 
tentative in the level scheme (Figure~\ref{fig:Pu240_levl_sche}). 

Applying the same method that was used in the case of band 2, the relative intensity and angular distribution 
coefficients for the transitions associated with band 3 were obtained (see Table~\ref{tab:240Pu-B3-gintns-agdscff}). 
When choosing the proper gates for generating certain spectra, extra caution was taken due to the weaker intensity 
and the more complex contaminations for this band, as described above. Compared with the typical values of $\gamma$-ray 
angular distribution coefficients ($A_2$ and $A_4$) given in Table~\ref{tab:angu_dist_coeff_values} 
of Chapter~\ref{chap:exp_techs}, the $A_2$ and $A_4$ coefficients for the transitions connecting bands 2 and 3 and 
the ones in band 3 (given in Table~\ref{tab:240Pu-B3-gintns-agdscff}) suggest their dipole and quadrupole nature, 
respectively. Hence, positive parity and even spins were assigned to the states of band 3 (the spins and parity for 
states above $20^{+}$ were again assigned based on the fact that these transitions represent the natural extension 
of the lower-spin sequence in this band and all in-band transitions possess $E2$ multipolarity). As will be shown later in the 
present work, the routhian plot of this band (Figure~\ref{fig:240Pu_routhian} in Sec.~\ref{subsec:Pu_ex_positv_parity_bd}) 
also suggests that the sequence of $\gamma$ 
rays observed in the present work (above the $6^{+}$ level) extrapolates well at lower angular frequency to the known 
$4^{+}$, $2^{+}$, and $0^{+}$ levels (the excitation energy of the $0^{+}$ state, $E_x$ is 860.7 $keV$). 
This band head was interpreted to be associated with a $\beta$ vibration in 
the decay work mentioned above~\cite{Parekh-PRC-26-2178-82}. The consistency of data points from the present work 
(at higher spin) with the ones from the decay study (at lower spin) supports, from another point of view, the assignment 
of spin and parity to this band proposed above. As in the case of band 2, the transitions connecting bands 2 
and 3 can be grouped into two types: $J^{+}{\rightarrow}(J-1)^{-}$ and $J^{+}{\rightarrow}(J+1)^{-}$. The latter ones are also much 
weaker than the former in intensity (see Table~\ref{tab:240Pu-B2-gintns-agdscff}). As can be seen in Figure~\ref{fig:Pu240_sp_4_band3}, 
the 290.6-$keV$ ($12^{+}{\rightarrow}13^{-}$) line, as an example of the weak $J^{+}{\rightarrow}(J+1)^{-}$ inter-band transitions, 
also faces strong contaminating $\gamma$ rays (293.5-$keV$ in band 1). These transitions were not observed clearly in the present 
work, and, hence, were taken as tentative in the level scheme. It is also interesting that band 3 was found to decay only to 
band 2. In other words, no linking transition between bands 1 and 3 was found, based on the observation that the 
maximum value of the relative intensity for virtual inter-band transitions connecting bands 1 and 3 was estimated to be 1 (the relative 
intensity of the 249.9-$keV$ line in band 1 defined as 1000) in the present work. 

\begin{table}
\begin{center}
\caption{THE EXCITATION ENERGIES ($E_x$) OF INITIAL STATES, ASSIGNED SPINS, $\gamma$-RAY ENERGIES ($E_{\gamma}$), RELATIVE 
$\gamma$-RAY INTENSITIES ($I_{\gamma}$) AND ANGULAR DISTRIBUTION COEFFICIENTS ($A_2$ AND $A_4$) FOR THE TRANSITIONS ASSOCIATED 
WITH BAND 3 IN $^{240}$Pu\label{tab:240Pu-B3-gintns-agdscff}}
\begin{tabular}{cccccc}
\hline \hline 
\multicolumn{6}{c}{Band 3 in $^{240}$Pu}\\
\hline 
$E_x$ ($keV$) & Assigned spin ($\hbar$) & $E_{\gamma}$ ($keV$) & $I_{\gamma}$ (rel.) & $A_{2}$ & $A_{4}$\\ 
\hline
1137.8 & $6^{+}{\rightarrow}4^{+}$ & 145.8(4) & & & \\
1322.8 & $8^{+}{\rightarrow}6^{+}$ & 185.0(3) & 1.2(8) & & \\
 & $8^{+}{\rightarrow}7^{-}$ & 445.2(4) & & & \\
1556.3 & $10^{+}{\rightarrow}8^{+}$ & 233.5(3) & 1.8(12) & & \\
 & $10^{+}{\rightarrow}9^{-}$ & 500.3(3) & 2.6(17) & -0.18(3) & 0.07(4)\\
 & $10^{+}{\rightarrow}11^{-}$ & 279.6(4) & & & \\
1829.4 & $12^{+}{\rightarrow}10^{+}$ & 273.2(3) & 3.1(17) & 0.18(2) & -0.12(4)\\
 & $12^{+}{\rightarrow}11^{-}$ & 552.7(4) & 2.8(15) & -0.23(4) & 0.03(3)\\
 & $12^{+}{\rightarrow}13^{-}$ & 290.6(4) & & & \\
2135.9 & $14^{+}{\rightarrow}12^{+}$ & 306.5(3) & 9(5) & 0.26(2) & -0.10(4)\\
 & $14^{+}{\rightarrow}13^{-}$ & 597.1(3) & 7(4) & -0.20(2) & 0.04(4)\\
 & $14^{+}{\rightarrow}15^{-}$ & 295.0(3) & 2.9(17) & & \\
2474.2 & $16^{+}{\rightarrow}14^{+}$ & 338.2(3) & 13(12) & 0.30(5) & -0.13(6)\\
 & $16^{+}{\rightarrow}15^{-}$ & 633.3(4) & 4(4) & -0.22(3) & 0.02(5)\\
 & $16^{+}{\rightarrow}17^{-}$ & 292.6(4) & & & \\
\hline 
\end{tabular}
\end{center}
\end{table}

\begin{table}
\begin{center}
\centerline{TABLE~\ref{tab:240Pu-B3-gintns-agdscff} (contd.)}
\begin{tabular}{cccccc}
\hline \hline 
$E_x$ ($keV$) & Assigned spin ($\hbar$) & $E_{\gamma}$ ($keV$) & $I_{\gamma}$ (rel.) & $A_{2}$ & $A_{4}$\\ 
\hline
2836.2 & $18^{+}{\rightarrow}16^{+}$ & 362.0(3) & 7(5) & 0.22(5) & -0.14(6)\\
 & $18^{+}{\rightarrow}17^{-}$ & 654.6(3) & 2.3(18) & & \\
3217.3 & $20^{+}{\rightarrow}18^{+}$ & 381.2(3) & 5(4) & 0.18(5) & -0.11(7)\\
 & $20^{+}{\rightarrow}19^{-}$ & 657.8(4) & 0.9(6) & & \\
3625.6 & $22^{+}{\rightarrow}20^{+}$ & 408.3(4) & & & \\
 & $22^{+}{\rightarrow}21^{-}$ & 652.8(4) & 0.8(4) & & \\
4062.5 & $24^{+}{\rightarrow}22^{+}$ & 436.9(5) & & & \\
4529.9 & $26^{+}{\rightarrow}24^{+}$ & 467.4(5) & & & \\
5029.0 & $28^{+}{\rightarrow}26^{+}$ & 499.1(5) & & & \\
5558.2 & $30^{+}{\rightarrow}28^{+}$ & 529.2(5) & & & \\
\hline 
\end{tabular}
\end{center}
\end{table}

\subsection{\label{subsec:240Pu_band4}Sequence(s) not assigned}
\begin{figure}
\begin{center}
\includegraphics[angle=0,width=0.20\columnwidth]{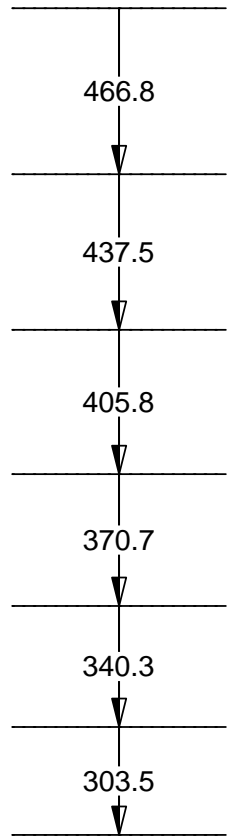}
\caption{A $\gamma$-ray sequence that was not assigned in present work.\label{fig:Pu240_unasgnd_band}}
\end{center}
\end{figure}

\begin{figure}
\begin{center}
\includegraphics[angle=270,width=\columnwidth]{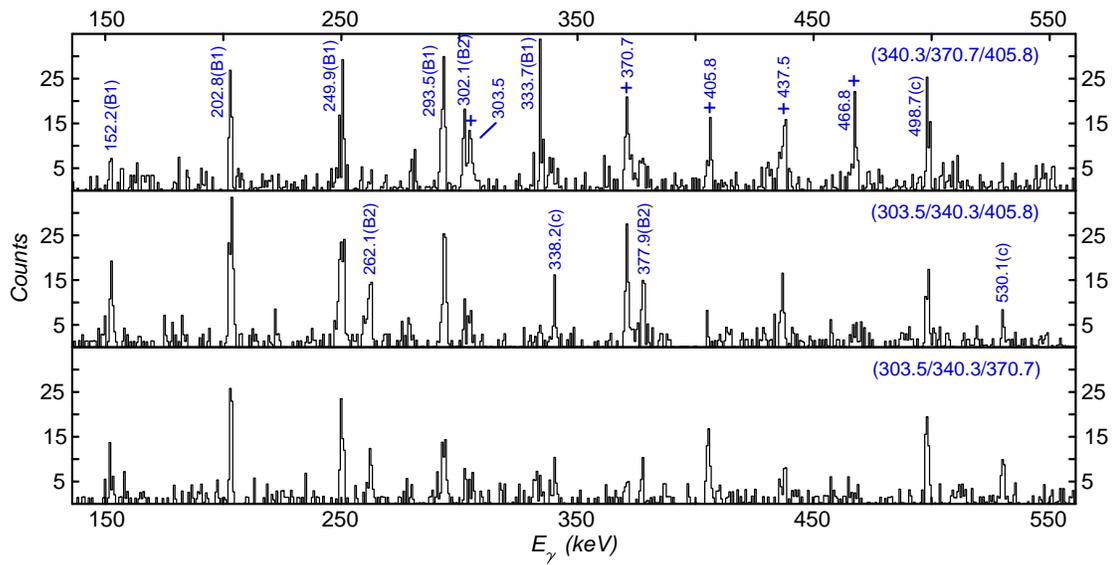}
\caption[Triple gated spectra supporting the observation of the new band that was not assigned in present 
work.]{Triple gated spectra supporting the observation of the new band that was not assigned in present work. 
The gates used to generate these spectra are written in the corresponding panels. 
``B1'', ``B2'', ``$+$'' and ``(c)'' symbols represent the transitions in band 1, in band 2, in this new band and 
the contaminated lines, respectively.\label{fig:Pu240_sp_4_band4}}
\end{center}
\end{figure}

In the present $^{240}$Pu data, another new band has been observed as illustrated in Figure~\ref{fig:Pu240_unasgnd_band}. 
Three supporting spectra in Figure~\ref{fig:Pu240_sp_4_band4} were generated by triple gating on three of the four 
transitions at 303.5-, 340.3-, 370.7- and 405.8-$keV$. The transitions used as gates in these spectra could not 
be in mutual coincidence if they were not part of this new band. The coincidence relations shown in the 
spectra of Figure~\ref{fig:Pu240_sp_4_band4} suggest that this new cascade is coincident with the transitions 
in $^{240}$Pu. Unfortunately, it was not possible to give any simple interpretation for this band. A thorough 
search was made in order to connect this cascade to the known states in $^{240}$Pu and in neighboring nuclei, 
but without success. Hence, this new band could not be assigned in present work, but, it will perhaps be identified 
in future experiments. 

\section{$^{242}$Pu data}
\begin{figure}
\begin{center}
\includegraphics[angle=0,width=0.92\columnwidth]{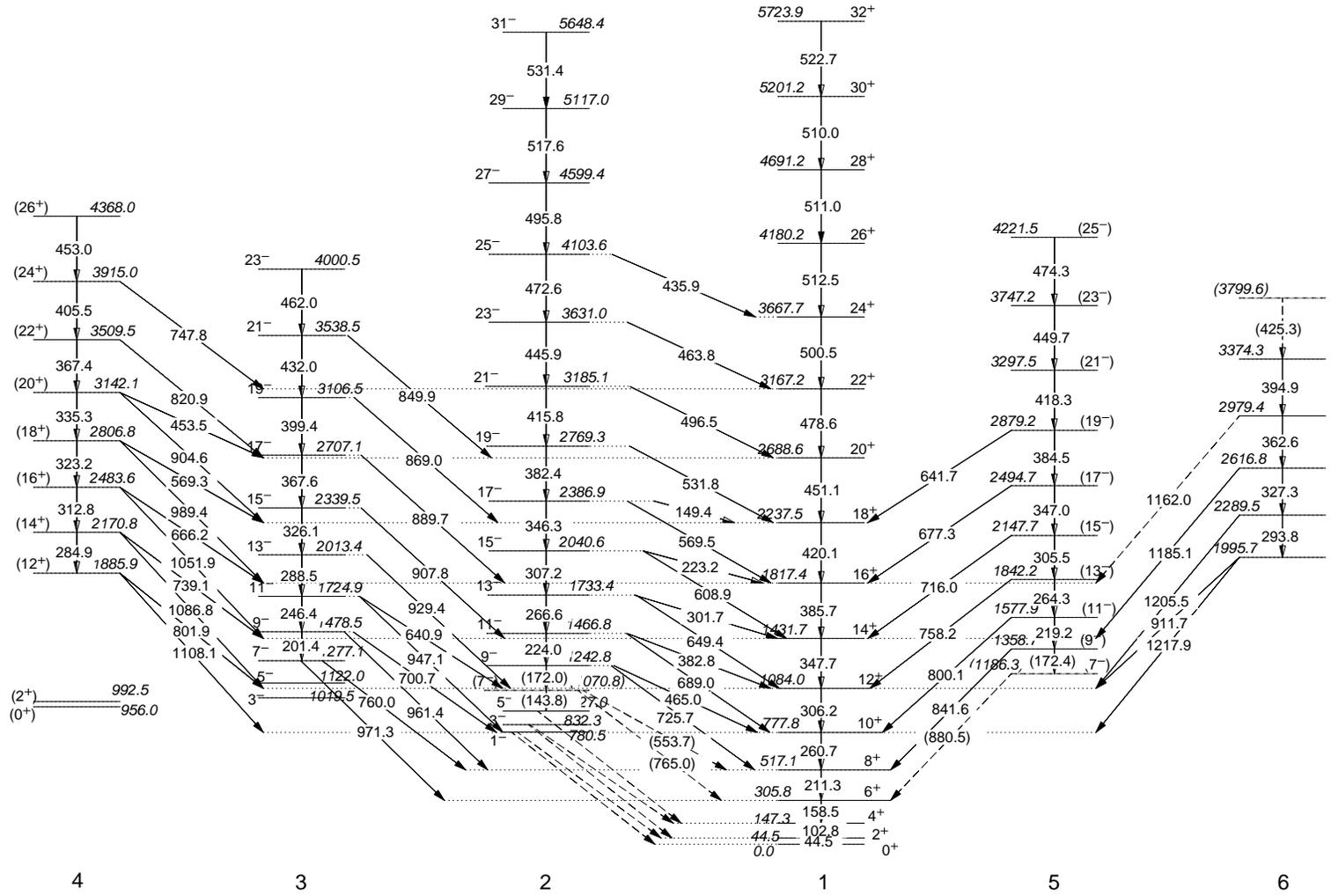}
\caption[Partial level scheme of $^{242}$Pu resulting from the present work.]{Partial 
level scheme of $^{242}$Pu resulting from the present work. See text for details.\label{fig:Pu242_levl_sche}}
\end{center}
\end{figure}

The level scheme of $^{242}$Pu resulting from the present work is presented in Figure~\ref{fig:Pu242_levl_sche}. 
The six bands, observed in our experiment, are labeled as bands 1 -- 6 and will be 
discussed in detail one by one below. The states and transitions in the $^{242}$Pu level scheme 
were labeled in a manner similar to that used in the $^{240}$Pu case. 

\begin{figure}
\begin{center}
\includegraphics[angle=0,width=\columnwidth]{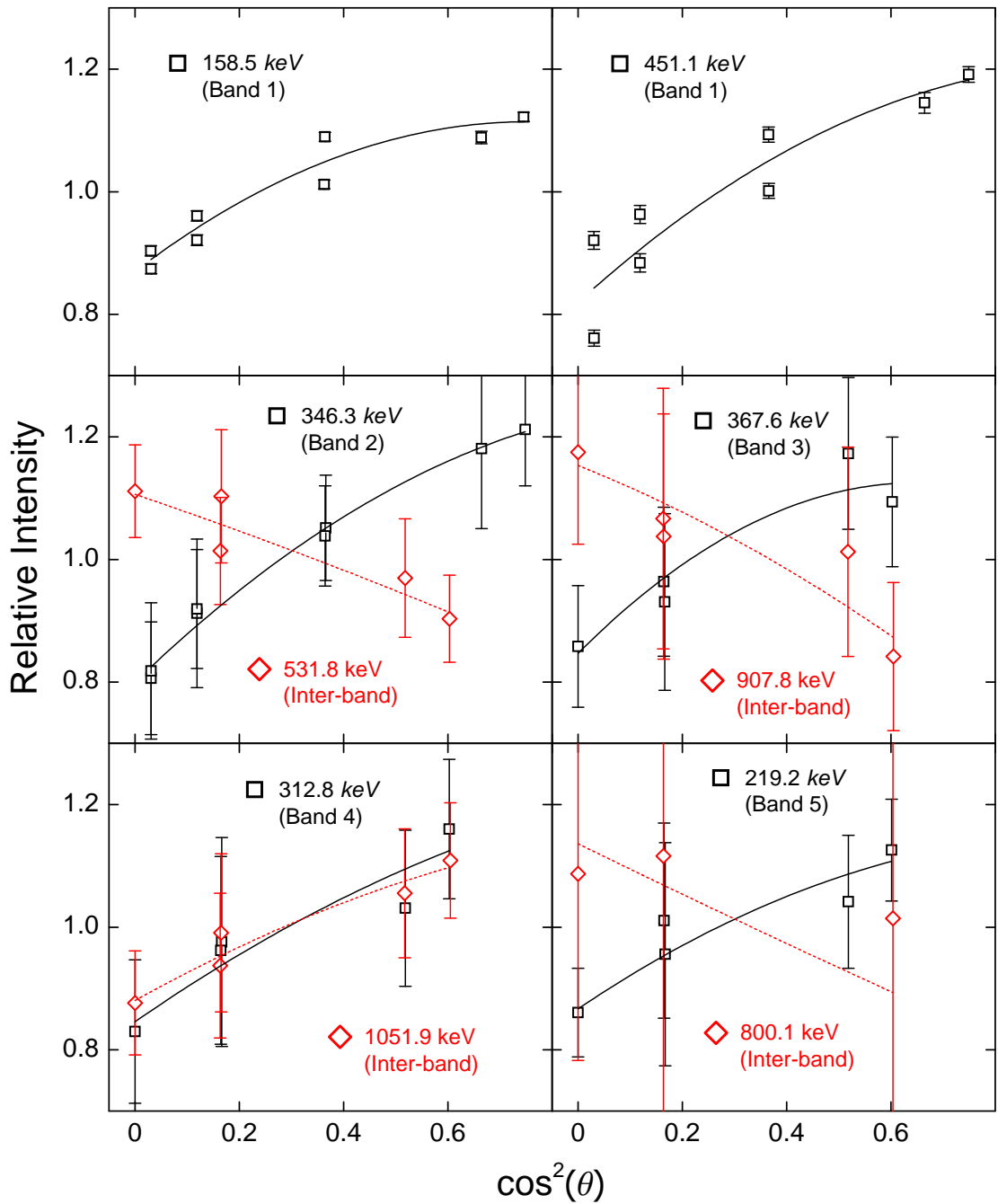}
\caption[Samples of angular distributions for transitions in $^{242}$Pu.]{Samples 
of angular distribution for transitions in $^{242}$Pu. The drawn curves (solid or dashed) represent 
the best fit of the data points. See text for details.\label{fig:242Pu_angu_dist_smp}}
\end{center}
\end{figure}

For computation of the relative intensities in the column ``$I_{\gamma}$'' of 
Tables~\ref{tab:242Pu-B1-gintns-agdscff}--\ref{tab:242Pu-B6-gintns-agdscff}, the 102.8-$keV$ ($4^{+}{\rightarrow}2^{+}$) transition 
in band 1 was taken as the reference, and its intensity was normalized to ``10000'' for convenience. Based on the 
relative $\gamma$-ray intensities obtained, the population of bands 2 -- 5 relative to band 1 in this experiment 
was estimated to be $7\%$, $2\%$, $1\%$ and $1\%$, respectively. It was found that band 6 was populated even less 
than band 5. A definite number for its population is not available due to the lack of statistics, 
but the corresponding upper limit was estimated to be $0.5\%$. 
Representative angular distributions for in-band and inter-band transitions associated with each band (except band 6) in 
$^{242}$Pu are compared in Figure~\ref{fig:242Pu_angu_dist_smp}, $\it{i.e.}$, the examples of the 158.5-$keV$ ($6^{+}{\rightarrow}4^{+}$) 
and 451.1-$keV$ ($20^{+}{\rightarrow}18^{+}$) lines in band 1, the 346.3-$keV$ ($17^{-}{\rightarrow}15^{-}$) $\gamma$ ray in band 2, the 
531.8-$keV$ ($19^{-}{\rightarrow}18^{+}$) transition linking bands 1 and 2, and the 367.6-$keV$ ($17^{-}{\rightarrow}15^{-}$) line in 
band 3, the 907.8-$keV$ ($15^{-}{\rightarrow}14^{+}$) transition linking bands 1 and 3, and the 312.8-$keV$ ($16^{+}{\rightarrow}14^{+}$) 
$\gamma$ ray in band 4, the 1051.9-$keV$ ($16^{+}{\rightarrow}14^{+}$) transition linking bands 1 and 4, the 219.2-$keV$ 
($11^{-}{\rightarrow}9^{-}$) line in band 5, and the 800.1-$keV$ ($11^{-}{\rightarrow}10^{+}$) transition linking bands 1 and 5. It 
was found in the analysis that the measured $A_2$ and $A_4$ coefficents for in-band and inter-band transitions in $^{242}$Pu 
(see Tables~\ref{tab:242Pu-B1-gintns-agdscff}--\ref{tab:242Pu-B6-gintns-agdscff}) are very close to the typical values expected for 
quadrupole and dipole $\gamma$ rays (see Table~\ref{tab:angu_dist_coeff_values} in Sec.~\ref{subsec:LS-AngDist} 
of Chapter~\ref{chap:exp_techs}), respectively. The only exception occurred in the case of $\gamma$ rays linking bands 1 and 4. 
The measured $A_2$ and $A_4$ coefficients for them, such as the 1051.9-$keV$ line (its angular distribution is illustrated in 
Figure~\ref{fig:242Pu_angu_dist_smp}), for example, suggest their quadrupole, rather than dipole, multipolarity. This exception 
will be discussed further in Sec.~\ref{subsec:242Pu_band4}. 

\subsection{\label{subsec:242Pu_band1}$^{242}$Pu band 1}
\begin{figure}
\begin{center}
\includegraphics[angle=270,width=\columnwidth]{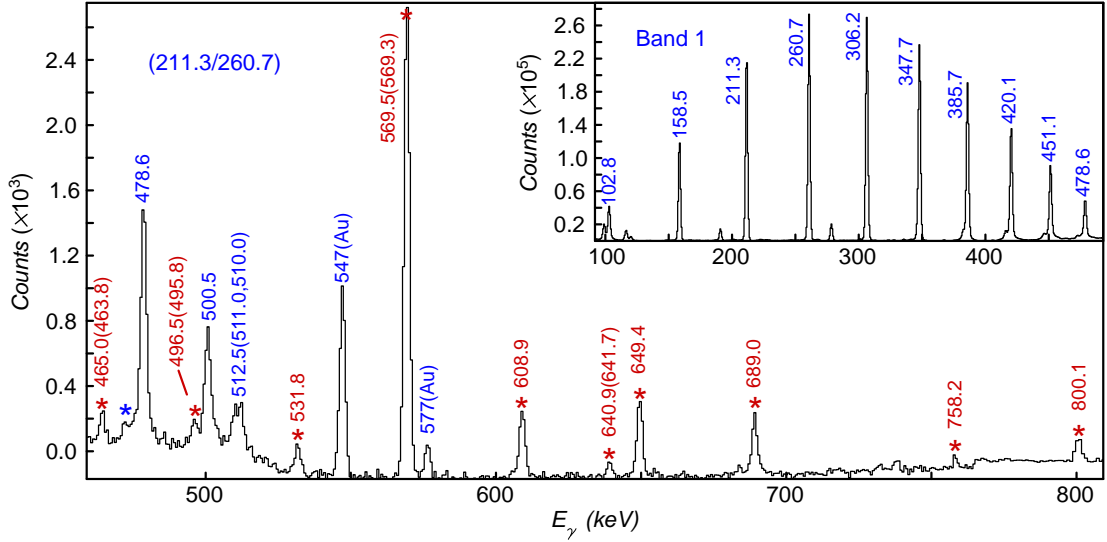}
\caption[Spectra representative of band 1 in $^{242}$Pu.]{Spectra 
representative of band 1 in $^{242}$Pu. The main panel is the spectrum gated on the 211.3-$keV$ 
and 260.7-$keV$ lines in band 1. Insert: the sum of spectra double gated on any two of the 11 in-band $\gamma$ rays 
(from 102.8-$keV$ to 500.5-$keV$) of band 1. The related transitions shown in the spectra are labeled in a 
manner similar to the one used in Figure~\ref{fig:Pu240_sp_4_band2}, except that the ``(c)'' symbols are replaced by 
the definite energy values of contaminants in parentheses, and, ``(Au)'' symbols denote the $\gamma$ rays from Coulomb 
excitation of the Au backings.\label{fig:Pu242_sp_4_band1}}
\end{center}
\end{figure}

\begin{figure}
\begin{center}
\includegraphics[angle=270,width=\columnwidth]{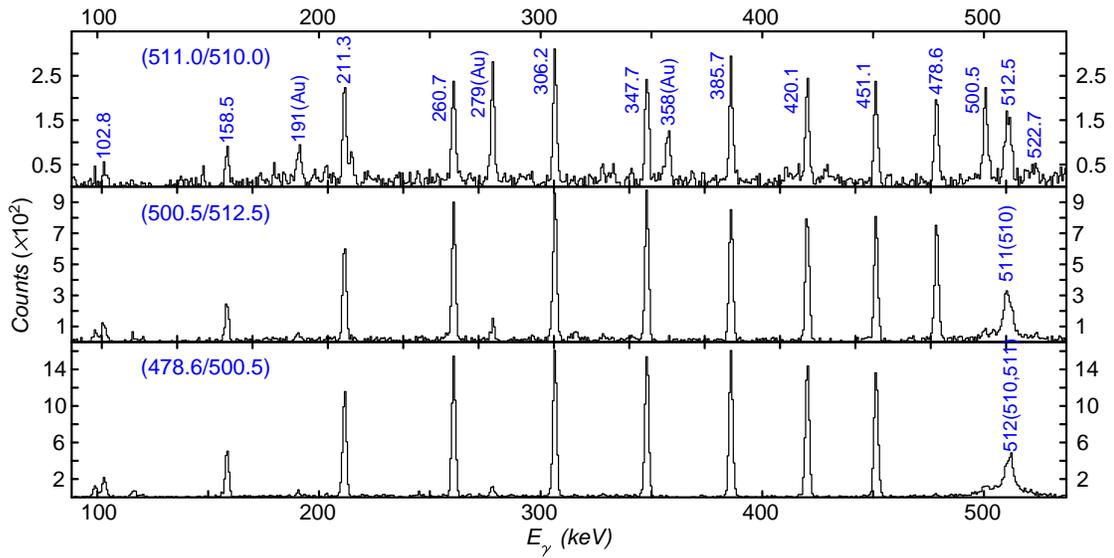}
\caption[Spectra highlighting the high-spin triplet in band 1 of $^{242}$Pu.]{Spectra 
highlighting the high-spin triplet in band 1 of $^{242}$Pu. The gates used to generate the three spectra 
(bottom, middle and top) are written in the corresponding panels. The related transitions shown in the spectra 
are labeled in the manner used in Figure~\ref{fig:Pu242_sp_4_band1}. See text for details.\label{fig:Pu242_sp_4_band1_tp}}
\end{center}
\end{figure}

Band 1 in Figure~\ref{fig:Pu242_levl_sche} is the yrast band. The states with spin up to $26{\hbar}$ in this band 
have been identified in several previous measurements~\cite{Spreng-PRL-51-1522-83,Cline-APPB-30-1291-99,Wiedenhover-PRL-83-2143-99}. 
In the present work, three additional transitions above the previously known $26^{+}$ level were observed, and these 
extend this band to spin $32\hbar$. All in-band transitions of band 1 below the $24^{+}$ state, except the 44.5-$keV$ 
($2^{+}{\rightarrow}0^{+}$) line, as well as some decay transitions from other bands in $^{242}$Pu to band 1 can be 
seen in Figure~\ref{fig:Pu242_sp_4_band1}. It is not surprising that the 44.5-$keV$ transition was 
suppressed, due to its high internal electron conversion probability. In the process of analysing spectra double gated on 
high-spin transitions in the present work, it was found that the $26^{+}{\rightarrow}24^{+}$ line in band 1, which was considered 
as the available highest-spin one with 510.0-$keV$ energy in the literature~\cite{Spreng-PRL-51-1522-83,Cline-APPB-30-1291-99}, 
is actually a triplet, $\it{i.e.}$, three transitions having almost degenerate energy. Since energy values of the three lines located 
in the same band are very close, $\it{i.e.}$, 510.0-, 511.0- and 512.5-$keV$, they have not been separated before as the experiments were 
mostly carried out with thin targets and Doppler broadening contributed significantly. The three transitions in this high-spin 
triplet and an additional line (522.7-$keV$) above it as well as the transitions at lower spin in band 1 are confirmed 
here by three supporting coincidence spectra (Figure~\ref{fig:Pu242_sp_4_band1_tp}) generated with proper gating conditions. 
It can be seen that, for the peaks located at the energy value of this triplet ($\sim$ 511-$keV$), the width and the area are 
the largest in the bottom spectrum gated on the lowest-spin $\gamma$ rays (both below triplet), and are the smallest in the 
top spectrum gated on the highest-spin lines (both in triplet). In the middle spectrum, generated with one gate set on a 
transition below the triplet (500.5-$keV$) and the other one placed on a line in the triplet (512.5-$keV$), both the width 
and area values for the peak of interest are intermediate. Furthermore, careful coincidence scans through the triplet 
provide additional evidence for the ordering proposed in the scheme (Figure~\ref{fig:Pu242_levl_sche}). 
In addition, the higher-spin transition next to this triplet (522.7-$keV$) was observed most clearly in the top spectrum. 
As a result, it was concluded that there exists a triplet of $\gamma$ rays with energy about 511 $keV$ in this band. 
The relative intensity of each line in this triplet, obtained from fitting the peaks of interest in the spectra of 
Figure~\ref{fig:Pu242_sp_4_band1_tp}, is given in Table~\ref{tab:242Pu-B1-gintns-agdscff}. The order in spin of the 
three transitions reflects the measured intensities. 

The intensities of the three low-lying transitions, $\it{i.e.}$, the 102.8-$keV$ ($4^{+}{\rightarrow}2^{+}$), the 158.5-$keV$ 
($6^{+}{\rightarrow}4^{+}$) and the 211.3-$keV$ ($8^{+}{\rightarrow}6^{+}$) transitions, were obtained from the analysis of the total 
projection of the data. Using the method applied in the $^{240}$Pu case, the relative intensity of the 260.7-$keV$ 
($10^{+}{\rightarrow}8^{+}$) transition was then obtained after getting the ratio of intensity between the 211.3-$keV$ and 
260.7-$keV$ lines in the spectrum double gated on the 102.8-$keV$ and 158.5-$keV$ $\gamma$ rays. Similarly, the intensities 
of the 306.2-$keV$ transition and all others in the sequence were acquired from the appropriate spectra. The values of 
angular distribution coefficients ($A_2$ and $A_4$) for the in-band transitions of this band, obtained by analyzing the sum 
of angular spectra double gated on pairs of the 12 in-band lines (from 102.8-$keV$ to 511.0-$keV$), can be found in 
Table~\ref{tab:242Pu-B1-gintns-agdscff}. The spins and parity of states up to $24^{+}$ in this band, established in 
the work of Refs.~\cite{Spreng-PRL-51-1522-83,Cline-APPB-30-1291-99}, were confirmed in the present work. As the 
512-511-510-$keV$ sequence appears to be the natural extension of the yrast band, $E2$ multipolarity is assigned 
to these transitions as well although $A_2$ and $A_4$ coefficients could not be extracted in the view of the spectral 
complexity discussed above. 

\begin{table}
\begin{center}
\caption{THE EXCITATION ENERGIES ($E_x$) OF INITIAL STATES, ASSIGNED SPINS, $\gamma$-RAY ENERGIES ($E_{\gamma}$), RELATIVE 
$\gamma$-RAY INTENSITIES ($I_{\gamma}$) AND ANGULAR DISTRIBUTION COEFFICIENTS ($A_2$ AND $A_4$) FOR THE TRANSITIONS ASSOCIATED 
WITH BAND 1 IN $^{242}$Pu\label{tab:242Pu-B1-gintns-agdscff}}
\begin{tabular}{cccccc}
\hline \hline 
\multicolumn{6}{c}{Band 1 in $^{242}$Pu}\\
\hline 
$E_x$ ($keV$) & Assigned spin ($\hbar$) & $E_{\gamma}$ ($keV$) & $I_{\gamma}$ (rel.) & $A_{2}$ & $A_{4}$\\ 
\hline
147.3 & $4^{+}{\rightarrow}2^{+}$ & 102.8(2) & 10000(397) & 0.30(3) & -0.12(5)\\
305.8 & $6^{+}{\rightarrow}4^{+}$ & 158.5(2) & 7474(441) & 0.19(3) & -0.10(5)\\
517.1 & $8^{+}{\rightarrow}6^{+}$ & 211.3(2) & 10643(654) & 0.21(4) & -0.05(5)\\
777.8 & $10^{+}{\rightarrow}8^{+}$ & 260.7(2) & 9064(830) & 0.22(4) & -0.04(5)\\
1084.0 & $12^{+}{\rightarrow}10^{+}$ & 306.2(2) & 6716(815) & 0.19(3) & -0.06(3)\\
1431.7 & $14^{+}{\rightarrow}12^{+}$ & 347.7(2) & 4730(487) & 0.18(3) & -0.06(3)\\
1817.4 & $16^{+}{\rightarrow}14^{+}$ & 385.7(2) & 2899(319) & 0.23(5) & -0.08(7)\\
2237.5 & $18^{+}{\rightarrow}16^{+}$ & 420.1(2) & 1631(172) & 0.18(4) & -0.11(6)\\
2688.6 & $20^{+}{\rightarrow}18^{+}$ & 451.1(2) & 975(113) & 0.30(7) & -0.09(10)\\
3167.2 & $22^{+}{\rightarrow}20^{+}$ & 478.6(2) & 488(80) & 0.26(6) & -0.12(7)\\
3667.7 & $24^{+}{\rightarrow}22^{+}$ & 500.5(2) & 241(57) & 0.32(10) & -0.12(14)\\
4180.2 & $26^{+}{\rightarrow}24^{+}$ & 512.5(3) & 76(19) & & \\
4691.2 & $28^{+}{\rightarrow}26^{+}$ & 511.0(5) & 39(12) & & \\
5201.2 & $30^{+}{\rightarrow}28^{+}$ & 510.0(7) & 12(4) & & \\
5723.9 & $32^{+}{\rightarrow}30^{+}$ & 522.7(4) & 3.4(9) & & \\
\hline 
\end{tabular}
\end{center}
\end{table}

\subsection{\label{subsec:242Pu_band2}$^{242}$Pu band 2}
\begin{figure}
\begin{center}
\includegraphics[angle=270,width=\columnwidth]{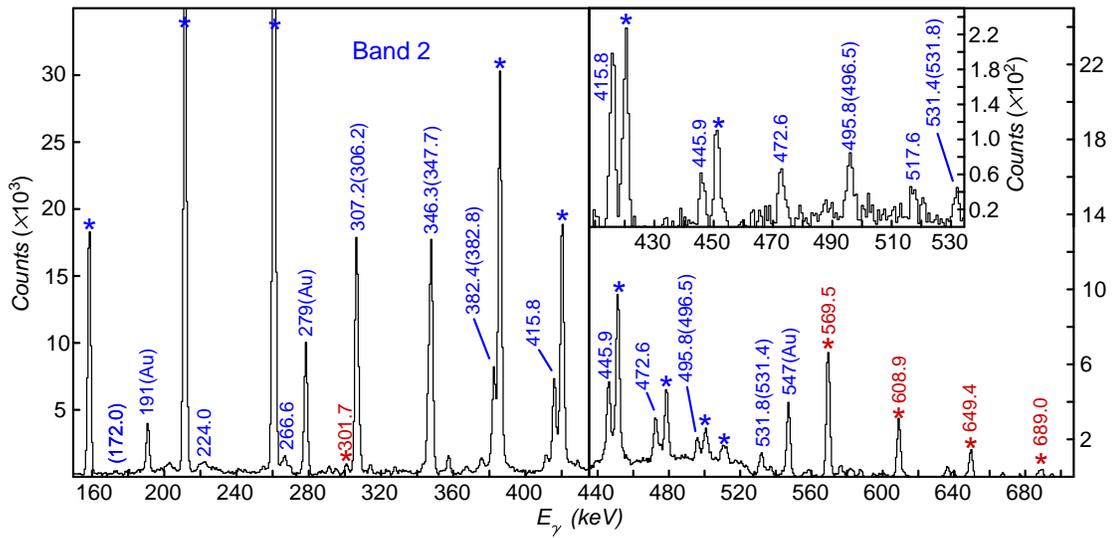}
\caption[Spectra representative of band 2 in $^{242}$Pu.]{Spectra 
representative of band 2 in $^{242}$Pu. Main panel: the sum of spectra double gated on any two of 
the 10 in-band $\gamma$ rays (from 224.0-$keV$ to 517.6-$keV$) of band 2. Insert: the sum of spectra gated on 
any two of the 445.9-$keV$, 472.6-$keV$ and 495.8-$keV$ lines in band 2. The related transitions are 
labeled in the manner used in Figure~\ref{fig:Pu242_sp_4_band1}.\label{fig:Pu242_sp_4_band2}}
\end{center}
\end{figure}

Band 2 in Figure~\ref{fig:Pu242_levl_sche}, observed in previous measurements~\cite{Elze-NPA-187-545-72,Wiedenhover-PRL-83-2143-99}, 
has been associated with a $K^{\pi}=0^{-}$ octupole vibration. It consists of 15 transitions. All of the 12 transitions 
above the $7^{-}$ level can be seen in Figure~\ref{fig:Pu242_sp_4_band2}. The inserted spectrum in this figure, generated with 
the gating conditions placed on the 445.9-$keV$ $\gamma$ ray and all other in-band lines above it, is presented here in order to highlight the 
high-spin (19 -- $31\hbar$) part of this band. The $7^{-}$ level ($E_x=1070.8~keV$) was not identified in the 
literature~\cite{Elze-NPA-187-545-72,Wiedenhover-PRL-83-2143-99} because of insufficient statistics. In the present work, the 
172.0-$keV$ ($9^{-}{\rightarrow}7^{-}$) transition, observed in Figure~\ref{fig:Pu242_sp_4_band2}, is still too weak to determine its 
energy precisely, therefore, this transition as well as the associated $7^{-}$ state are indicated as tentative in the level scheme 
(Figure~\ref{fig:Pu242_levl_sche}). The two bottom (below the $5^{-}$ level) $\gamma$ rays were not observed here due to the high internal 
electron conversion probability, but the associated states, $\it{i.e.}$, $1^{-}$, $3^{-}$ and $5^{-}$, have been established in 
earlier work~\cite{Elze-NPA-187-545-72,Haustein-PRC-19-2332-79}. Gamma rays associated with the deexcitation of this 
band to the yrast sequence (band 1) can be grouped into two types: $J^{-}{\rightarrow}(J-1)^{+}$ 
and $J^{-}{\rightarrow}(J+1)^{+}$, with the latter being much weaker in intensity (see Table~\ref{tab:242Pu-B2-gintns-agdscff}). 
These inter-band transitions are also indicated in Figure~\ref{fig:Pu242_sp_4_band2}. Additionally, as 
can be seen in the level scheme, the $(J+1)^{+}$ levels in band 1 are located, either, lower in energy ($J{\leq}21\hbar$) than, 
or, very close (${\Delta}E_x<100\;keV$; $J>21\hbar$) to the $J^{-}$ states in band 2. Thus, no linking transitions between 
bands 1 and 2 in an opposite direction, like the $(J+1)^{+}{\rightarrow}J^{-}$ linking $\gamma$ rays identified 
in the $^{240}$Pu case, were observed in $^{242}$Pu. 

\begin{table}
\begin{center}
\caption{THE EXCITATION ENERGIES ($E_x$) OF INITIAL STATES, ASSIGNED SPINS, $\gamma$-RAY ENERGIES ($E_{\gamma}$), RELATIVE 
$\gamma$-RAY INTENSITIES ($I_{\gamma}$) AND ANGULAR DISTRIBUTION COEFFICIENTS ($A_2$ AND $A_4$) FOR THE TRANSITIONS ASSOCIATED 
WITH BAND 2 IN $^{242}$Pu\label{tab:242Pu-B2-gintns-agdscff}}
\begin{tabular}{cccccc}
\hline \hline 
\multicolumn{6}{c}{Band 2 in $^{242}$Pu}\\
\hline 
$E_x$ ($keV$) & Assigned spin ($\hbar$) & $E_{\gamma}$ ($keV$) & $I_{\gamma}$ (rel.) & $A_{2}$ & $A_{4}$\\ 
\hline
1070.8 & $7^{-}{\rightarrow}5^{-}$ & 143.8(10) & 279(884) & & \\
 & $7^{-}{\rightarrow}6^{+}$ & 765.0(2) & 279(41) & & \\
 & $7^{-}{\rightarrow}8^{+}$ & 553.7(2) & 280(397) & & \\
1242.8 & $9^{-}{\rightarrow}7^{-}$ & 172.0(5) & 23(19) & & \\
 & $9^{-}{\rightarrow}8^{+}$ & 725.7(2) & 60(14) & -0.29(11) & 0.11(9)\\
 & $9^{-}{\rightarrow}10^{+}$ & 465.0(3) & 46(25) & & \\
1466.8 & $11^{-}{\rightarrow}9^{-}$ & 224.0(3) & 36(24) & 0.26(9) & -0.10(16)\\
 & $11^{-}{\rightarrow}10^{+}$ & 689.0(2) & 92(17) & -0.24(8) & -0.08(10)\\
 & $11^{-}{\rightarrow}12^{+}$ & 382.8(2) & 24(19) & & \\
1733.4 & $13^{-}{\rightarrow}11^{-}$ & 266.6(2) & 134(61) & 0.34(20) & -0.07(25)\\
 & $13^{-}{\rightarrow}12^{+}$ & 649.4(2) & 121(21) & -0.22(10) & 0.06(5)\\
 & $13^{-}{\rightarrow}14^{+}$ & 301.7(2) & 39(26) & & \\
2040.6 & $15^{-}{\rightarrow}13^{-}$ & 307.2(2) & 357(114) & 0.24(10) & -0.05(7)\\
 & $15^{-}{\rightarrow}14^{+}$ & 608.9(2) & 128(22) & -0.15(10) & 0.03(5)\\
 & $15^{-}{\rightarrow}16^{+}$ & 223.2(4) & 30(17) & & \\
\hline 
\end{tabular}
\end{center}
\end{table}

\begin{table}
\begin{center}
\centerline{TABLE~\ref{tab:242Pu-B2-gintns-agdscff} (contd.)}
\begin{tabular}{cccccc}
\hline \hline 
$E_x$ ($keV$) & Assigned spin ($\hbar$) & $E_{\gamma}$ ($keV$) & $I_{\gamma}$ (rel.) & $A_{2}$ & $A_{4}$\\ 
\hline
2386.9 & $17^{-}{\rightarrow}15^{-}$ & 346.3(2) & 638(322) & 0.34(2) & -0.08(3)\\
 & $17^{-}{\rightarrow}16^{+}$ & 569.5(2) & 175(54) & -0.17(8) & 0.02(4)\\
 & $17^{-}{\rightarrow}18^{+}$ & 149.4(4) & 9(10) & & \\
2769.3 & $19^{-}{\rightarrow}17^{-}$ & 382.4(3) & 97(26) &  0.19(8) & -0.08(10)\\
 & $19^{-}{\rightarrow}18^{+}$ & 531.8(2) & 53(9) & -0.22(11) & -0.01(14)\\
3185.1 & $21^{-}{\rightarrow}19^{-}$ & 415.8(2) & 124(43) & 0.40(14) & -0.06(8)\\
 & $21^{-}{\rightarrow}20^{+}$ & 496.5(2) & 34(6) & -0.29(17) & 0.02(21)\\
3631.0 & $23^{-}{\rightarrow}21^{-}$ & 445.9(2) & 105(52) & 0.19(7) & -0.04(11)\\
 & $23^{-}{\rightarrow}22^{+}$ & 463.8(3) & 14(5) & & \\
4103.6 & $25^{-}{\rightarrow}23^{-}$ & 472.6(3) & 77(45) & 0.20(10) & -0.12(16)\\
 & $25^{-}{\rightarrow}24^{+}$ & 435.9(3) & 4(2) & & \\
4599.4 & $27^{-}{\rightarrow}25^{-}$ & 495.8(4) & & & \\
5117.0 & $29^{-}{\rightarrow}27^{-}$ & 517.6(4) & & & \\
5648.4 & $31^{-}{\rightarrow}29^{-}$ & 531.4(4) & & & \\
\hline 
\end{tabular}
\end{center}
\end{table}

As was done in the analysis of $^{240}$Pu data, the ratios of peak areas between the transitions with 
known intensities and those to be derived were extracted from the appropriate coincidence spectra. Thus, the relative intensities for 
the transitions associated with band 2 were obtained by, first, establishing the inter-band $\gamma$ rays, and subsequently, 
the in-band transitions (see the related discussion in Sec.~\ref{subsec:240Pu_band2}). For example, the relative intensity of the 
569.5-$keV$ ($17^{-}{\rightarrow}16^{+}$) inter-band line was acquired by analyzing the summed spectra double gated on the transitions 
below the $16^{+}$ state in band 1. Then, the relative intensity of the 346.3-$keV$ ($17^{-}{\rightarrow}15^{-}$) $\gamma$ ray in band 2 
was obtained by the analysis of the summed spectra double gated on the transitions above the $17^{-}$ state in band 2. 

The spins and parity of states in this band, proposed originally in Ref.~\cite{Wiedenhover-PRL-83-2143-99}, were adopted in the 
present work (Figure~\ref{fig:Pu242_levl_sche}). The assignment of the states up to the $25^{-}$ level was confirmed by the measured 
angular distribution coefficients for the transitions linking band 1 to band 2 (indicative of dipole multipolarity) and the ones 
in band 2 (indicative of quadrupole multipolarity), see Table~\ref{tab:242Pu-B2-gintns-agdscff}, while the assignment of spins and parity 
of the $27^{-}$ level and the states above it is supported by the fact that all in-band transitions are associated with an $E2$ multipolarity 
(referring to what has been done for the high-spin states in band 1). It is worth to note that the gating conditions used to generate 
appropriate spectra (double-gated), in order to study the angular distributions of the weak inter-band transitions, were placed 
on one transition in both bands. 

\subsection{\label{subsec:242Pu_band3}$^{242}$Pu band 3}
\begin{figure}
\begin{center}
\includegraphics[angle=270,width=\columnwidth]{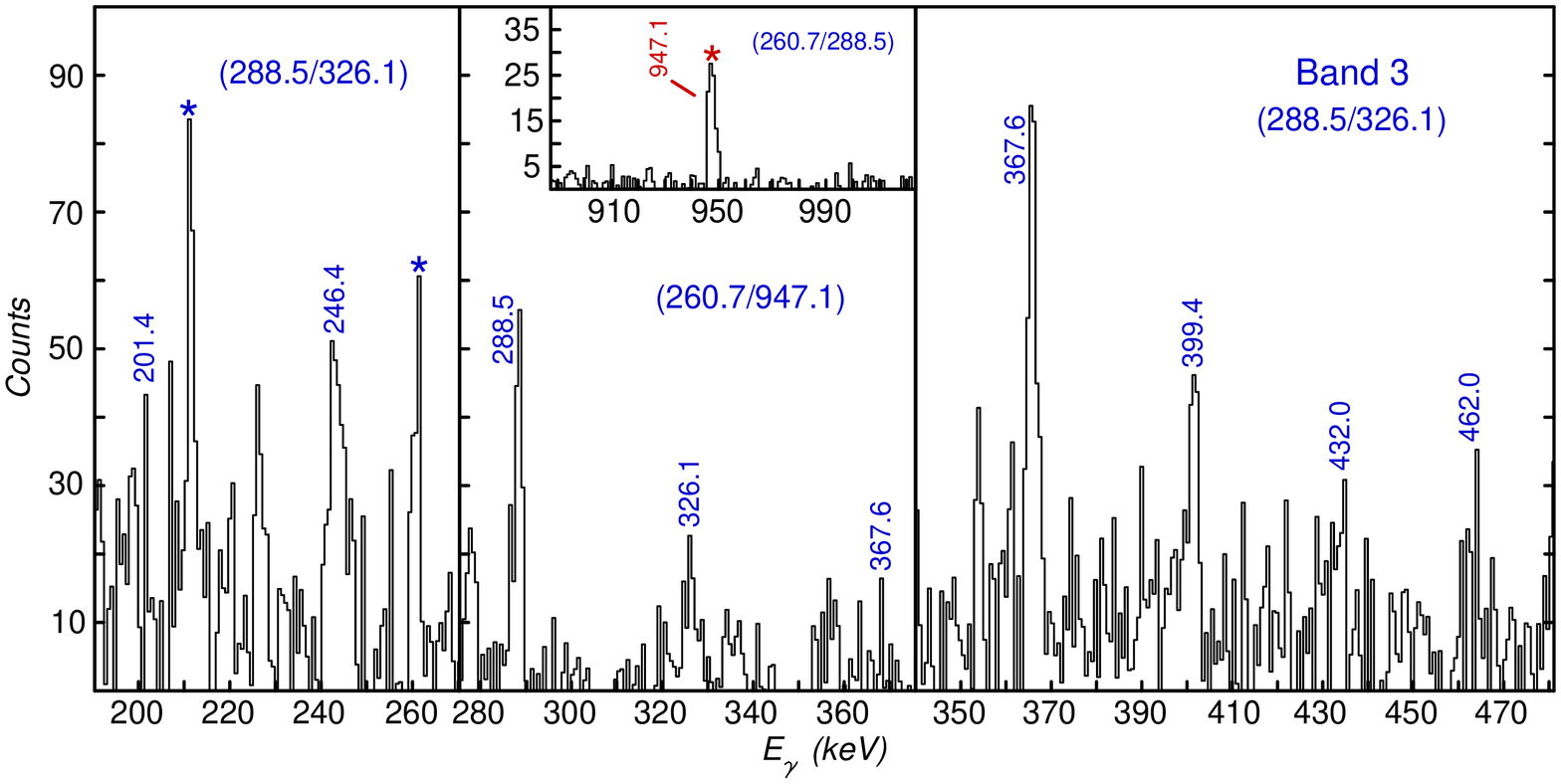}
\caption[Spectra representative of band 3 in $^{242}$Pu.]{Spectra 
representative of band 3 in $^{242}$Pu. The gates used to generate these spectra 
are written in the corresponding panels, respectively. The related transitions shown in the spectra are 
labeled in the manner used in Figure~\ref{fig:Pu242_sp_4_band1}.\label{fig:Pu242_sp_4_band3}}
\end{center}
\end{figure}

Due to their weak population and the presence of rather complex contaminations, the experimental evidence for 
bands 3 -- 6 in $^{242}$Pu cannot be provided through a single sample spectrum. Hence, for the supporting 
coincidence spectra shown below (Figures~\ref{fig:Pu242_sp_4_band3} -- \ref{fig:Pu242_sp_4_band6}), a figure generally 
consists of several panels. In each panel, one portion of an appropriate spectrum is illustrated with the purpose of 
highlighting a part of the related transitions only. The combination of all panels in a figure is then used to present 
all of the transitions observed in the band. 

Band 3 in Figure~\ref{fig:Pu242_levl_sche} consists of ten transitions. The $7^{-}$ level and the states above it have been 
established for the first time in the present work. The seven in-band lines above the $7^{-}$ level as well as some 
inter-band transitions between bands 1 and 3 are indicated in Figure~\ref{fig:Pu242_sp_4_band3}. The two lowest-spin 
$\gamma$ rays ($5^{-}{\rightarrow}3^{-}$ and $7^{-}{\rightarrow}5^{-}$) were suppressed because of high internal electron 
conversion probabilities. However, the associated $3^{-}$ and $5^{-}$ states were established in previous work, 
more than two decades ago~\cite{Elze-NPA-187-545-72}. This band decays to band 1 (yrast band) only. In other words, no 
linking transition between this band and any sequence other than band 1, such as band 2, for example, was found. This is based 
on the observation that the upper limit for the relative intensity of any virtual inter-band transition connecting this band 
with any other, except band 1, was determined to be 5 (the relative intensity of the 102.8-$keV$ line in band 1 is taken 
as 10000). The $\gamma$ rays associated with the deexcitation of band 3 to 1 can also be grouped into two types: 
$J^{-}{\rightarrow}(J-1)^{+}$ and $J^{-}{\rightarrow}(J+1)^{+}$. Transitions with the same initial state in these two 
types have comparable intensities, as seen in Table~\ref{tab:242Pu-B3-gintns-agdscff}. 

The intensities and angular distributions for transitions associated with band 3 were studied in the method also 
used in the case of band 2. The results of the analysis are summarized in Table~\ref{tab:242Pu-B3-gintns-agdscff}. 
The in-band and inter-band transitions (between bands 1 and 3) are indicative of quadrupole and dipole 
multipolarity, respectively, based on the measured $A_2$ and $A_4$ coefficients. As a result, negative parity 
and odd spins were assigned to the states of this band (the spins and parity for states above $17^{-}$ were assigned 
based on the fact that these transitions appear to represent a natural extension of the lower-spin sequence in 
this band and all in-band transitions possess $E2$ multipolarity. This is similar to the reasoning use several times 
above). Further, the routhian plot of this band, indicated in Figure~\ref{fig:242Pu_routhian_oth} 
(in Sec.~\ref{subsec:Pu242_other_bds}), was obtained later using 
the method described in Sec.~\ref{subsec:LabTrnsfBody} of Chapter~\ref{chap:theoriBkgd}. This plot also suggests 
that the sequence of $\gamma$ rays observed in the present work (above the $7^{-}$ level) extrapolates well at lower 
angular frequency to the known $5^{-}$, $3^{-}$ levels ($E_x=$ 1019.5 $keV$ for the $3^{-}$ state). The consistency 
of data points from the present work (at higher spin) with the ones from the previous work~\cite{Elze-NPA-187-545-72} 
(at lower spin) supports, from another point of view, the above assignment of spin and parity. 

\begin{table}
\begin{center}
\caption{THE EXCITATION ENERGIES ($E_x$) OF INITIAL STATES, ASSIGNED SPINS, $\gamma$-RAY ENERGIES ($E_{\gamma}$), RELATIVE 
$\gamma$-RAY INTENSITIES ($I_{\gamma}$) AND ANGULAR DISTRIBUTION COEFFICIENTS ($A_2$ AND $A_4$) FOR THE TRANSITIONS ASSOCIATED 
WITH BAND 3 IN $^{242}$Pu\label{tab:242Pu-B3-gintns-agdscff}}
\begin{tabular}{cccccc}
\hline \hline 
\multicolumn{6}{c}{Band 3 in $^{242}$Pu}\\
\hline 
$E_x$ ($keV$) & Assigned spin ($\hbar$) & $E_{\gamma}$ ($keV$) & $I_{\gamma}$ (rel.) & $A_{2}$ & $A_{4}$\\ 
\hline
1277.1 & $7^{-}{\rightarrow}6^{+}$ & 971.3(5) & 53(18) & -0.27(16) & 0.06(19)\\
 & $7^{-}{\rightarrow}8^{+}$ & 760.0(5) & 114(156) & & \\
1478.5 & $9^{-}{\rightarrow}7^{-}$ & 201.4(3) & 24(9) & 0.19(11) & -0.07(14)\\
 & $9^{-}{\rightarrow}8^{+}$ & 961.4(4) & 35(10) & -0.21(9) & 0.02(5)\\
 & $9^{-}{\rightarrow}10^{+}$ & 700.7(4) & 80(73) & & \\
1724.9 & $11^{-}{\rightarrow}9^{-}$ & 246.4(2) & 10(4) & 0.27(17) & -0.13(11)\\
 & $11^{-}{\rightarrow}10^{+}$ & 947.1(2) & 24(6) & -0.23(31) & 0.04(39)\\
 & $11^{-}{\rightarrow}12^{+}$ & 640.9(3) & 49(34) & -0.35(23) & 0.10(13)\\
2013.4 & $13^{-}{\rightarrow}11^{-}$ & 288.5(2) & 11(10) & 0.22(14) & -0.10(18)\\
 & $13^{-}{\rightarrow}12^{+}$ & 929.4(2) & 34(6) & -0.4(2) & -0.04(27)\\
2339.5 & $15^{-}{\rightarrow}13^{-}$ & 326.1(2) & 15(9) & 0.3(3) & -0.04(39)\\
 & $15^{-}{\rightarrow}14^{+}$ & 907.8(3) & 21(5) & -0.3(2) & -0.05(26)\\
2707.1 & $17^{-}{\rightarrow}15^{-}$ & 367.6(2) & 30(13) & 0.20(12) & -0.15(17)\\
 & $17^{-}{\rightarrow}16^{+}$ & 889.7(4) & 11(3) & & \\
\hline 
\end{tabular}
\end{center}
\end{table}

\begin{table}
\begin{center}
\centerline{TABLE~\ref{tab:242Pu-B3-gintns-agdscff} (contd.)}
\begin{tabular}{cccccc}
\hline \hline 
$E_x$ ($keV$) & Assigned spin ($\hbar$) & $E_{\gamma}$ ($keV$) & $I_{\gamma}$ (rel.) & $A_{2}$ & $A_{4}$\\ 
\hline
3106.5 & $19^{-}{\rightarrow}17^{-}$ & 399.4(3) & & & \\
 & $19^{-}{\rightarrow}18^{+}$ & 869.0(4) & 10(3) & & \\
3538.5 & $21^{-}{\rightarrow}19^{-}$ & 432.0(3) & & & \\
 & $21^{-}{\rightarrow}20^{+}$ & 849.9(4) & & & \\
4000.5 & $23^{-}{\rightarrow}21^{-}$ & 462.0(4) & & & \\
\hline 
\end{tabular}
\end{center}
\end{table}

\subsection{\label{subsec:242Pu_band4}$^{242}$Pu band 4}
\begin{figure}
\begin{center}
\includegraphics[angle=270,width=\columnwidth]{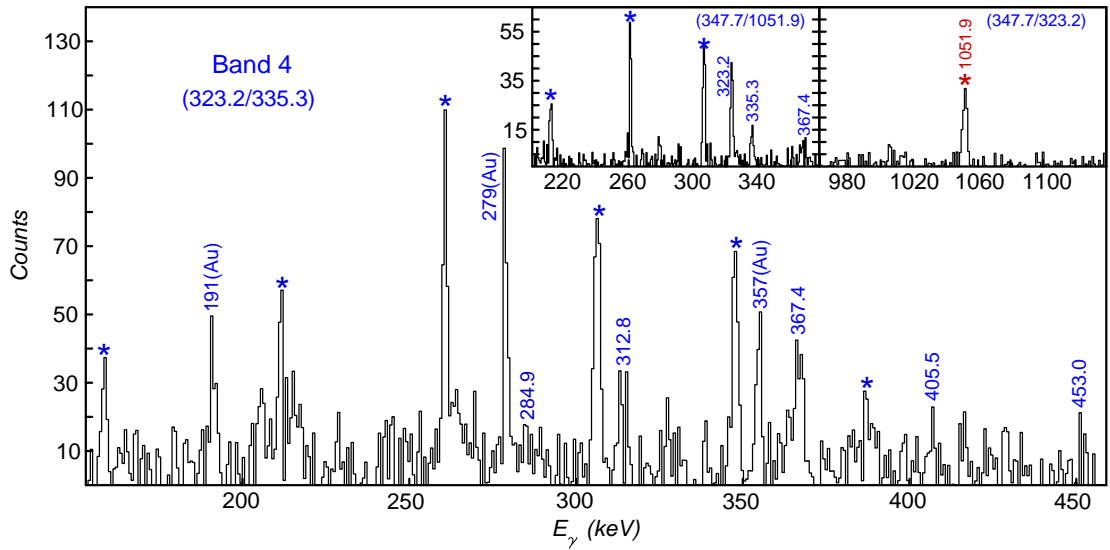}
\caption[Spectra representative of band 4 in $^{242}$Pu.]{Spectra 
representative of band 4 in $^{242}$Pu. The gates used to generate these spectra 
are written in the corresponding panels. The transitions are 
labeled in the manner used in Figure~\ref{fig:Pu242_sp_4_band1}.\label{fig:Pu242_sp_4_band4}}
\end{center}
\end{figure}

Band 4 in Figure~\ref{fig:Pu242_levl_sche} is the second positive-parity band in $^{242}$Pu. The $12^{+}$ level and the 
states above it have been established here for the first time. The seven in-band lines above the $12^{+}$ level 
as well as some inter-band transitions between bands 1 and 4 are indicated in Figure~\ref{fig:Pu242_sp_4_band4}. The 
$\gamma$ rays below the $12^{+}$ state were not observed in the present work due to the lack of statistics 
and/or the high internal electron conversion probability, but the two bottom levels ($0^{+}$ and $2^{+}$) have been 
reported in the literature~\cite{Maher-PRL-25-302-70,Maher-PRC-5-1380-72}. It was found that this band is connected with 
band 1 only. The linking transitions between the two bands can be grouped into two types: $J^{+}{\rightarrow}(J-2)^{+}$ and 
$J^{+}{\rightarrow}(J)^{+}$, with the latter being weaker in intensity (see Table~\ref{tab:242Pu-B4-gintns-agdscff}). 
The maximum value of the relative intensity for virtual inter-band transitions connecting this band with any other, 
except band 1, was estimated to be 3 (the relative intensity of the 102.8-$keV$ line in band 1 was defined as 10000). 

The results of data analysis for this band are summarized in Table~\ref{tab:242Pu-B4-gintns-agdscff}. 
The measured $A_2$ and $A_4$ coefficients for the in-band transitions of band 4 are indicative of their quadrupole nature. 
It is worth noting that the measured $A_2$ and $A_4$ values for some of the transitions connecting band 1 with 4, for example, 
the 1051.9-$keV$ (proposed $16^{+}{\rightarrow}14^{+}$) line, indicate their quadrupole rather than dipole character. In 
other words, the angular distributions for these inter-band $\gamma$ rays linking bands 1 and 4 
($J^{+}{\rightarrow}(J-2)^{+}$) are distinctly different from the ones of transitions connecting band 1 with bands 2 or 3. 
Rather, they are close to those for in-band $\gamma$ rays, as illustrated in Figure~\ref{fig:242Pu_angu_dist_smp}. As a result, 
positive parity and even spins, $\it{i.e.}$, $12^{+}$, $14^{+}$,..., $26^{+}$, were assigned to the states of band 4 (the spin and 
parity for the $22^{+}$ level and the states above it were assigned based on the fact that these transitions represent a 
natural extension of the lower-spin sequence in this band and all in-band transitions possess $E2$ multipolarity.). This 
assignment is further supported by the fact that all of the measured $A_2$ and $A_4$ coefficients for transitions associated 
with band 4 (Table~\ref{tab:242Pu-B4-gintns-agdscff}) can be explained well with the spins and parity assigned above. 
As will be shown later, the routhian plot of this band (Figure~\ref{fig:242Pu_routhian_pp} in Sec.~\ref{subsec:Pu_ex_positv_parity_bd}) 
also suggests that the sequence of $\gamma$ rays observed here extrapolates at lower angular frequency 
to the known $2^{+}$, $0^{+}$ levels ($K=0$; $E_x=956.0~keV$ for the $0^{+}$ state)~\cite{Maher-PRL-25-302-70,Maher-PRC-5-1380-72}. 
Nevertheless, the $4^{+}$, $6^{+}$, $8^{+}$ and $10^{+}$ levels have not been established yet; $\it{i.e.}$, the sequence of $\gamma$ rays 
in this band cannot be rigorously followed to low spin. Therefore, all of the spins and parity 
assigned above to the states of this band are given as tentative in the level scheme (Figure~\ref{fig:Pu242_levl_sche}). 

\begin{table}
\begin{center}
\caption{THE EXCITATION ENERGIES ($E_x$) OF INITIAL STATES, ASSIGNED SPINS, $\gamma$-RAY ENERGIES ($E_{\gamma}$), RELATIVE 
$\gamma$-RAY INTENSITIES ($I_{\gamma}$) AND ANGULAR DISTRIBUTION COEFFICIENTS ($A_2$ AND $A_4$) FOR THE TRANSITIONS ASSOCIATED 
WITH BAND 4 IN $^{242}$Pu\label{tab:242Pu-B4-gintns-agdscff}}
\begin{tabular}{cccccc}
\hline \hline 
\multicolumn{6}{c}{Band 4 in $^{242}$Pu}\\
\hline 
$E_x$ ($keV$) & Assigned spin ($\hbar$) & $E_{\gamma}$ ($keV$) & $I_{\gamma}$ (rel.) & $A_{2}$ & $A_{4}$\\ 
\hline
1885.9 & $12^{+}{\rightarrow}10^{+}$ & 1108.1(4) & 8(3) & & \\
 & $12^{+}{\rightarrow}12^{+}$ & 801.9(2) & 12(4) & & \\
2170.8 & $14^{+}{\rightarrow}12^{+}$ & 284.9(3) & 15(8) & 0.23(9) & -0.09(7)\\
 & $14^{+}{\rightarrow}12^{+}$ & 1086.8(2) & 12(4) & 0.15(10) & -0.06(11)\\
 & $14^{+}{\rightarrow}14^{+}$ & 739.1(2) & 10(3) & -0.23(11) & -0.08(12)\\
2483.6 & $16^{+}{\rightarrow}14^{+}$ & 312.8(3) & 16(7) & 0.27(15) & -0.05(20)\\
 & $16^{+}{\rightarrow}14^{+}$ & 1051.9(2) & 19(4) & 0.21(7) & -0.04(9)\\
 & $16^{+}{\rightarrow}16^{+}$ & 666.2(4) & 9(3) & & \\
2806.8 & $18^{+}{\rightarrow}16^{+}$ & 323.2(2) & 20(7) & 0.13(17) & -0.15(22)\\
 & $18^{+}{\rightarrow}16^{+}$ & 989.4(2) & 24(5) & & \\
 & $18^{+}{\rightarrow}18^{+}$ & 569.3(3) & & & \\
3142.1 & $20^{+}{\rightarrow}18^{+}$ & 335.3(2) & 6(3) & 0.27(16) & -0.15(21)\\
 & $20^{+}{\rightarrow}18^{+}$ & 904.6(2) & 13(3) & & \\
 & $20^{+}{\rightarrow}20^{+}$ & 453.5(3) & & & \\
\hline 
\end{tabular}
\end{center}
\end{table}

\begin{table}
\begin{center}
\centerline{TABLE~\ref{tab:242Pu-B4-gintns-agdscff} (contd.)}
\begin{tabular}{cccccc}
\hline \hline 
$E_x$ ($keV$) & Assigned spin ($\hbar$) & $E_{\gamma}$ ($keV$) & $I_{\gamma}$ (rel.) & $A_{2}$ & $A_{4}$\\ 
\hline
3509.5 & $22^{+}{\rightarrow}20^{+}$ & 367.4(2) & & & \\
 & $22^{+}{\rightarrow}20^{+}$ & 820.9(3) & & & \\
3915.0 & $24^{+}{\rightarrow}22^{+}$ & 405.5(3) & & & \\
 & $24^{+}{\rightarrow}22^{+}$ & 747.8(3) & & & \\
4368.0 & $26^{+}{\rightarrow}24^{+}$ & 453.0(3) & & & \\
\hline 
\end{tabular}
\end{center}
\end{table}

\subsection{\label{subsec:242Pu_band5_6}$^{242}$Pu bands 5 and 6}
\begin{figure}
\begin{center}
\includegraphics[angle=270,width=\columnwidth]{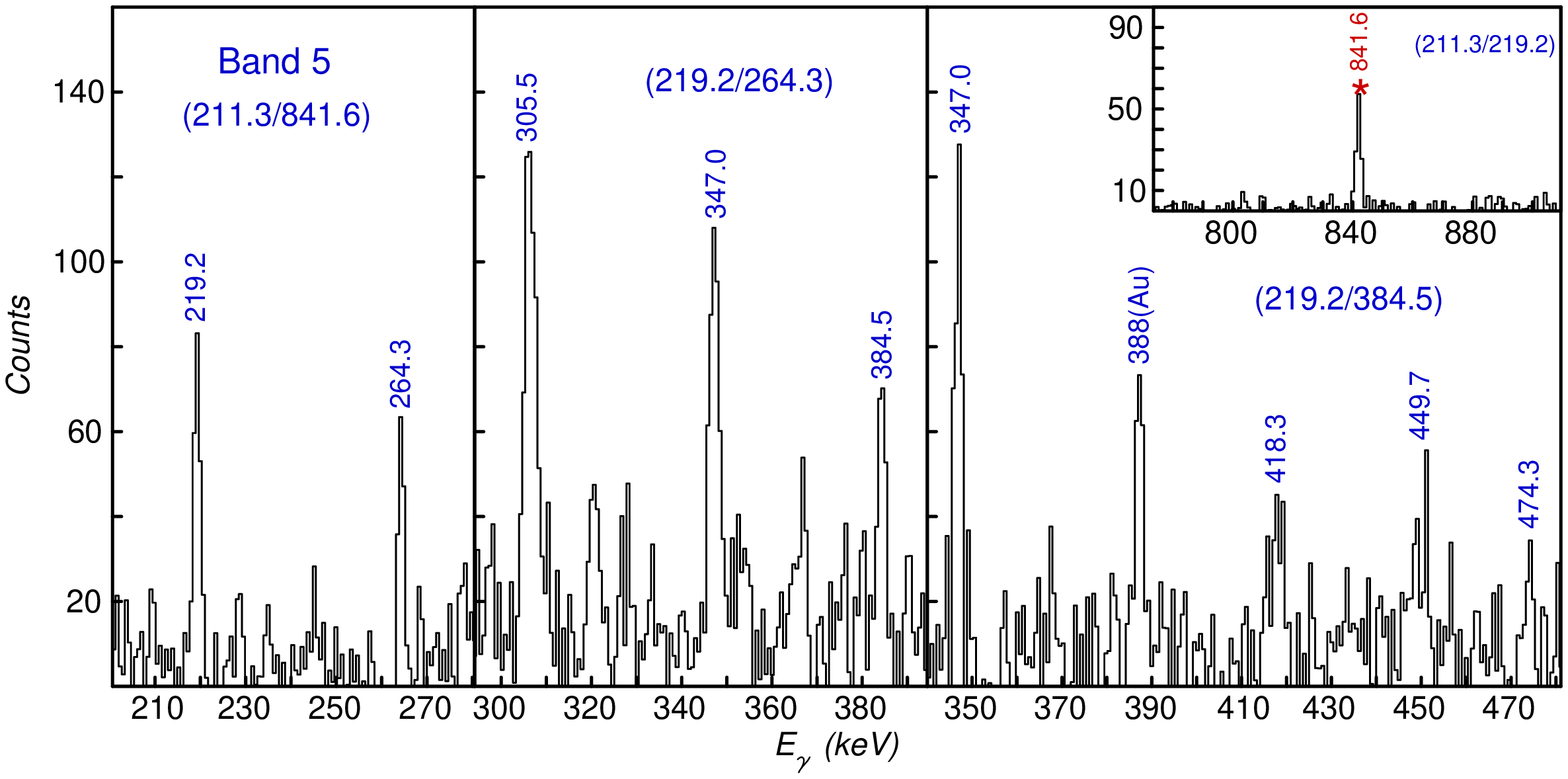}
\caption[Spectra representative of band 5 in $^{242}$Pu.]{Spectra 
representative of band 5 in $^{242}$Pu. The gates used to generate these spectra 
are written in the corresponding panels. The transitions are 
labeled in the manner used in Figure~\ref{fig:Pu242_sp_4_band1}.\label{fig:Pu242_sp_4_band5}}
\end{center}
\end{figure}

\begin{figure}
\begin{center}
\includegraphics[angle=270,width=\columnwidth]{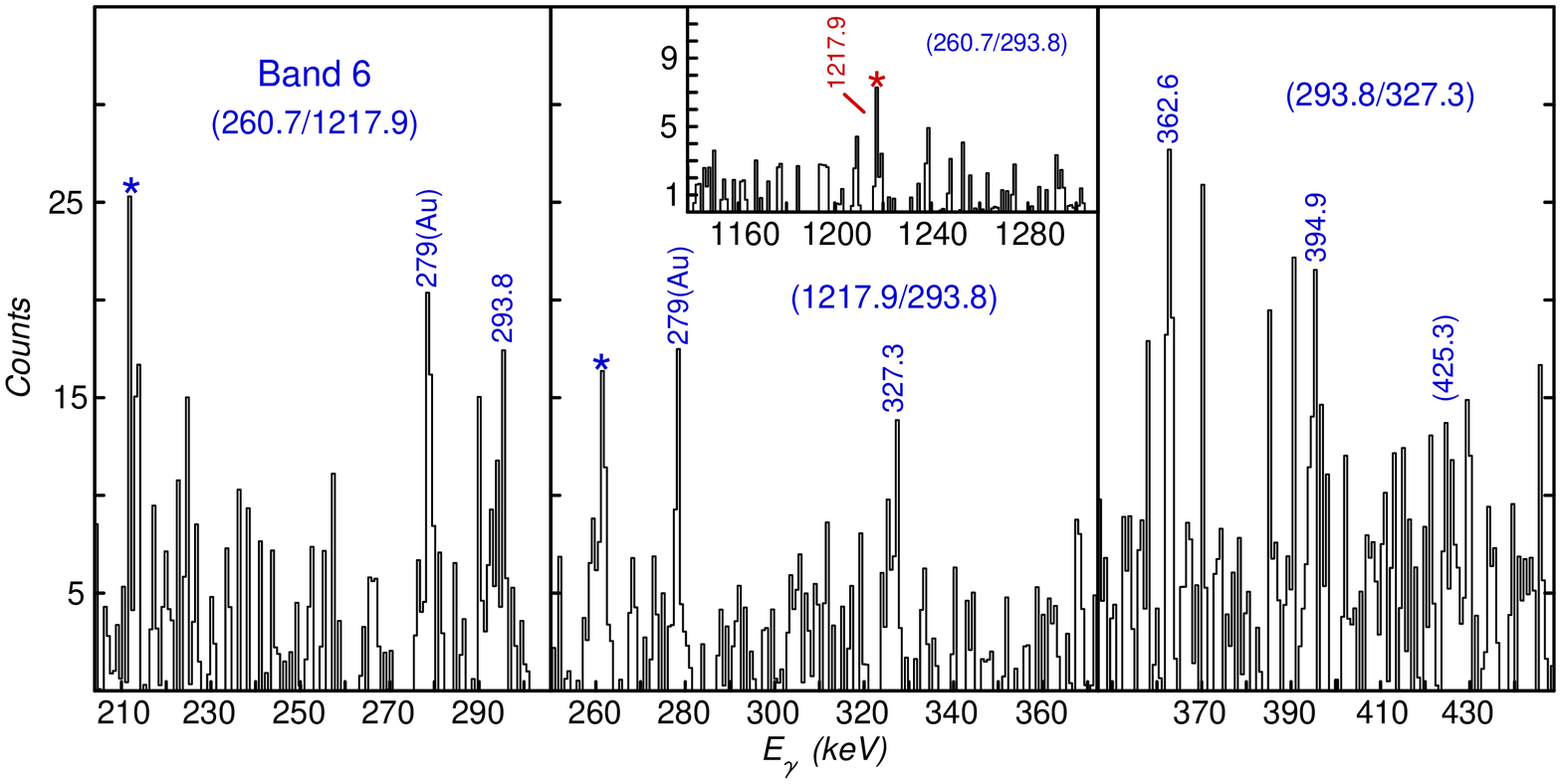}
\caption[Spectra representative of band 6 in $^{242}$Pu.]{Spectra 
representative of band 6 in $^{242}$Pu. The gates used to generate these spectra 
are written in the corresponding panels. The transitions are 
labeled in the manner used in Figure~\ref{fig:Pu242_sp_4_band1}.\label{fig:Pu242_sp_4_band6}}
\end{center}
\end{figure}

The $\gamma$-ray sequences of bands 5 and 6, shown in Figure~\ref{fig:Pu242_levl_sche}, are both new structures observed 
here for the first time. The eight transitions in band 5 and the five $\gamma$ rays in band 6, as well as some associated 
inter-band transitions, are indicated in Figures~\ref{fig:Pu242_sp_4_band5} and \ref{fig:Pu242_sp_4_band6}. In the spectra, 
the 172.4-$keV$ ($9^{-}{\rightarrow}7^{-}$) in-band and the 880.5-$keV$ ($7^{-}{\rightarrow}6^{+}$) inter-band transitions, 
associated with band 5, and the 425.3-$keV$ in-band and the 1162.0-$keV$ inter-band transitions, associated with band 6, 
were not identified firmly because of insufficient statistics. Thus, these lines as well as the associated states 
were assigned as tentative in the level scheme (Figure~\ref{fig:Pu242_levl_sche}). None of the low-spin (below the $7^{-}$ 
level for band 5 and below the 1995.7-$keV$ level for band 6) $\gamma$ rays was observed because of the lack of statistics 
and/or the high internal conversion probability. As was the case for bands 2 -- 4, bands 5 and 6 are linked to band 1 only. 
The upper limit for the relative intensity of virtual inter-band transitions, connecting bands 5 or 6 with any 
sequence other than band 1, was estimated to be 3 (the relative intensity of the 102.8-$keV$ line in band 1 is again 
taken as 10000). 

Because of the poor statistics for band 5 in the data, the measured relative intensities are small quantities with relatively 
large errors, and the angular distribution coefficients (the errors are also relatively large) could be extracted for only 
a small portion of the related transitions (4 of 16). The intensities and the angular distribution coefficients for 
transitions associated with band 6 were not determined due to even poorer statistics. Despite the large uncertainties, the 
measured $A_2$ and $A_4$ coefficients for the in-band and inter-band transitions in band 5 (Table~\ref{tab:242Pu-B5-gintns-agdscff}) 
indicate either quadrupole or dipole character depending on the case. Therefore, the levels in band 5 were assigned negative 
parity and odd spins ($7^{-}$, $9^{-}$, ..., $25^{-}$). Again, the spins and parity for the $15^{-}$ level and the states above it were 
assigned based on the continuation of the band through $E2$ transitions. Neither for band 5 or band 6, could the sequence be followed 
to low spin. In addition, none of the bandheads identified so far in the literature~\cite{Natnl-Nucl-Data-Cnt-242Pu} 
could be considered as a reasonable extrapolation at lower angular frequency of the sequence of $\gamma$ rays observed in the 
the present work. 
%in the resulting routhian plot (shown later in Figure~\ref{fig:242Pu_routhian_oth} in Sec.~\ref{subsec:Pu242_other_bds}). 
Therefore, all spin and parity 
values assigned above to the states of bands 5 are given as tentative in the level scheme and no quantum numbers are assigned 
to the levels of band 6. 

\begin{table}
\begin{center}
\caption{THE EXCITATION ENERGIES ($E_x$) OF INITIAL STATES, ASSIGNED SPINS, $\gamma$-RAY ENERGIES ($E_{\gamma}$), RELATIVE 
$\gamma$-RAY INTENSITIES ($I_{\gamma}$) AND ANGULAR DISTRIBUTION COEFFICIENTS ($A_2$ AND $A_4$) FOR THE TRANSITIONS ASSOCIATED 
WITH BAND 5 IN $^{242}$Pu\label{tab:242Pu-B5-gintns-agdscff}}
\begin{tabular}{cccccc}
\hline \hline 
\multicolumn{6}{c}{Band 5 in $^{242}$Pu}\\
\hline 
$E_x$ ($keV$) & Assigned spin ($\hbar$) & $E_{\gamma}$ ($keV$) & $I_{\gamma}$ (rel.) & $A_{2}$ & $A_{4}$\\ 
\hline
1186.3 & $7^{-}{\rightarrow}6^{+}$ & 880.5(5) & & & \\
1358.7 & $9^{-}{\rightarrow}7^{-}$ & 172.4(4) & 24(16) & & \\
 & $9^{-}{\rightarrow}8^{+}$ & 841.6(5) & 26(12) & -0.15(19) & 0.06(24)\\
1577.9 & $11^{-}{\rightarrow}9^{-}$ & 219.2(3) & 43(16) & 0.22(13) & -0.07(16)\\
 & $11^{-}{\rightarrow}10^{+}$ & 800.1(2) & 36(7) & -0.3(7) & 0.01(90)\\
1842.2 & $13^{-}{\rightarrow}11^{-}$ & 264.3(3) & 26(16) & 0.27(11) & -0.11(12)\\
 & $13^{-}{\rightarrow}12^{+}$ & 758.2(3) & 17(4) & & \\
2147.7 & $15^{-}{\rightarrow}13^{-}$ & 305.5(3) & 25(56) & & \\
 & $15^{-}{\rightarrow}14^{+}$ & 716.0(4) & 14(4) & & \\
2494.7 & $17^{-}{\rightarrow}15^{-}$ & 347.0(3) & 23(57) & & \\
 & $17^{-}{\rightarrow}16^{+}$ & 677.3(5) & 7(3) & & \\
2879.2 & $19^{-}{\rightarrow}17^{-}$ & 384.5(3) & 33(57) & & \\
 & $19^{-}{\rightarrow}18^{+}$ & 641.7(5) & & & \\
3297.5 & $21^{-}{\rightarrow}19^{-}$ & 418.3(3) & 27(56) & & \\
3747.2 & $23^{-}{\rightarrow}21^{-}$ & 449.7(4) & & & \\
4221.5 & $25^{-}{\rightarrow}23^{-}$ & 474.3(4) & & & \\
\hline 
\end{tabular}
\end{center}
\end{table}

\begin{table}
\begin{center}
\caption{THE EXCITATION ENERGIES ($E_x$) OF INITIAL STATES, ASSIGNED SPINS, $\gamma$-RAY ENERGIES 
($E_{\gamma}$) FOR THE TRANSITIONS ASSOCIATED WITH BAND 6 IN $^{242}$Pu\label{tab:242Pu-B6-gintns-agdscff}}
\begin{tabular}{ccc}
\hline \hline 
\multicolumn{3}{c}{Band 6 in $^{242}$Pu}\\
\hline 
$E_x$ ($keV$) & Assigned spin ($\hbar$) & $E_{\gamma}$ ($keV$)\\ 
\hline
1995.7 & $11^{-}{\rightarrow}10^{+}$ & 1217.9(3)\\
 & $11^{-}{\rightarrow}12^{+}$ & 911.7(3)\\
2289.5 & $13^{-}{\rightarrow}11^{-}$ & 293.8(3)\\
 & $13^{-}{\rightarrow}12^{+}$ & 1205.5(3)\\
2616.8 & $15^{-}{\rightarrow}13^{-}$ & 327.3(3)\\
 & $15^{-}{\rightarrow}14^{+}$ & 1185.1(4)\\
2979.4 & $17^{-}{\rightarrow}15^{-}$ & 362.6(3)\\
 & $17^{-}{\rightarrow}16^{+}$ & 1162.0(4)\\
3374.3 & $19^{-}{\rightarrow}17^{-}$ & 394.9(4)\\
3799.6 & $21^{-}{\rightarrow}19^{-}$ & 425.3(4)\\
\hline 
\end{tabular}
\end{center}
\end{table}

\section{$^{238}$Pu data}
\begin{figure}
\begin{center}
\includegraphics[angle=0,width=0.53\columnwidth]{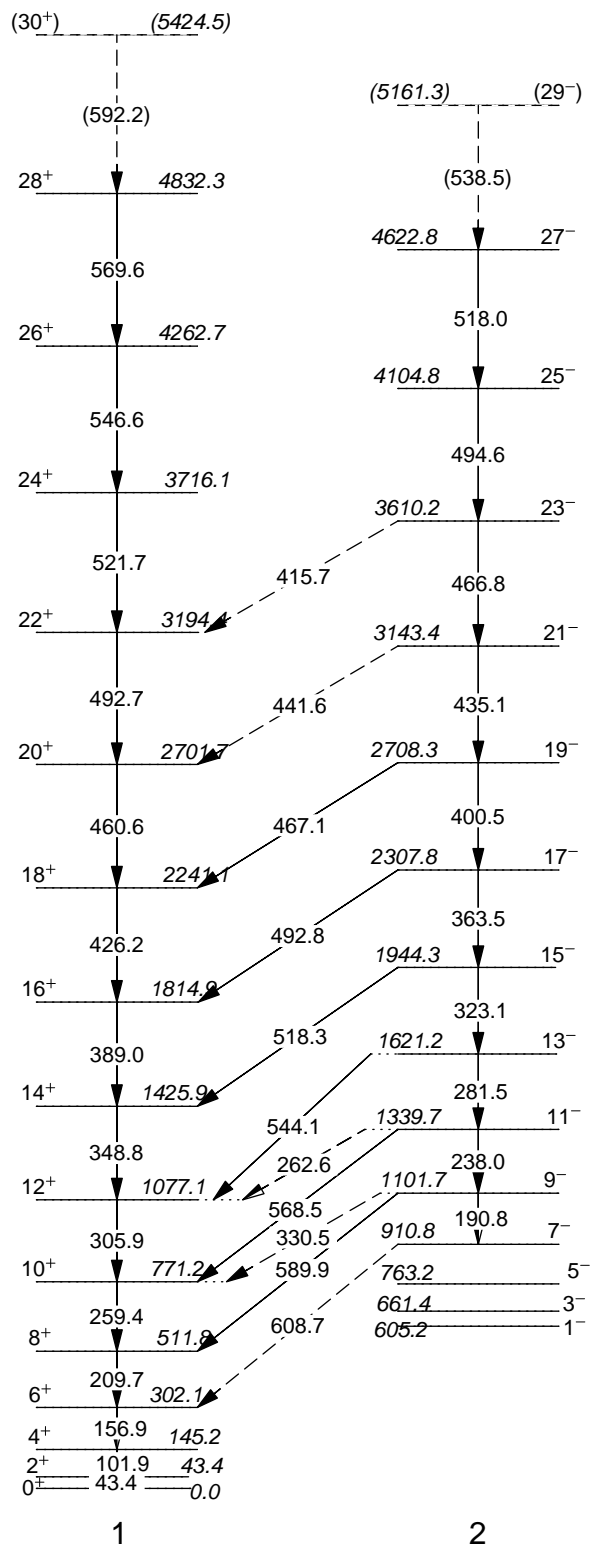}
\caption[Partial level scheme of $^{238}$Pu resulting from present work.]{Partial 
level scheme of $^{238}$Pu. See text for details.\label{fig:Pu238_levl_sche}}
\end{center}
\end{figure}

The level scheme of $^{238}$Pu resulting from this work is given in Figure~\ref{fig:Pu238_levl_sche}. 
As pointed out above, the $^{238}$Pu data was acquired from the weak, single-neutron transfer channel in the reaction 
of a $^{207}$Pb beam with a $^{239}$Pu target. The ratio of the number of events associated with $^{238}$Pu to those of 
$^{239}$Pu was roughly $1/8$. As a result, it should come as no surprise that only two bands were observed in $^{238}$Pu. 
These are labeled as bands 1 and 2 in Figure~\ref{fig:Pu238_levl_sche}. 

\begin{figure}
\begin{center}
\includegraphics[angle=0,width=0.70\columnwidth]{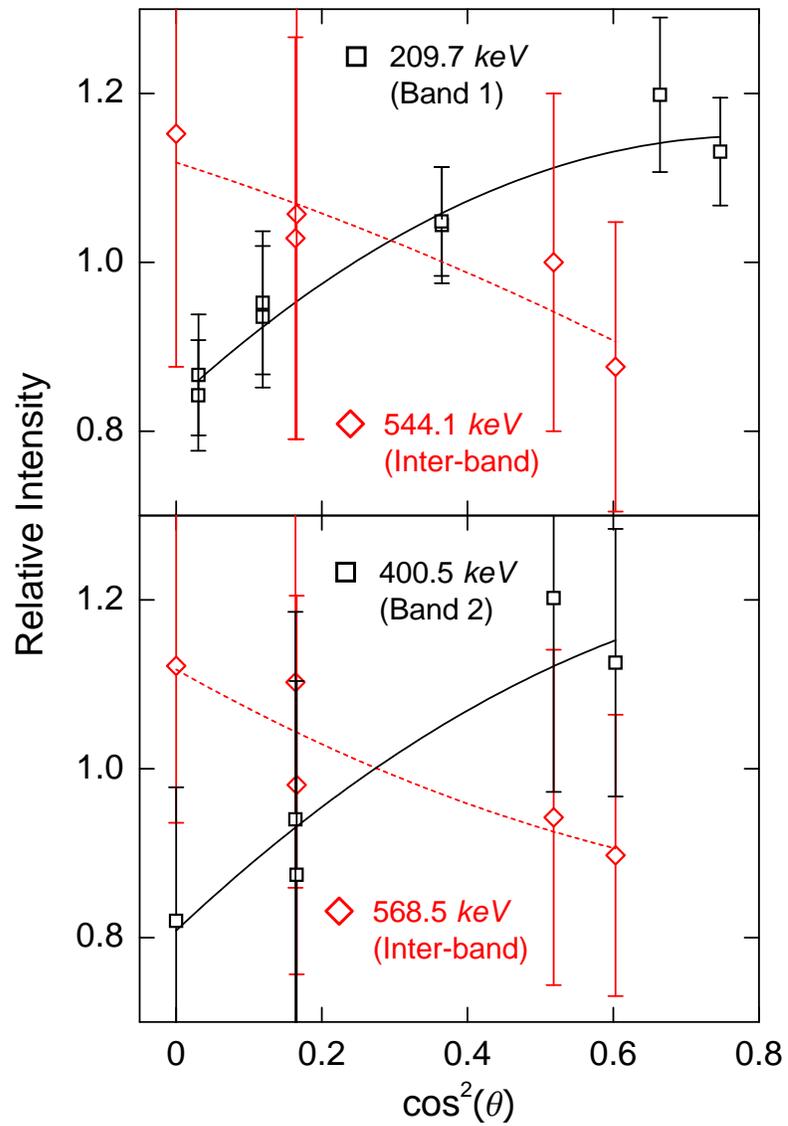}
\caption[Samples of angular distributions for transitions in $^{238}$Pu.]{Samples 
of angular distributions for transitions in $^{238}$Pu. The curves (solid or dashed) represent 
the best fit of the data points. See text for details.\label{fig:238Pu_angu_dist_smp}}
\end{center}
\end{figure}

As described in Sec.~\ref{sec:Pu_exp_datana}, triple coincidence gates, including one placed on one of the strongest 
transitions in $^{208}$Pb, the reaction partner of $^{238}$Pu, were always used to enhance the channel of interest. 
For calculating the relative intensities in the column ``$I_{\gamma}$'' of Tables~\ref{tab:238Pu-B1-gintns-agdscff} 
and \ref{tab:238Pu-B2-gintns-agdscff}, the 101.9-$keV$ ($4^{+}{\rightarrow}2^{+}$) transition in band 1 was taken 
as the reference, and its intensity was normalized to ``1000'' for convenience. Based on the measured relative 
$\gamma$-ray intensities, the population of band 2 relative to band 1 was determined to be ${\sim}10\%$, while the 
upper limit for the population of any other possible bands, except bands 1 or 2, was estimated to be $2\%$. 
Representative angular distributions for in-band and inter-band transitions associated with each 
band in $^{238}$Pu are compared in Figure~\ref{fig:238Pu_angu_dist_smp}. The examples include the 209.7-$keV$ 
($8^{+}{\rightarrow}6^{+}$) line in band 1, the 400.5-$keV$ ($19^{-}{\rightarrow}17^{-}$) $\gamma$ ray in 
band 2 and the 544.1-$keV$ ($13^{-}{\rightarrow}12^{+}$) and 568.5-$keV$ ($11^{-}{\rightarrow}10^{+}$) transitions linking 
bands 1 and 2. It was found in the analysis that the measured $A_2$ and $A_4$ coefficients for in-band and inter-band 
transitions in $^{238}$Pu are very close to the typical values expected for quadrupole and dipole $\gamma$ rays. 

\subsection{\label{subsec:238Pu_band1}$^{238}$Pu band 1}
\begin{figure}
\begin{center}
\includegraphics[angle=270,width=\columnwidth]{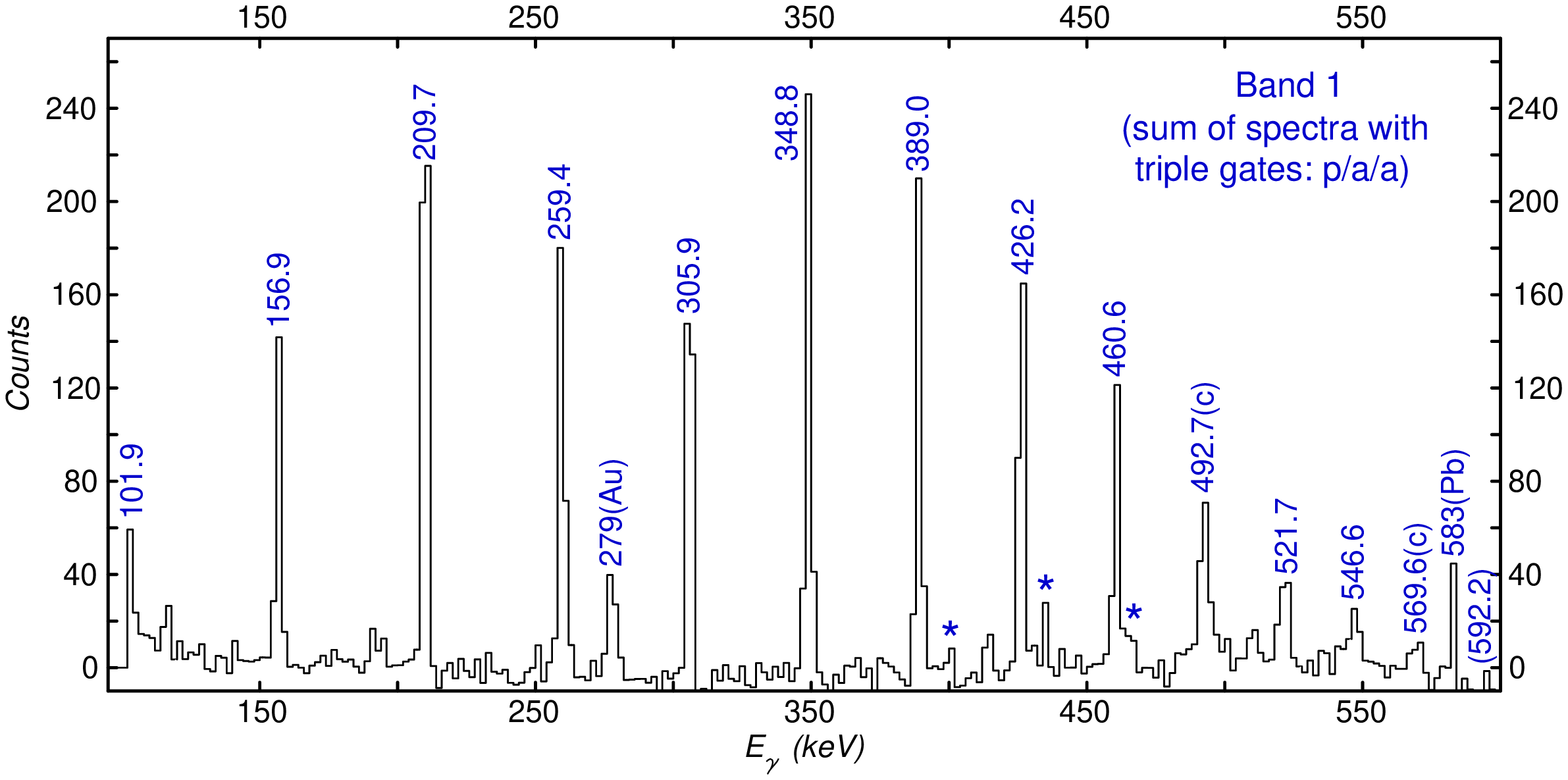}
\caption[Spectrum representative of band 1 in $^{238}$Pu.]{Spectrum representative of band 1 
in $^{238}$Pu. The triple coincidence gates are set on two of the 10 in-band transitions 
(from 209.7-$keV$ to 546.6-$keV$) of band 1 in $^{238}$Pu, $\it{i.e.}$, ``a'', and one of the strongest lines (583-$keV$ or 
2611-$keV$) in $^{208}$Pb, $\it{i.e.}$, ``p''. The transitions are labeled in a 
manner similar to that used in Figure~\ref{fig:Pu240_sp_4_band2}, except that ``(Au)'' and ``(Pb)''symbols denote the 
lines of Au and Pb, respectively.\label{fig:Pu238_sp_4_band1}}
\end{center}
\end{figure}

Band 1 in Figure~\ref{fig:Pu238_levl_sche} is the yrast band. It has been delineated up to a spin of $26{\hbar}$ in 
Refs.~\cite{Stoyer-90-thesis,Devlin-PRC-47-2178-93}. Here, this band was extended by two additional transitions. 
The sum of spectra, given in Figure~\ref{fig:Pu238_sp_4_band1}, was generared with the gating conditions placed on two of 
the in-band transitions and one of the strongest lines in $^{208}$Pb (either 583 or 2611 $keV$). 
This band consists of fifteen transitions. All in-band transitions of band 1 are seen in Figure~\ref{fig:Pu238_sp_4_band1} 
except the 43.4-$keV$ ($2^{+}{\rightarrow}0^{+}$) line, which is suppressed due to the low detection efficiency at low 
energy and the high internal conversion probability. Some of the transitions in band 2 can be seen in this figure as well 
because of the connections between the two sequences (see Figures~\ref{fig:Pu238_sp_4_band1} and \ref{fig:Pu238_levl_sche}). 
They will be discussed below. 

The intensities of the three low-lying transitions, $\it{i.e.}$, the 101.9-$keV$ ($4^{+}{\rightarrow}2^{+}$), the 156.9-$keV$ 
($6^{+}{\rightarrow}4^{+}$) and the 209.7-$keV$ ($8^{+}{\rightarrow}6^{+}$) $\gamma$ rays, were extracted from the total 
projection (single gated on one of the two lines in $^{208}$Pb). Using the method applied in $^{240}$Pu and $^{242}$Pu, 
the relative intensity of the 259.4-$keV$ ($10^{+}{\rightarrow}8^{+}$) transition was then obtained from the comparison 
of the peak area of the 209.7-$keV$ line with that of the 259.4-$keV$ $\gamma$ ray in the spectrum triple gated on the 
101.9-$keV$ and 156.9-$keV$ transitions together with one of the two $^{208}$Pb lines. The intensities of the 305.9-$keV$ 
$\gamma$ ray and all other lines located above it were obtained from appropriate spectra in the same manner. 
The $A_2$ and $A_4$ coefficients for the in-band transitions, given in Table~\ref{tab:238Pu-B1-gintns-agdscff}, 
were acquired at several angles from the sum of spectra generated with all available triple gates. 

The spins and parity of the states up to $24^{+}$, established in previous measurements~\cite{Stoyer-90-thesis,Devlin-PRC-47-2178-93}, 
were confirmed by the measured $A_2$ and $A_4$ coefficients of those associated $\gamma$ rays here. As the transitions 
above the $24^{+}$ state form a natural extension of the lower-spin sequence in this band, the spins and parity for states 
above $24^{+}$ were assigned accordingly, even though the information of $A_2$ and $A_4$ coefficients for these high-spin 
lines is not available in the present work. 

\begin{table}
\begin{center}
\caption{THE EXCITATION ENERGIES ($E_x$) OF INITIAL STATES, ASSIGNED SPINS, $\gamma$-RAY ENERGIES ($E_{\gamma}$), RELATIVE 
$\gamma$-RAY INTENSITIES ($I_{\gamma}$) AND ANGULAR DISTRIBUTION COEFFICIENTS ($A_2$ AND $A_4$) FOR THE TRANSITIONS ASSOCIATED 
WITH BAND 1 IN $^{238}$Pu\label{tab:238Pu-B1-gintns-agdscff}}
\begin{tabular}{cccccc}
\hline \hline 
\multicolumn{6}{c}{Band 1 in $^{238}$Pu}\\
\hline 
$E_x$ ($keV$) & Assigned spin ($\hbar$) & $E_{\gamma}$ ($keV$) & $I_{\gamma}$ (rel.) & $A_{2}$ & $A_{4}$\\ 
\hline
145.2 & $4^{+}{\rightarrow}2^{+}$ & 101.9(5) & 1000(133) & & \\
302.1 & $6^{+}{\rightarrow}4^{+}$ & 156.9(5) & 1252(247) & 0.25(13) & -0.2(2)\\
511.8 & $8^{+}{\rightarrow}6^{+}$ & 209.7(5) & 1549(338) & 0.24(3) & -0.11(4)\\
771.2 & $10^{+}{\rightarrow}8^{+}$ & 259.4(5) & 1636(364) & 0.24(6) & -0.17(10)\\
1077.1 & $12^{+}{\rightarrow}10^{+}$ & 305.9(5) & 1346(347) & 0.16(3) & -0.13(4)\\
1425.9 & $14^{+}{\rightarrow}12^{+}$ & 348.8(5) & 1117(302) & 0.24(5) & -0.14(7)\\
1814.9 & $16^{+}{\rightarrow}14^{+}$ & 389.0(5) & 888(252) & 0.18(11) & -0.08(17)\\
2241.1 & $18^{+}{\rightarrow}16^{+}$ & 426.2(5) & 569(166) & 0.33(5) & -0.15(8)\\
2701.7 & $20^{+}{\rightarrow}18^{+}$ & 460.6(5) & 469(142) & 0.24(4) & -0.04(6)\\
3194.4 & $22^{+}{\rightarrow}20^{+}$ & 492.7(5) & 260(96) & 0.22(4) & -0.07(6)\\
3716.1 & $24^{+}{\rightarrow}22^{+}$ & 521.7(5) & 202(81) & 0.20(9) & -0.15(11)\\
4262.7 & $26^{+}{\rightarrow}24^{+}$ & 546.6(5) & 141(61) & & \\
4832.3 & $28^{+}{\rightarrow}26^{+}$ & 569.6(6) & 79(38) & & \\
5424.5 & $30^{+}{\rightarrow}28^{+}$ & 592.2(6) & 20(14) & & \\
\hline 
\end{tabular}
\end{center}
\end{table}

\subsection{\label{subsec:238Pu_band2}$^{238}$Pu band 2}
\begin{figure}
\begin{center}
\includegraphics[angle=270,width=\columnwidth]{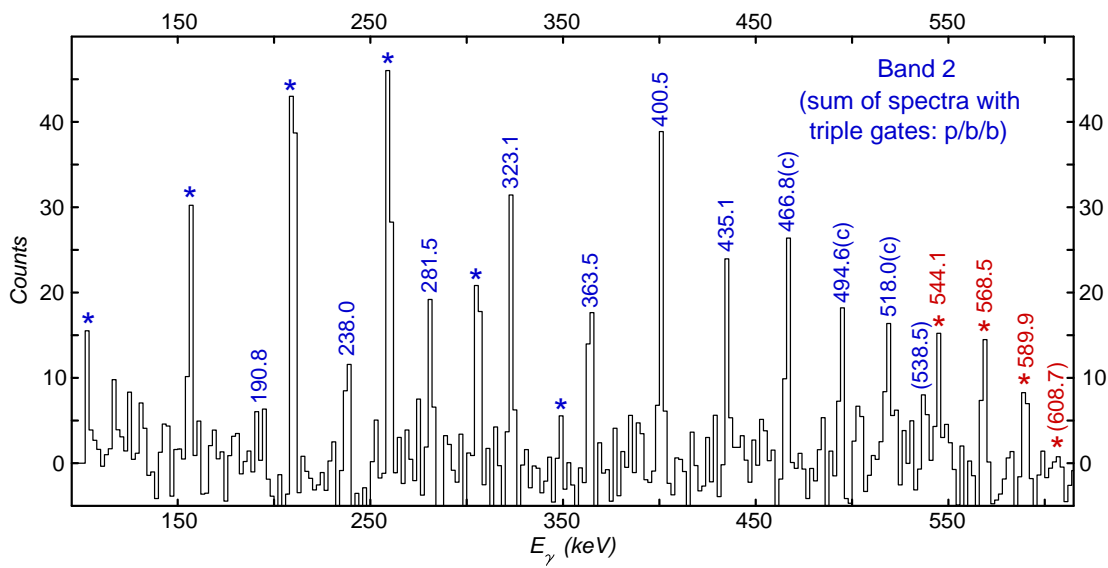}
\caption[Spectrum representative of band 2 in $^{238}$Pu.]{Spectrum representative of band 2 in $^{238}$Pu. 
The triple coincidence gates are placed on two of 9 in-band transitions 
(from 238.0-$keV$ to 518.0-$keV$) in band 2, $\it{i.e.}$, ``b'', and one of the strongest lines (583-$keV$ 
or 2611-$keV$) in $^{208}$Pb, $\it{i.e.}$, ``p''. The transitions are labeled in a manner used in 
Figure~\ref{fig:Pu238_sp_4_band1}.\label{fig:Pu238_sp_4_band2}}
\end{center}
\end{figure}

Band 2 in Figure~\ref{fig:Pu238_levl_sche} has been associated with the $K^{\pi}=0^{-}$ octupole vibration in 
Refs.~\cite{Asaro-UCRL-9566-50-60,Bjornholm-PR-130-2000-63,Bengtson-NPA-159-249-70,Lorenz-LAEATRS-261-86}, but the $7^{-}$ 
level and the states above it had not been established before. This band consists of fourteen transitions. All of the eleven transitions 
above the $7^{-}$ level can be seen in Figure~\ref{fig:Pu238_sp_4_band2}. In this summed triple-gated spectrum, the 538.5-$keV$ 
($29^{-}{\rightarrow}27^{-}$) transition is hard to confirm due to the lack of statistics combined with a sizeable Doppler shift 
and/or broadening. As a result, it was assigned on a tentative basis in the level scheme. The three transitions below 
the $7^{-}$ level were not observed because of the high internal conversion probability, but the associated $1^{-}$, 
$3^{-}$ and $5^{-}$ states have been established in several previous 
measurements~\cite{Asaro-UCRL-9566-50-60,Bjornholm-PR-130-2000-63,Bengtson-NPA-159-249-70,Lorenz-LAEATRS-261-86}. 
The $7^{-}$ level ($E_x=910.8~keV$), of which the energy was unknwon in the literature, has been identified for the first time in 
the present work, taking advantage of the observation of the 190.8-$keV$ ($9^{-}{\rightarrow}7^{-}$) transition in this band (see 
Figures~\ref{fig:Pu238_sp_4_band2} and \ref{fig:Pu238_levl_sche}). The $\gamma$ rays linking this sequence to the 
yrast band (band 1) are indicated in Figure~\ref{fig:Pu238_sp_4_band2}, and, they can also be grouped into two types: 
$J^{-}{\rightarrow}(J-1)^{+}$ and $J^{-}{\rightarrow}(J+1)^{+}$. The transitions of the latter type, $\it{i.e.}$, the 262.6- and 
330.5-$keV$ lines, are hard to verify in the spectra resulting from the above analysis, because of their low intensities, 
therefore, they are taken as tentative in the level scheme. This is also the case for some weak $J^{-}{\rightarrow}(J-1)^{+}$ 
inter-band transition, such as the 415.7-$keV$ line, for example. It is also worth noting that no linking transitions 
between bands 1 and 2 in an opposite direction, like the $(J+1)^{+}{\rightarrow}J^{-}$ lines seen in $^{240}$Pu, were observed in 
$^{238}$Pu, though the pattern of the difference in energy between the states of bands 1 and 2 at high spin in $^{238}$Pu 
is quite similar to that in $^{240}$Pu. The maximum value of the relative intensity for virtual $(J+1)^{+}{\rightarrow}J^{-}$ 
transitions between bands 1 and 2 was estimated to be 5 (the relative intensity of the 101.9-$keV$ line in band 1 is taken as 1000). 

\begin{table}
\begin{center}
\caption{THE EXCITATION ENERGIES ($E_x$) OF INITIAL STATES, ASSIGNED SPINS, $\gamma$-RAY ENERGIES ($E_{\gamma}$), RELATIVE 
$\gamma$-RAY INTENSITIES ($I_{\gamma}$) AND ANGULAR DISTRIBUTION COEFFICIENTS ($A_2$ AND $A_4$) FOR THE TRANSITIONS ASSOCIATED 
WITH BAND 2 IN $^{238}$Pu\label{tab:238Pu-B2-gintns-agdscff}}
\begin{tabular}{cccccc}
\hline \hline 
\multicolumn{6}{c}{Band 2 in $^{238}$Pu}\\
\hline 
$E_x$ ($keV$) & Assigned spin ($\hbar$) & $E_{\gamma}$ ($keV$) & $I_{\gamma}$ (rel.) & $A_{2}$ & $A_{4}$\\ 
\hline
910.8 & $7^{-}{\rightarrow}6^{+}$ & 608.7(5) & 99(25) & & \\
1101.7 & $9^{-}{\rightarrow}7^{-}$ & 190.8(6) & 84(26) & & \\
 & $9^{-}{\rightarrow}8^{+}$ & 589.9(5) & 140(33) & -0.4(2) & 0.1(2)\\
 & $9^{-}{\rightarrow}10^{+}$ & 330.5(6) & 49(16) & & \\
1339.7 & $11^{-}{\rightarrow}9^{-}$ & 238.0(6) & 130(44) & 0.21(2) & -0.09(2)\\
 & $11^{-}{\rightarrow}10^{+}$ & 568.5(6) & 175(51) & -0.20(14) & 0.05(17)\\
1621.2 & $13^{-}{\rightarrow}11^{-}$ & 281.5(6) & 163(63) & 0.25(5) & -0.13(9)\\
 & $13^{-}{\rightarrow}12^{+}$ & 544.1(6) & 119(53) & -0.26(16) & -0.02(22)\\
1944.3 & $15^{-}{\rightarrow}13^{-}$ & 323.1(5) & 150(66) & 0.22(6) & -0.10(8)\\
 & $15^{-}{\rightarrow}14^{+}$ & 518.3(5) & 86(43) & & \\
2307.8 & $17^{-}{\rightarrow}15^{-}$ & 363.5(5) & 127(61) & 0.4(3) & -0.05(45)\\
 & $17^{-}{\rightarrow}16^{+}$ & 492.8(5) & 58(58) & & \\
2708.3 & $19^{-}{\rightarrow}17^{-}$ & 400.5(5) & 107(52) & 0.31(16) & -0.1(2)\\
 & $19^{-}{\rightarrow}18^{+}$ & 467.1(5) & 41(69) & & \\
\hline 
\end{tabular}
\end{center}
\end{table}

\begin{table}
\begin{center}
\centerline{TABLE~\ref{tab:238Pu-B2-gintns-agdscff} (contd.)}
\begin{tabular}{cccccc}
\hline \hline 
$E_x$ ($keV$) & Assigned spin ($\hbar$) & $E_{\gamma}$ ($keV$) & $I_{\gamma}$ (rel.) & $A_{2}$ & $A_{4}$\\ 
\hline
3143.4 & $21^{-}{\rightarrow}19^{-}$ & 435.1(5) & 100(49) & 0.14(18) & -0.09(24)\\
 & $21^{-}{\rightarrow}20^{+}$ & 441.6(5) & 38(20) & & \\
3610.2 & $23^{-}{\rightarrow}21^{-}$ & 466.8(5) & 82(28) & 0.4(2) & -0.2(3)\\
 & $23^{-}{\rightarrow}22^{+}$ & 415.7(5) & 33(19) & & \\
4104.8 & $25^{-}{\rightarrow}23^{-}$ & 494.6(6) & 53(23) & 0.5(3) & -0.1(5)\\
4622.8 & $27^{-}{\rightarrow}25^{-}$ & 518.0(7) & 35(16) & & \\
5161.3 & $29^{-}{\rightarrow}27^{-}$ & 538.5(7) & & & \\
\hline 
\end{tabular}
\end{center}
\end{table}

As in the $^{240}$Pu and $^{242}$Pu cases, the relative intensities of the inter-band transitions 
were extracted first, and, those of the in-band lines were obtained subsequently. Further, the values of $A_2$ and $A_4$ 
coefficients for the transitions associated with band 2 were also derived. The rules of choosing 
proper coincidence gates are very similar to those applied in the analyses for band 2 in $^{240}$Pu 
and in $^{242}$Pu, except that triple (an additional gate placed on one of the strongest transitions in $^{208}$Pb) 
instead of double gates were applied. The results of the above analysis are summarized in Table~\ref{tab:238Pu-B2-gintns-agdscff}. 
The measured $A_2$ and $A_4$ coefficients for the inter-band transitions between bands 1 and 2 and the in-band ones of 
band 2 indicate their dipole and quadrupole nature, respectively. Hence, negative parity and odd spins were assigned to 
the levels of band 2 (the spins and parity for states above $25^{-}$ were adopted under the assumption that these 
transitions form the natural extension of the $E2$ sequence). 
As will be shown later, the resulting routhian plot (Figure~\ref{fig:238Pu_routhian} in Sec.~\ref{subsec:Pus_condense_interpt}) 
also suggests that the sequence of 
$\gamma$ rays observed in the present work (above the $7^{-}$ level) extrapolates well at lower angular frequency to 
the known bandhead sequence of ``$5^{-}$, $3^{-}$, $1^{-}$'' states ($E_x=$ 605.2 $keV$ for the $1^{-}$ level). The 
consistency of the data points from the present work (at higher spin) with the ones from the previous 
work~\cite{Asaro-UCRL-9566-50-60,Bjornholm-PR-130-2000-63,Bengtson-NPA-159-249-70,Lorenz-LAEATRS-261-86} 
(at lower spin) supports, from another point of view, the assignment of spins and parity to this band. 

\section{\label{sec:Pu-dis-intpt}Discussion and interpretation}
Making use of the method described in Sec.~\ref{subsec:LabTrnsfBody} of Chapter~\ref{chap:theoriBkgd}, the experimental 
results for the three even-even Pu isotopes were used to extract essential physical quantities such as routhians and 
alignments, for example. In order to make meaningful comparisons between the intrinsic properties in different nuclei, 
common Harris parameters, $J_0=65{\hbar}^{2}{MeV}^{-1}$ and $J_1=369{\hbar}^{4}{MeV}^{-3}$, were adopted for 
all nuclei compared in the present work. These parameters have proved suitable for most nuclei in the 
region~\cite{Wiedenhover-Proc-NS98,Wiedenhover-PRL-83-2143-99,AbuSaleem-PRC-70-024310-04}. In fact, these values keep 
the aligned spins $i_{x}$ of the yrast bands in the even-even nuclei close to zero at all frequencies below the 
first band crossing ($\it{i.e.}$, backbending; see Sec.~\ref{subsec:CSMintro} in Chapter~\ref{chap:theoriBkgd}). 

\subsection{\label{subsec:Pu_yrast_bds}Yrast bands}
Before discussing these measured alignments, it is necessary to briefly introduce the important concept of ``Pauli blocking''. 
In a rotating odd-even nucleus, as described in Sec.~\ref{subsec:PairInteract} of Chapter~\ref{chap:theoriBkgd}, 
the Coriolis force tends to break a pair of nucleons in an orbital of interest and align the individual angular 
momenta with the rotation axis. However, the presence of an aligned odd nucleon in the same orbital will prevent the 
occupation of this orbital, because of the Pauli exclusion principle: once the pair is broken, one of the nucleons 
will need to occupy a higher-lying orbital. In other words, this alignment will require more energy and will occur 
at much higher frequency. Hence, the alignment curve will remain constant in the frequency region where a sudden increase 
of the angular momentum occurs in the yrast bands in the neighboring even-even nuclei. 
In other words, the alignment is blocked. The phenomenon has been predicted and reproduced well by Cranked Shell Model (CSM) 
calculations. A more detailed discussion of the subject can be found in Refs.~\cite{AbuSaleem-02-thesis,AbuSaleem-PRC-70-024310-04}. 

\begin{figure}
\begin{center}
\includegraphics[angle=0,width=0.80\columnwidth]{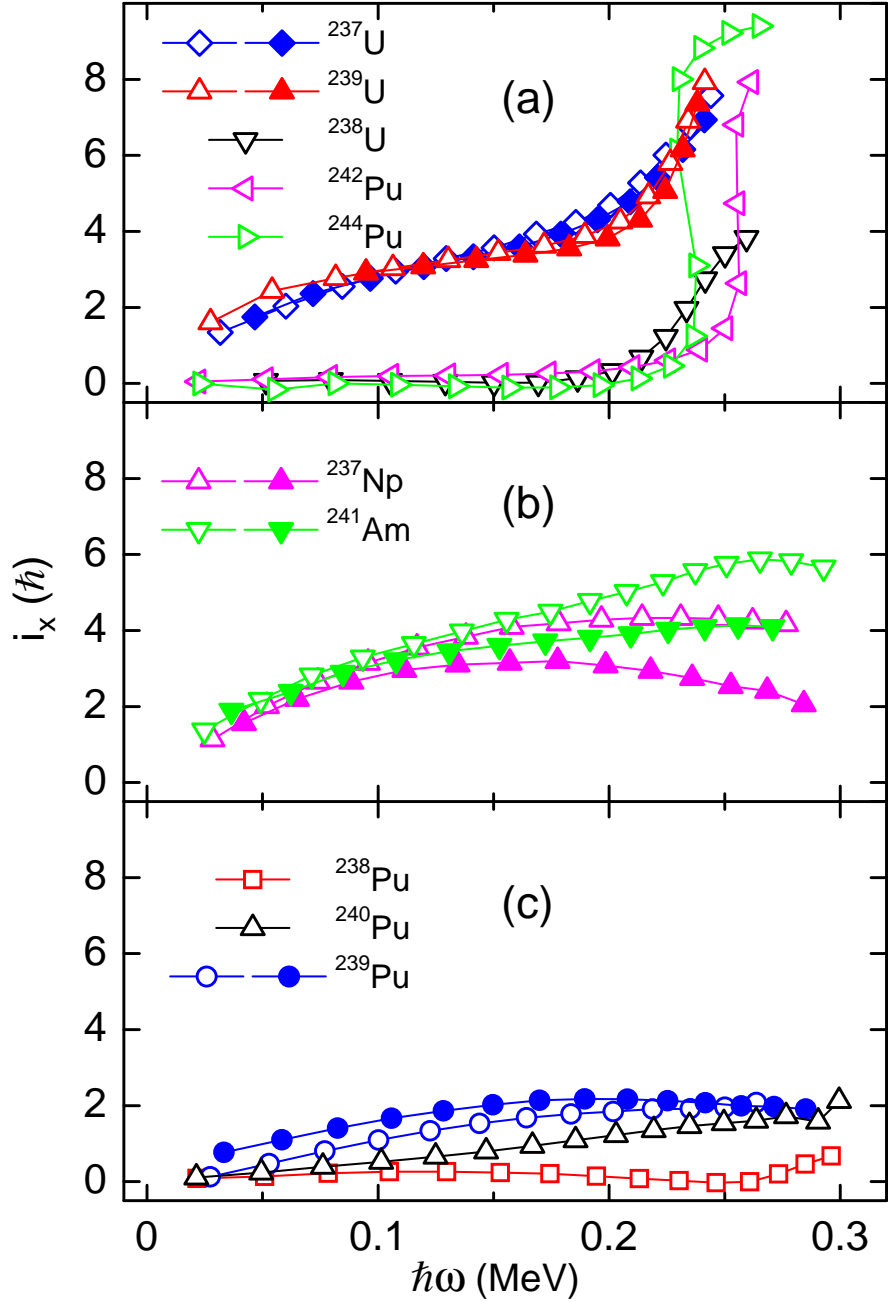}
\caption[The alignments of the yrast bands in $^{238,240,242,244}$Pu, $^{237,238,239}$U, $^{237}$Np and $^{241}$Am.]{The aligned 
spins $i_{x}$ of the yrast bands in the $^{238,240,242,244}$Pu isotopes and in the neighboring even-even 
$^{238}$U, odd-neutron $^{237,239}$U and odd-proton $^{237}$Np and $^{241}$Am nuclei. See text for details.\label{fig:Pu_U_Np_Am_compare_aligns}}
\end{center}
\end{figure}

Figure~\ref{fig:Pu_U_Np_Am_compare_aligns} (a), (b) and (c) compares the aligned spins $i_{x}$ as a function of 
the rotational frequency $\hbar\omega$ for the yrast bands (band 1) in the $^{238}$Pu, 
%(built on $i_{13/2}$ proton orbital) 
$^{240}$Pu and $^{242}$Pu isotopes with those seen 
%for the bands built on $i_{13/2}$ proton or $j_{15/2}$ neutron orbital 
in the neighboring even-even $^{244}$Pu~\cite{Wiedenhover-PRL-83-2143-99} and $^{238}$U~\cite{Ward-NPA-600-88-96} nuclei. 
Comparisons are also provided with the bands built on the $j_{15/2}$ neutron orbital in $^{237,239}$U~\cite{Zhu-PLB-618-51-05} 
and on the $i_{13/2}$ proton orbital in $^{237}$Np and $^{241}$Am~\cite{AbuSaleem-02-thesis,AbuSaleem-PRC-70-024310-04}. 

As can be seen in Figure~\ref{fig:Pu_U_Np_Am_compare_aligns} (a), a strong alignment occurs in $^{242,244}$Pu at a frequency 
of $\hbar\omega{\sim}0.25\;MeV$. In the case of $^{244}$Pu, the data show that the gain in angular momentum 
is essentially $10\hbar$. 
The alignment in the yrast band of $^{238}$U exhibits a pattern similar to those of $^{242,244}$Pu, but the rate of 
increase is smaller, probably reflecting differences in interaction strength between the aligned and ground configurations 
at the crossing point. In contrast, the bands built on the $j_{15/2}$ neutron orbital in the odd-$A$ (odd-neutron) neighboring 
nuclei, such as $^{237,239}$U, all exhibit alignments which follow those of the even-even cores, $\it{i.e.}$, no Pauli blocking occurs. 
The difference of $i_x$ at low frequency between the even-even and odd-$A$ nuclei, discussed above, simply reflects the 
contribution to the total spin due to the alignment with the rotation axis of the odd neutron. 

In contrast, the alignments for the bands built on odd $i_{13/2}$ quasi-protons in $^{237}$Np and $^{241}$Am, 
illustrated in Figure~\ref{fig:Pu_U_Np_Am_compare_aligns} (b), do not exhibit any sudden gain within the 
$0.20\;MeV<\hbar\omega<0.30\;MeV$ frequency range, $\it{i.e.}$, these alignments are blocked. 

All of the above observations naturally lead to the conclusion that the gain in alignment 
(backbending) seen in these actinide nuclei is due to the alignment of a pair of $i_{13/2}$ quasi-protons. This settles a 
longstanding debate in the literature~\cite{Dudek-PRC-26-1708-82,Dudek-PScr-T5-171-83,Spreng-PRL-51-1522-83} about 
whether $i_{13/2}$ quasi-protons, or, $j_{15/2}$ quasi-neutrons are involved. Further detailed discussions about the 
impact of the proton $i_{13/2}$ and neutron $j_{15/2}$ orbitals on the alignment patterns in the actinide region are 
presented in Refs.~\cite{AbuSaleem-02-thesis,AbuSaleem-PRC-70-024310-04}. 

\begin{figure}
\begin{center}
\includegraphics[angle=0,width=\columnwidth]{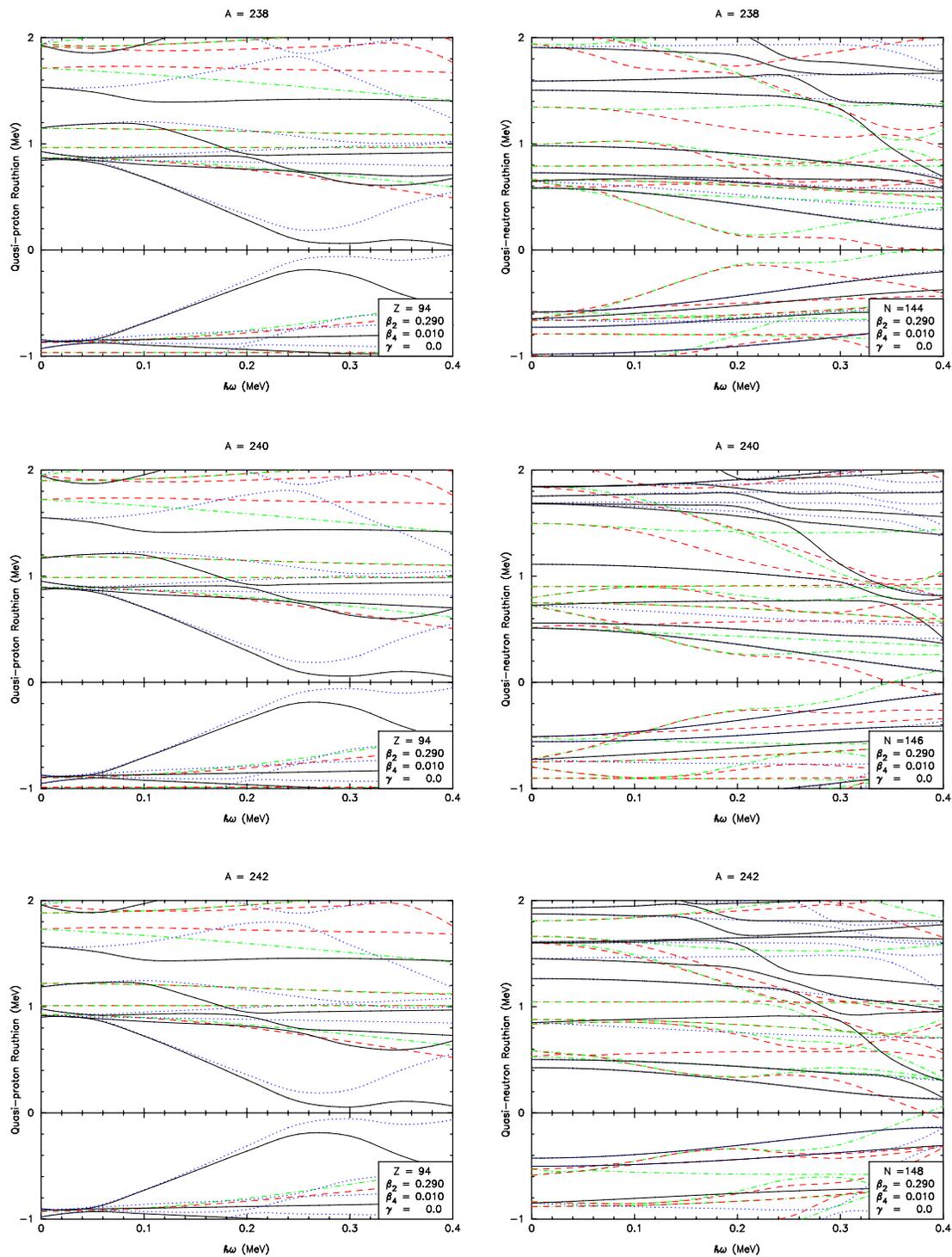}
\caption[The calculated quasi-proton and -neutron routhians for $^{238,240,242}$Pu.]{The quasi-proton and quasi-neutron 
routhians from the Cranked Shell Model (CSM) calculations described in the text.\label{fig:Pus_CSM_routhan}}
\end{center}
\end{figure}

With the purpose of interpreting the above observations of alignment, the standard Cranked Shell Model (CSM) calculations 
with the Warsaw-Lund code using a Woods-Saxon potential were performed for the $^{238,240,242}$Pu isotopes. The universal 
deformation parameters, $\beta_{2}=0.29$, $\beta_{4}=0.01$, $\gamma=0^{\circ}$, and a pairing strength 
($\it{i.e.}$, the $G_p$ and $G_n$ factors for protons and neutrons, calculated from the fit of a large number of 
data~\cite{AbuSaleem-02-thesis}) at $90\%$ of the full value were adopted in 
the present calculations. CSM calculations with these values of the parameters have been found 
to give the most satisfactory agreement with the data for most nuclei in the region~\cite{AbuSaleem-02-thesis}. 
The gain in alignment $\delta{i_x}$ and the frequency $\hbar\omega_{c}$ where the 
alignment occurs for the $i_{13/2}$ quasi-protons in $^{238,240,242}$Pu, extracted from the resulting quasi-proton 
routhian plots (at the left in Figure~\ref{fig:Pus_CSM_routhan}), are almost identical: $\delta{i_x}=7.75\hbar$ and 
$\hbar\omega_{a}=0.28\;MeV$. Hence, in the $^{242}$Pu case, the measured $\delta{i_x}$ ($\sim$ 8 -- $10\hbar$) and 
$\hbar\omega_{a}$ ($\sim$ $0.25\;MeV$) values for the $i_{13/2}$ quasi-protons were reproduced well by the 
CSM calculations. 

The performed CSM calculations (see the plots at the right in Figure~\ref{fig:Pus_CSM_routhan}) also indicate 
that the alignment of $j_{15/2}$ quasi-neutrons is expected in the same frequency range. As discussed above, 
the data do not provide much direct evidence for such neutron alignments. This issue remains 
unresolved at this time, although it has been suggested in Ref.~\cite{AbuSaleem-02-thesis} that a rather small 
neutron gap may have as a consequence a gradual, hardly visible alignment of the $j_{15/2}$ neutrons. 

Because of the successful interpretation of the $^{242,244}$Pu and $^{238}$U data illustrated in 
Figure~\ref{fig:Pu_U_Np_Am_compare_aligns} (a) and (b) using the concepts of $i_{13/2}$ quasi-proton alignments 
and Pauli blocking, the experimental $i_x$ curves for the yrast bands in $^{238,239,240}$Pu become rather striking: 
there is no sign of any sudden gain in alignment throughout the entire rotation frequency range 
($0.02\;MeV<\hbar\omega<0.30\;MeV$). This observation indicates that the expected $i_{13/2}$ quasi-proton 
alignment is absent or, at least, severely delayed in frequency in $^{238,239,240}$Pu. It is worth pointing 
out that this is not a small effect. For example, the absence of a sudden alignment in $^{240}$Pu continues 
for at least five transitions beyond the point where it occurs in $^{242}$Pu. To the best of our knowledge, 
the same strong $i_{13/2}$ quasi-proton alignment is predicted to occur at the same frequency 
($\hbar\omega{\sim}0.25\;MeV$) by all available 
calculations~\cite{Dudek-PRC-26-1708-82,Dudek-PScr-T5-171-83,Spreng-PRL-51-1522-83}. The striking difference 
in pattern was revealed for the first time in the previous work of Wiedenh{\"o}ver $\it{et~al.}$~\cite{Wiedenhover-PRL-83-2143-99}. 
A natural explanation was suggested therein in terms of strong octupole correlations in $^{238,239,240}$Pu, which 
would have a considerable impact on their intrinsic structure and result in the absence or delay of the strong 
proton alignment seen in the heavier Pu and neighboring U nuclei. 

\subsection{\label{subsec:Pu_octupl_bds}One-phonon octupole bands}
\begin{figure}
\begin{center}
\includegraphics[angle=270,width=0.80\columnwidth]{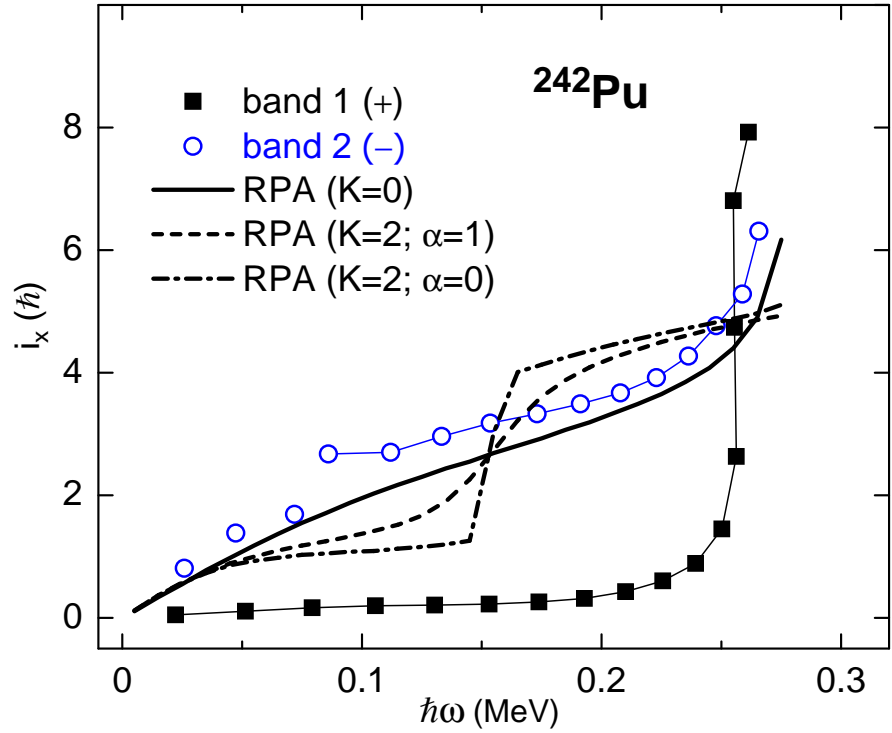}
\caption[Spin alignments as a function of rotational frequency for bands 1 and 2 in $^{242}$Pu.]{Spin alignments 
as a function of rotational frequency for bands 1 and 2 in $^{242}$Pu. The $+$ and $-$ symbols in 
parentheses represent the parity of band. The data points connected by thin lines were obtained 
in the present work, while the separated points were adopted from the literature. The drawn thick 
lines are obtained from the cranked RPA calculations with different $K$ and $\alpha$ quantum numbers. 
See text for details.\label{fig:242Pu_alignment_np}}
\end{center}
\end{figure}

\begin{figure}
\begin{center}
\includegraphics[angle=270,width=0.80\columnwidth]{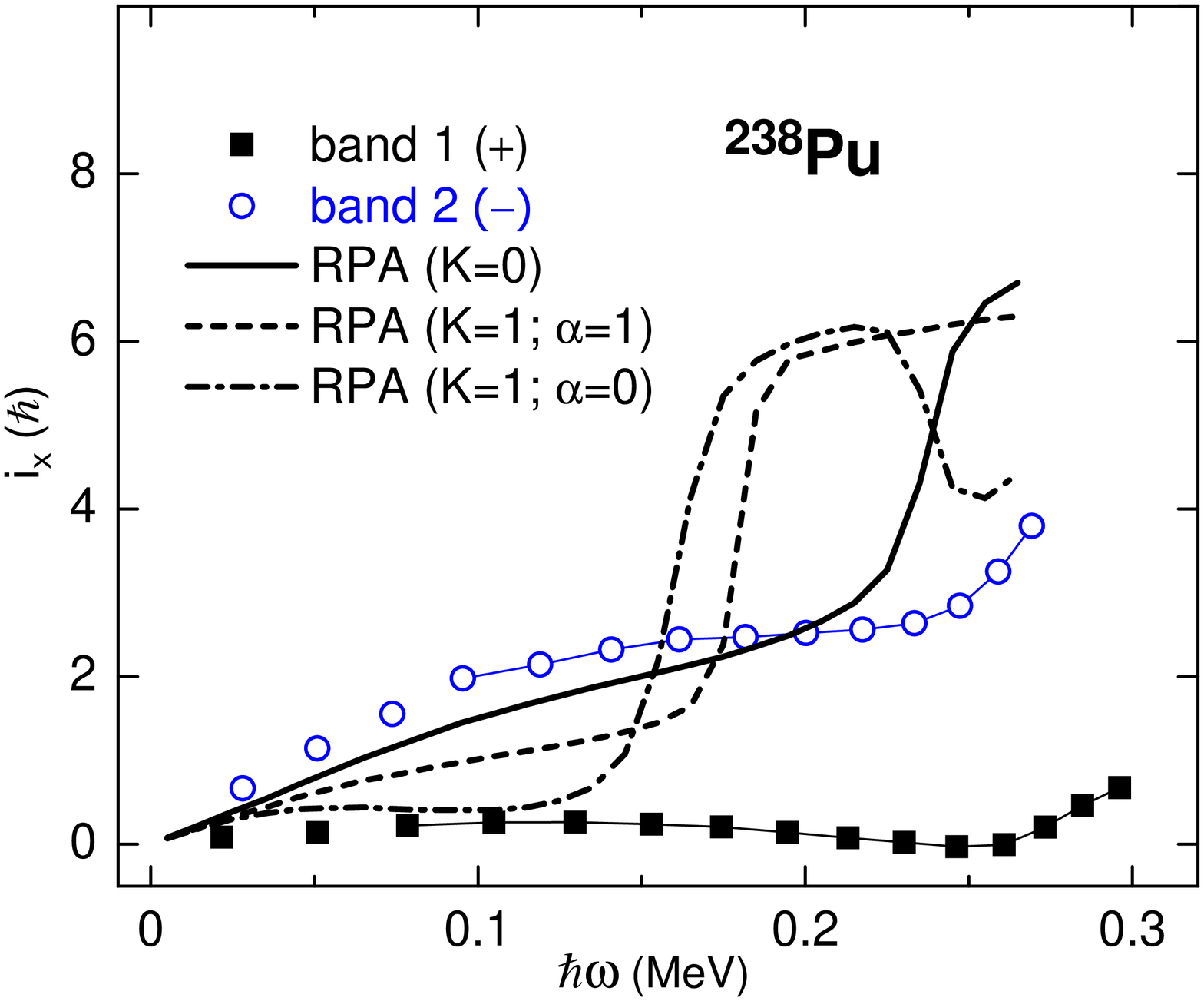}
\caption[Spin alignments as a function of rotational frequency for bands 1 and 2 in $^{238}$Pu.]{Spin alignments 
as a function of rotational frequency for bands 1 and 2 in $^{238}$Pu from the data and from the RPA calculations. 
See text for details.\label{fig:238Pu_alignment}}
\end{center}
\end{figure}

\begin{figure}
\begin{center}
\includegraphics[angle=270,width=0.80\columnwidth]{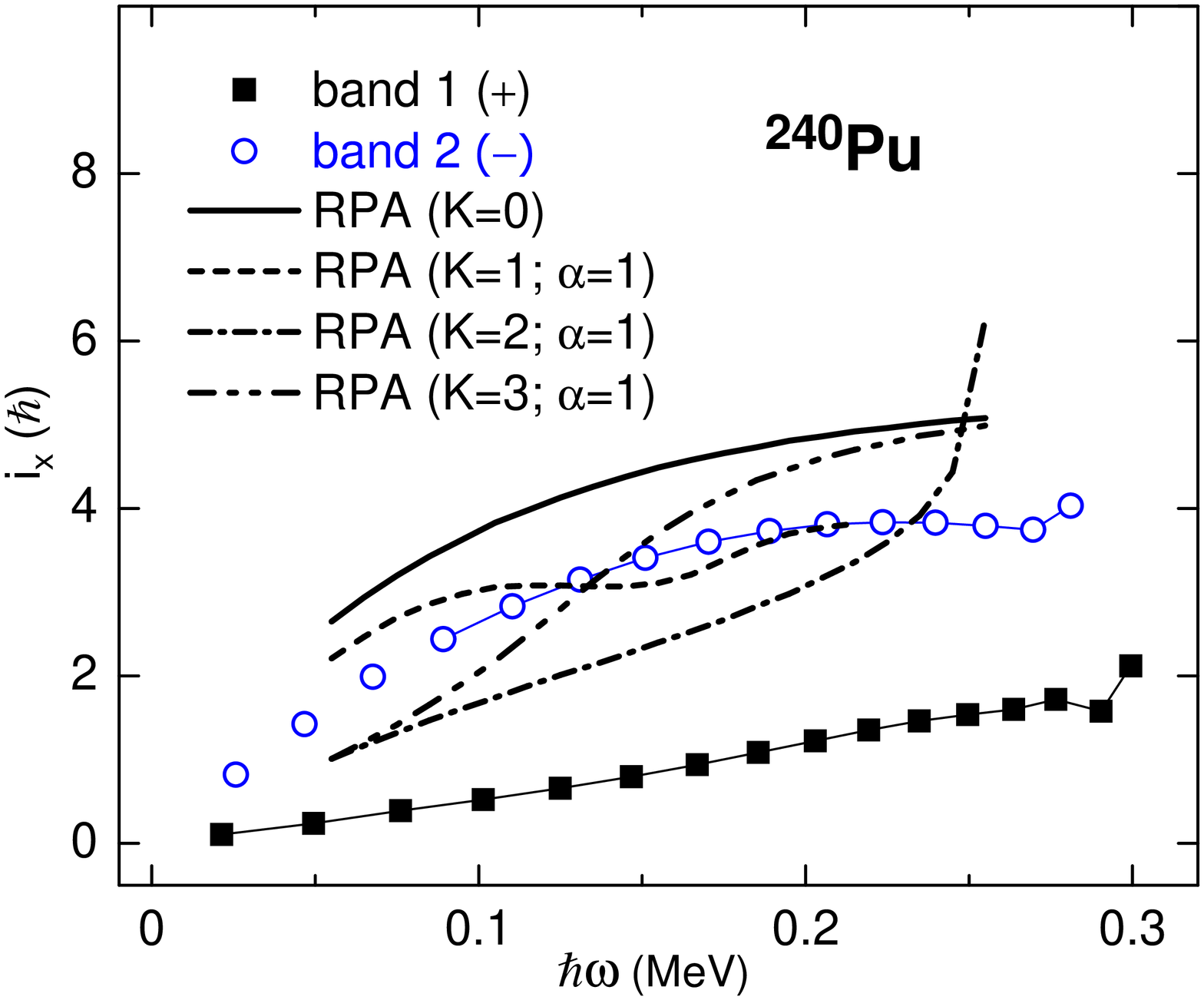}
\caption[Spin alignments as a function of rotational frequency for bands 1 and 2 in $^{240}$Pu.]{Spin alignments 
as a function of rotational frequency for bands 1 and 2 in $^{240}$Pu from the data and from the RPA calculations. 
See text for details.\label{fig:240Pu_alignment_np}}
\end{center}
\end{figure}

The alignments of the negative-parity bands (band 2) in $^{238,240,242}$Pu, which have been associated with a one-phonon octupole 
vibration, are illustrated in Figures~\ref{fig:242Pu_alignment_np}, \ref{fig:238Pu_alignment} and \ref{fig:240Pu_alignment_np}. The 
indicated patterns are consistent with those seen in their respective yrast (positive-parity) bands. 

In $^{242}$Pu, the aligned spin $i_x$ of the octupole band (band 2) gradually increases until an upbending occurs 
at $\hbar\omega{\sim}0.25\;MeV$ just as in the yrast band. The difference of $i_x$ between the yrast and octupole bands 
gradually rises with increasing frequency, until it becomes almost constant at 3$\hbar$ over wide 
range of frequencies ($0.10\;MeV<\hbar\omega<0.25\;MeV$). The value of $3\hbar$ corresponds to what would be expected 
for an octupole phonon. This negative-parity band upbends at essentially the same high frequency 
($\hbar\omega{\sim}0.25\;MeV$), $\it{i.e.}$, the alignment processes occurring in the core are reflected in band 2 
as well. 

In order to understand the observations, a microscopic model that is able to describe well different types of vibrations 
is necessary. As introduced in Sec.~\ref{subsec:Reft_asym_theory} of this chapter, the calculations with the Random 
Phase Approximation (RPA) model based on the cranked Nilsson potential, $\it{i.e.}$, the cranked RPA calculations, can 
account well for the interplay between octupole vibrations and collective quasi-particle excitations under the stress 
of rotation. Such calculations have proved to reproduce quite successfully the experimental observables in many 
nuclei of the region, such as $^{238}$U~\cite{Ward-NPA-600-88-96}, $^{248}$Cm~\cite{Hackman-PRC-57-R1056-98} 
and $^{232}$Th~\cite{AbuSaleem-02-thesis}, for example. Hence, similar 
calculations for the three Pu nuclei of interest were carried out by T. Nakatsukasa of RIKEN for the present work. 

As seen in Figure~\ref{fig:242Pu_alignment_np}, the cranked RPA calculations reproduce the alignment in band 2 of 
$^{242}$Pu rather well: the magnitude of the alignment and its evolution with spin are accounted for, justifying 
the $K=0$ octupole assignment. 

In $^{238}$Pu, the aligned spin of the octupole band, and, hence, the difference $\Delta{i}_{x}$ between the yrast 
and octupole bands is close to zero at low frequency and gradually increases with increasing frequency, until it 
becomes flat around $3\hbar$ in the medium spin region ($0.10\;MeV<\hbar\omega<0.25\;MeV$). Thus, the data indicate 
that this negative-parity band is associated with an octupole vibration. Around $\hbar\omega=0.25\;MeV$, there is 
indication of the start of an upbend. However, the experimental data do not allow to fully delineate the character 
of the alignment, although it seems to mirror the beginning of the upbend seen in the yrast sequence at somewhat 
higher frequency. As seen in Figure~\ref{fig:238Pu_alignment}, the cranked RPA calculations with $K=0$ do not 
reproduce the characteristics of this band as well as was the case in $^{242}$Pu, especially at high frequency 
($\hbar\omega>0.20\;MeV$). 

The most striking pattern exhibited by the alignments is seen in $^{240}$Pu. It has been pointed out earlier that the 
$i_x$ curve of its yrast band is unique among those of the yrast bands in $^{238-244}$Pu isotopes; $\it{i.e.}$, it forms 
an almost straight line with small and positive slope and no sign of a sudden alignment gain. At the same time, 
the alignment of its octupole band grows gradually from a small number (${\sim}0.5\hbar$) at $\hbar\omega=0.02\;MeV$ 
until it reaches the maximum of about $4\hbar$ at $\hbar\omega{\sim}0.20\;MeV$, before remaining almost constant at 
higher frequencies. As a result, the value of $\Delta{i}_{x}$ of the negative-parity band relative to the yrast 
(positive-parity) sequence remains between 2 to 3$\hbar$ in the medium spin region, and, even decreases slightly 
with the increasing angular frequency for $\hbar\omega>0.20\;MeV$. It is worth pointing out that the corresponding 
$\Delta{i}_{x}$ in $^{220}$Ra and $^{222}$Th~\cite{Smith-PRL-75-1050-95,Cocks-NPA-645-61-99}, two of the best 
examples of nuclei with stable octupole deformation, also decreases from about $3\hbar$ at $\hbar\omega{\sim}0.10\;MeV$ 
to a smaller value at higher frequencies ($\hbar\omega>0.22\;MeV$). Hence, this characteristic behavior in $\Delta{i}_{x}$ 
can be viewed as a first similarity with observations made in nuclei understood in terms of octupole deformation. 
As can be seen in Figure~\ref{fig:240Pu_alignment_np}, none of the cranked RPA calculations can reproduce 
the alignment of band 2. This observation is consistent with that seen at high spin in $^{238}$Pu, 
and is presumably due to the impact of strong octupole correlations. 

As described in Sec.~\ref{subsec:Pu_yrast_bds}, it was seen in the data that the $i_{13/2}$ quasi-proton 
alignment is absent or at least severely delayed in frequency in the $^{240}$Pu and $^{238}$Pu yrast bands. 
This surprising phenomenon, suggested to be due to the impact of strong octupole 
correlations~\cite{Wiedenhover-PRL-83-2143-99}, has actually been predicted in theoretical work by Frauendorf and 
Pashkevich~\cite{Frauendorf-PLB-141-23-84}. The CSM calculations therein indicate that, with the presence of 
octupole deformation, the pronounced backbending is absent at the frequency where it is observed in reflection-symmetric 
nuclei, and is calculated to occur at higher frequency. As shown above, 
the observed alignments in the octupole bands of $^{240}$Pu and $^{238}$Pu are not reproduced well by 
the cranked RPA calculations, which have proved to do well for ``normal'' octupole vibrations, for example 
those in $^{242}$Pu (in the present work), $^{238}$U~\cite{Ward-NPA-600-88-96} and $^{232}$Th~\cite{AbuSaleem-02-thesis}. 
These facts point to the additional stength of octupole correlations in $^{240}$Pu and $^{238}$Pu. To 
firmly establish this pattern of increased octupole correlations and to learn further about their impact in the two Pu nuclei, 
evidence other than the alignment properties discussed thus far is required. 

\begin{figure}
\begin{center}
\includegraphics[angle=270,width=0.80\columnwidth]{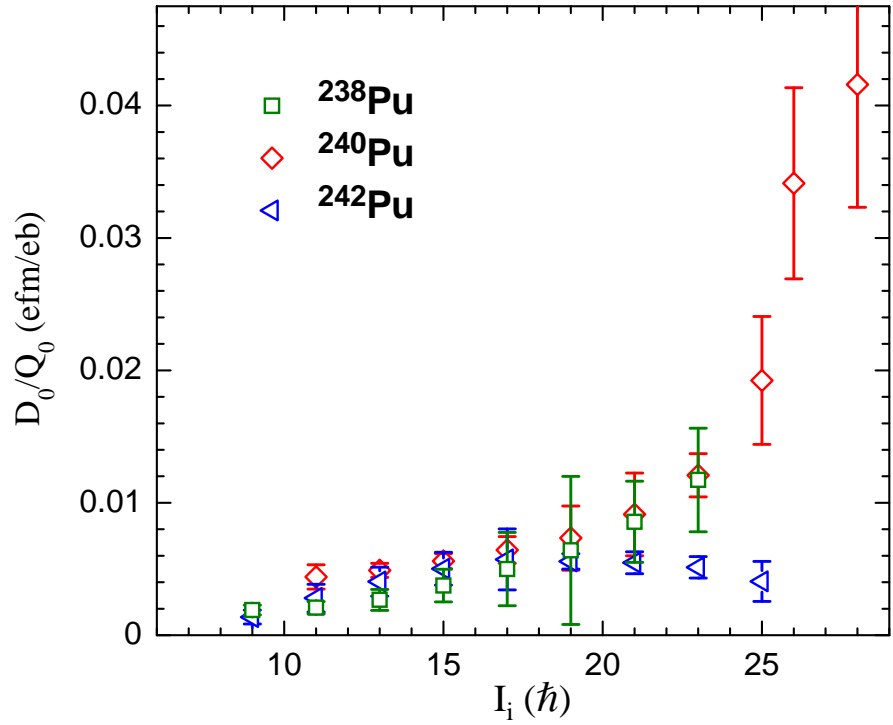}
\caption[The ratios between the dipole and quadrupole moments in $^{238,240,242}$Pu.]{The ratios 
between the dipole and quadrupole moments extracted from the intensity balance between transitions linking 
the yrast and octupole bands and the in-band transitions in the octupole band.\label{fig:Pu_isotopes_DQ}}
\end{center}
\end{figure}

Further evidence of the increased importance of octupole correlations 
in $^{240}$Pu and $^{238}$Pu comes from the ratios of dipole and quadrupole moments which can be extracted from 
the branching ratios between in-band and out-of-band transitions in bands 2 in the three Pu isotopes. 
Although such ratios were already reported and discussed in Ref.~\cite{Wiedenhover-PRL-83-2143-99}, 
the data from the present work extend to higher spin, and, in many instances, are of better accuracy. 
As seen clearly in Figure~\ref{fig:Pu_isotopes_DQ}, the ratios of moments ${D_0}/{Q_0}$ in the three even-even Pu 
isotopes are close and grow gradually with increasing spin, up to angular momenta lower than $19\hbar$, but, they behave 
differently for higher spin values (${>}19\hbar$). In $^{238}$Pu and $^{240}$Pu, the ${D_0}/{Q_0}$ ratios keep rising, while 
the values become essentially constant in $^{242}$Pu. In other words, the inter-band $E1$ decays become increasingly 
competitive with the in-band $E2$ transitions, as the spin increases. The effect is most pronounced for $^{240}$Pu. 
With the assumption that the $Q_0$ moments are constant, as would be expected for a rotational band, the resulting 
increase of $D_0$ moment in $^{238,240}$Pu reflects the enhacement of octupole collectivity and this observation 
can be viewed as a second indication that enhanced octupole correlations indeed play a major role in these two nuclei. 

\begin{figure}
\begin{center}
\includegraphics[angle=270,width=0.90\columnwidth]{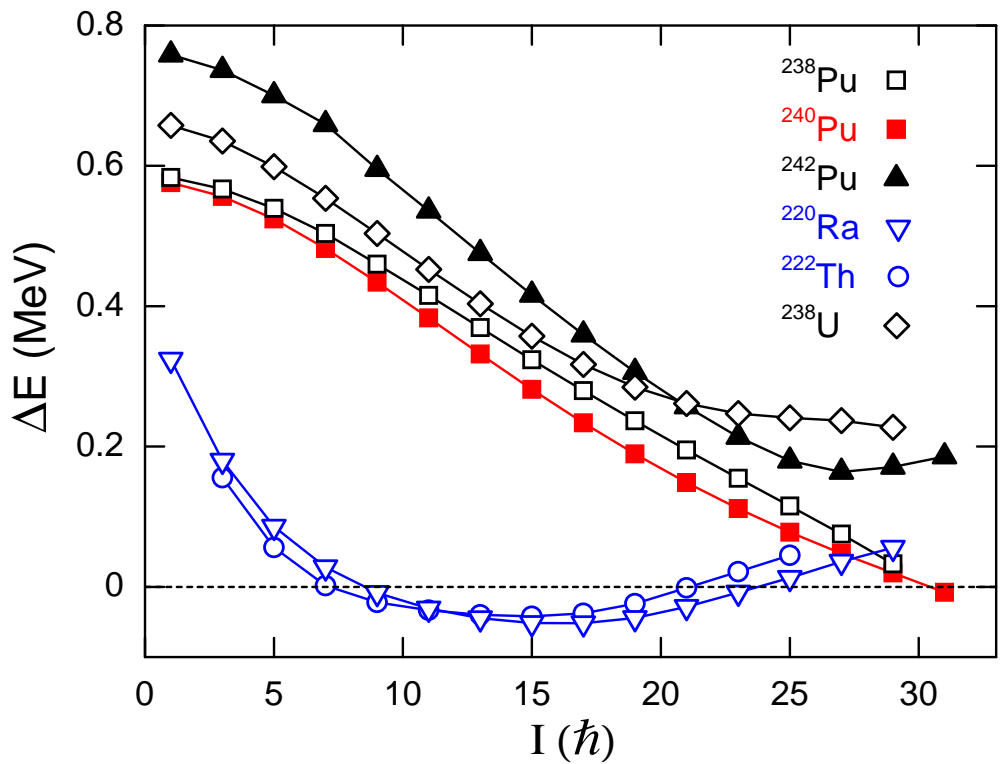}
\caption[The plots of $\Delta{E}$ for several even-even actinide nuclei.]{The plots of the energy difference between 
the yrast and one-phonon bands $\Delta{E}$ as a function of spin $I$ for several even-even actinide nuclei. 
See text for details.\label{fig:Pu_Ra_Th_U_DelE}}
\end{center}
\end{figure}

In fact, the energy difference between the yrast and octupole bands $\Delta{E}$, which is defined as 
$\Delta{E}=E_{-}(I)-(E_{+}(I+1)+E_{+}(I-1))/2$, is a measure of the extent to which the positive-parity levels 
of band 1 and the negative-parity levels of band 2 can be viewed as merging into a single rotational band. 
As can be seen in Figure~\ref{fig:Pu_Ra_Th_U_DelE}, at the highest spins the situation in $^{238}$Pu and 
$^{240}$Pu is rather similar to that seen in the ``octupole-deformed'' $^{220}$Ra and $^{222}$Th nuclei: 
the measured $\Delta{E}$ values are very close to zero. In contrast, the negative-parity band in $^{242}$Pu 
exhibits a different trajectory: it never merges with the yrast band and its $\Delta{E}$ pattern has similarities with 
that seen in $^{238}$U. 

More importantly, at spins ${\geq}25\hbar$ in $^{240}$Pu, a key sequence of $E1$ transitions, linking the 
alternating-parity states that are well interleaved in energy, $\it{i.e.}$, the ``zig-zag'' structure, is 
observed (see the level scheme of $^{240}$Pu and the discussion in Sec.~\ref{subsec:240Pu_band2}). These inter-band transitions 
are seen to go ``both ways'', $\it{i.e.}$, they go not only from band 2 to band 1, such as the $27^{-}{\rightarrow}26^{+}$ 
line, but also from band 1 to band 2, see the $26^{+}{\rightarrow}25^{-}$ $\gamma$ ray, for example. This 
unusual decay mode has been established for the first 
time in the present work (see Sec.~\ref{subsec:240Pu_band2}). In $^{238}$Pu, no such $E1$ transitions connecting the 
interleaved levels in the yrast (positive-parity) and octupole (negative-parity) bands at high spin were seen. 
In $^{242}$Pu, the levels $(I+1)^{+}$ and $I^{-}$ with $I{\geq}25\hbar$ in the two bands with opposite 
parity have almost degenerate energies, and no inter-band lines were observed. Assuming a constant moment 
$Q_0=11.6\;eb$ adopted from a previous measurement of the $B(E2)$ rate for the $2^{+}{\rightarrow}0^{+}$ in the yrast 
band~\cite{Bemis-PRC-8-1466-73}, the key values of $D_0$ for some of those linking transitions 
can be derived. Despite the relatively large errors, the $D_0$ values at high spin become quite large: $D_0{\geq}0.2\;efm$ 
for $I{\geq}25\hbar$ in $^{240}$Pu, and correspond to $B(E1)$ strength ${\geq}2{\times}10^{-3}$ W.u. Such values 
are much larger than the $B(E1)$ strengths that are usually observed ($<10^{-4}$ W.u.). They are of the same order as 
those observed in the light Ra and Th isotopes~\cite{Schuler-PLB-174-241-86}, which are among the best examples of 
octupole rotors~\cite{Butler-RMP-68-349-96}. 

From the discussion above, it is clear that, at high spin, the yrast (positive-parity) and the negative-parity 
bands in $^{240}$Pu appear to merge into a single octupole rotational band ($I^{+}$, $(I+1)^{-}$, $(I+2)^{+}$, ...), 
which is widely viewed as one of the signatures for octupole deformed nuclei. This is the first observation of 
this unique feature in actinide nuclei located outside of the predicted boundary of the ``island'' of octupole 
deformation~\cite{Nazarewicz-NPA-429-269-84,Sheline-PLB-197-500-87}, illustrated in Figure~\ref{fig:Actinide_chart} 
in Sec.~\ref{subsec:Motiv_regn_octu}. It provides the most direct indication for the fact that the octupole correlations 
are enhanced considerably in $^{240}$Pu by the rotation of nucleus and, perhaps, to the point where static octupole 
deformation develops at sufficiently high angular momentum~\cite{Jolos-PRC-49-R2301-94,Jolos-NPA-587-377-95}. 

%This is perhaps the case in $^{238}$Pu, too. 

\subsection{\label{subsec:Pu_ex_positv_parity_bd}Excited positive-parity bands in $^{240}$Pu and $^{242}$Pu}
The three bottom levels in band 3 of $^{240}$Pu were originally associated with a collective 
quadrupole vibration, mostly a $\beta$ vibration with preserved axial symmetry~\cite{Parekh-PRC-26-2178-82}, 
although a triaxial $\gamma$ vibration with two phonons combining to a $0^{+}$ state was proposed as well. 
However, it was pointed out by Hoogduin $\it{et~al.}$~\cite{Hoogduin-PLB-384-43-96} that a triaxial $\gamma$ 
band would be expected to manifest itself at higher excitation energy. In the same work~\cite{Hoogduin-PLB-384-43-96}, 
another new experimental observation was provided: band 3 in $^{240}$Pu does not decay by $E0$ transitions to band 1, 
unlike every other first excited even-spin, positive-parity band in the neighboring nuclei. Such strong $E0$ 
linking lines have been widely associated with $\beta$ bands~\cite{Davydov-NP-60-529-64}. Furthermore, studies of 
$(p,t)$ transfer reactions, carried out in the 1970s~\cite{Maher-PRL-25-302-70,Maher-PRC-5-1380-72}, indicated 
that, while in most actinide nuclei (Th, U, Cm, ...) the $(p,t)$ strength to excited $0^{+}$ states is concentrated 
on a single state, this is not the case in $^{240}$Pu; rather, the strength is split on two excited states: 
the $0_{2}^{+}$ level at 862 $keV$ and the $0_{3}^{+}$ level at 1091 $keV$. Thus, from this discussion it results 
that band 3 cannot readily be associated with a $\beta$ vibration, either. 

\begin{figure}
\begin{center}
\includegraphics[angle=270,width=0.80\columnwidth]{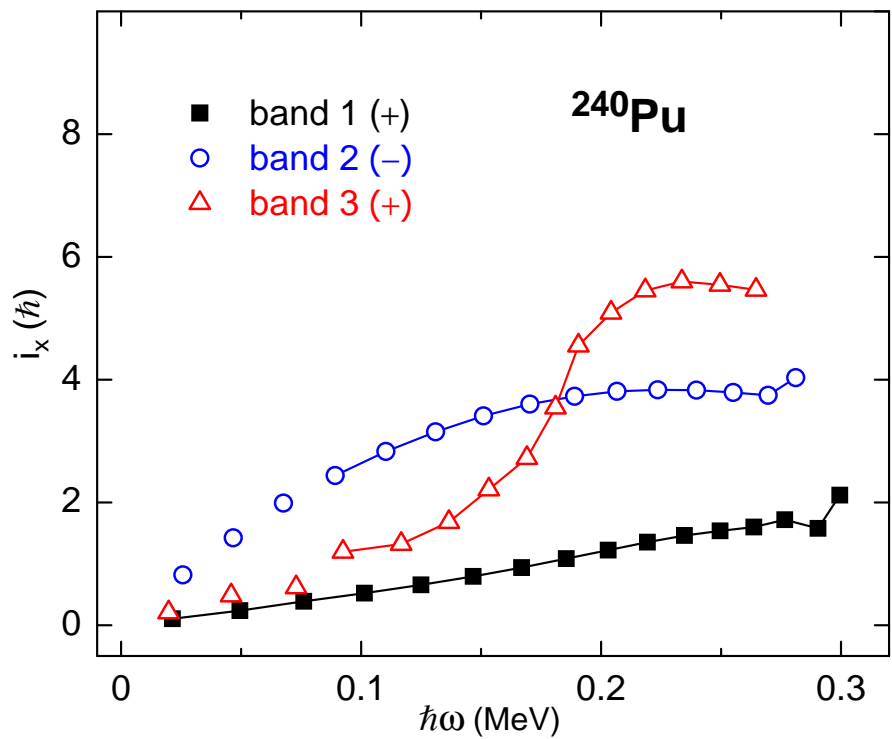}
\caption{The aligned spins obtained from the present data for bands 1, 2 and 3 in $^{240}$Pu.\label{fig:240Pu_alignment}}
\end{center}
\end{figure}

\begin{figure}
\begin{center}
\includegraphics[angle=270,width=0.80\columnwidth]{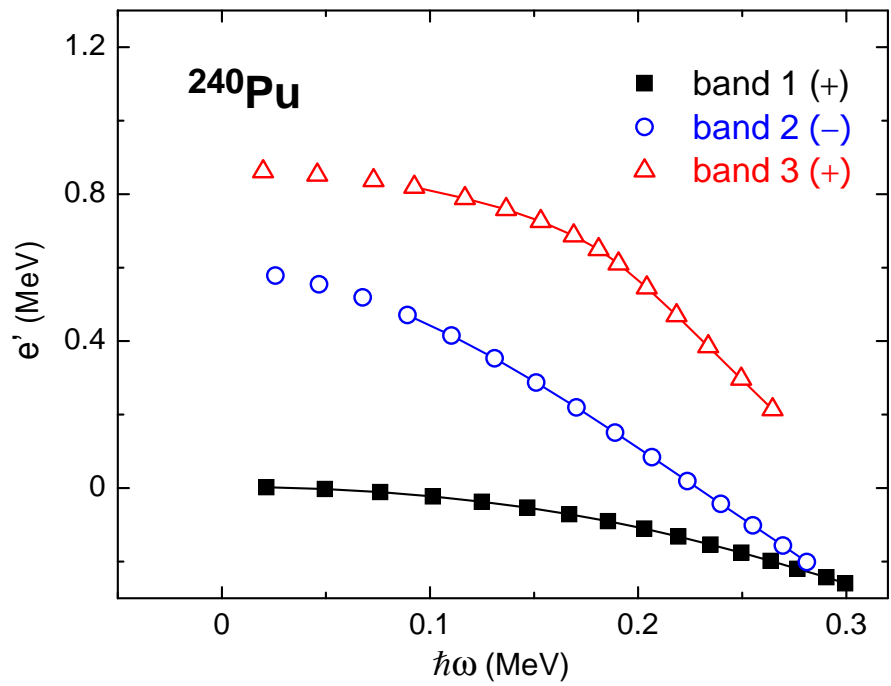}
\caption{The routhians obtained from the present data for bands 1, 2 and 3 in $^{240}$Pu.\label{fig:240Pu_routhian}}
\end{center}
\end{figure}

As can be seen in the level scheme (Figure~\ref{fig:Pu240_levl_sche}), the sequence of $\gamma$ rays at higher spin built 
on the first excited $0^{+}$ state in $^{240}$Pu ($\it{i.e.}$, the in-band transitions of band 3) has been observed for the first 
time in the present work. Experimentally, this band only decays via $E1$ transitions to the negative-parity 
band (band 2). This is a very surprising and, to our knowledge, unique result. It is the only band 
of its kind in all three Pu isotopes studied here. Hence, the excited positive-parity band (band 3) sets 
the $^{240}$Pu nucleus apart from all the others. The aligned spin $i_x$ of band 3 in $^{240}$Pu is also interesting. As illustrated 
in Figure~\ref{fig:240Pu_alignment}, it starts from a low value ($<1\;\hbar$), but at the point where a $3\hbar$ 
alignment is achieved in the negative-parity band (band 2), $\it{i.e.}$, $\hbar\omega{\sim}0.20\;MeV$, a gain in alignment of 
$\sim$ $6\hbar$ has occurred in the excited positive-parity band. Furthermore, the $i_x$ value of band 3 remains essentially 
constant at $6\hbar$ for higher frequencies. The routhian plot for this band (see Figure~\ref{fig:240Pu_routhian}) changes 
its slope sharply at the point where the sudden gain in alignment happens. 

\begin{figure}
\begin{center}
\includegraphics[angle=270,width=0.80\columnwidth]{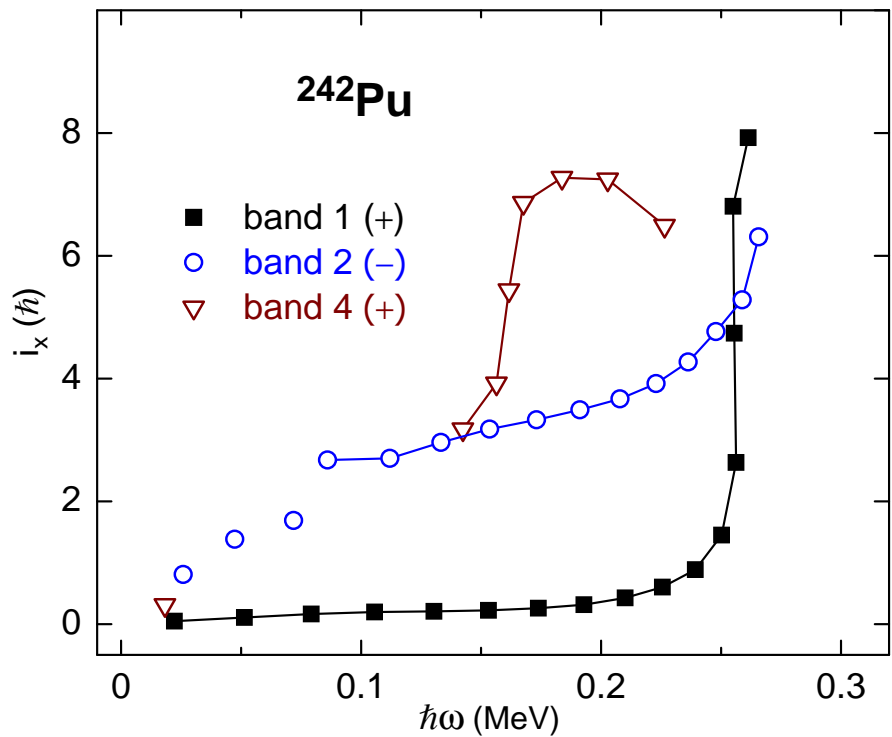}
\caption{The aligned spins obtained from the present data for bands 1, 2 and 4 in $^{242}$Pu.\label{fig:242Pu_alignment_pp}}
\end{center}
\end{figure}

\begin{figure}
\begin{center}
\includegraphics[angle=270,width=0.80\columnwidth]{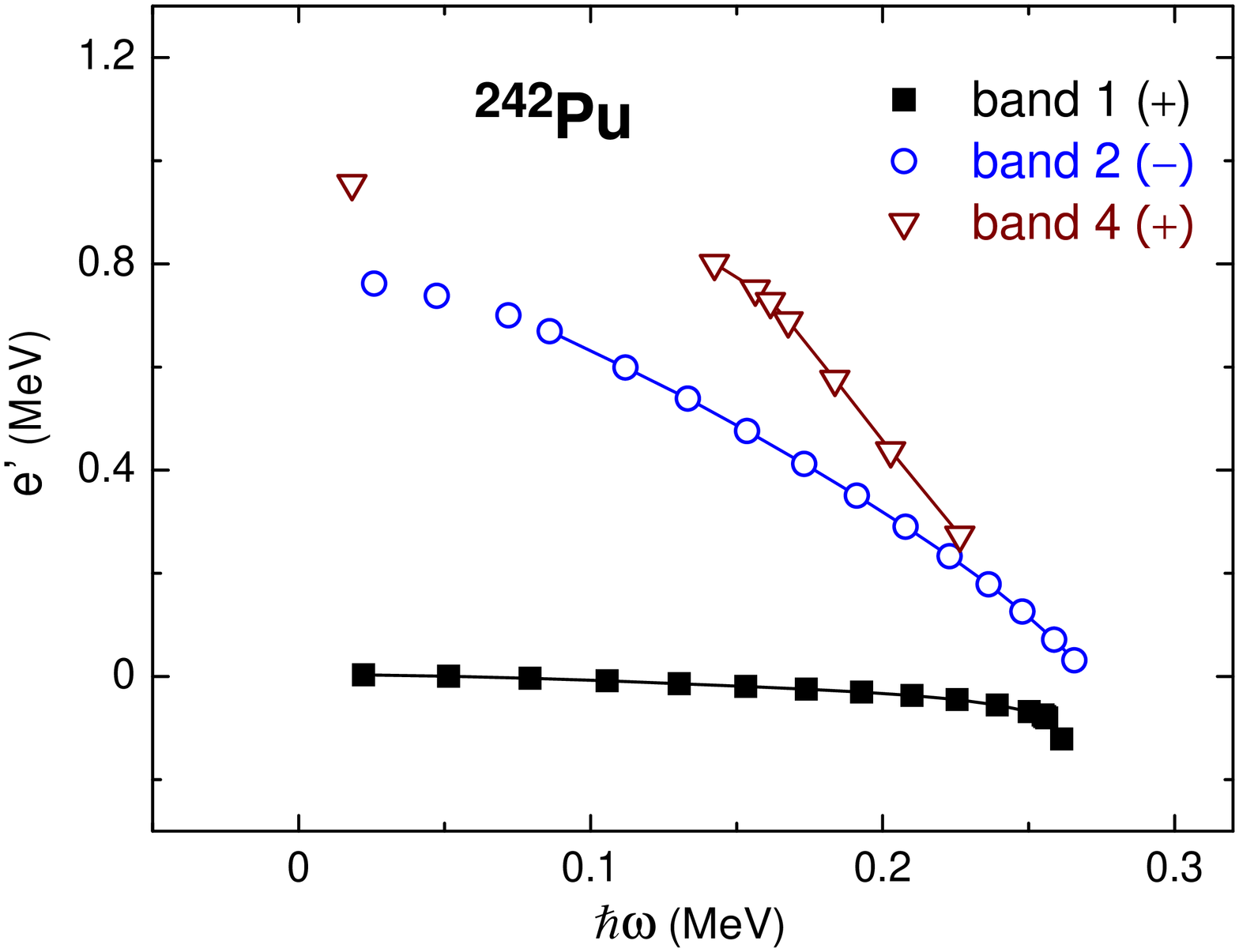}
\caption{The routhians obtained from the present data for bands 1, 2 and 4 in $^{242}$Pu.\label{fig:242Pu_routhian_pp}}
\end{center}
\end{figure}

The first excited positive-parity band (band 4) in $^{242}$Pu exhibits a markedly different behavior. Experimentally, 
it only decays to the yrast band (instead of to the negative-parity band in the $^{240}$Pu case). Only a few states 
were identified, and this band was not traced down to low spin due to the lack of statistics. It can be seen 
clearly in Figure~\ref{fig:242Pu_alignment_pp} that a gain in alignment of $\sim$ $8\hbar$ occurs around $\hbar\omega=0.16\;MeV$. 
Thus, this band does not have any similarity with band 3 in $^{240}$Pu, and its exact nature has not been thoroughly 
investigated here. A rather natural interpretation of this band might be that it starts out as a $\beta$ vibration 
at low spin (where the $0_{2}^{+}$ and $2_{2}^{+}$ states are known from earlier work~\cite{Maher-PRL-25-302-70,Maher-PRC-5-1380-72}), 
which is crossed at higher frequency by a two quasi-particles excitation, possibly a quasi-neutron excitation. 
As can be seen in Figure~\ref{fig:Pus_CSM_routhan}, at least one possible crossing is calculated to occur in the 
routhian diagram below $\hbar\omega=0.2\;MeV$. On the other hand, an extrapolation of the routhian of band 4 to 
higher frequency in Figure~\ref{fig:242Pu_routhian_pp} would suggest that this band would cross band 1 at the 
frequency where the upbend occurs in latter. This observation would associate band 4 with a two quasi-protons 
excitation. Clearly, further elucidation of this issue will require a new data set where band 4 would be expanded 
to both higher and lower spin states. 

In summary, the evidence for very strong octupole correlations in $^{240}$Pu (and possibly also in $^{238}$Pu) that 
has emerged in the present work includes: (a) the absence or severe delay of quasi-proton alignment in the yrast band 
(see Sec.~\ref{subsec:Pu_yrast_bds}); (b) the rising of the $D_0$ moment with increasing spin throughout the measured 
range of angular momentum (see Sec.~\ref{subsec:Pu_octupl_bds}); (c) the sequence consisting of interleaved levels 
with alternating spin and parity, which are connected by strong $E1$ transitions, $\it{i.e.}$, the ``zig-zag'' pattern 
(see Sec.~\ref{subsec:Pu_octupl_bds}); and, (d) an excited positive-parity band (band 3) that decays solely to the 
octupole band (band 2) via $E1$ transitions (see Sec.~\ref{subsec:Pu_ex_positv_parity_bd}). In addition, there is 
further evidence for the strength of octupole correlations in $^{238}$Pu and $^{240}$Pu from $\alpha$ 
decay~\cite{Sheline-PRC-61-057301-00}. The hindrance factors (HFs) of $\alpha$ decay from the $0^{+}$ ground 
states of $^{238}$Pu and $^{240}$Pu to the low-lying $1^{-}$ levels in the daughter nuclei are found to be small. 
In fact, these two nuclei form a minimum in HFs for the Pu isotopes, which can be related to large octupole correlations, 
according to Ref.~\cite{Sheline-PRC-61-057301-00}. Indeed, this observation of low HFs 
is one of the criteria for octupole deformation, which have been established by a large number of experiments 
(see Ref.~\cite{Butler-RMP-68-349-96}). All of these observations underline the need for a new interpretation. 

\subsection{\label{subsec:Pus_condense_interpt}Interpretation of observables using ``Octupole Condensation''}
As discussed earlier in this section, the RPA calculations appear to be inappropriate to deal with cases 
associated with strong octupole correlations, such as is the case here with $^{240}$Pu and, perhaps, $^{238}$Pu. 

\begin{figure}
\begin{center}
\includegraphics[angle=270,width=0.90\columnwidth]{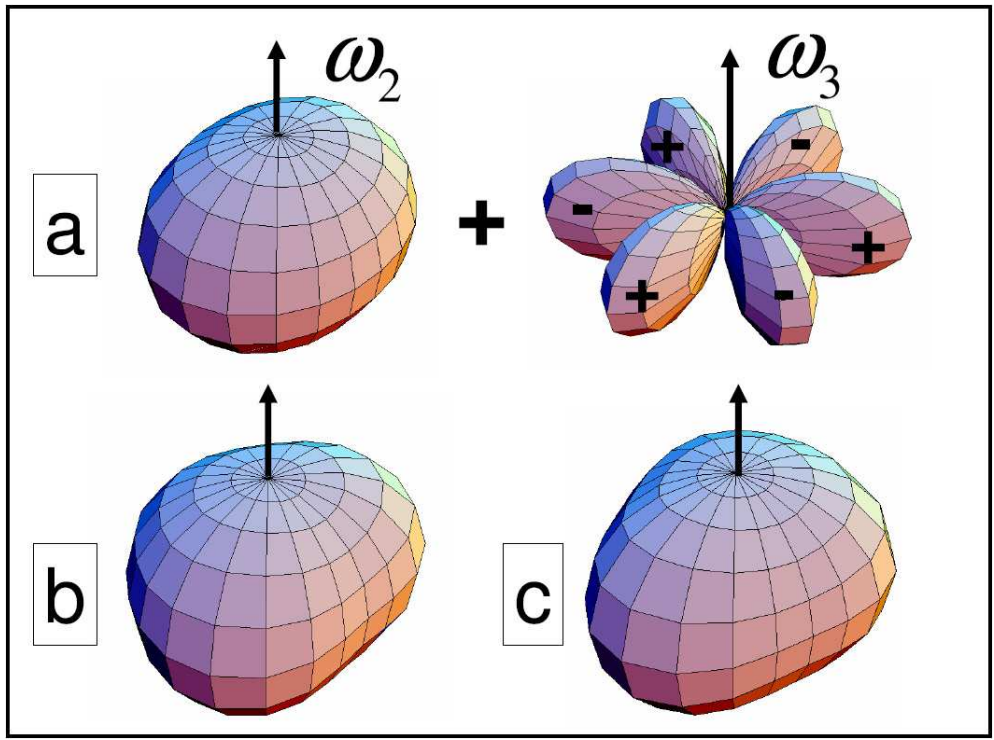}
\caption[An octupole wave traveling over the surface of a quadrupole-deformed nucleus.]{An octupole 
wave traveling over the surface of a quadrupole-deformed nucleus. In c), the octupole wave has 
turned by $90^{\circ}$ as compared to b). See text for details. 
Taken from Ref.~\cite{Frauendorf-condestn-unpublsh-07}.\label{fig:Octu_condens_illust}}
\end{center}
\end{figure}

\begin{figure}
\begin{center}
\includegraphics[angle=0,width=0.52\columnwidth]{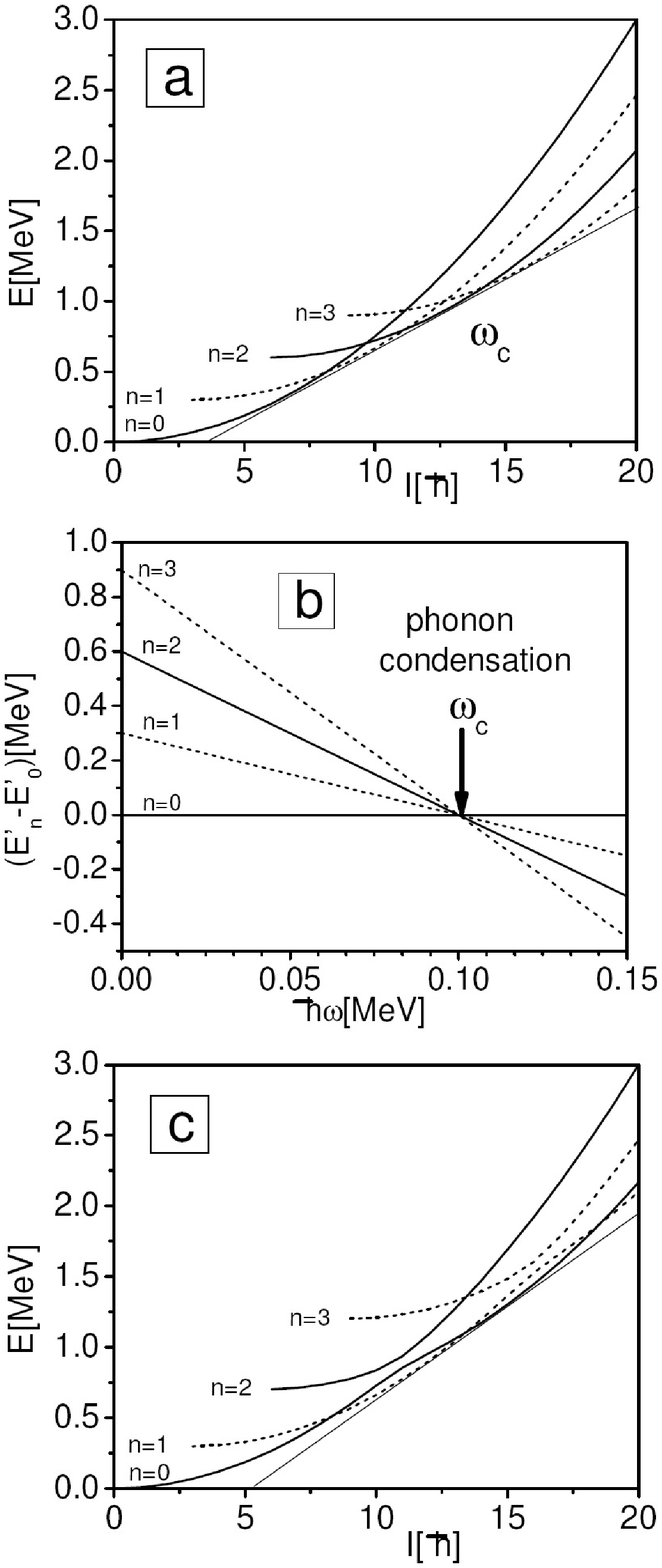}
\caption[Illustration of the concept of octupole condensation.]{Top panel, (a): Energies of aligned octupole 
multiphonon bands. Full lines show $\pi=+$ and dashed lines $\pi=-$ states. Middle panel, (b): Energy in the 
rotating frame (routhian) relative to the zero-phonon band. Bottom panel, (c): As top, but, assuming an 
interaction between the phonons and an anharmonicity. See text for details. Taken from 
Ref.~\cite{Frauendorf-condestn-unpublsh-07}.\label{fig:Octu_condens_concpt}}
\end{center}
\end{figure}

Recently, a new mechanism has been proposed by S. Frauendorf, entitled ``condensation 
of rotational-aligned octupole phonons''~\cite{Frauendorf-condestn-unpublsh-07}. In this work, first it is 
assumed that the quadrupole deformed nucleus is a rigid rotor with the moment of inertia, ${\cal J}$, and 
that the octupole vibration is harmonic with frequency ${\Omega}_3$. Furthermore, it is postulated there 
is no interaction between the octupole phonons and the quadrupole deformed potential of the nucleus. 
The picture that then results is given in Figure~\ref{fig:Octu_condens_illust}: while the prolate nucleus 
rotates with a frequency $\omega_2$, an ``octupole'' wave travels over its surface with a frequency 
$\omega_3$ ($=\Omega_3/3$). Frozen ``images'' of the superposition of the two modes resemble in a striking 
way an octupole deformed nucleus. The running of the wave then gives the appearance of an octupole 
nucleus rotating or vibrating. The energy of the nucleus assuming this new collective mode is 
$E_n(I)=\hbar{\Omega}_3(n+1/2)+(I-ni)^2/(2{\cal J})$, where $I=ni+\omega{\cal J}$, $\omega$ is the angular 
velocity, $n$ is the number of octupole bosons, each boson carrying $3\hbar$ angular momentum ($i=3\hbar$). 
The resulting energy spectrum is plotted as a function of $I$ in Figure~\ref{fig:Octu_condens_concpt} (a). 
At $I_n=\Omega_3{\cal J}/i+i(n+1/2)$, it becomes energetically favorable to increase the spin $I$ through 
exciting an aligned phonon rather than by a further increase of the angular velocity 
$\omega$ ($=\omega_2$). Thus, as shown in Figure~\ref{fig:Octu_condens_concpt} (b), a number of band 
crossings will occur. At each crossing, a band with a higher number of phonons crosses the yrast band. 
The routhians of these multiphonon states $E^{\prime}_{n}-E^{\prime}_{0}$ 
($E^{\prime}_{n}=E_n-\omega{I}$), illustrated in Figure~\ref{fig:Octu_condens_concpt} (b), become zero 
at the same point $\omega=\omega_{c}={\Omega}_3/i$, $\it{i.e.}$, at a critical angular frquency. In this language, 
boson condensation occurs at $\omega_{c}$, the point where $\omega_2$ also equals $\omega_3$. 

The figure given above (Figure~\ref{fig:Octu_condens_concpt} (a)) is an idealized one. In reality, the 
anharmonicities of the octupole mode and the coupling to the quadrupole mode~\cite{Bohr-98-book} lead to an 
interaction between states with different phonon numbers. Due to parity conservation, the states with even $n$ (or with 
odd $n$) only mix with others with even $n$ (or with odd $n$). The interaction causes a repulsion 
between crossing bands of the same parity. Figure~\ref{fig:Octu_condens_concpt} (c) illustrates the resulting 
pattern of condensation when moderate anharmonicities are included. 

\begin{figure}
\begin{center}
\includegraphics[angle=270,width=0.80\columnwidth]{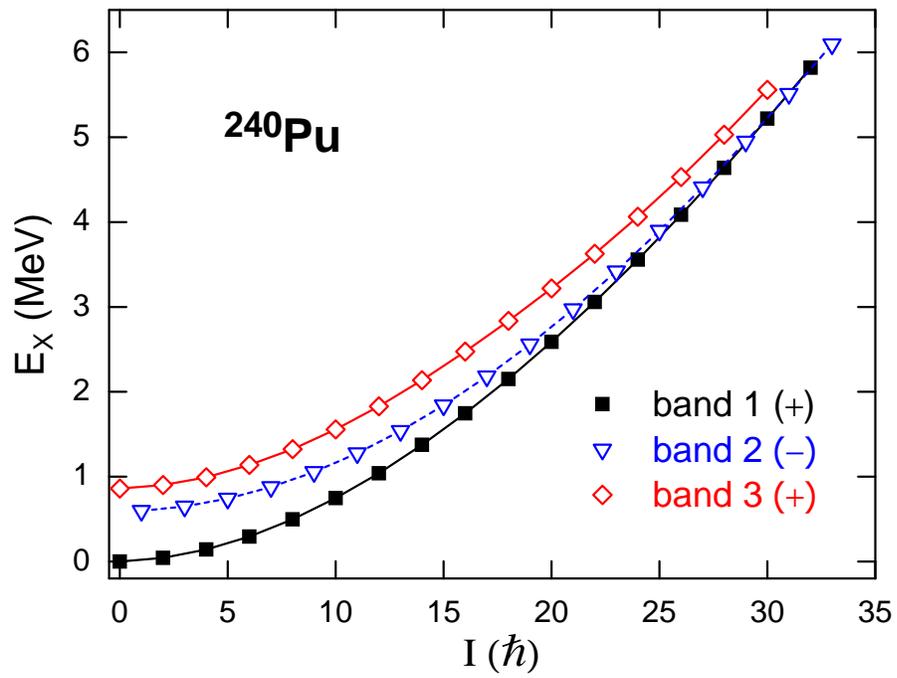}
\caption[The excitation energies of states in bands 1, 2 and 3 of $^{240}$Pu.]{The excitation energies of states 
in bands 1, 2 and 3 of $^{240}$Pu. See text for details.\label{fig:240Pu_ex_ener}}
\end{center}
\end{figure}

Assuming that bands 1, 2 and 3 can be assimilated with $n=0$, $n=1$ and $n=2$ phonon bands, 
the experimental excitation energies $E_x$ (Figure~\ref{fig:240Pu_ex_ener}) and routhians 
$e^{\prime}$ (Figure~\ref{fig:240Pu_routhian}) of the three bands in $^{240}$Pu can be 
compared with Figures~\ref{fig:Octu_condens_concpt} (b) and (c), respectively. In Figure~\ref{fig:240Pu_routhian}, 
the one-phonon octupole band (band 2) crosses the yrast band (band 1) at a frequency of $\sim$ 0.28 $MeV$. Due 
to the lack of statistics, the proposed two-phonon band (band 3) was not extended sufficiently to observe the crossing 
with the yrast band. However, if one extrapolates its trajectory while keeping the slope unchanged, such a crossing 
would not occur at the same frequency as that where bands 1 and 2 cross, but rather around 0.35 $MeV$. The disagreement 
between the observation and the ideal situation of octupole condensation, shown in Figures~\ref{fig:Octu_condens_concpt} (b), 
presumably reflects the involved anharmonicities. The excitation energies in $^{240}$Pu 
(Figure~\ref{fig:240Pu_ex_ener}) exhibit essentially the expected pattern at low-spin ($0<I<8\hbar$) as shown in 
Figures~\ref{fig:Octu_condens_concpt} (c), $\it{i.e.}$, the prediction for octupole condensation with band 
interaction. Unfortunately, the predicted full mixing between $n=0$ and $n=2$ phonons was 
not reached, even at the highest spin ($\sim$ $33\hbar$), in the present work. Based on the difference of the crossing 
frequency of $n=0$ and $n=1$ phonons with that of $n=0$ and $n=2$ phonons, the impact of anharmonicities can be 
quantified approximately. According to Ref.~\cite{Frauendorf-condestn-unpublsh-07}, 
at the point of condensation, the relation $E_n=\omega_c{I}$ applies for the $n$-phonon state. 
%In a nucleus, $\it{i.e.}$, a realistic small system, the critical frequency $\omega_c$ 
%fluctuates instead of being a constant. 
In the $^{240}$Pu case, $\omega_c$ was extracted 
to be 0.28 $MeV$ and 0.35 $MeV$ for the $n=1$ and $n=2$ states, respectively. As pointed out above, the aligned 
spins for these two states are about $3\hbar$ ($n=1$) and $6\hbar$ ($n=2$), as would be expected for one- and 
two-phonon states. Hence, $E_2$ can be estimated to be $2.5E_1$, rather than the idealized value of $2E_1$. 
The discrepancy in $E_2$ directly points to the strengh of the involved anharmonicities. It should be noted also 
that the presence of sizable anharmonicities is also seen in the fact that the excitation energy of $0^{+}$ state 
of band 3 (861 $keV$) is significantly lower than twice the excitation energy of the band 2 bandhead (597 $keV$). 

\begin{figure}
\begin{center}
\includegraphics[angle=0,width=\columnwidth]{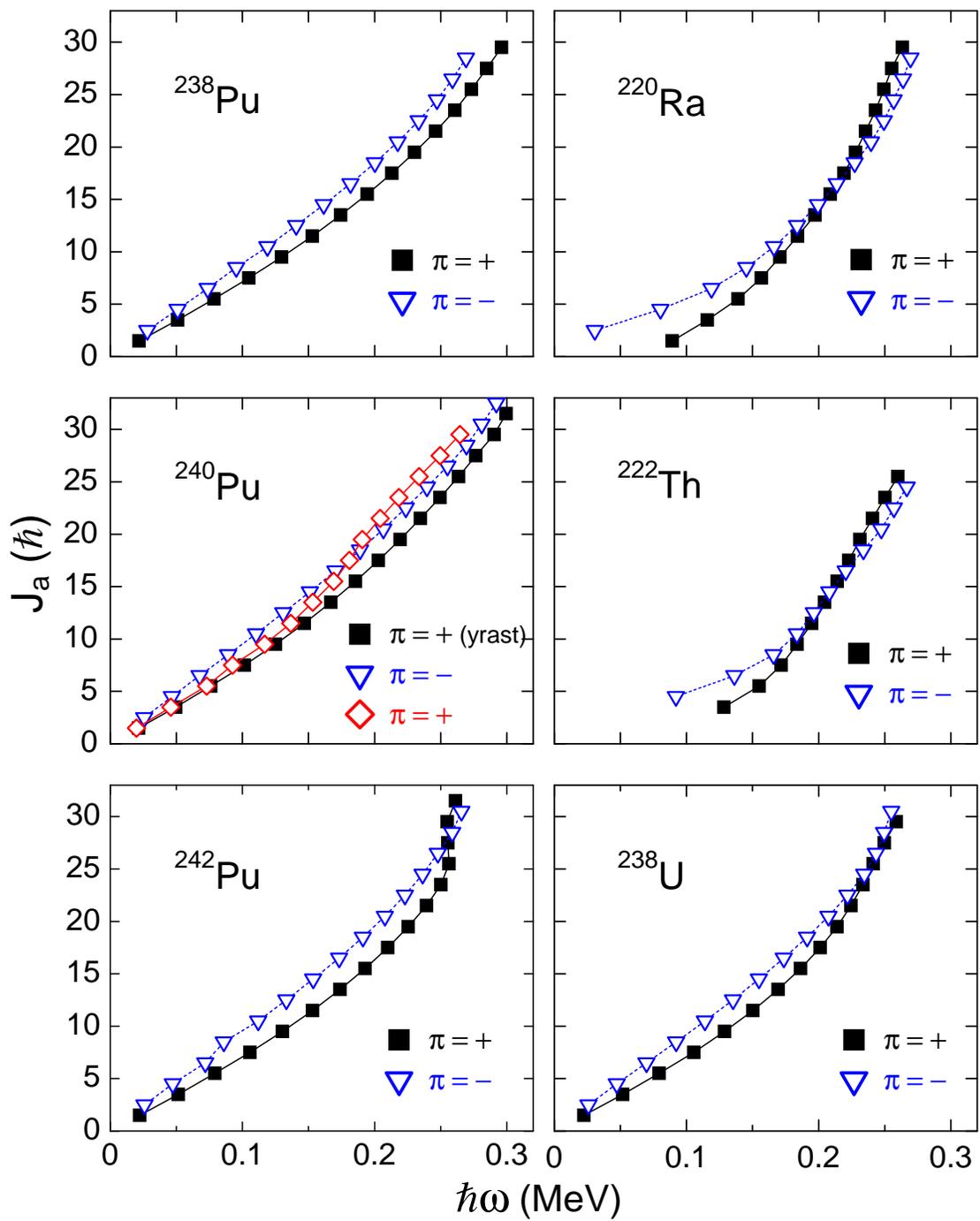}
\caption[The plots of the angular momentum for the yrast and octupole bands in several even-even actinide nuclei.]{The 
plots of the angular momentum $J_a=I-1/2$ as a function of angular frequency $\hbar\omega$ for the yrast and octupole bands 
in several even-even actinide nuclei. See text for details.\label{fig:Pu_Ra_Th_U_alt_spin}}
\end{center}
\end{figure}

The impact of octupole condensation also shows up in the plots of the energy difference $\Delta{E}$ as a 
function of spin $I$ (Figure~\ref{fig:Pu_Ra_Th_U_DelE} in Sec.~\ref{subsec:Pu_octupl_bds}) and of the 
spin $J_a$ as a function of the angular frequency $\hbar\omega$ 
(Figure~\ref{fig:Pu_Ra_Th_U_alt_spin}). Following the discussion of Ref.~\cite{Frauendorf-condestn-unpublsh-07} on 
octupole condensation in $^{220}$Ra and $^{222}$Th, the one-phonon band crosses the zero-phonon band before 
it feels much of an interaction with the two-phonon band (see Figure~\ref{fig:Octu_condens_concpt}). At the crossing, 
the energy difference between the yrast and one-phonon bands, $\it{i.e.}$, $\Delta{E}=E_{-}(I)-(E_{+}(I+1)+E_{+}(I-1))/2$, 
changes sign. At higher frequency the two-phonon band encounters the zero-phonon one, and the two mix and exchange 
character. The level repulsion attenuates the growth of $-\Delta E$, and this quantity starts decreasing 
when the $\pi=+$ band has become predominantly of two-phonon character. At the crossing between the two- and 
one-phonon bands, $\Delta E$ changes sign again. According the same rule, the one- and three-phonon bands are 
mixed at higher spin and the next change of sign of $\Delta E$ would occur at the crossing of three- and 
two-phonon bands. In the $^{240}$Pu case, presumably because of weaker octupole correlations, 
only the first change of sign of $\Delta E$ (the crossing of one- and zero-phonon bands) 
is seen at the highest spins, while the zero- and two-phonon bands have not yet started their mixing. It is 
worth pointing out that the observation of the yrast and one-phonon bands not merging, but crossing has been 
interpreted satisfactorily for the first time by this concept of octupole condensation. 
The indication of octupole condensation in the angular momentum $J_a$ ($=I-1/2$) was also discussed using 
the example of $^{220}$Ra and $^{222}$Th in Ref.~\cite{Frauendorf-condestn-unpublsh-07}. 
As can be seen in Figure~\ref{fig:Pu_Ra_Th_U_alt_spin}, the one-phonon band starts with three units 
of angular momentum more than the zero-phonon band at the same $\hbar\omega$, as expected for one 
aligned octupole phonon. The difference decreases, when the two-phonon state, which carries six units 
of angular momentum, starts mixing into the zero-phonon band. The two observed bands have equal angular momentum 
at the frequency of full mixing where $\Delta E$ starts rising. Near the crossing of the two- and one-phonon bands at 
$I{\sim}24\hbar$ ($\hbar\omega{\sim}0.25\;MeV$), where the mixing is small, the angular momentum difference is 
$-3\hbar$. It is also expected that the difference of $J_a$ will be zero again when the full mixing between 
the one- and three-phonon states is reached (see Figure~\ref{fig:Octu_condens_concpt}). The $J_a$ functions 
for the three bands of $^{240}$Pu observed in the present work exhibit the expected pattern characteristic of 
octupole condensation in the stage before the occurrence of full mixing between the zero- and two-phonon states. The 
difference of $J_a$ between the zero- and one-phonon bands remains $3\hbar$ at $\hbar\omega{\sim}0.28\;MeV$, and, 
then, starts decreasing because of the mixing of zero- and two-phonon states. This difference would reach zero at a 
frequency larger than $0.30\;MeV$ (the highest frequency in the observation) as shown by the extrapolation of the $J_a$ 
curves of the zero- and one-phonon bands. The frequency where the difference of $J_a$ may become $-3\hbar$ in $^{240}$Pu 
(as seen in $^{220}$Ra and $^{222}$Th) is expected to be well beyond the measured range in the present work. 
In addition, the two- and one-phonon bands cross at $\hbar\omega{\sim}0.20\;MeV$ as expected. 
Therefore, both the $\Delta E$ and $J_a$ behaviors support the onset of octupole condensation in $^{240}$Pu. 

It is also pointed out in Ref.~\cite{Frauendorf-condestn-unpublsh-07} that, with the anharmonicities included, the 
$I^{+}{\rightarrow}(I-1)^{-}$ inter-band transitions have equal strength with the $I^{-}{\rightarrow}(I-1)^{+}$ ones, 
rather than being suppressed (in the case of excluding the anharmonicities). The consistency between this prediction 
and the observation of the ``zig-zag'' pattern of inter-band lines only in $^{240}$Pu provides further 
confidence for the validity of the interpretation with octupole condensation. 

Finally, it is worth pointing out as well that interband transitions are allowed only between states of bands 
differing by only one octupole phonon. This accounts for the experimental observations that the levels of band 3 
solely decay to those of band 2 and that any deexcitation towards band 1 was beyond the detection limit of 
the measurements. 

As seen in both Figures~\ref{fig:Pu_Ra_Th_U_DelE} (in Sec.~\ref{subsec:Pu_octupl_bds}) and \ref{fig:Pu_Ra_Th_U_alt_spin}, 
the situation in $^{242}$Pu is 
very different. The first excited positive-parity sequence (band 4) in $^{242}$Pu has been identified as a 
quasi-particle excitation, rather than a two-phonon octupole band. At the highest spins (or frequency) in the present work, 
the $\Delta{E}$ values remain almost constant at ${\sim}0.2\;MeV$. The $J_a$ curves of the yrast and one-phonon bands 
cross at $\hbar\omega{\sim}0.25\;MeV$, but the reason for this crossing is the alignment of a pair of $i_{13/2}$ 
quasi-protons in the yrast band and the crossing has no similarity in pattern with those of phonon states. 
Both $\Delta E$ and $J_a$ values in $^{242}$Pu behave in an almost identical manner with those in $^{238}$U, a well-known 
octupole vibrator (see Figures~\ref{fig:Pu_Ra_Th_U_DelE} and \ref{fig:Pu_Ra_Th_U_alt_spin}). 
In summary, the different scenarios in $^{240}$Pu and $^{242}$Pu indicate that the two-phonon octupole and the $i_{13/2}$ two 
quasi-protons bands are rather close in energy and compete with each other. In $^{240}$Pu, the two-phonon octupole 
band is lower. The yrast (zero-phonon) band interacts with the two-phonon band. This effect results in the absence 
or delay of the $i_{13/2}$ quasi-protons alignment. In $^{242}$Pu, the $i_{13/2}$ two quasi-protons band is at lower 
excitation energy. It interacts with the yrast band, and a strong alignment occurs. Unfortunately, the two-phonon 
octupole band could not be seen in this case. 

\begin{figure}
\begin{center}
\includegraphics[angle=270,width=0.80\columnwidth]{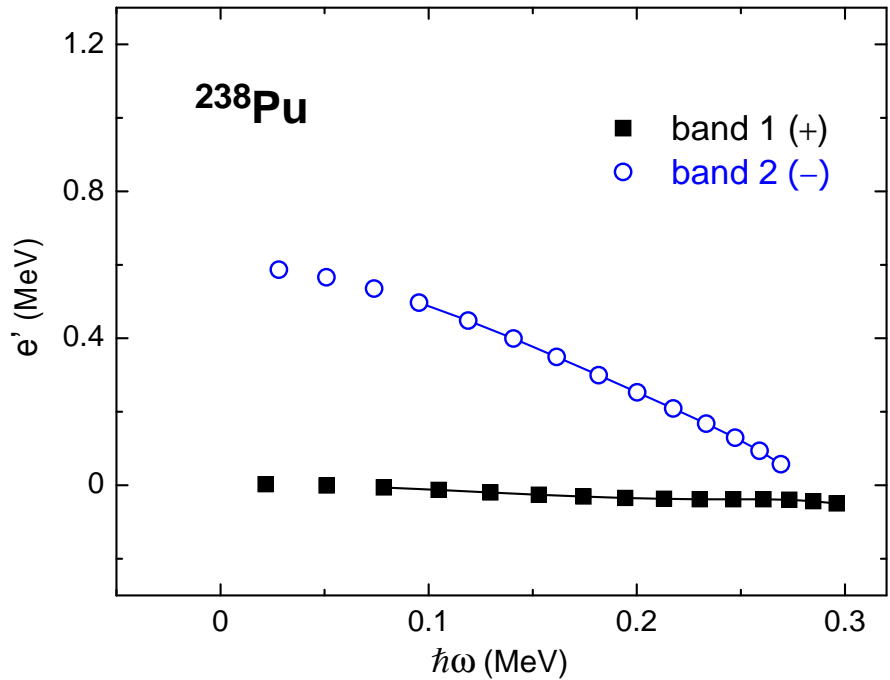}
\caption{The routhians obtained from the present data for bands 1 and 2 in $^{238}$Pu.\label{fig:238Pu_routhian}}
\end{center}
\end{figure}

Compared with the observations in $^{240}$Pu and $^{242}$Pu, the strength of octupole correlations in $^{238}$Pu seems moderate. 
In its routhian diagram (Figure~\ref{fig:238Pu_routhian}), due to the lack of statistics, the one-phonon octupole band (band 2) 
was not extended sufficiently to observe the crossing with the yrast band (band 1). However, the extrapolattion of its trajectory 
suggests that band 2 would cross the yrast band in a manner similar to that seen in the $^{240}$Pu case, but, at a higher 
frequency ($\sim$ $0.30\;MeV$). 
The $\Delta E$ values (see Figure~\ref{fig:Pu_Ra_Th_U_DelE} in Sec.~\ref{subsec:Pu_octupl_bds}) keep decreasing and get very 
close to, but do not reach, zero at the highest observed spins. The difference of $J_a$ (see Figure~\ref{fig:Pu_Ra_Th_U_alt_spin}) 
between the yrast band and the one-phonon one remains constant ($\sim$ $3\hbar$) at $\hbar\omega>0.20\;MeV$; $\it{i.e.}$, no sign 
of a decrease is seen. On the other hand, the proton alignment seen in $^{242}$Pu is absent or delayed, as well. No 
inter-band transition was observed at high spin. It is possible that the two quasi-particle band and the two-phonon 
octupole band are located at almost the same energy, although neither of them is seen in the present work. The impact 
from both of these two bands on the yrast band could possibly result in the observed moderate pattern in $^{238}$Pu, 
$\it{e.g.}$, the milder upbend of the alignment at higher frequency for the yrast and one-phonon octupole bands 
(see Figure~\ref{fig:238Pu_alignment} in Sec.~\ref{subsec:Pu_octupl_bds}). 

\subsection{\label{subsec:Pu242_other_bds}Other bands in $^{242}$Pu}
\begin{figure}
\begin{center}
\includegraphics[angle=270,width=0.80\columnwidth]{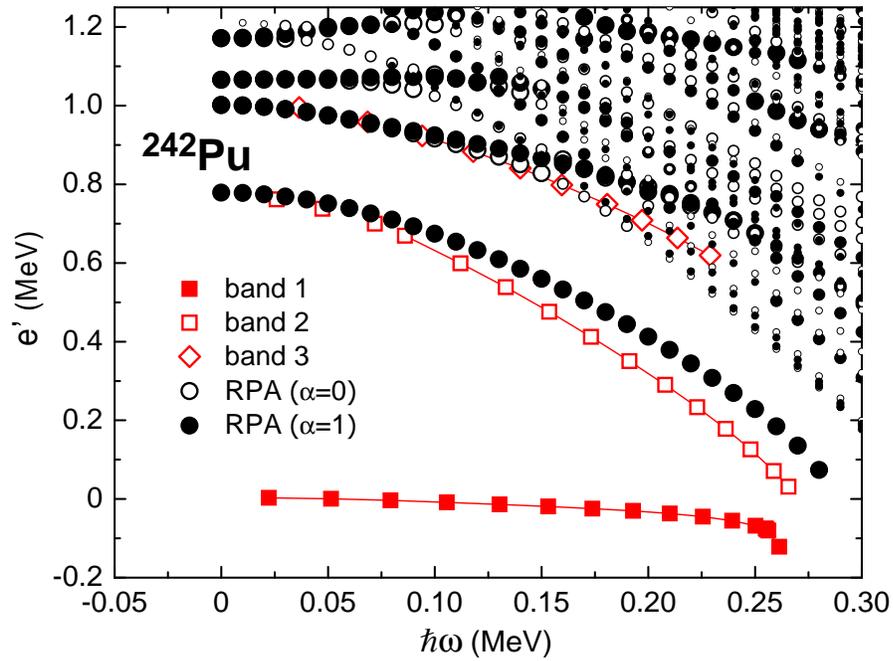}
\caption[The routhians of bands 1, 2 and 3 from the present data and the negative-parity RPA solutions for $^{242}$Pu.]{The 
routhians of bands 1, 2 and 3 from the present data and the cranked RPA calculations for the negative-parity 
bands in $^{242}$Pu. The $K$ quantum numbers of the octupole multiplets used in the RPA calculations from low to high 
in energy in the figure are: 0, 3, 2, 1. The open and filled circles denote that $\alpha$ (signature) $=0$ and $=1$, 
respectively. The size of the circle represents the strength of collectivity, $\it{i.e.}$, large - highly collective, 
medium - moderately collectively, and, small - non-collective.\label{fig:242Pu_routhian_oth}}
\end{center}
\end{figure}

\begin{figure}
\begin{center}
\includegraphics[angle=270,width=0.80\columnwidth]{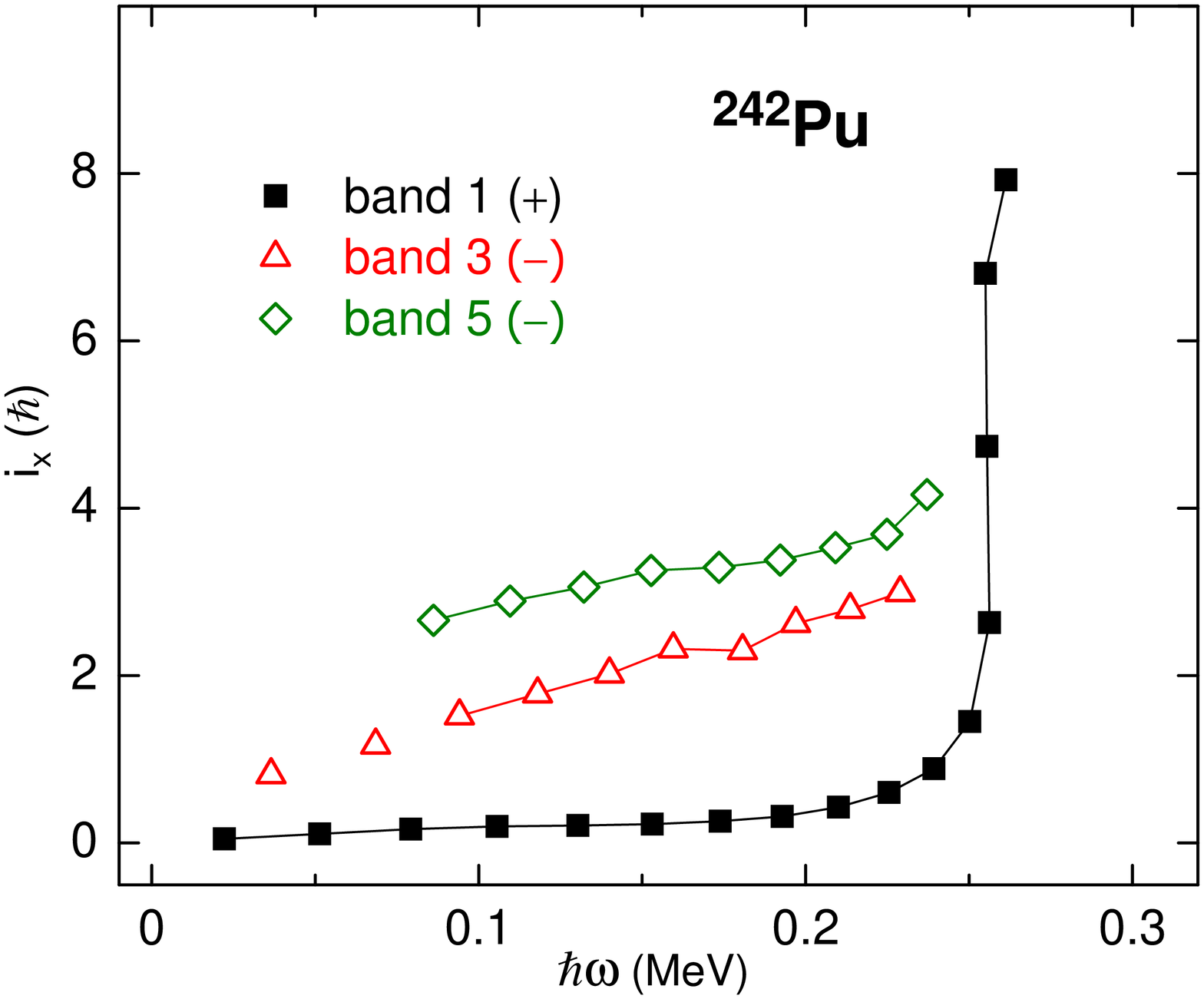}
\caption{The aligned spins obtained from the present data for bands 1, 3 and 5 in $^{242}$Pu.\label{fig:242Pu_alignment_oth}}
\end{center}
\end{figure}

Band 3 only decays to the yrast band experimentally. As can be seen Figure~\ref{fig:242Pu_routhian_oth}, the cranked RPA 
calculations for the negative-parity bands in $^{242}$Pu reproduced this band reasonably well. Based on the experimental observables, 
such as the routhian (Figure~\ref{fig:242Pu_routhian_oth}) and the aligned spin (Figure~\ref{fig:242Pu_alignment_oth}), 
for example, and the comparisons with calculations, band 3 can probably be associated with an octupole vibration with 
$K=3$ and $\alpha=1$. Due to the large uncertainty on spin and parity value in band 5 and the unknown properties of band 6, these 
two bands were not compared with the RPA calculations and no further interpretation is given in the present work. 

\section{Conclusions and outlook}

In the present work, the measured properties of $^{238,240,242}$Pu and the comparison with neighboring nuclei indicate 
that octupole correlations in $^{240}$Pu are very strong (although weaker than those in two of the best examples of 
``octupole rotors'', $^{220}$Ra and $^{222}$Th). The octupole correlations appear to be enhanced with increasing spin and 
become even sufficient to give rise at high spin to structures similar to those traditionally associated with the 
rotation of an octupole deformed nucleus. Octupole correlations in $^{238}$Pu seem to be substantial as well; $\it{i.e.}$, 
sufficient to delay the proton alignment. The octupole correlations appear weakest in $^{242}$Pu, and do not affect 
the occurrence of the proton alignment. 

The cranked RPA calculations reproduced reasonably well all experimental observables associated with weak octupole correlations 
(octupole vibrators). The behavior of $^{240}$Pu can be understood by invoking octupole condensation. 

It would be interesting to see if the yrast and one-phonon bands in $^{240}$Pu behave in the expected manner; $\it{i.e.}$, in 
a way similar to that seen in $^{220}$Ra and $^{222}$Th, as predicted by the theory. This would require data at spins 
beyond the measured range in the present work.

%
% Chapter 5
%

%
% Chapter 5
%

\chapter{SUMMARY AND PERSPECTIVE}
The purpose of this work was two-fold: explore the nuclear shape 
associated with new bands discovered in $^{163}$Tm and provide new data to 
investigate the strength of octupole correlations in the Pu isotopes. 
These two objectives have now been met. 

In $^{163}$Tm, two excited bands were shown to be associated with quasi-particle 
excitations in a triaxial minimum rather than with the wobbling mode. This 
minimum was measured to correspond to a larger deformation than that 
characterizing the ground state. The data can be reproduced reasonably 
well by calculations which also provide an explanation for the 
presence of wobbling in the Lu isotopes and its absence in the neighboring 
isotopic chains of Ta, Hf and now also Tm. 

The results obtained thus far lead naturally to a number of open questions 
deserving attention in the future. First, it was recognized in the present 
work that the measured quadrupole transition moments are significantly 
smaller than those calculated. This observation is not limited to $^{163}$Tm: 
quadrupole moments smaller than those predicted have also been found in the Lu 
isotopes where wobbling is present. The origin of this discrepancy between 
experiment and theory remains to be clarified. Furthermore, specifically for 
$^{163}$Tm, the data presented here indicate that the $\gamma$-ray intensity 
responsible for the feeding of the two triaxial strongly deformed bands 
originates from states associated with larger deformation than that of the 
two bands themselves. The exact nature of these feeding levels warrants 
further attention, but would presumably require a detector system of larger 
resolving power than Gammasphere in order to extract the weak $\gamma$ rays 
from the quasicontinuum of unresolved transitions. Finally, a systematic 
search for wobbling excitations in this region of the nuclear chart needs 
to continue in order to validate the present theoretical understanding in 
terms of specific shell gaps. 

The new data available for the even-even $^{238-242}$Pu isotopes have clarified the 
role of octupole correlations in these nuclei and have led to a 
satisfactory interpretation in terms of octupole phonon condensation. 
Specifically, the special role of octupole correlations in $^{240}$Pu is 
highlighted by the data. Not only has a pattern expected for the rotation 
of an octupole-deformed nucleus been experimentally established at high 
spin, but a seldom observed mode of deexcitation has been uncovered for a 
rotational band built on the first excited $0^{+}$ level:  the decay 
proceeds solely towards the octupole band. 

The picture that emerges is one where strong octupole correlations are 
interpreted as rotation-induced condensation of octupole phonons having 
their angular momentum aligned with the rotational axis. The phonon 
condensate co-rotates with the prolate deformed nucleus. This new 
collective motion leads naturally to oscillations in the energy difference 
between the lowest positive- and negative-parity bands. Within this picture, 
the two excited bands observed here in $^{240}$Pu are associated with one- and 
two-phonon excitations and the data extend to sufficiently high frequency 
to observe the first of the crossings predicted by theory. The $^{238}$Pu data 
can be interpreted within the same framework. The extended level 
structure delineated in $^{242}$Pu is the only one to exhibit the type of 
excitations seen in other well-deformed actinide nuclei of the region, 
indicating that the octupole strength weakens considerably as the number 
of neutrons increases. 

While a satisfactory picture emerges, additional work is highly desirable 
in this instance as well. If the rotational sequences, especially in 
$^{240}$Pu, could be extended to higher spin, it would be possible to verify the 
theoretical predictions of a crossing between the yrast and the two-phonon 
bands. Also, expanding the level structure of $^{240}$Pu would be of value. In
particular, information on the excitation energy and the behavior with 
rotational frequency of quasi-particle excitations would provide further 
information on the strength of octupole correlations as well as on the 
interactions between the various modes. Exploration of other 
nuclei, not only in the actinide region, but also in other parts of the 
nuclear chart would also be important in order to assess whether this 
exotic collective mode is a rather general phenomenon. 

Just as in the case of $^{163}$Tm, progress demands the availability of a 
detection system with the intrinsic ability to detect weaker $\gamma$-ray 
transitions. Indeed, such a system, GRETA, 
an array of Ge detectors with tracking capability, is currently 
being developed. This array is expected to exceed the resolving 
power of Gammasphere by at least two orders of magnitude.

%
% Appendix
%

%\appendix

%\include{appendix}

%
% Back stuff
%

\backmatter
\bibliographystyle{apsrev}
\bibliography{XWangThesisBibli}

\end{document}